\documentclass[preprint]{aastex}
\usepackage{hyperref}
\usepackage{xcolor}
\usepackage{graphicx}
\usepackage{booktabs}
\usepackage{gensymb}
\pdfoutput=1
\let\oldbibliography\thebibliography
\renewcommand{\thebibliography}[1]{%
  \oldbibliography{#1}%
  \setlength{\itemsep}{0pt}%
}
 {\typeout{Couldn't patch the command}}

\begin{document}

\title{KFPA Examinations of Young STellar Object Natal Environments (KEYSTONE): \\ Hierarchical Ammonia Structures in Galactic Giant Molecular Clouds}

\author{Jared Keown\altaffilmark{1}, James Di Francesco\altaffilmark{1,2}, Erik Rosolowsky\altaffilmark{3}, Ayushi Singh\altaffilmark{4,5}, Charles Figura\altaffilmark{6}, Helen Kirk\altaffilmark{2,1}, L. D. Anderson\altaffilmark{7}, Michael Chun-Yuan Chen\altaffilmark{1}, Davide Elia\altaffilmark{8}, Rachel Friesen\altaffilmark{9}, Adam Ginsburg\altaffilmark{10}, A. Marston\altaffilmark{11}, Stefano Pezzuto\altaffilmark{8}, Eugenio Schisano\altaffilmark{8}, Sylvain Bontemps\altaffilmark{12}, Paola Caselli\altaffilmark{13}, Hong-Li Liu\altaffilmark{14, 15}, Steven Longmore\altaffilmark{16}, Fr\'ed\'erique Motte\altaffilmark{17}, Philip C. Myers\altaffilmark{18}, Stella S. R. Offner\altaffilmark{19}, Patricio Sanhueza\altaffilmark{20}, Nicola Schneider\altaffilmark{21}, Ian Stephens\altaffilmark{18}, James Urquhart\altaffilmark{22,23} \\ and the KEYSTONE collaboration}

\email{jkeown@uvic.ca}

\altaffiltext{1}{Department of Physics and Astronomy, University of Victoria, Victoria, BC, V8P 5C2, Canada} 

\altaffiltext{2}{NRC Herzberg Astronomy and Astrophysics, 5071 West Saanich Road, Victoria, BC, V9E 2E7, Canada}

\altaffiltext{3}{Department of Physics, University of Alberta, Edmonton, AB, Canada}

\altaffiltext{4}{Department of Astronomy \& Astrophysics, University of Toronto, 50 St. George Street, Toronto, Ontario, Canada M5S 3H4}

\altaffiltext{5}{Canadian Institute for Theoretical Astrophysics, University of Toronto, 60 St. George St., Toronto, Ontario, Canada, M5S 3H8}

\altaffiltext{6}{Mathematics, Computer Science, and Physics, Wartburg College, Waverly, IA, USA}

\altaffiltext{7}{Department of Physics and Astronomy, West Virginia University, Morgantown, WV 26506, USA}

\altaffiltext{8}{INAF-IAPS, via del Fosso del Cavaliere 100, I-00133 Roma, Italy 0000-0002-9120-5890}

\altaffiltext{9}{National Radio Astronomy Observatory, Charlottesville, VA 22903, USA}

\altaffiltext{10}{National Radio Astronomy Observatory, 1003 Lopezville Rd, Socorro, NM 87801, USA}

\altaffiltext{11}{ESAC/ESA, Camino Bajo del Castillo s/n, 28691 Villanueva de la Ca{\~n}ada, Spain}

\altaffiltext{12}{OASU/LAB-UMR5804, CNRS, Université Bordeaux, 33615 Pessac, France}

\altaffiltext{13}{Max-Planck-Institut f\"{u}r extraterrestrische Physik, Giessenbachstrasse 1, D-85748, Garching, Germany}

\altaffiltext{14}{Department of Physics, The Chinese University of Hong Kong, Shatin, N.T., Hong Kong}

\altaffiltext{15}{Departamento de Astronom\'ia, Universidad de Concepci\'on, Av. Esteban Iturra s/n, Distrito Universitario, 160-C, Chile}

\altaffiltext{16}{Astrophysics Research Institute, Liverpool John Moores University, 146 Brownlow Hill, Liverpool L3 5RF, UK}

\altaffiltext{17}{Univ. Grenoble Alpes, CNRS, IPAG, 38000 Grenoble, France}

\altaffiltext{18}{Harvard-Smithsonian Center for Astrophysics, 60 Garden Street, Cambridge, MA 02138, USA}

\altaffiltext{19}{Department of Astronomy, The University of Texas, Austin, TX 78712, USA}

\altaffiltext{20}{National Astronomical Observatory of Japan, National Institutes of Natural Sciences, 2-21-1 Osawa, Mitaka, Tokyo 181-8588, Japan}

\altaffiltext{21}{I. Physik. Institut, University of Cologne, Zülpicher Str. 77, 50937 Cologne, Germany}

\altaffiltext{22}{Centre for Astrophysics and Planetary Science, University of Kent, Canterbury, CT2 7NH, UK}

\altaffiltext{23}{Max-Planck-Institut f\"{u}r Radioastronomie, Auf dem H\"{u}gel 69, D-53121 Bonn, Germany}



\begin{abstract}
We present initial results from the K-band focal plane array Examinations of Young STellar Object Natal Environments (KEYSTONE) survey, a large project on the 100-m Green Bank Telescope mapping ammonia emission across eleven giant molecular clouds at distances of $0.9-3.0$ kpc (Cygnus X North, Cygnus X South, M16, M17, MonR1, MonR2, NGC2264, NGC7538, Rosette, W3, and W48).  This data release includes the NH$_3$ (1,1) and (2,2) maps for each cloud, which are modeled to produce maps of kinetic temperature, centroid velocity, velocity dispersion, and ammonia column density.  Median cloud kinetic temperatures range from $11.4\pm2.2$ K in the coldest cloud (MonR1) to $23.0\pm6.5$ K in the warmest cloud (M17).  Using dendrograms on the NH$_3$ (1,1) integrated intensity maps, we identify 856 dense gas clumps across the eleven clouds.  Depending on the cloud observed, $40-100\%$ of the clumps are aligned spatially with filaments identified in H$_2$ column density maps derived from SED-fitting of dust continuum emission.  A virial analysis reveals that 523 of the 835 clumps ($\sim63\%$) with mass estimates are bound by gravity alone.  We find no significant difference between the virial parameter distributions for clumps aligned with the dust-continuum filaments and those unaligned with filaments.  In some clouds, however, hubs or ridges of dense gas with unusually high mass and low virial parameters are located within a single filament or at the intersection of multiple filaments.  These hubs and ridges tend to host water maser emission, multiple 70$\mu$m-detected protostars, and have masses and radii above an empirical threshold for forming massive stars. 




\end{abstract}

\section{Introduction}

The ubiquity of filaments in star-forming environments was first revealed by continuum observations of nearby ($<$300 pc), low-mass star-forming molecular clouds, which showed that filaments are present in both quiescent \cite[][]{Miville_2010, Ward-Thompson_2010} and active \citep[][]{Andre_2010, Menshchikov_2010} star-forming regions.  These results suggest filaments are created during the molecular cloud formation process prior to the onset of star formation, likely as a result of turbulence \citep{Vazquez_2006, Smith_2014, Smith2_2014, Federrath_2016} and magnetic fields \citep{Hennebelle_2013, Palmeirim_2013, Seifried_2015}.  Furthermore, prestellar cores, the arguably gravitationally bound structures that likely collapse to form stars, are predominantly found along filaments \citep{Konyves_2010, Konyves_2015, Marsh_2016}. These results provide evidence that the formation and gravitational collapse of filaments is related to the core and star formation processes in low-mass star-forming environments.  

Although the study of nearby molecular clouds undoubtedly provides us with a close-up view of the star formation process, such clouds are not representative of the most productive star-forming engines in our Galaxy due to their low abundance of O- and B-type stars and clusters.  To observe large samples of high-mass stars ($>$8 M$_{\odot}$) and stellar clusters, we must probe giant molecular clouds (GMCs) at distances typically $>$300 pc from our Solar system.  While these distant environments require higher spatial resolution and sensitivity, they are more indicative of the majority of clouds in the Galaxy.  Similar to nearby clouds, filamentary networks of dense gas are also prevalent throughout GMCs and have been found to be spatially correlated with signposts of high-mass star formation \citep[e.g.,][]{Nguyen_2011, Hill_2012, Motte_2018_2}.  In particular, massive young stellar objects (MYSOs) and embedded stellar clusters appear to be preferentially located at the intersections of multiple filaments seen in dust continuum observations \citep{Myers_2009, Schneider_2010_cyg, Schneider_2012, Hennemann_2012, Li_2016, Motte_2017}.  The combination of the pervasiveness of filaments throughout molecular clouds with the finding that clusters form at the intersections of multiple filaments motivates the idea that mass flow along filaments provides the localized high-density conditions necessary to form stellar clusters and the MYSOs that form within them \citep{Kirk_2013, Friesen_2013, Henshaw_2013, Schneider_2010_cyg, Fukui_2015, Motte_2017}.

While dust continuum emission provides a detailed look at the distribution of dense cores and filaments within molecular clouds, it does not provide the gas velocity dispersion measurements required to understand whether or not those structures are gravitationally bound.  Rather, observations of dense gas emission from molecules such as NH$_3$ (ammonia) and N$_2$H$^+$ (diazenylium) are necessary to probe core and filament kinematics.  These tracers provide an advantage over commonly observed carbon-based molecules (e.g., CO) for tracing dense gas because they suffer less from freeze-out onto dust grains at the high densities within dense cores \citep[see, e.g.,][]{DiFrancesco_2007} and they are also typically optically thin with Gaussian-like profiles that allow an easier interpretation of kinematics.  In addition, the hyperfine splitting of ammonia emission provides a convenient method for obtaining optical depths.  Since the relative heights of the NH$_3$ hyperfine structures are well known in the optically thin limit, optical depths and excitation temperatures can easily be determined by measuring the intensities of the hyperfine components \citep{Ho_1983}.  Furthermore, observations of multiple NH$_3$ transitions allow a kinetic gas temperature to be calculated from the relative intensities of the central hyperfine groups in each transition.  This line strength relationship serves as a proxy for the distribution of populations within each excited state \citep{Ho_1979}, i.e., the kinetic energy over the observed portion of the cloud. 

The combination of dense gas kinematics and temperatures with continuum observations provides a way to measure the virial stability of dense cores and filaments \citep[e.g.,][]{Friesen_2016, Kirk_2017, Keown_2017}, the dissipation of turbulence from clouds and filaments to cores \citep[``transition to coherence,''][]{Pineda_2010, Chen_2018}, and the flow of gas along or onto filaments \citep[e.g.,][]{Schneider_2010_cyg, Kirk_2013, Friesen_2013, Henshaw_2013}.  Such measurements can also be used to determine if dense structures associated with filament intersections are susceptible to gravitational collapse.  If so, the structures may be the precursors of future stellar clusters, further linking filament intersections to the star formation process in GMCs. 

Recent large surveys have set out to investigate the connection between dense gas kinematics and star formation by observing ammonia emission throughout different regions of the Galaxy.  The Green Bank Ammonia Survey (GAS) mapped NH$_3$ emission throughout the nearby Gould Belt molecular clouds ($d <$ 500 pc) where A$_v$ $>$ 7 \citep[e.g.,][]{Friesen_2017, Kirk_2017, Keown_2017, Redaelli_2017, Kerr_2019, Chen_2018}.  The Galactic plane, which typically excludes nearby ($<$ 3 kpc) GMCs, has been mapped in ammonia by the Radio Ammonia Mid-Plane Survey \citep[RAMPS; covering $10^\circ < l < 40^\circ$, $-0.5^\circ < b < +0.5^\circ$;][]{Hogge_2018} and the H$_2$O Southern Galactic Plane Survey \citep[HOPS; covering $-70^\circ > l > 30^\circ$, $-0.5^\circ < b < +0.5^\circ$;][]{Purcell_2012}.  Similarly, \cite{Urquhart_2011, Urquhart_2015} observed ammonia and water maser emission from $\sim600$ massive young stellar objects and ultra-compact \ion{H}{2} regions as part of the Red MSX Source Survey.  While these surveys trace the kinematics of the most quiescent and extreme environments in the Galaxy, they do not cover the nearest GMCs producing massive stars.

Here, we present KFPA Examinations of Young STellar Object Natal Environments (KEYSTONE, PI: J. Di Francesco), a large project on the Green Bank Telescope (GBT) that has mapped NH$_3$ emission in eleven GMCs at intermediate distances (0.9 kpc $< d <$ 3.0 kpc) using the K-band Focal Plane Array (KFPA) receiver and VEGAS spectrometer on the GBT.  KEYSTONE targeted GMCs observable from Green Bank that are part of the Herschel OB Young Stars Survey \citep[HOBYS,][]{Motte_2010}, which mapped dust continuum emission in all GMCs out to 3 kpc using the \textit{Herschel Space Observatory}.  This sample of molecular cloud complexes presented in \cite{Motte_2017} (see also \cite{Schneider_2011}) gives a complete view of high-mass star formation at distances less than 3 kpc. This sample notably contains the Cygnus X molecular complex 
\citep{Hennemann_2012, Schneider_2016}, the M16/M17 complex \citep{Hill_2012, Tremblin_2013, Tremblin_2014}, the Monoceres complex \citep{Didelon_2015, Rayner_2017}, Rosette \citep{Motte_2010, DF_2010, Schneider_2010, Schneider_2012}, W48 \citep{Nguyen_2011, Rygl_2014}, the W3/KR140 complex \citep{Rivera-Ingraham_2013, Rivera-Ingraham_2015}, NGC7538 \citep{Fallscheer_2013} plus southern regions not presented here \citep{Hill_2012_2, Minier_2013, Tige_2017}. Thus, KEYSTONE provides the kinematic counterpart to the HOBYS survey that is required to understand the relationship between dense gas dynamics and massive stars.

This paper, which is the first KEYSTONE publication, provides an initial look at the NH$_3$ (1,1) and (2,2) emission maps observed in each region, catalogs each region's dense gas clumps, estimates the virial stability of those clumps, and compares the spatial distribution of the clumps to the positions of filaments and protostars identified in \textit{Herschel} observations.  Dendrograms, tree-diagrams that identify intensity peaks in a map and determine their hierarchical structure, are used to select dense gas clumps in each cloud.  The top-level structures in the dendrogram hierarchy are often called ``leaves,'' a term that we use synonymously with ``clumps'' throughout this paper.  In \S~2, we describe our GBT observations and data reduction techniques, along with the archival data that were retrieved for our analysis.  In \S~3, we outline the methods used to model the NH$_3$ data, identify NH$_3$ structures, derive their stability parameters, and compare their spatial distributions to those of dust continuum filaments.  In \S~4, we estimate the cloud weight pressure and turbulent pressure exerted on the NH$_3$ structures.  We conclude with a summary of the paper in \S~5 and a discussion of future analyses using the KEYSTONE data in \S~6.

\section{Observations and Data Reduction}

\subsection{Targets}
Table \ref{Table_regions} lists the eleven clouds observed by KEYSTONE and their distances.  Here, we provide a brief overview of each cloud.  For more detailed comparisons between the clouds, see the review by \cite{Motte_2017}.

\begin{deluxetable}{cccccccc}
\tabletypesize{\footnotesize}
\tablewidth{0pt}
\tablecolumns{8}
\tablecaption{KEYSTONE Target GMCs}
\tablehead{\colhead{Region} & R.A. & Dec. & \colhead{Distance} & Total Mass & Total Area & \colhead{Footprints\tablenotemark{a}} & Completeness \\
   & (J2000) & (J2000) & (kpc) & (M$_\odot$) & (pc$^2$) & Observed & $A_V > 10$}
\startdata
W3 & 02:23:22.140 & +61:36:17.432 & 2.0 $\pm$ 0.1\tablenotemark{b} & 1.0E5 & 2.7E3 & 26+ & $98\%$\\
Mon R2 & 06:08:25.657 & $-$06:14:32.812 & 0.9 $\pm$ 0.1\tablenotemark{c, d} & 4.9E3 & 1.4E2 & 5+ & $100\%$\\
Mon R1 & 06:32:32.294 & +10:27:13.335 & 0.9 $\pm$ 0.1\tablenotemark{c, e} & 8.8E3 & 1.4E2 & 5 & $98\%$\\
Rosette & 06:33:38.530 & +04:29:10.771 & 1.4 $\pm$ 0.1\tablenotemark{c} & 3.2E4 & 7.2E2 &  15 & $85\%$\\
NGC2264 & 06:40:41.339 & +09:25:42.177 & 0.9 $\pm$ 0.1\tablenotemark{e} & 1.0E4 & 2.2E2 & 8+ & $99\%$\\
M16 & 18:18:38.140 & $-$13:39:30.050 & 1.8 $\pm$ 0.5\tablenotemark{f} & 8.6E4 & 5.6E2 &  5 & $52\%$\\
M17 & 18:19:35.479 & $-$16:19:09.088 & 2.0 $\pm$ 0.1\tablenotemark{g} & 5.0E5 & 1.9E3 &  9 & $31\%$\\
W48 & 19:00:52.657 & +01:41:55.338 & 3.0\tablenotemark{h} & 1.4E6 & 8.8E3 & 13+ & $60\%$\\
Cygnus X South & 20:33:42.800 & +39:35:41.356 & 1.4 $\pm$ 0.1\tablenotemark{i} & 2.2E5 & 2.3E3 & 43 & $84\%$\\
Cygnus X North & 20:37:14.998 & +41:56:04.742 & 1.4 $\pm$ 0.1 \tablenotemark{i} & 2.7E5 & 3.3E3 & 36+ & $82\%$\\
NGC7538 & 23:14:50.333 & +61:29:04.744 & 2.7 $\pm$ 0.1\tablenotemark{j} & 9.3E4 & 1.4E3 & 17 & $100\%$\\

\enddata
\tablecomments{The right ascensions and declinations listed are the mid-point of the entire mapped area.  The total mass  and total area are calculated as the sum of all H$_2$ column density and area, respectively, mapped in each cloud by \textit{Herschel}.  The completeness represents the percentage of pixels with $A_V > 10$ in the \textit{Herschel} H$_2$ column density maps that were observed by KEYSTONE.  We assumed an extinction conversion factor of N$_{H2}$ / $A_V = 0.94\times10^{21}$ \citep{Bohlin_1978}.  The completion percentages for M16, M17, and W48 account for the RAMPS intended coverage of those regions.  }
\tablenotetext{a}{Each footprint is $10'\times10'$. A `+' denotes that a partially completed tile was also observed in that region.}
\tablenotetext{b}{\cite{Hachisuka_2006}} 
\tablenotetext{c}{\cite{Schlafly_2014}} 
\tablenotetext{d}{\cite{Lombardi_2011}} 
\tablenotetext{e}{\cite{Baxter_2009}} 
\tablenotetext{f}{\cite{Bonatto_2006}} 
\tablenotetext{g}{\cite{Xu_2011}} 
\tablenotetext{h}{\cite{Rygl_2010}} 
\tablenotetext{i}{\cite{Rygl_2012}} 
\tablenotetext{j}{\cite{Moscadelli_2009}} 

\label{Table_regions}
\end{deluxetable}

\subsubsection{W3}
W3 is part of a larger complex located in the Perseus spiral arm that also includes the W4 and W5 molecular clouds \citep{Megeath_2008}.  The W3 Main, W3(OH), and AFGL 333 regions on the eastern edge of W3 all show signatures of high-mass star formation that may have been triggered by superbubbles from previous generations of star formation \citep{Oey_2005}.  W3 Main is a particularly popular source for high-mass star formation studies due to its array of \ion{H}{2} regions \citep{Colley_1980, Tieftrunk_1997} powered by a cluster of OB stars \citep{Megeath_1996, Ojha_2004}.  For instance, \cite{Tieftrunk_1998} used NH$_3$ (1,1) and (2,2) observations of W3 Main and W3(OH) to show that the stellar clusters are littered with cold dense gas clumps.  More recently, \cite{Nakano_2017} mapped the AFGL 333 ridge in NH$_3$ and found evidence for triggered star formation at the edges of the ridge but quiescent (non-triggered) formation in the ridge center.  Similarly, \cite{Rivera-Ingraham_2011} argued that both triggered and quiescent star formation are required to explain the YSO population detected in the cloud.  More recent large-scale \textit{Herschel} mapping of W3 by \cite{Rivera-Ingraham_2013, Rivera-Ingraham_2015} suggested that the triggered star formation was a result of ``convergent constructive feedback,'' which involves massive stars serving as triggers for subsequent star formation by funneling gas onto a central massive structure.  

In this paper, we present the observations of the southwestern half of W3, which includes the small \ion{H}{2} region KR 140 \citep{Kallas_1980}, as a separate region that we named W3-west.  

\subsubsection{Mon R2}
Monoceros R2 (Mon R2) is the most distant member of the larger Orion-Monoceros molecular cloud complex, which also includes the Orion A and Orion B clouds.  \cite{Wilson_2005} contend that Mon R2 and the Orion clouds share a common origin, as evidenced by the alignment of spurs in their CO emission with the Vela supershell.  Mon R2 hosts a central reflection nebula with a high stellar volume density ($\sim$ 9000 stars pc$^{-3}$), including several B-type stars \citep{Carpenter_1997}.  \cite{Didelon_2015} estimated that the size of the four main \ion{H}{2} regions in Mon R2 range from 0.1 pc for the central ultra-compact \ion{H}{2} region, which they suggest is undergoing pressure-driven large-scale collapse, to 0.8 pc for the most extended classical \ion{H}{2} region.  Previous NH$_3$ mapping by \cite{Willson_1981} and \cite{Montalban_1990} have shown that the \ion{H}{2} regions are surrounded by dense gas clumps with masses of $1-65$ M$_\odot$ and kinetic temperatures of $15-30$ K.  Moreover, recent \textit{Herschel} dust continuum and C$^{18}$O observations by \cite{Rayner_2017} showed that the gas and dust in Mon R2 has a distinct hub-spoke geometry, with a central hub of protostars and dense cores that may be fed by several connected filaments.  The column density probability distribution function from the \textit{Herschel} observations also shows two power-law tails, suggesting both turbulent- and gravity-dominated regimes in Mon R2 \citep{Schneider_2015, Pokhrel_2016}. 

A \textit{Herschel}-derived sample of 177 dense cores in MonR2 was published by \cite{Rayner_2017}. Their masses span 0.084 M$_\odot$ to 24 M$_\odot$ and their radii span 0.023 pc to 0.3 pc.  Of the 177 dense cores identified by \cite{Rayner_2017}, 29 ($\sim 16 \%$) were found to be protostellar and eleven had masses $>10$ M$_\odot$.

\subsubsection{Mon R1 and NGC 2264}
The Monoceros OB1 (Mon OB1) GMC includes NGC 2264, one of the most massive star clusters ($\sim$ 1400 members) within 1 kpc of our position in the Galaxy \citep{Dahm_2008, Teixeira_2012, Rapson_2014}.  Initial CO and CS mapping of the region revealed several outflows associated with the cluster \citep[e.g.,][]{Margulis_1986, Wolf_1995}.  Six Herbig-Haro objects have also been detected within this region \citep{Adams_1979, Walsh_1992, Wang_2003}. Ammonia mapping by \cite{Lang_1980} and \cite{Pagani_1987} revealed that the dense gas in NGC 2264 is comprised of two components, each $\sim$ 0.9 pc in diameter and separated by 0.9 pc, with kinetic temperatures of $\sim$ 20 K.  In addition, \cite{Peretto_2006} used more recent observations of dust continuum and molecular line emission to show that several massive clumps in NGC 2264 indicate infall motions and may comprise an intermediate mode of massive star formation.  

Just north of NGC 2264 is a more quiescent region of dense gas where a collection of Class 0/I and II objects are forming \citep{Rapson_2014}.  We henceforth refer to this northern region as ``Mon R1,'' which it has been referred to in previous literature \citep{Kutner_1974, Ogura_1984}.  Large-scale CO mapping covering NGC 2264 and Mon R1 by \cite{Oliver_1996} revealed that the kinematics of the region are dominated by the Perseus and Local spiral arms.  

\subsubsection{Rosette}
The Rosette complex is located in the Monoceros constellation south in declination from Mon OB1, NGC 2264, and Mon R2 \citep{RZ_2008}.  The cloud's emission is dominated by NGC 2244, its central OB association of 70 high-mass stars that has created a large \ion{H}{2} region \citep{Wang_2008}.  Rosette has been mapped extensively in CO \citep{Blitz_1980, Blitz_1986, Schneider_1998, Heyer_2006}, which revealed outflows from the massive proto-binary AFGL 961 \citep{Castelaz_1985}.  Large-scale \textit{Herschel} dust continuum mapping by \cite{DF_2010} revealed 473 dense clumps throughout Rosette, 371 being starless and 102 being protostellar, which includes 6 protostellar massive dense cores and 3 prestellar massive dense cores with masses between 20 M$_\odot$ and 40 M$_\odot$ \citep{Motte_2010}.  \cite{Schneider_2010} also used the \textit{Herschel} observations to show a negative temperature gradient, positive density gradient, and age sequence (more evolved to younger) as distance from the NGC 2244 cluster increases, highlighting the influence of the OB association upon the star formation in the cloud.  In addition, \cite{Schneider_2012} note that the massive stars and infrared clusters discovered in Rosette tend to align with the intersections of dust-identified filaments, providing compelling evidence that massive star formation occurs at the sites of filament mergers.

\subsubsection{M16}
M16, which is also known as the Eagle Nebula, is an \ion{H}{2} region located in the Sagittarius spiral arm \citep{Oliveira_2008}.  The cloud's structure and temperature are influenced by the open cluster NGC 6611 at its center, which contains 52 OB stars \citep{Evans_2005}.  For example, \cite{Hill_2012} used \textit{Herschel Space Observatory} dust continuum mapping to show there is a clear dust temperature gradient moving away from the NGC 6611 cluster.  \cite{Tremblin_2014} also show that the dust-derived column density probability distribution function in M16 has a second peak at high densities, which they attribute to a compressed zone of gas caused by an expanding shell of ionized gas from NGC 6611.  In the south of M16 are the famous ``Pillars of Creation'' or ``elephant trunks'' imaged with the \textit{Hubble Space Telescope} \citep{Hester_1996} and with \textit{Herschel} \citep{Hill_2012, Tremblin_2013}.  The morphology of the Pillars is caused by the ionizing radiation from the central OB stars in M16 \citep{White_1999, Williams_2001, Gritschneder_2010}.  In addition, recent CO mapping of M16 by \cite{Nishimura_2017} revealed a 10 pc diameter cavity of molecular gas near NGC 6611, providing further evidence of the cluster's impact on the star formation in the GMC.  

\subsubsection{M17}
M17 (the Omega Nebula) is located south in declination from M16 by an angular separation of 2.5$\degree$ \citep{Oliveira_2008}.  \cite{Elmegreen_1979} used CO mapping, however, to show that M17 and M16 form a continuous molecular cloud structure despite their large angular separation, which is a conclusion supported by recent near-infrared imaging \citep{Comeron_2019}.  Similar to M16, M17 has a central \ion{H}{2} region created by an open cluster (NGC 6618) of 53 OB stars \citep{Hoffmeister_2008}.  While much of the literature is focused on mapping the molecular gas \citep[e.g.,][]{Thronson_1983, Stutzki_1988, Stutzki_1990, PB_2015} and dust continuum \citep[e.g.,][]{Gatley_1979, Povich_2009} of the M17SW region near NGC 6618, the whole of M17 has recently been mapped in $^{12}$CO, $^{13}$CO, and C$^{18}$O by \cite{Nishimura_2018} and in $^{12}$CO, $^{13}$CO, HCO$^+$ and HCN by Nguyen Luong et al. (2019, submitted).  M17SW has also been mapped in NH$_3$ by \cite{Lada_1976} and \cite{Guesten_1988}, which revealed several distinct velocity components in the dense gas and kinetic temperatures of $30-100$ K.

\subsubsection{W48}
At 3 kpc \citep{Rygl_2010}, W48 is the most distant HOBYS and thus KEYSTONE target.  \textit{Herschel} observations of the cloud by \cite{Nguyen_2011} revealed numerous \ion{H}{2} regions with extended warm dust emission.  The IRDC G035.39$-$00.33 region in the north of W48 was also found to host 13 high-mass ($M ~>$ 20 M$_\odot$), compact (diameters of $0.1-0.2$ pc), and dense ($2-20 \times$10$^{5}$ cm$^{-3}$) massive dense cores that could be the precursors of massive stars \citep{Nguyen_2011}.  \cite{Liu_2018} used dust polarization and NH$_3$ measurements to show that these clumps are likely supported against gravitational collapse by magnetic fields and turbulence. Similarly, \cite{Pillai_2011} used interferometric observations of G35.20$-$1.74 in the east of W48 to show that the cores there were also massive ($\sim 9-250$ M$_\odot$), dense ($>$ 10$^{5}$ cm$^{-3}$), cold ($< 20$ K), and highly deuterated ([NH$_2$D/NH$_3$] $> 10 \%$), which suggest they are on the verge of forming protoclusters.  With several methanol maser emission line detections \citep{Slysh_1995, Minier_2000, Sugiyama_2008, Surcis_2012}, which are a signpost of massive stars, it is clear that W48 is an interesting testbed for high-mass star formation studies.  

\subsubsection{Cygnus X}
The Cygnus X molecular cloud complex is one of the most active star-forming regions in the nearby Galaxy \citep{Schneider_2016}.  It hosts over 1800 protostars \citep{Kryukova_2014} and is a favored target for studies of high-mass star formation due to its high concentration of OB associations \citep[e.g.,][]{Hanson_2003, Comeron_2012, Wright_2014}.  The OB associations range in age and size from the young proto-globular cluster Cyg OB2 \citep{Knodlseder_2000, Wright_2014}, harboring nearly one hundred O-stars, to the slightly older and smaller Cyg OB1, OB3, and OB9 \citep{Uyaniker_2001}.  It has been mapped extensively in a variety of molecular gas tracers \citep{Schneider_2006, Schneider_2010, Wilson_1990, Csengeri_2011, Csengeri_2011_2, Duarte-Cabral_2013, Duarte-Cabral_2014, Dobashi_2014, Schneider_2016, Pillai_2012}, dust continuum \citep{Motte_2007, Bontemps_2010, Hennemann_2012}, and dust polarization \citep[e.g.,][]{Ching_2017}. 

Although previous papers have treated Cygnus X as a single complex \citep{Schneider_2006, Rygl_2012}, our observations split the Cygnus X cloud into a North and South region.  The choice to treat Cygnus X North and South as separate regions in our analysis is motivated by the observations of \cite{Kryukova_2014}, which showed that each have distinct luminosity functions and morphological differences indicative of dissimilar star-forming environments. Cygnus X North contains DR21, the massive ridge where a slew of massive stars are forming, including the high-mass core DR21(OH) \citep[e.g.,][]{Mangum_1991, Csengeri_2011_2}.  Previous observations of the DR21 \ion{H}{2} region by \cite{Guilloteau_1983} mapped the region in NH$_3$ (1,1), (2,2), (3,3), and (4,4), which revealed absorption in the (1,1) and (2,2) emission that indicates high excitation temperatures $\geq$ 100 K.  The southern section of Cygnus X is home to DR15, a cluster of $\sim$ 200 protostars that sits atop a filamentary pillar extended over 10 pc to the south \citep{Rivera_Galvez_2015}. 

\subsubsection{NGC 7538}
NGC 7538 is a GMC associated with the Perseus spiral arm \citep{Kun_2008}.  It harbors several bright \ion{H}{2} regions, most notably around the IRS $1-11$ sources in its center \citep{Werner_1979, Mallick_2014}.  Strong outflows have been observed throughout the cloud \citep{Campbell_1984, Scoville_1986, Sandell_2005, Qui_2011}, one of which has signatures of a massive ($\sim 40$ M$_\odot$) accreting Class 0 protostar \citep{Sandell_2003}.  Dust continuum observations covering the IRS $1-11$ sources by \cite{Reid_2005} showed that the bright IR sources are surrounded by massive cold clumps.  \textit{Herschel} observations by \cite{Fallscheer_2013} that covered a wider field-of-view revealed an evacuated ring structure in the east of NGC 7538, with a string of cold clumps detected along the ring's edge.  \citeauthor{Fallscheer_2013} also detected 13 massive ($M ~>$ 40 M$_{\odot}$) and cold ($T ~<$ 15 K) clumps that may be starless or contain embedded Class 0 sources, further highlighting the high-mass star-forming potential of the cloud.   

Previous ammonia observations in NGC 7538 have been focused primarily on IRS 1, which has shown a slew of rare emission features such as: maser emission in H$_2$O, the nonmetastable $^{14}$NH$_3$ (10,6), (10,8), (9,8), and (9,6) transitions, and $^{15}$NH$_3$ (3,3) \citep{Johnston_1989, Hoffman_2011, Hoffman_2012} as well as vibrationally excited ammonia \citep{Schilke_1990}.

\subsection{GBT NH$_3$ Data}

Data were obtained as part of the KEYSTONE (KFPA Examinations of Young STellar Object Natal Environments) survey, a large project on the GBT that mapped NH$_3$, HC$_5$N, HC$_7$N, HNCO, H$_2$O, CH$_3$OH, and CCS emission across eleven GMCs at distances between 0.9 kpc and 3 kpc.  Observations were conducted between 2016 October and 2019 March for a total of 356.25 observing hours, including overheads.  Table \ref{Table_transitions} summarizes all observed transitions along with their rest frequencies.  The eleven GMCs observed by KEYSTONE were selected from the HOBYS survey.  The observing strategy for KEYSTONE targeted all filamentary structures where $A_V$ $>$ 10 mag in the HOBYS column density maps (see Ladjelate et al., in preparation), which is slightly higher than that used in the GAS survey \citep[$A_V$ $>$ 7;][]{Friesen_2017}.  Due to the large amount of foreground and/or background contamination along the line of sight to some of the clouds, this extinction threshold does not have much physical meaning but rather is meant to highlight the densest regions in each cloud. The KEYSTONE observations also exclude parts of M16, M17, and W48 that will be mapped with the GBT by the Radio Ammonia Mid-Plane Survey \citep[RAMPS,][]{Hogge_2018}.  

Observations were made with the GBT's K-band Focal Plane Array, which has seven beams arranged in the shape of a hexagon with beam centers separated by $\sim95\arcsec$ on the sky.  Following the observational setup used in GAS, each cloud was segmented into $10' \times 10'$ tiles that were observed using on-the-fly mapping and frequency-switching for 11 seconds of on+off integration time (5.5 seconds on-source and 5.5 seconds at reference frequency) per beam (i.e., a total of 77 seconds when summing over all seven beams) for each resolution element.  The $10' \times 10'$ tiles were scanned using on-the-fly mapping, covering the observed region in the Right Ascension and Declination directions.  The row separation ($\sim13\arcsec$) and spectrometer dump cadence ensured that each resolution element in the map was sampled by $>3$ samples in both directions, ensuring Nyquist sampling.  Each tile took $\sim$ 1.3 hours to complete, with 1 to 3 tiles observed per session.  Table \ref{Table_regions} lists the number of tiles completed for each cloud.  The survey's completeness, defined as the percentage of the HOBYS maps with $A_V$ $>$ 10 mag observed by KEYSTONE, ranged from $31\%$ for M17 to $100\%$ for MonR2 and NGC7538 (see Table \ref{Table_regions}).  

\begin{figure}[ht]
\epsscale{0.6}
\plotone{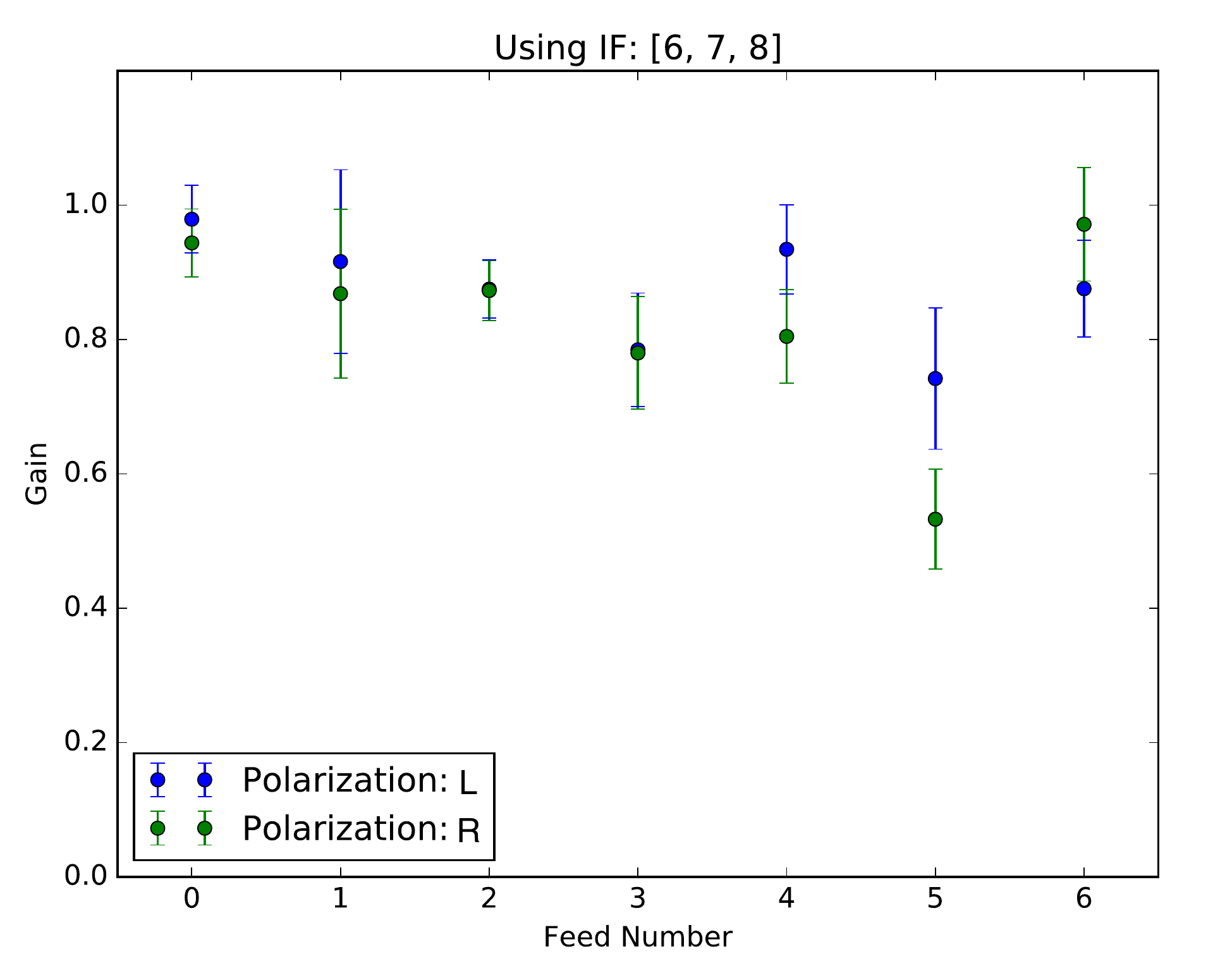}
\caption{Beam gains for the NH$_3$ (1,1) spectral windows (IFs 6, 7, and 8) averaged over all Moon observations for each feed and polarization.}
\label{beam_gains}
\end{figure}

The telescope's pointing and focus were aligned before mapping each tile to account for changes in the optical performance due to, e.g., temperature- and weather-dependent structural deformations.  The KFPA receiver's noise diodes were used to measure the off-source system temperatures for each observing session, which are also temperature- and weather-dependent. Since each of the KFPA's beams has an independent response (i.e., gain), the Moon was observed at least once per session for flux density calibration, if available.  The Moon's large angular size compared to the size of the KFPA beam allowed for beam gains to be calculated from single on-source and off-source observations during each observing session.  Figure \ref{beam_gains} shows the beam gains for the NH$_3$ (1,1) spectral windows (IFs 6, 7, and 8) averaged over all observations of the Moon for each polarization.  Table \ref{gains} displays the final beam gains used for flux density calibration, along with the standard deviation for each average.


The GBT's VEGAS backend was configured with eight spectral windows, each 23.44 MHz wide.  All five NH$_3$ transitions (1,1) up to (5,5), along with HC$_7$N $(9-8)$ and CH$_3$OH ($10_1-9_2$) $A^-$, were observed in seven of the windows across all seven of the KFPA beams.  The eighth VEGAS window covered H$_2$O ($6_{16}-5_{23}$), HC$_5$N ($8-7$), HC$_7$N ($19-18$), HNCO ($1_{0,1}-0_{0,0}$), CH$_3$OH ($12_2-11_1$) $A^-$, and CCS ($2_0-1_0$) in only the central KFPA beam.  The GBT beam has a FWHM of 32$\arcsec$ at the NH$_3$ (1,1) rest frequency. This VEGAS configuration is the same as that used by the RAMPS survey \citep{Hogge_2018}.

In this paper, we present the NH$_3$ (1,1) and (2,2) emission. Other lines will be presented in future KEYSTONE papers.  We also identify H$_2$O ($6_{16}-5_{23}$) maser emission by eye to include in figures presented in Section 4, but leave the full presentation of those data and a more thorough maser identification technique to White et al. (in prep.).  The NH$_3$ data were reduced using \texttt{gbtpipe}\footnote{\url{https://github.com/GBTSpectroscopy/gbtpipe}}, a Python-packaged version of the standard GBT reduction pipeline.  The data were calibrated and output as 3D FITS spectral cubes, with R.A., Dec., and spectral frequency comprising each axis.  The on-the-fly observations were mapped to a grid of square pixels with width of 8.8$\arcsec$, which corresponds to $\sim3.5$ pixels per FWHM beam of the NH$_3$ (1,1) line.  The spectrum corresponding to each spatial pixel was determined using a weighted average of on-the-fly integrations from all seven beams of the K-band Focal Plane array, including those samples with separations less than one FWHM beam size away from a given map pixel.  The weighting scheme is a Gaussian-tapered Bessel function, as described in \cite{Friesen_2017} following \cite{mangum}.  This procedure results in data cubes with a resolution of $32''$ and the dense sampling from the mapping strategy and multiple receiver feeds produces high-quality maps without discernible scanning patterns in the image or Fourier domain. 


The pipeline also subtracts a first-order polynomial fit to the channels on the edges of each scan prior to gridding to remove any shape in the spectral baselines introduced by, e.g., instrumental effects.  The pixel size of the final data cubes is 8.8$\arcsec$, with a spectral resolution of 5.7 kHz, or 0.07 km s$^{-1}$.  

To remove any remaining shape in the spectral baselines, we perform an additional round of per-pixel baseline fitting similar to the method described in \cite{Hogge_2018}.  Namely, a sliding window with a width of 31 channels is used to calculate a ``local'' standard deviation for every channel in a spectrum.  For the 15 channels at each end of the spectrum, the first and last 31 channels are used as the ``local'' windows, while all other channels are at the center of their ``local'' window.  From the standard deviation distribution of all ``local'' windows, the central channels belonging to the lowest two quintiles are used for the baseline fit.  Thus, channels that belong to an emission line or noise spike are excluded from the baseline fit due to their high ``local'' standard deviation relative to the non-emission-line channels in the spectrum. Next, polynomials up to a third order are fit to the selected channels.  A reduced chi-squared value is then calculated for each of the best-fit polynomials against the full spectrum.  Finally, the polynomial with the lowest reduced chi-squared value is subtracted from the original spectrum.  This baseline subtraction technique is publicly available\footnote{\url{https://github.com/GBTAmmoniaSurvey/keystone}}, along with the full KEYSTONE data reduction code base.  The final baseline-subtracted NH$_3$ (1,1) and (2,2) data cubes are publicly available\footnote{\url{https://doi.org/10.11570/19.0074}}.

\begin{figure}[ht]
\epsscale{1.0}
\plottwo{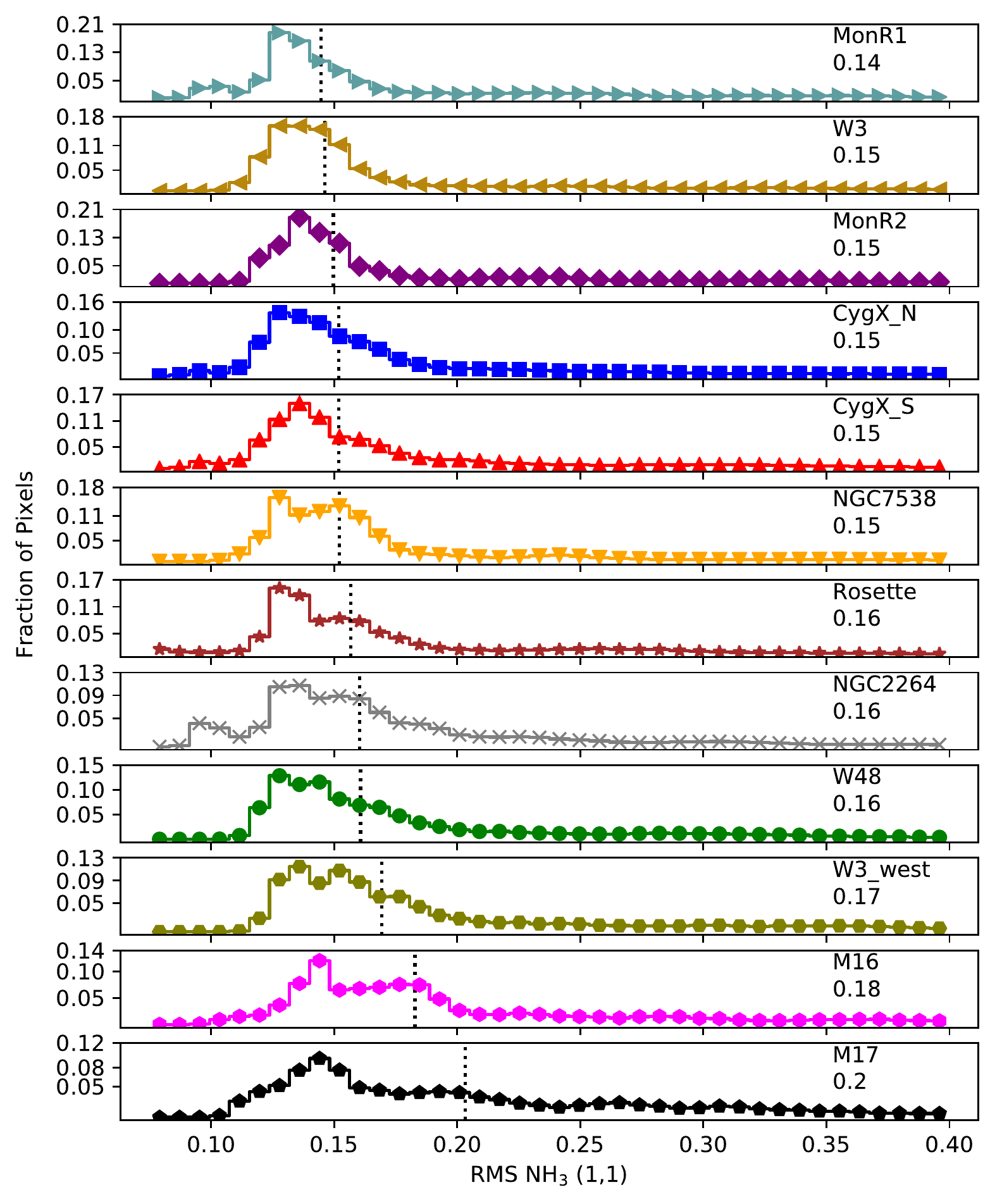}{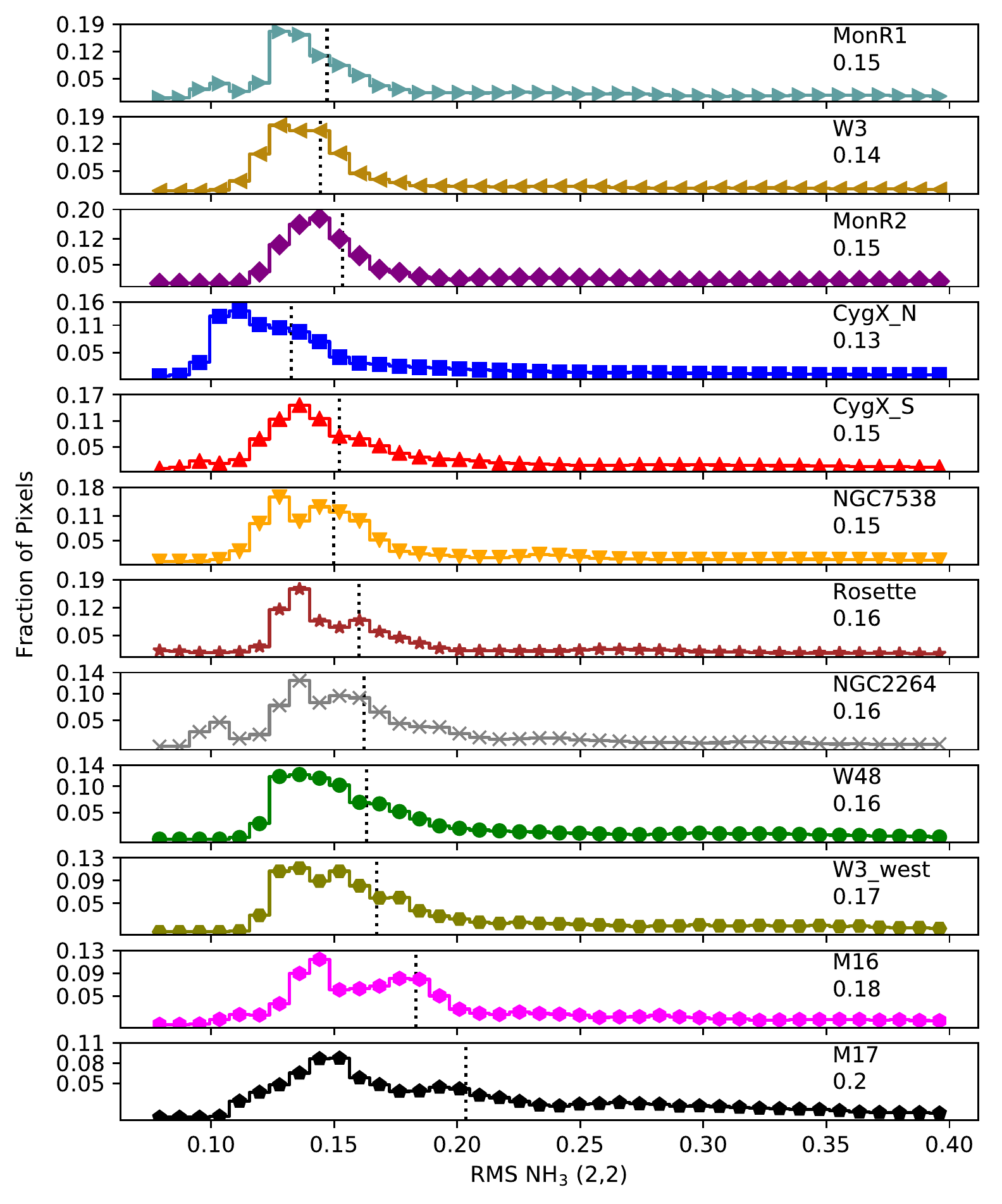}
\caption{Histograms of the RMS noise for the NH$_3$ (1,1) (left) and NH$_3$ (2,2) (right) maps of each cloud.  The median of the distribution is displayed as a vertical dotted line, with the corresponding value shown in the upper right corner of each panel. The clouds are ordered from top to bottom by increasing median NH$_3$ (1,1) RMS noise.}
\label{rms_hists}
\end{figure}

The system temperatures for the observations were typically $40-50$ K, with a median of $\sim$ 43 K.  Figure \ref{rms_hists} shows histograms of the RMS noise for the NH$_3$ (1,1) and (2,2) maps of each cloud.  We calculate the RMS using the channels in the best-fit model from our line-fitting procedure (see Section 3.1) with brightness lower than 0.0125 K.  While the medians of the RMS distributions range from 0.13 K to 0.2 K, most of the distributions have a peak below 0.15 K.  M17 and M16 have slightly higher noise than the other regions since they are the lowest declination sources observed ($\sim -16\degree$ and $-13\degree$, respectively). 

\begin{deluxetable}{cccc}
\tablewidth{0pt}
\tablecolumns{4}
\tablecaption{KEYSTONE Observed Transitions}
\tablehead{\colhead{Molecule} & \colhead{Transition} & \colhead{Rest Frequency\tablenotemark{a}} & \colhead{Number of Beams}\\
   & & (MHz) }
\startdata
HC$_5$N & ($8-7$) & 21301.26 & 1\\
HC$_7$N & ($19-18$) & 21431.93 & 1\\
CH$_3$OH & ($12_2-11_1$) $A^-$ & 21550.34 & 1\\ 
HNCO & ($1_{0,1}-0_{0,0}$) & 21981.4706(1) & 1\\  
H$_2$O & ($6_{16}-5_{23}$) & 22235.08 & 1\\
CCS & ($2_0-1_0$) & 22344.030 & 1\\
CH$_3$OH & ($10_1-9_2$) $A^-$ & 23444.78 & 7\\ 
NH$_3$ & (1,1) & 23694.4955 & 7\\  
NH$_3$ & (2,2) & 23722.6336 & 7\\  
NH$_3$ & (3,3) & 23870.1296 & 7\\
HC$_5$N & ($9-8$) & 23963.9010 & 7 \\    
NH$_3$ & (4,4) & 24139.35 & 7\\  
NH$_3$ & (5,5) & 24532.92 & 7\\      

\enddata
\tablenotetext{a}{Accessed from \cite{Lovas_2004}}

\label{Table_transitions}
\end{deluxetable}

\begin{deluxetable}{ccc}
\tablewidth{0pt}
\tablecolumns{3}
\tablecaption{Beam Gains}
\tablehead{\colhead{Beam} & \colhead{Polarization L} & \colhead{Polarization R}}
\startdata
0	&	0.979	(0.050)	&	0.944	(0.051) \\
1	&	0.916	(0.137)	&	0.868	(0.126) \\
2	&	0.875	(0.043)	&	0.873	(0.044) \\
3	&	0.785	(0.084)	&	0.780	(0.084) \\
4	&	0.934	(0.066)	&	0.805	(0.070) \\
5	&	0.742	(0.105)	&	0.533	(0.074) \\
6	&	0.876	(0.072)	&	0.972	(0.085) \\
\enddata
\tablecomments{Average beam gains with one-$\sigma$ variations shown in parentheses.}
\label{gains}
\end{deluxetable}

\subsection{\textit{Herschel} Dust Continuum Data}
\textit{Herschel Space Observatory} Level 2.5 data products at 70 $\mu$m, 160 $\mu$m, 250 $\mu$m, 350 $\mu$m, and 500 $\mu$m for each KEYSTONE region were downloaded from the European Space Agency Herschel Science Archive\footnote{\url{http://archives.esac.esa.int/hsa/whsa/}}.  These maps were originally observed by the Herschel OB Young Stars Survey \citep{Motte_2010} and have spatial resolutions of 8.4$\arcsec$, 13.5$\arcsec$, 18.2$\arcsec$, 24.9$\arcsec$, and 36.3$\arcsec$, respectively.  Although the HOBYS team has released dense core and protostar catalogs for MonR2 \citep{Rayner_2017}, W3 (Rivera-Ingraham et al., submitted), Cygnus X North (Bontemps et al., in preparation),  NGC 6334 \citep{Tige_2017}, and NGC 6537 (Russeil et al. 2019, in press), no catalogs have been yet released for many of the clouds targeted by KEYSTONE. In this paper, we use the \textit{Herschel} $160 - 500~\mu$m maps to estimate the H$_2$ column densities and masses of structures identified in the KEYSTONE observations (see Section 3.4).  Additionally, we use the 70 $\mu$m maps to identify embedded protostars in each cloud (see Section 3.6). 



\begin{figure}[ht]
\epsscale{0.8}
\plotone{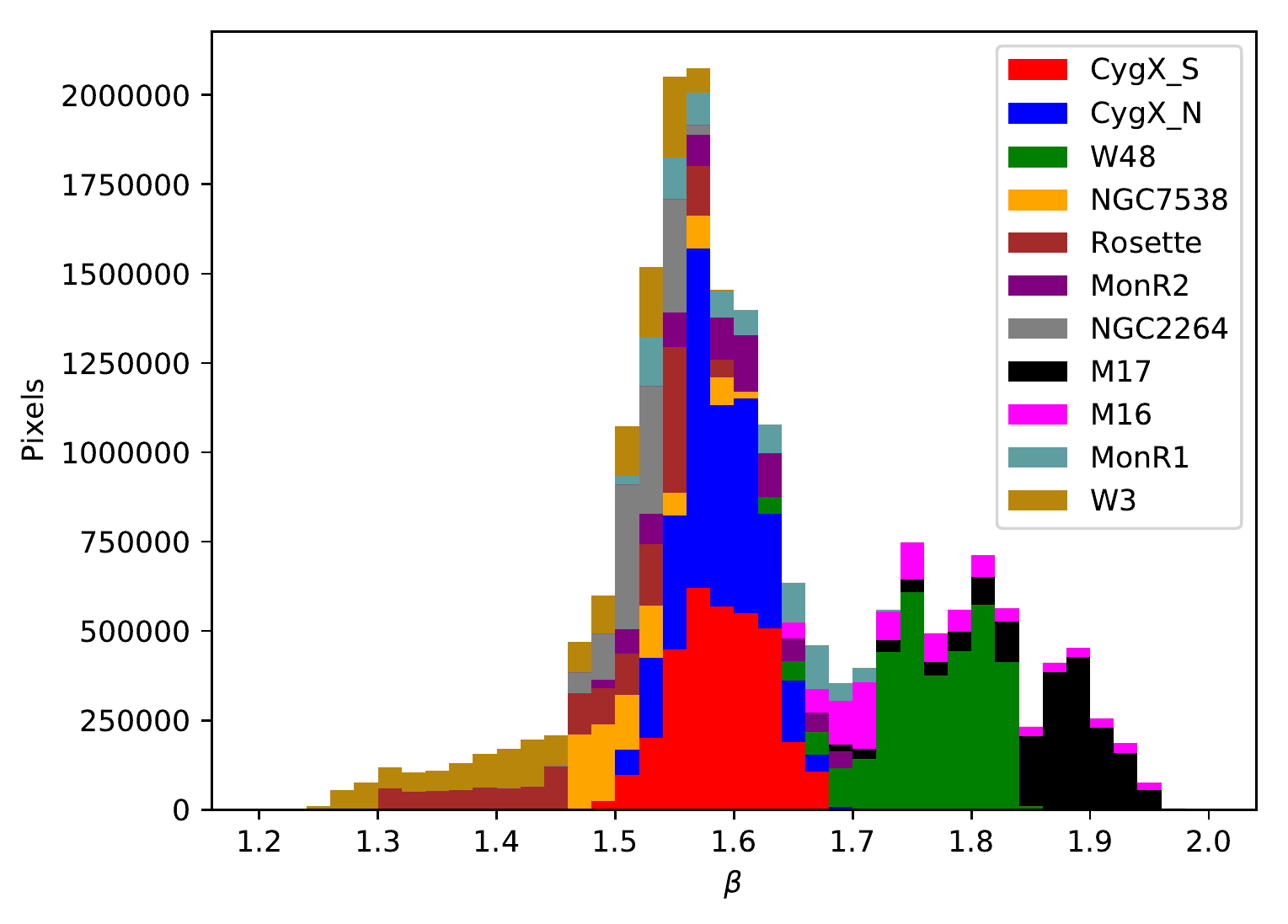}
\caption{Stacked histogram of the dust emissivity, $\beta$, used for SED fitting of each pixel in the \textit{Herschel} dust continuum maps for all clouds observed by KEYSTONE.  The $\beta$ values are from \textit{Planck}-derived dust models \citep{Planck_2014} that have been resampled to the same pixel grid as the \textit{Herschel} data.}
\label{beta}
\end{figure}

To estimate H$_2$ column densities for each region, spectral energy distributions (SEDs) were created by combining the $160 - 500 ~\mu$m maps for each observed pixel.  Full details of the SED-fitting method are described in Singh et al. (2019, in preparation), but are similar to the method applied in all HOBYS papers (see, e.g., Ladjelate et al. in preparation).  Here, we provide a brief summary of the process: first, a zero-level offset was added to the 160 $\mu$m map based on \textit{Planck} observations to account for background continuum emission not included in the \textit{Herschel} data.  The \textit{Herschel} Level 2.5 products for $250-500 ~\mu$m already have this offset applied, so no additional offsets were added to those maps.  Next, all maps were convolved to a resolution of 36.3$\arcsec$ and aligned to the same pixel grid as the 500 $\mu$m map.  SEDs were then assembled on a pixel-by-pixel basis and a modified blackbody model of the form $I_{\lambda} = B_{\lambda}(T_D) \kappa_{\lambda} \Sigma$ was fit to the data, where $I_{\lambda}$ is the surface brightness of the emission, $B_{\lambda}(T_D)$ is the Planck blackbody function at dust temperature $T_D$, and $\kappa_{\lambda}$ is the dust opacity defined as $\kappa_{\lambda}$ = 0.1($\lambda/300 \mu$m)$^{-\beta}$ cm$^2$/g following \cite{Hildebrand_1983} and assuming a gas-to-dust ratio of 100.  The dust emissivity, $\beta$, varies between 1.2 and 2.0 for each pixel and is based on \textit{Planck}-derived dust models \citep{Planck_2014} that were resampled to the same pixel grid as the \textit{Herschel} maps.  A stacked histogram showing the $\beta$ distribution across all pixels used for SED fitting in each region is displayed in Figure \ref{beta}.  The $\beta$ distributions vary from cloud to cloud, with the highest values observed in clouds close to the Galactic plane such as W48, M17, and M16.  We used \textit{Planck}-derived values of $\beta$ because the \textit{Herschel} data include only the portion of the dust SED close to the intensity peak. The position of the intensity peak is a function of both $T_D$ and $\beta$ \citep[e.g., $\lambda_{peak} = 1.493/T_{D}(3+\beta)$ for a modified blackbody,][]{Elia_2016} and it is not possible to remove the degeneracy between these two parameters unless data at longer wavelengths, where $I_\nu\propto\lambda^{-\beta}$, are used.  Since the \textit{Planck} data include observations down to 850 $\mu$m, they are more capable of constraining $\beta$ than the \textit{Herschel} data.

The gas surface mass density, $\Sigma$, and dust temperature were left as free parameters during the fitting procedure.  The resulting best-fit model's $\Sigma$ was converted to H$_2$ column density, $N(H_2)$, using $\Sigma=\mu_{H} m_H N(H_{2})$, where $\mu_{H}$=2.8 is the mean molecular weight per hydrogen molecule, which assumes the relative mass ratios of hydrogen, helium, and metals are 0.71, 0.27, and 0.02, respectively \citep[see, e.g., Appendix A in][]{Kauffmann_2008}, and $m_H$ is the mass of a hydrogen atom.  The SED fitting procedure failed to converge for a small fraction of pixels where the dust continuum emission was saturated.  The percentage of affected pixels for the KEYSTONE clouds affected are: M17 ($0.09 \%$), W48 ($0.008 \%$ of pixels), Cygnus X North ($0.006 \%$ of pixels), NGC7538 ($0.01 \%$ of pixels), W3 ($0.01 \%$ of pixels), and MonR2 ($0.001 \%$ of pixels).  For the affected pixels, we replace their values with the median column density of the ten closest pixels with reliable SED fits.  As such, this is likely a lower limit to the true column density for those pixels.  Any dendrogram-identified leaf (see Section 3.3) that overlaps with one of these affected pixels is also flagged in all catalogs and analyses.  The number of leaves in each cloud that include affected pixels are: M17 (1 of 38 leaves), W48 (1 of 100 leaves), Cygnus X North (2 of 200 leaves), NGC7538 (1 of 73 leaves), and W3 (2 of 84 leaves).

The main difference between the column densities derived in this paper and those of the HOBYS collaboration (Ladjelate et al. in preparation) involve the assumptions on $\beta$.  Specifically, the HOBYS column density maps assume $\beta=2$ for all pixels while we use $1.2 \leq \beta \leq 2.0$ based on \textit{Planck} dust models that constrain $\beta$ on large spatial scales \citep{Planck_2014}.  Our lower values of $\beta$ result in comparatively lower column densities in our maps.  For instance, the HOBYS team has released the H$_2$ column density maps and core/protostar catalog for MonR2 \citep{Rayner_2017}.  We find that the \cite{Rayner_2017} column densities are on average a factor of $\sim2.5$ higher than those derived in this paper.  Although the higher column densities in the HOBYS maps would lead to larger structure masses in our analysis, we discuss in Section 3.4 that the method used to convert the column densities into structure masses is likely a larger source of uncertainty than the $\beta$ assumption.  Moreover, we also recovered 22 of the 28 ($\sim79\%$) protostars identified by \cite{Rayner_2017}, with the six discrepant sources located in the central MonR2 hub that is bright at 70 $\mu$m.  This suggests that our protostar extraction is likely confusion limited in bright hubs, but can efficiently recover sources that are more isolated.  



\subsection{JCMT C$^{18}$O Data}   
C$^{18}$O $(3-2)$ data cubes observed by the HARP-ACSIS spectrometer on the James Clerk Maxwell Telescope (JCMT) were obtained from the JCMT Science Archive\footnote{\url{http://www.cadc-ccda.hia-iha.nrc-cnrc.gc.ca/en/}}, which is hosted by the Canadian Astronomy Data Centre.  Of the eleven clouds observed by KEYSTONE, six were found to have publicly available C$^{18}$O $(3-2)$ data cubes in the JCMT Science Archive: Cygnus X North, Cygnus X South, M16, M17, NGC7538, and W3.  The native spectral resolution of the C$^{18}$O $(3-2)$ cubes is $\sim$ 0.056 km s$^{-1}$ and the spatial resolution is 15.3$\arcsec$.  To match better the spatial and spectral resolution of our NH$_3$ observations and improve sensitivity, we smoothed the C$^{18}$O $(3-2)$ maps to a spatial resolution of 32$\arcsec$ and spectral resolution of 0.11 km s$^{-1}$.  In Section 4.3, we describe how Gaussian line fitting of these data cubes is used to estimate the external, turbulent pressure on the ammonia structures observed by KEYSTONE. 




\section{Analysis and Results}
\subsection{NH$_3$ Line Fitting}
The NH$_3$ (1,1) and (2,2) lines were used to estimate the excitation temperature ($T_{ex}$), kinetic gas temperature ($T_{K}$), centroid velocity ($V_{LSR}$), velocity dispersion ($\sigma$), and para-NH$_3$ column density ($N_{para-NH_3}$) for each pixel.  We adopted the line fitting method of the Green Bank Ammonia Survey (GAS) described in \cite{Friesen_2017}, which uses the \texttt{coldammonia} model in the \texttt{pyspeckit} Python package \citep{Ginsburg_2011} to generate model ammonia spectra under the assumptions of LTE and a single velocity component along the line of sight.  While most of the KEYSTONE spectra are well characterized by a single velocity component, we do see signs of multiple velocity components that are closely separated along the spectral axis in regions of W48 and M17.  For those spectra, our single velocity component fitting will produce a best-fit model that has a broadened line width to account for the larger width of the emission line features in the spectrum.  In a future KEYSTONE paper, we plan to implement a multiple velocity component fitting method that will robustly identify spectra with more than one velocity component and estimate better the line widths for those spectra (Keown et al., in preparation).

The GAS line-fitting pipeline\footnote{available at \url{http://gas.readthedocs.io/}} was applied to all pixels with NH$_3$ (1,1) signal-to-noise ratio (SNR) $>$ 3, where SNR is measured from the ratio of peak emission-line intensity to the rms of the off-line channels in the spectrum.  In addition to the minimum SNR threshold, pixels were excluded from our final parameter maps if they did not meet the following constraints on the best-fit model parameters and uncertainties: 

\begin{enumerate}
\item 5 K $< T_{K} <$ 40 K (outside this range, the NH$_3$ (1,1) and (2,2) lines cannot constrain $T_{K}$); 
\item 0.05 km s$^{-1}$ $< \sigma <$ 2.0 km s$^{-1}$ (below 0.05 km s$^{-1}$ is unrealistic since our channel width is only $\sim$ 0.07 km s$^{-1}$; above 2.0 km s$^{-1}$ is uncharacteristic of NH$_3$ (1,1) emission in the observed star-forming environments \citep[e.g.,][]{Pillai_2011, Olmi_2010} and likely indicates the presence of strong outflows or multiple velocity components along the line of sight);
\item $N_{para-NH_3}<10^{16}$ cm$^{-2}$ (above $10^{16}$ cm$^{-2}$ is uncharacteristic of NH$_3$ emission in the observed star-forming environments \citep[e.g.,][]{Olmi_2010});
\item $T_{K, err} <$ 5 K;
\item $\sigma_{err} <$ 2.0 km s$^{-1}$;
\item $V_{LSR, err} <$ 1 km s$^{-1}$;
\item $(\log{N_{para-NH_{3}}})_{err}<2$ ; 
\end{enumerate} where $4-7$ are included to cull fits that were unable to converge.

\begin{figure}[ht]
\epsscale{0.6}
\plotone{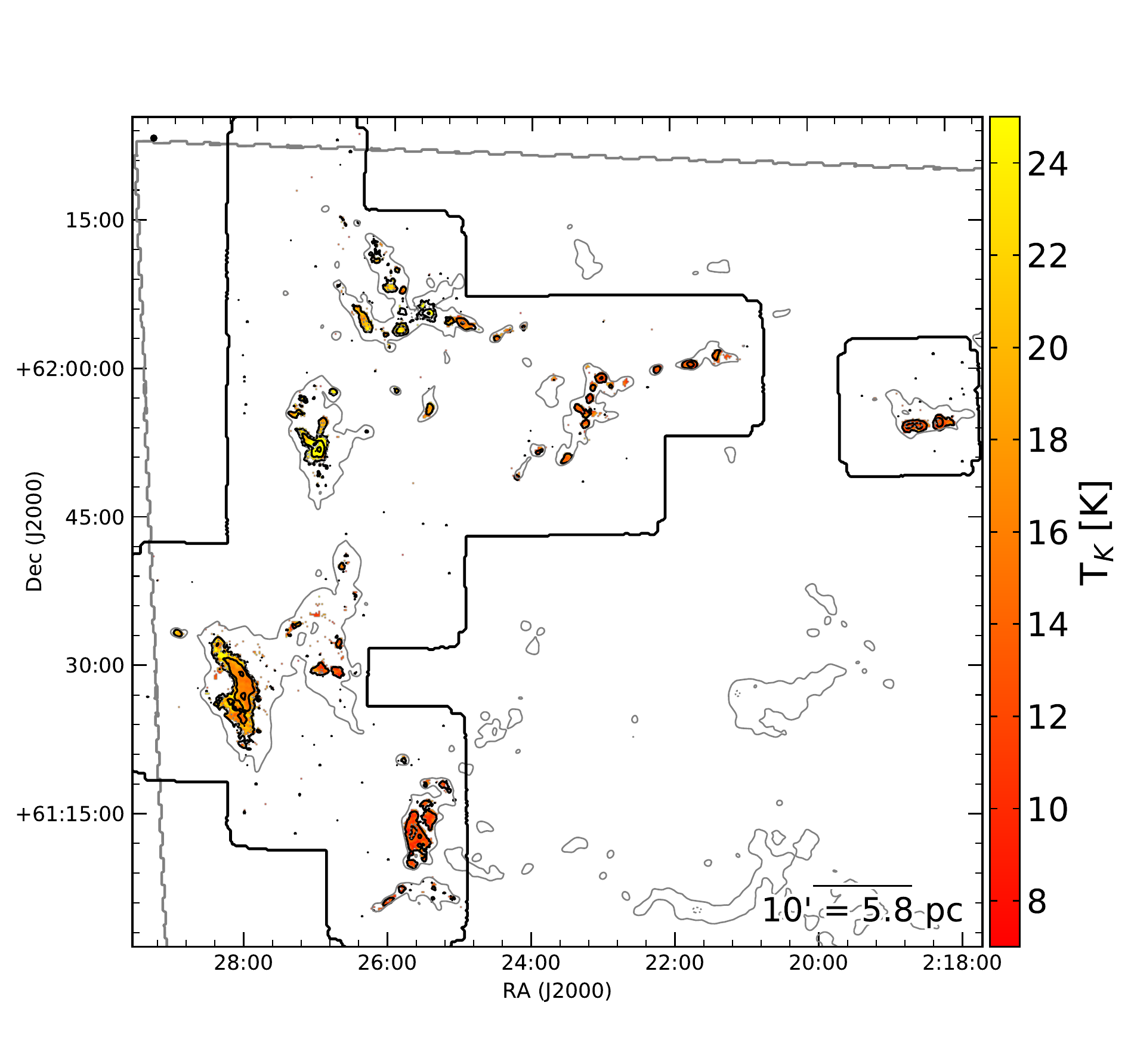}
\plotone{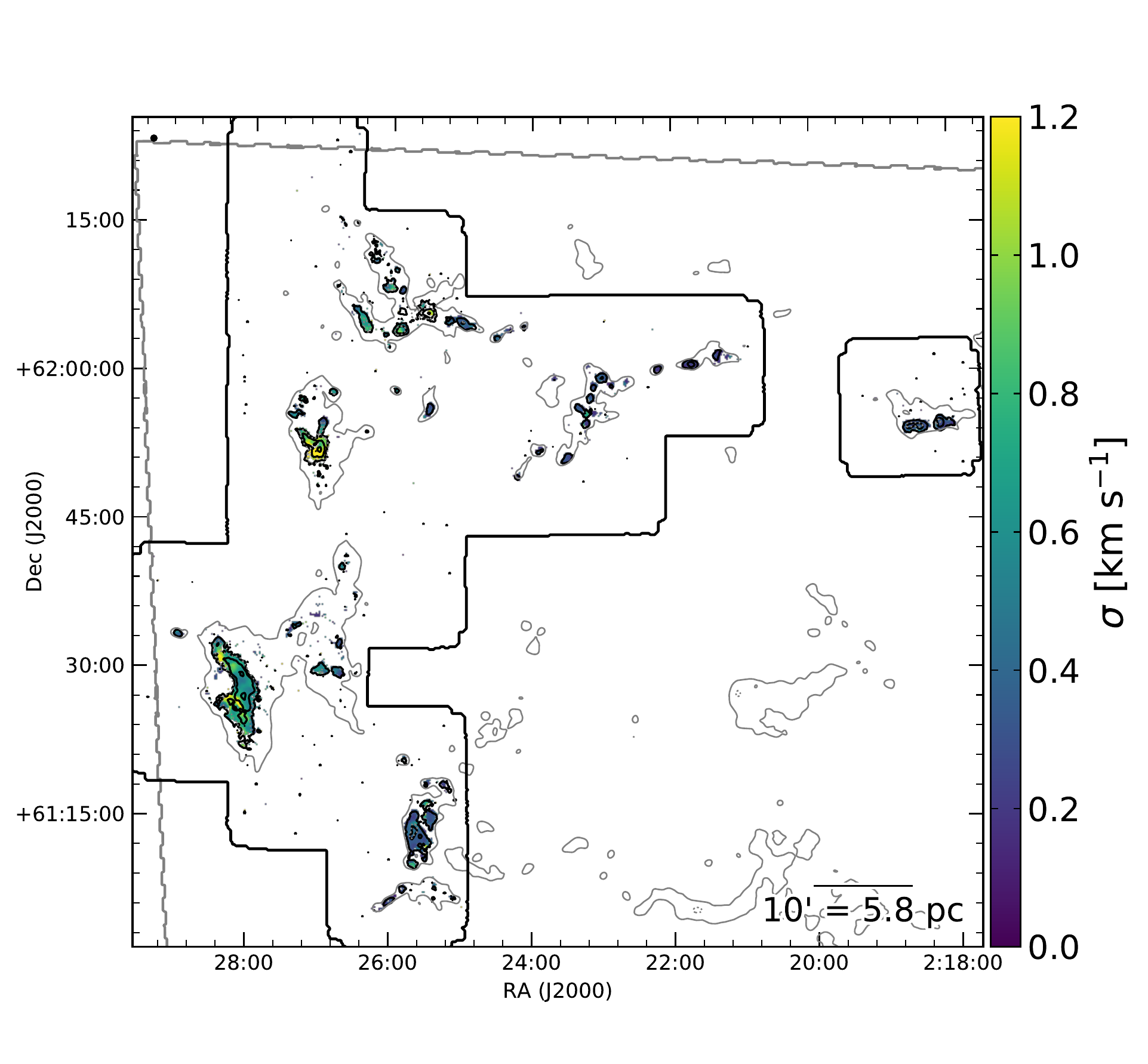}
\caption{Kinetic temperature (top) and NH$_3$ (1,1) velocity dispersion (bottom) derived from NH$_3$ (1,1) and (2,2) line fitting of the W3 observations.  The black contours show the NH$_3$ (1,1) integrated intensity at 1.0, 3.5, and 10 K km s$^{-1}$.  The solid and dotted grey contours outline H$_2$ column densities of $2.8\times10^{21}$ cm$^{-2}$ and $9.4\times10^{21}$ cm$^{-2}$, respectively, which are equivalent to a total extinction in the V band of A$_{V}=3$ mag and 10 mag. The 32$\arcsec$ beam size is shown as a black dot in the upper left corner of each plot. The thick black and gray lines outline the KEYSTONE and \textit{Herschel} mapping boundaries, respectively.}
\label{W3_params}
\end{figure}

\begin{figure}[ht]
\plotone{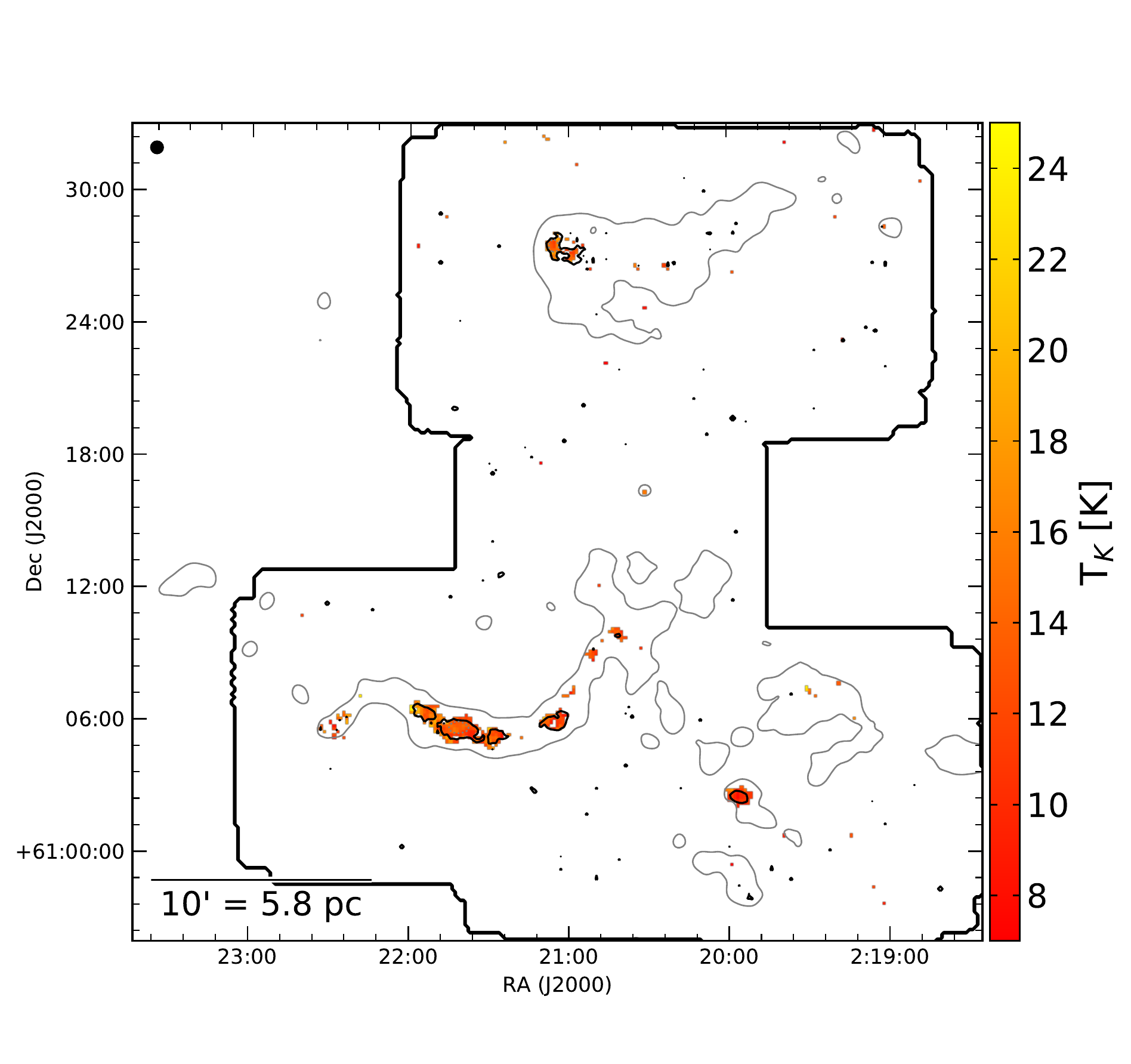}
\plotone{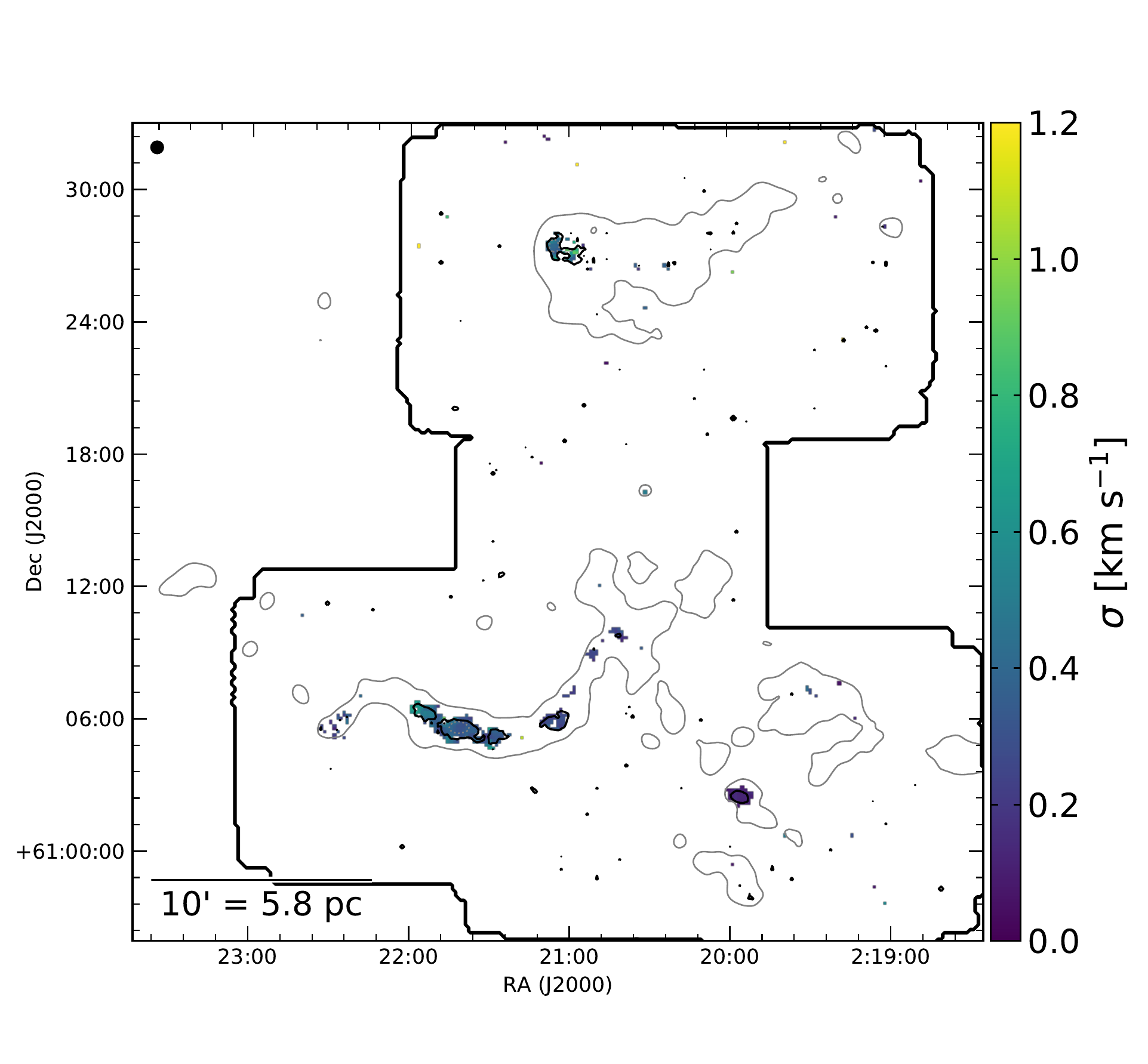}
\caption{Same as Figure \ref{W3_params} for W3-west. }
\label{W3_west_params}
\end{figure}

\begin{figure}[ht]
\epsscale{0.9}
\plotone{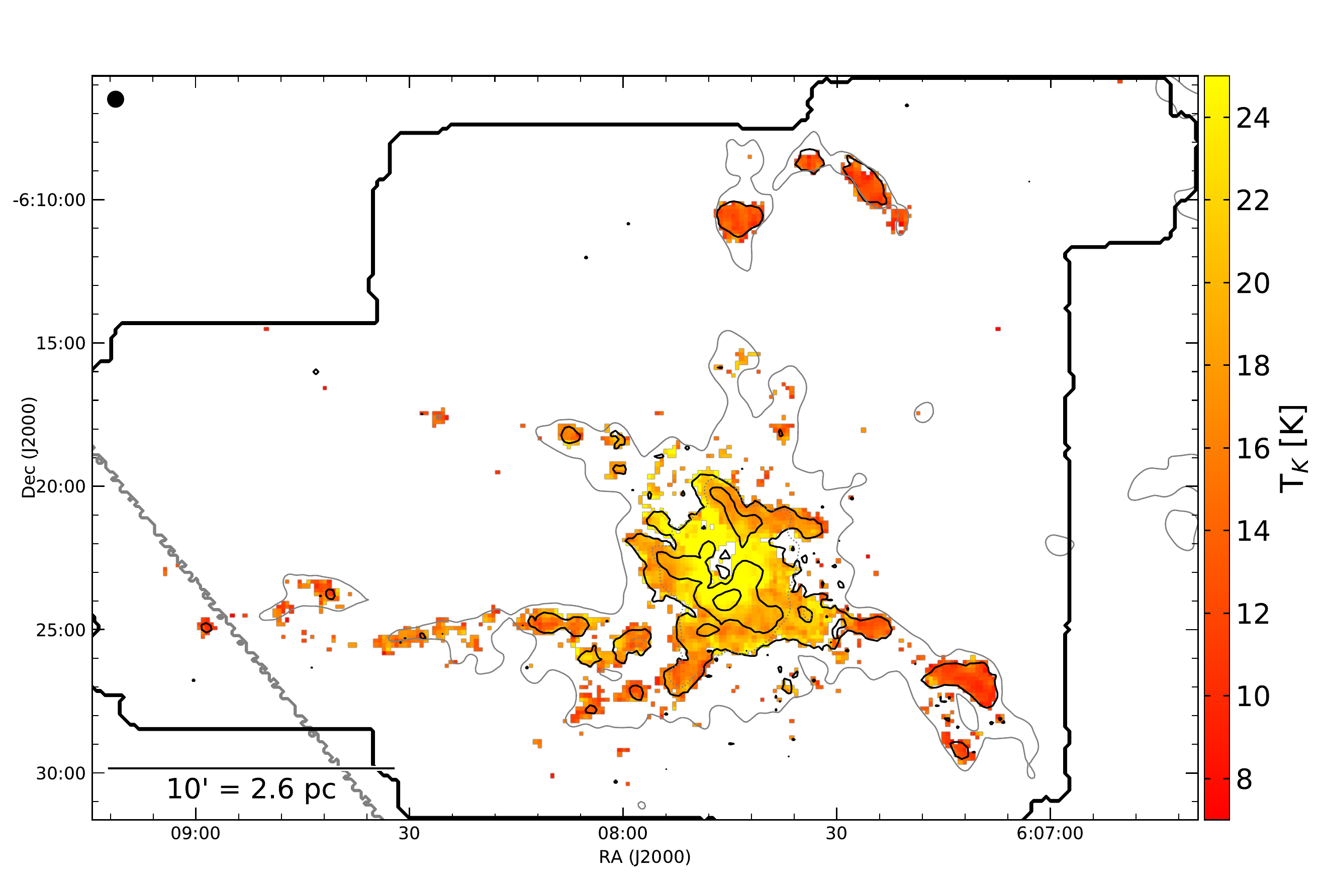}
\plotone{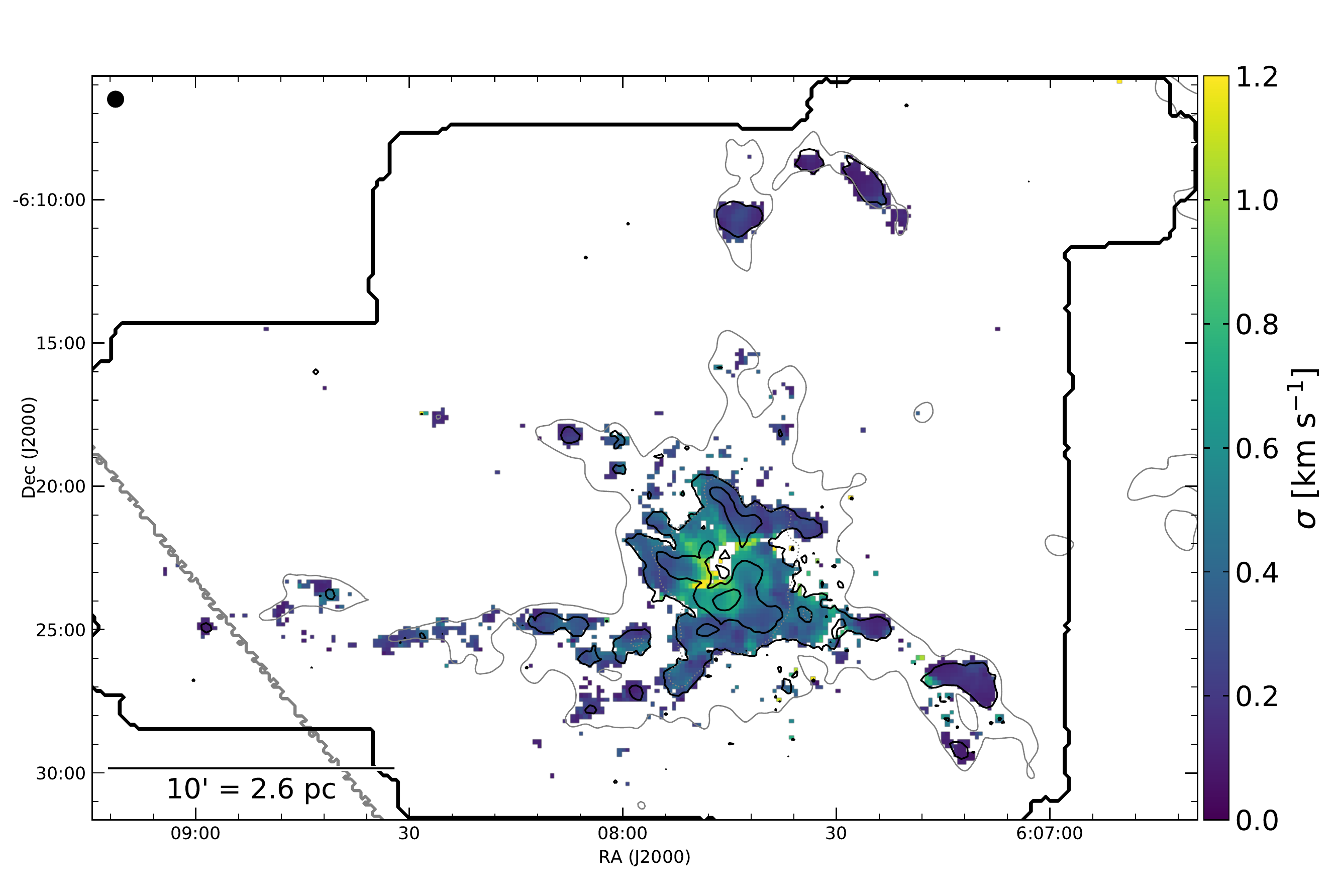}
\caption{Same as Figure \ref{W3_params} for MonR2. }
\label{MonR2_params}
\end{figure}

\begin{figure}[ht]
\epsscale{1.2}
\plottwo{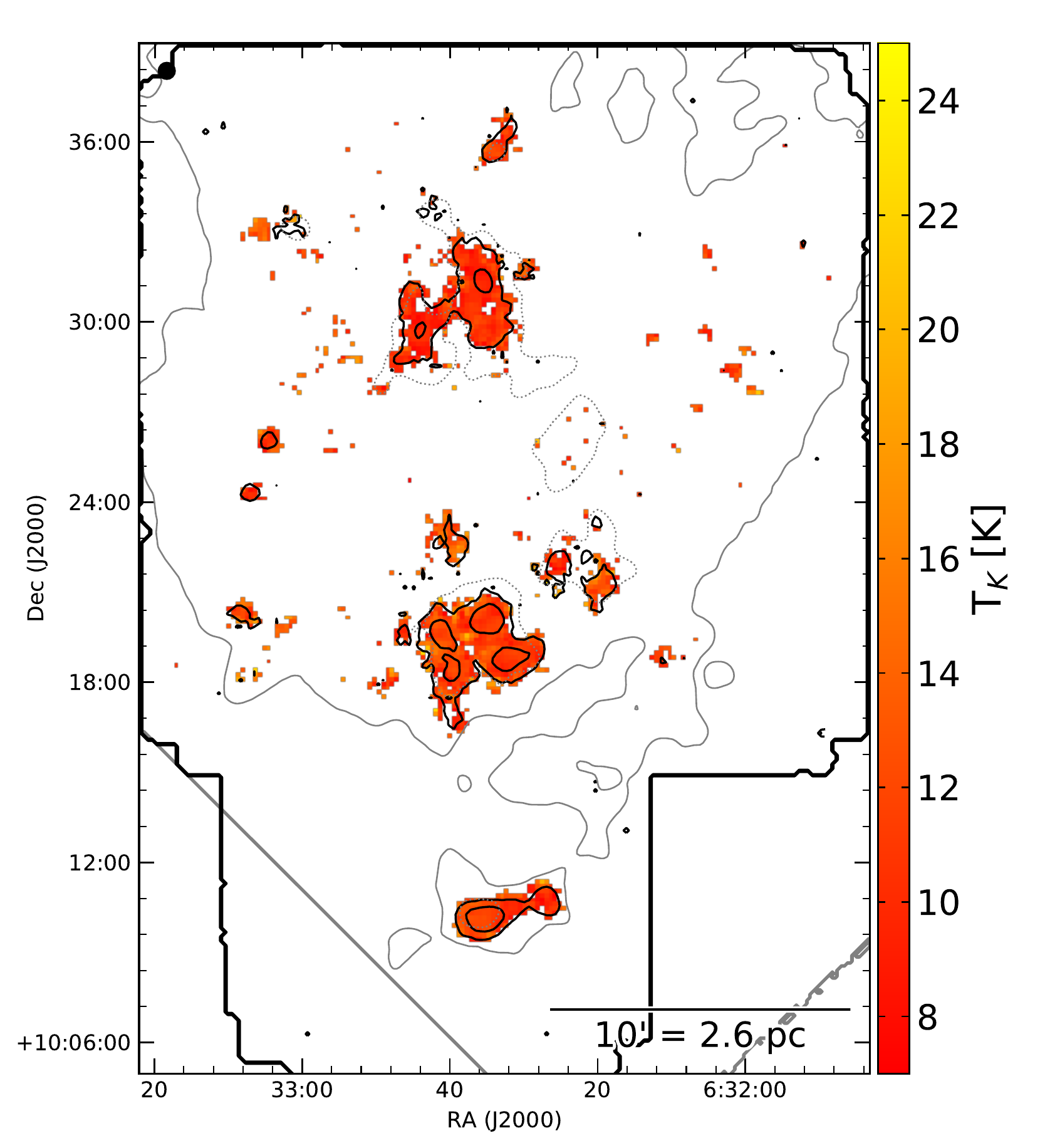}{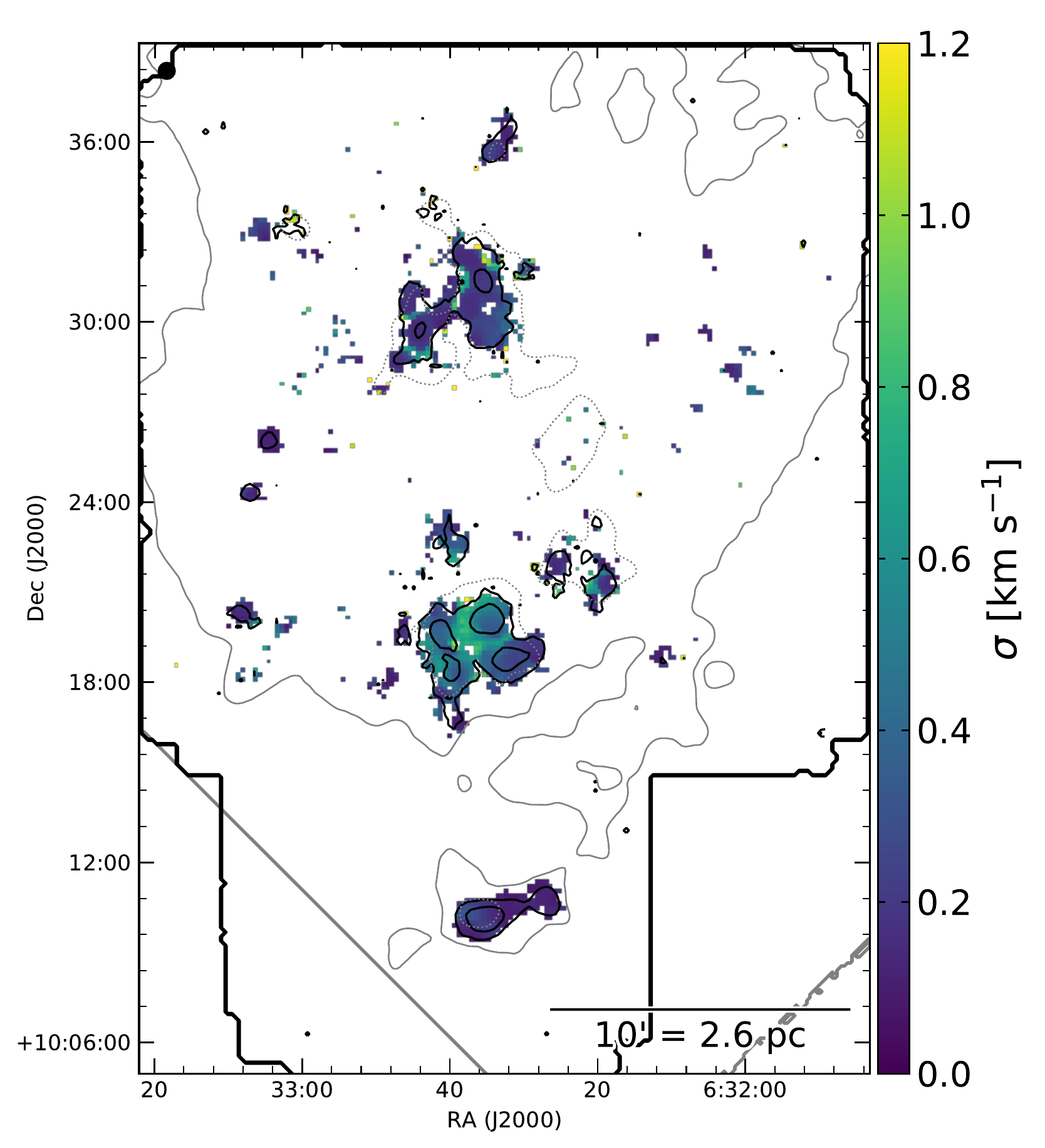}
\caption{Same as Figure \ref{W3_params} for MonR1. }
\label{MonR1_params}
\end{figure}

\begin{figure}[ht]
\epsscale{0.7}
\plotone{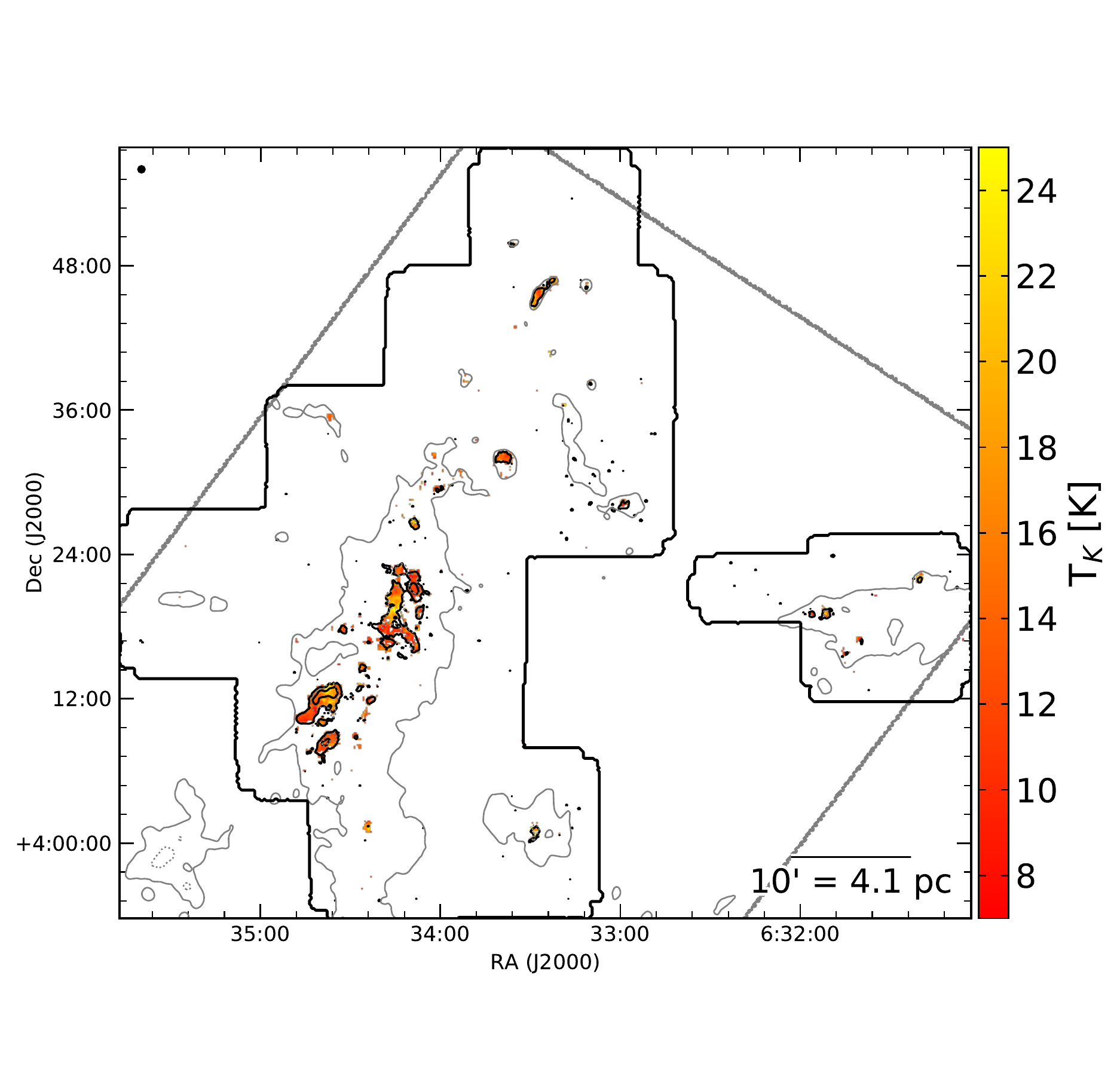}
\plotone{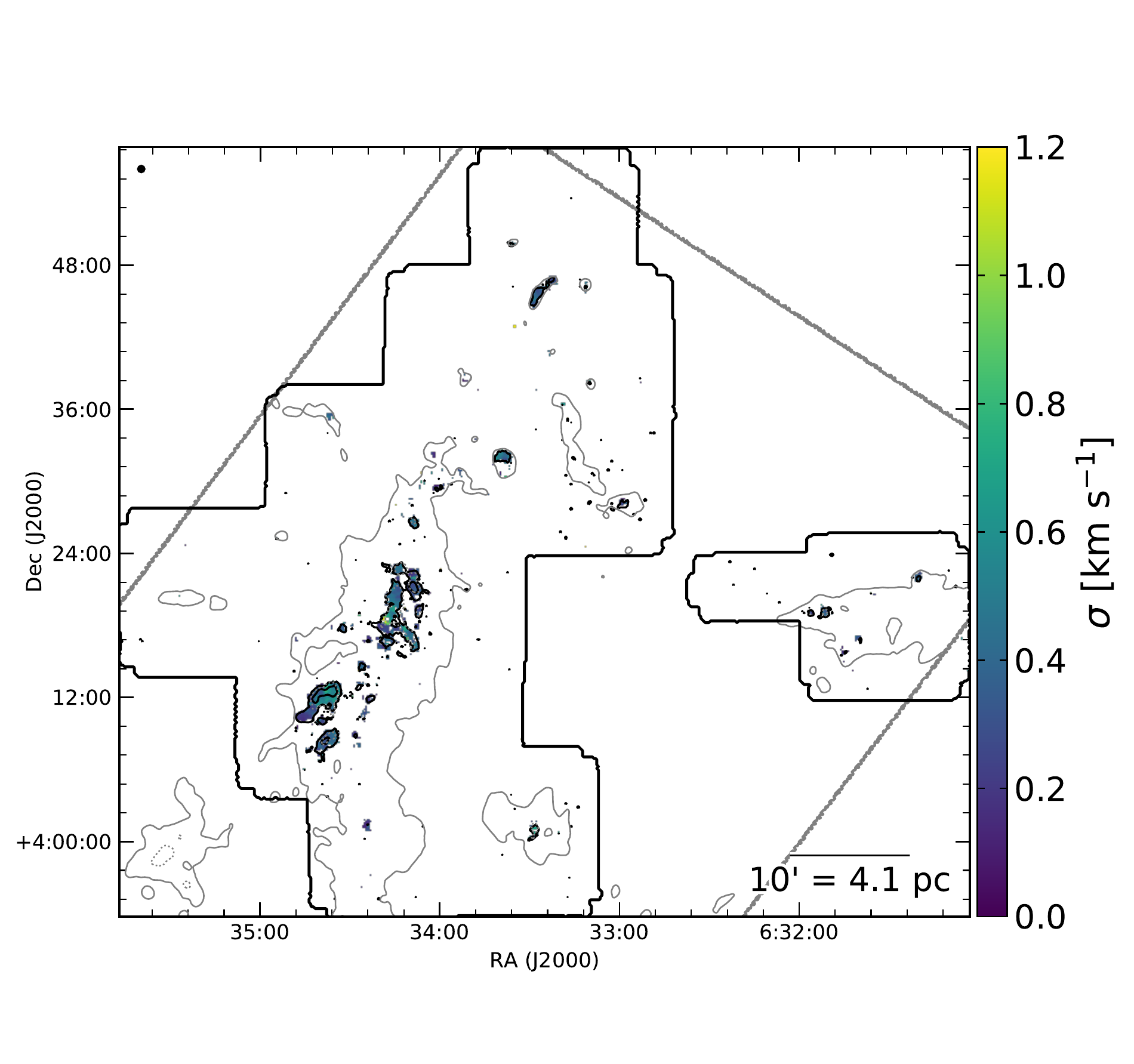}
\caption{Same as Figure \ref{W3_params} for Rosette. }
\label{Rosette_params}
\end{figure}

\begin{figure}[ht]
\epsscale{1.2}
\plottwo{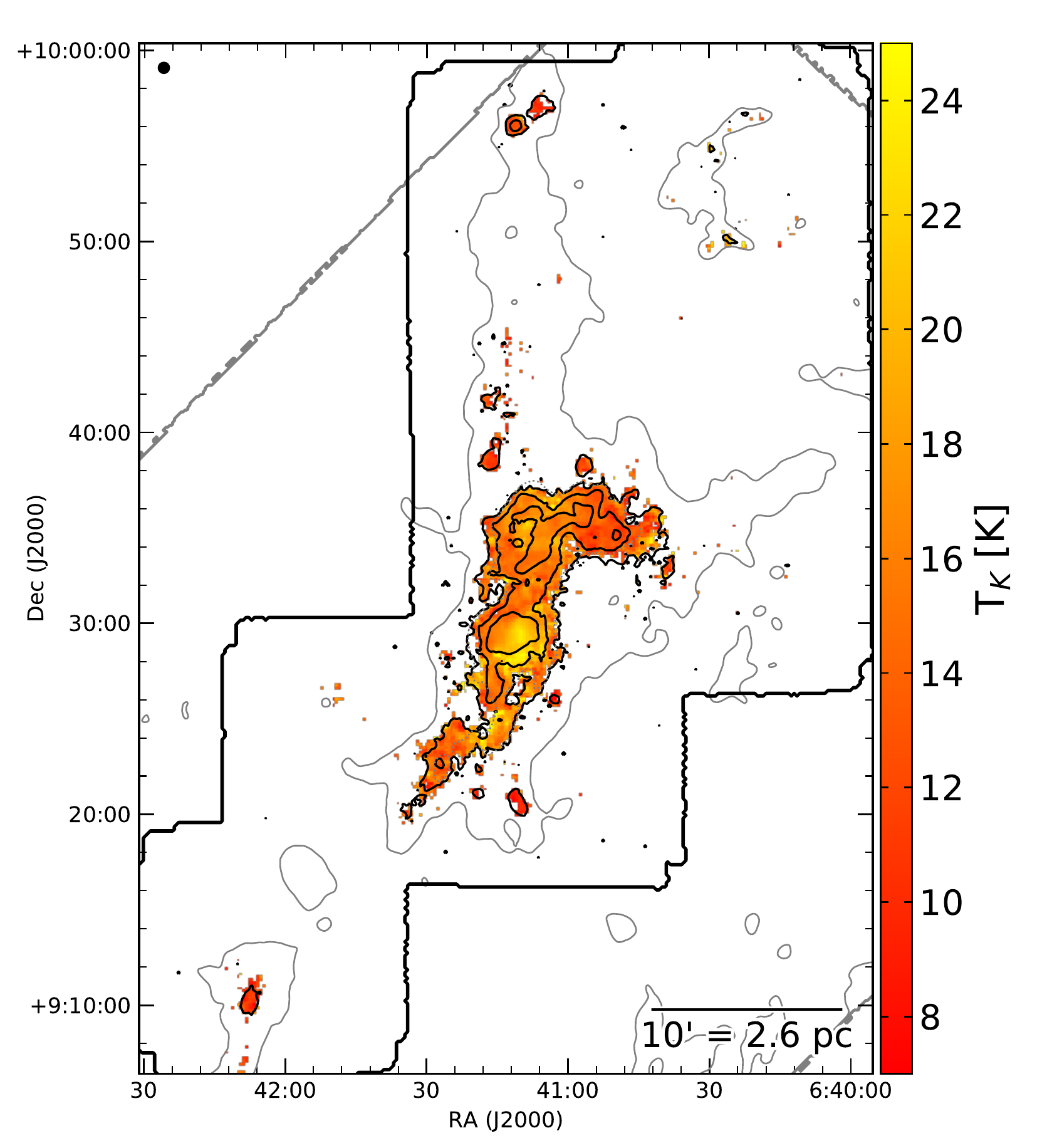}{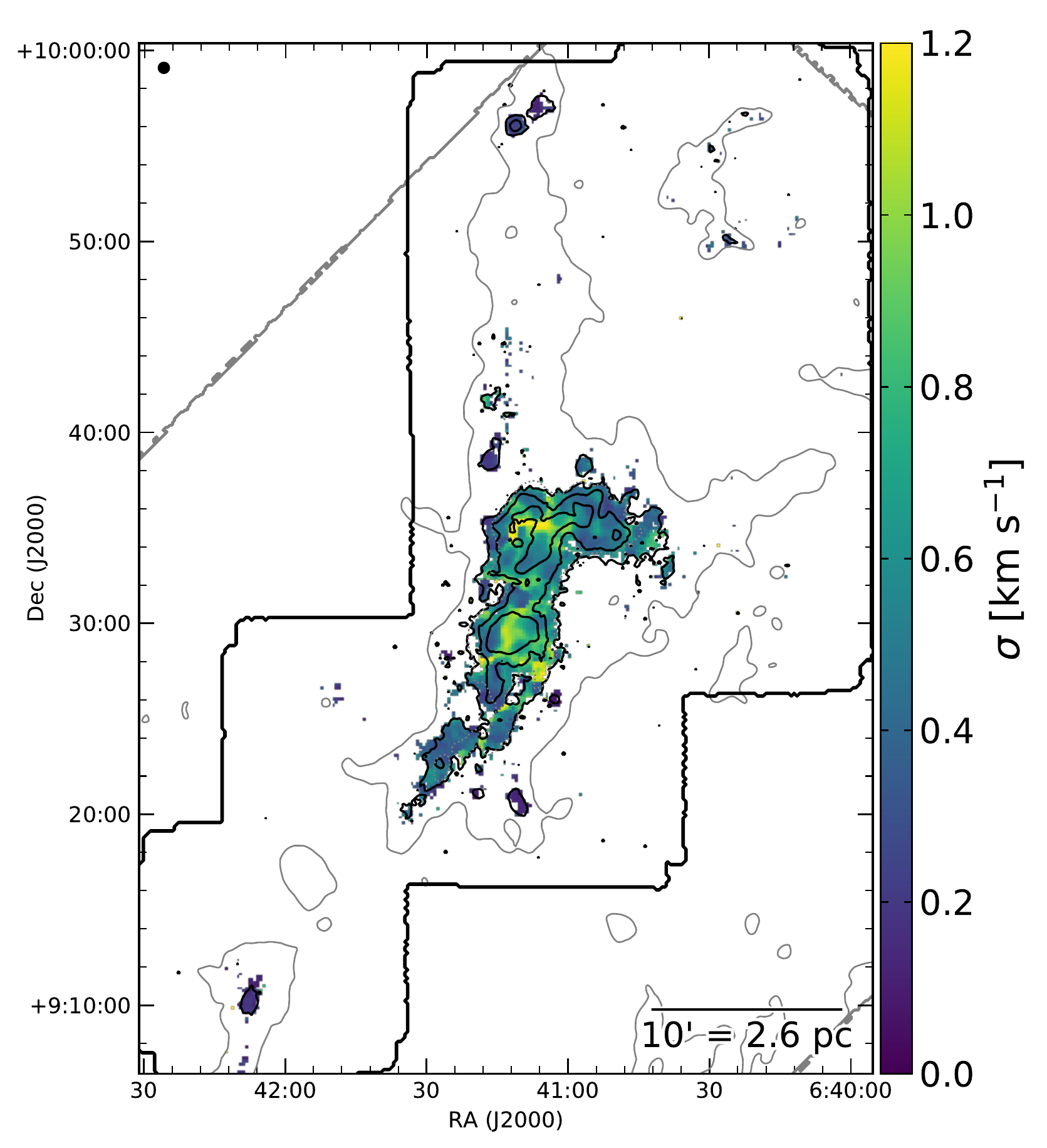}
\caption{Same as Figure \ref{W3_params} for NGC2264. }
\label{NGC2264_params}
\end{figure}

\begin{figure}[ht]
\plottwo{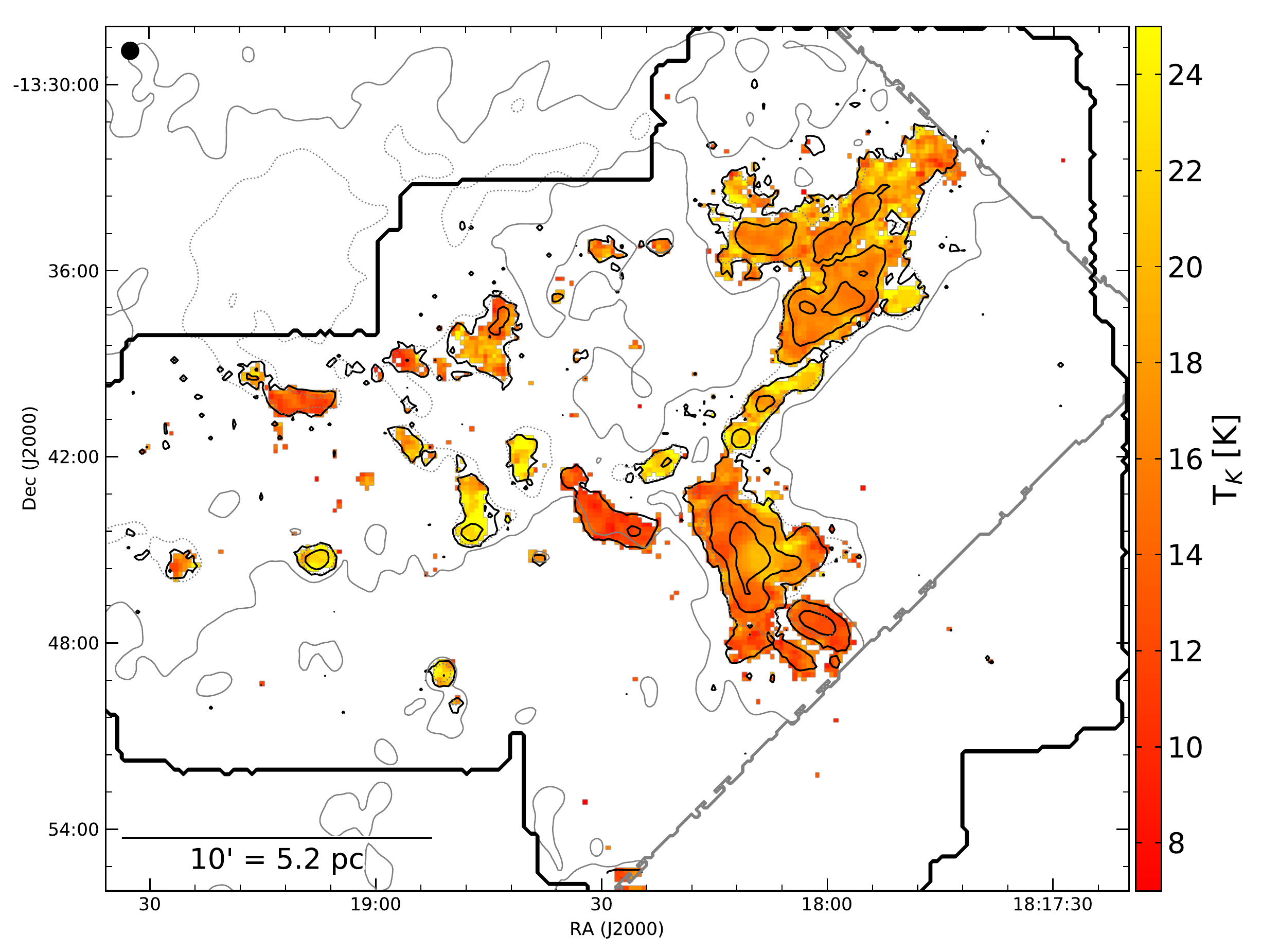}{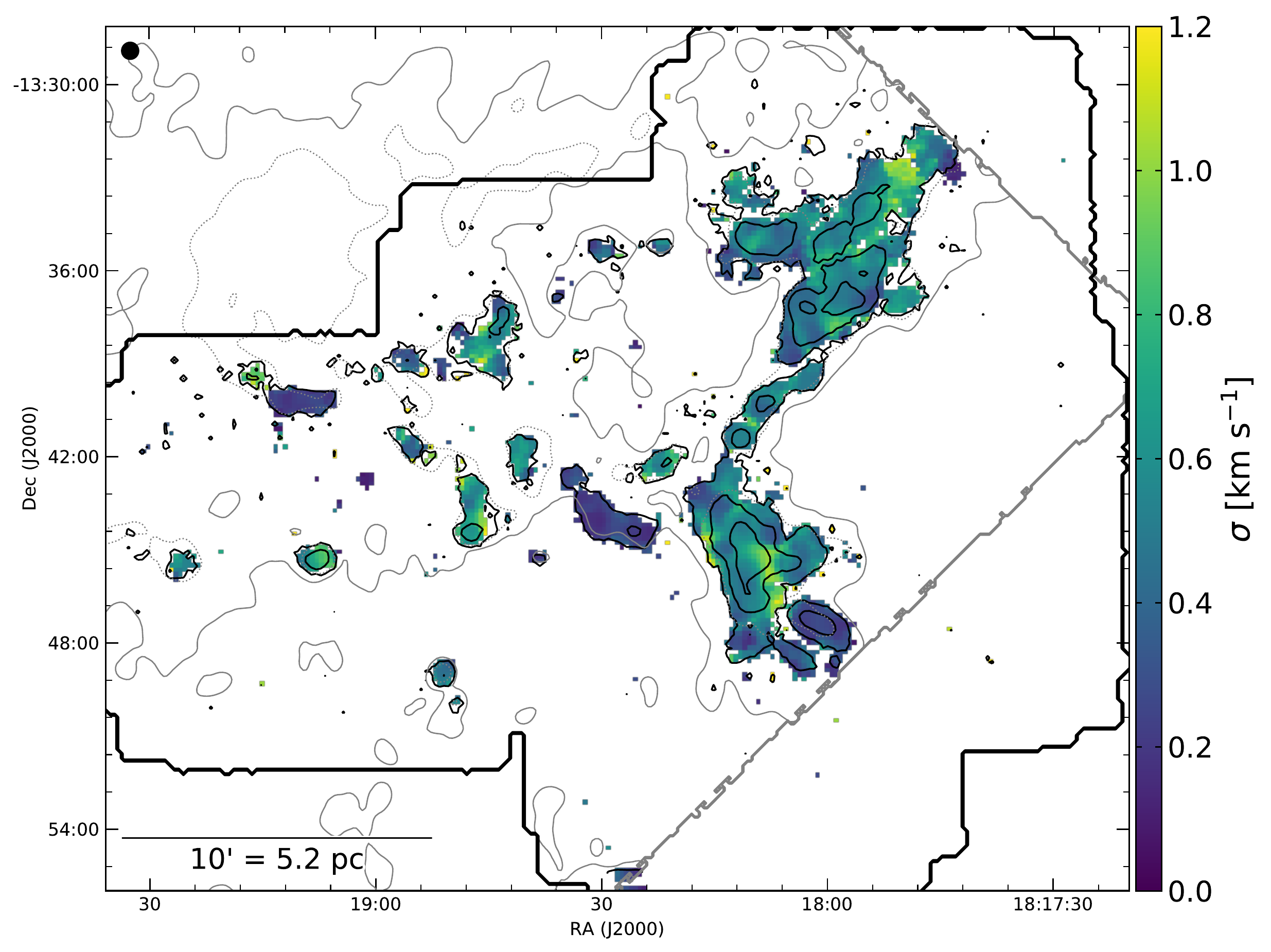}
\caption{Same as Figure \ref{W3_params} for M16. The solid and dotted grey contours outline H$_2$ column densities of $5.6\times10^{21}$ cm$^{-2}$ and $9.4\times10^{21}$ cm$^{-2}$, respectively, which are equivalent to a total extinction in the V band of A$_{V}=6$ mag and 10 mag.}
\label{M16_params}
\end{figure}

\begin{figure}[ht]
\plottwo{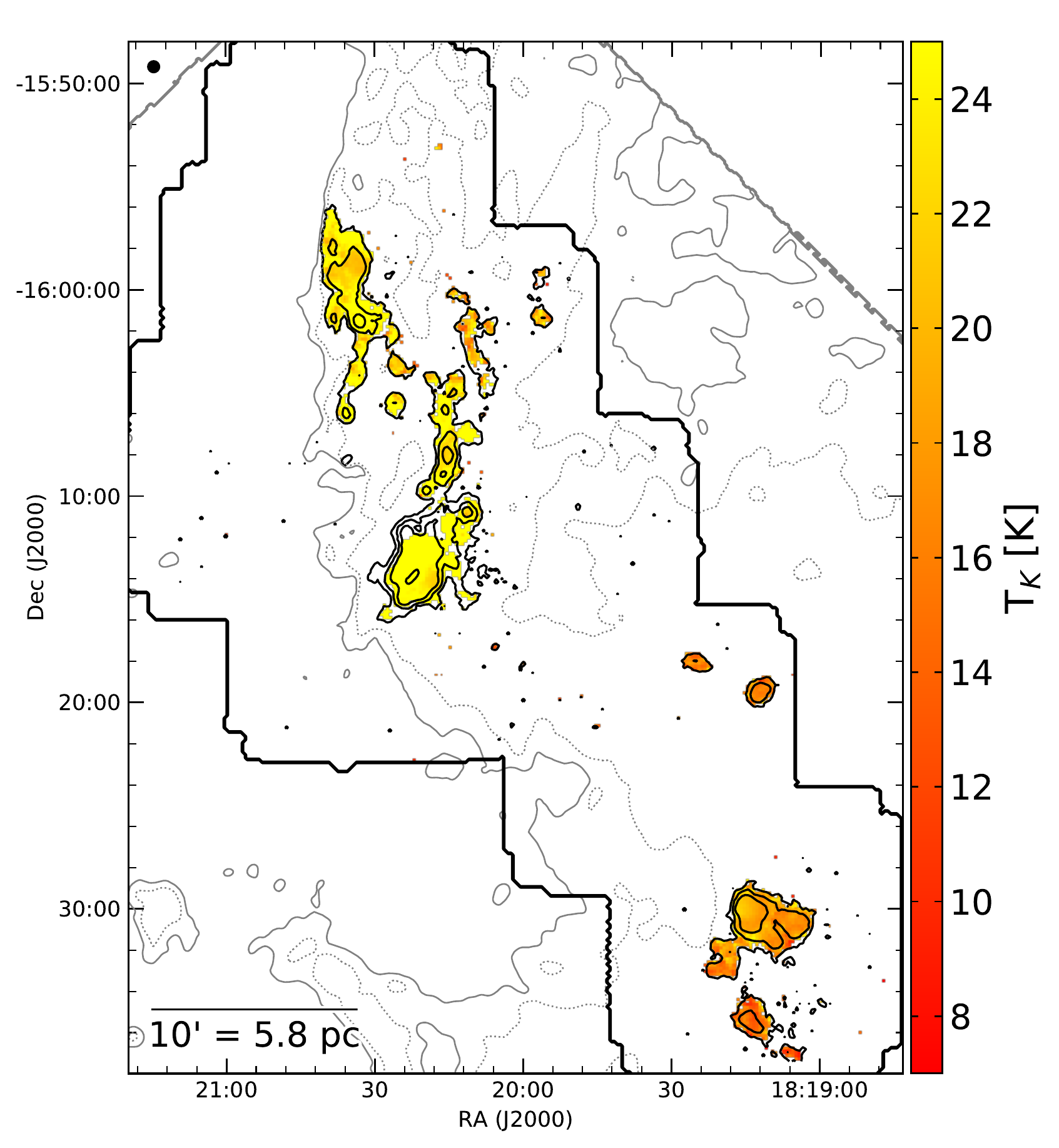}{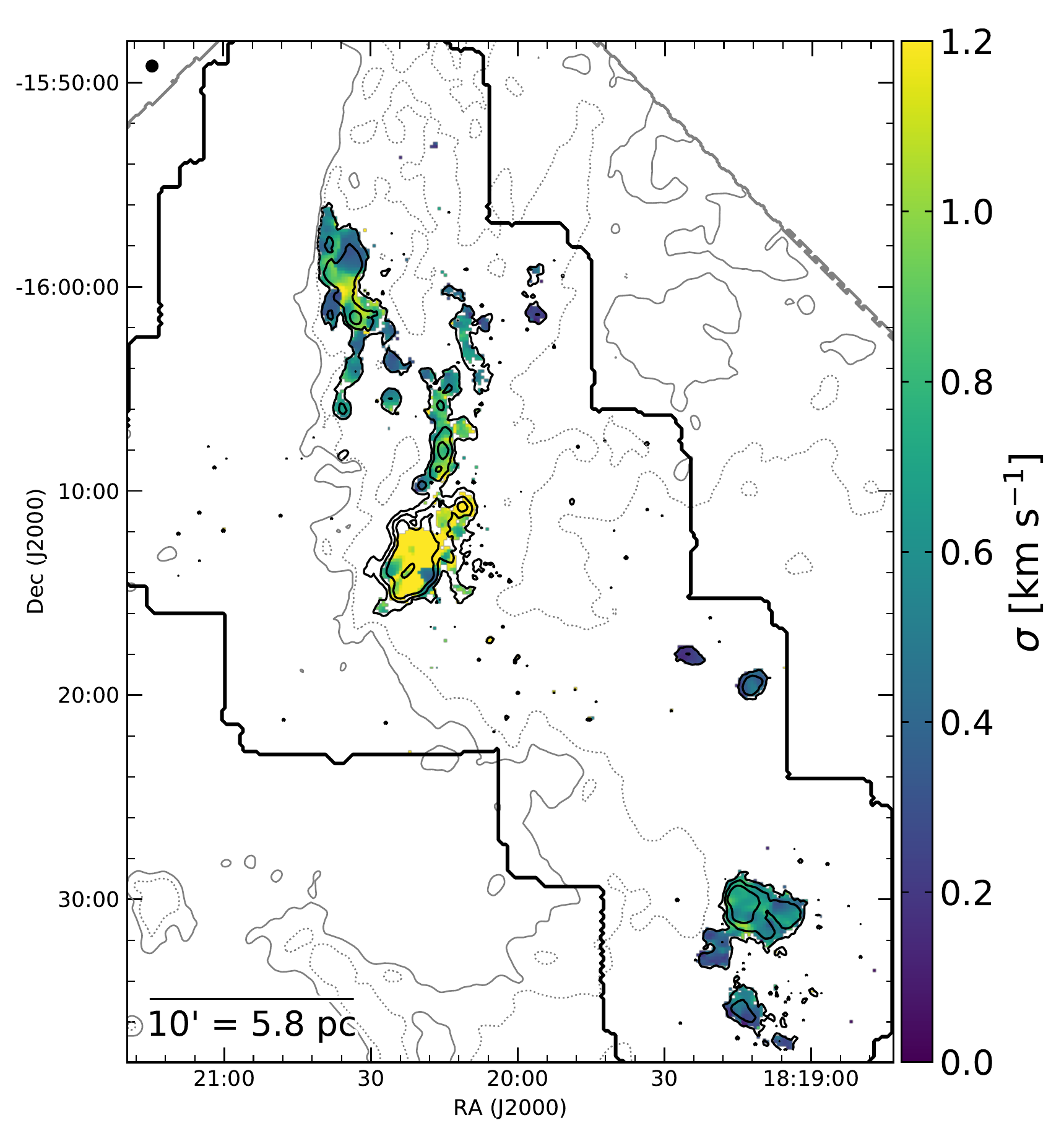}
\caption{Same as Figure \ref{W3_params} for M17. The solid and dotted grey contours outline H$_2$ column densities of $7.5\times10^{21}$ cm$^{-2}$ and $9.4\times10^{21}$ cm$^{-2}$, respectively, which are equivalent to a total extinction in the V band of A$_{V}=8$ mag and 10 mag.}
\label{M17_params}
\end{figure}

\begin{figure}[ht]
\epsscale{0.9}
\plotone{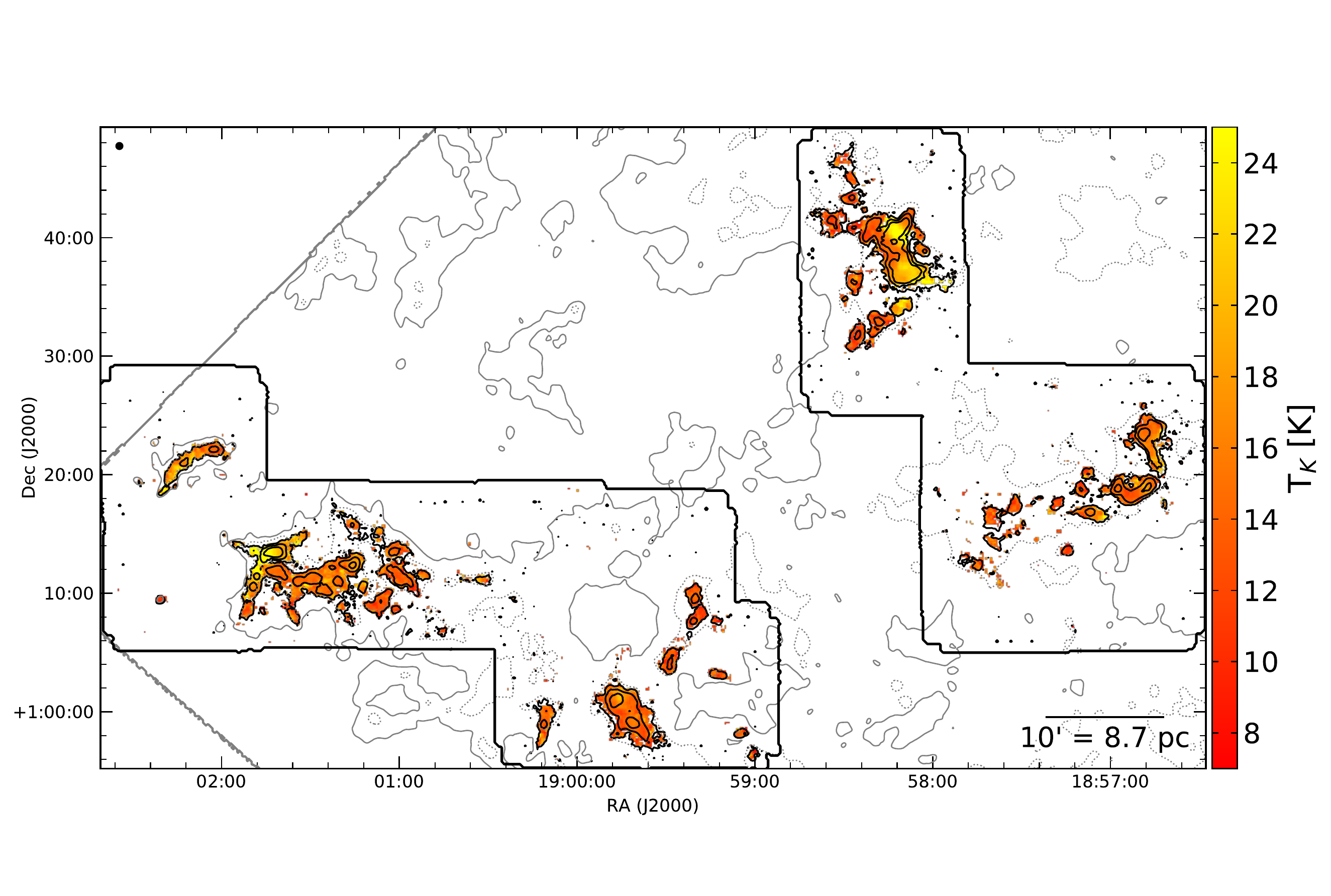}
\plotone{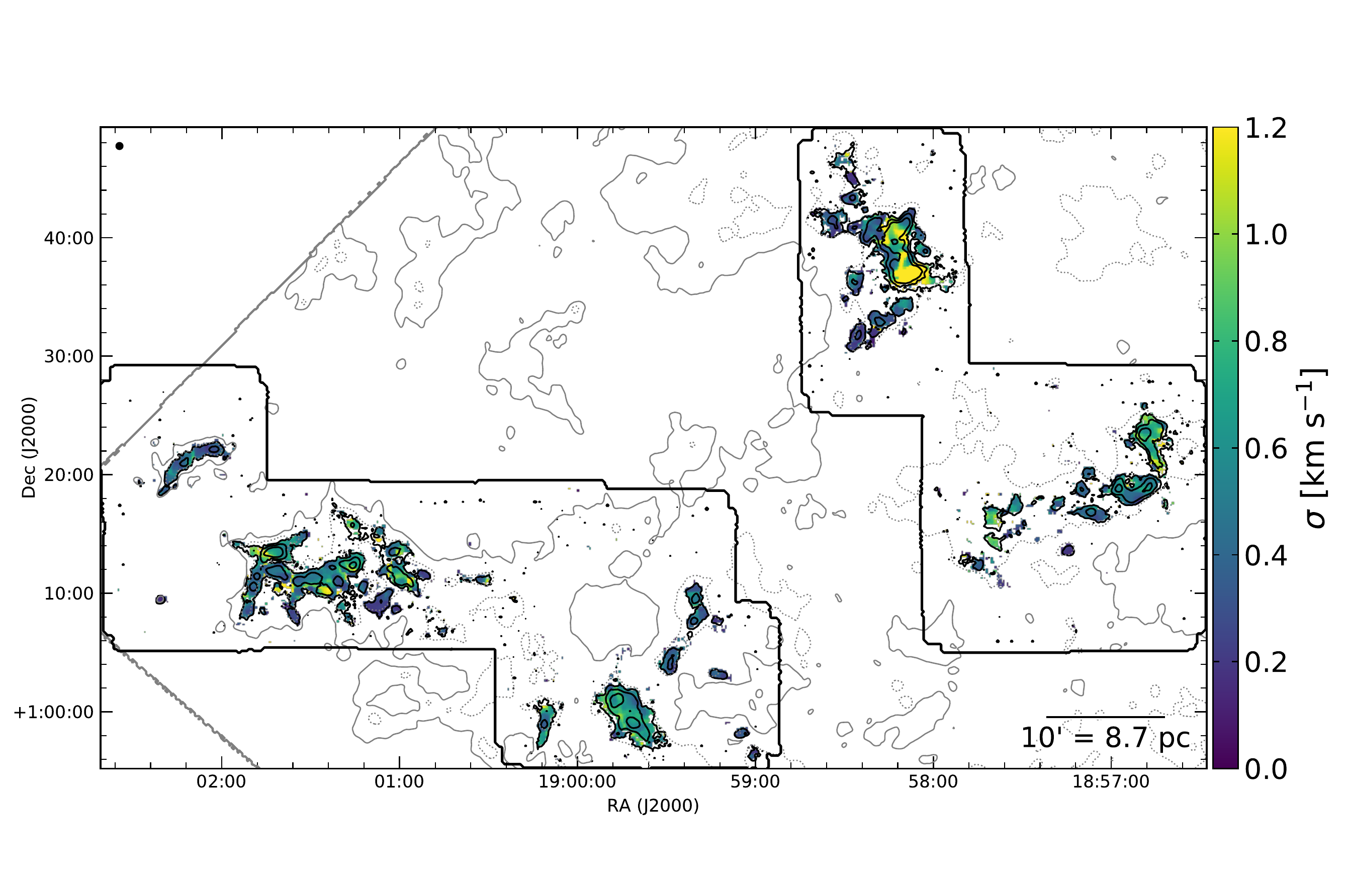}
\caption{Same as Figure \ref{M16_params} for W48. }
\label{W48_params}
\end{figure}

\begin{figure}[ht]
\epsscale{1.2}
\plottwo{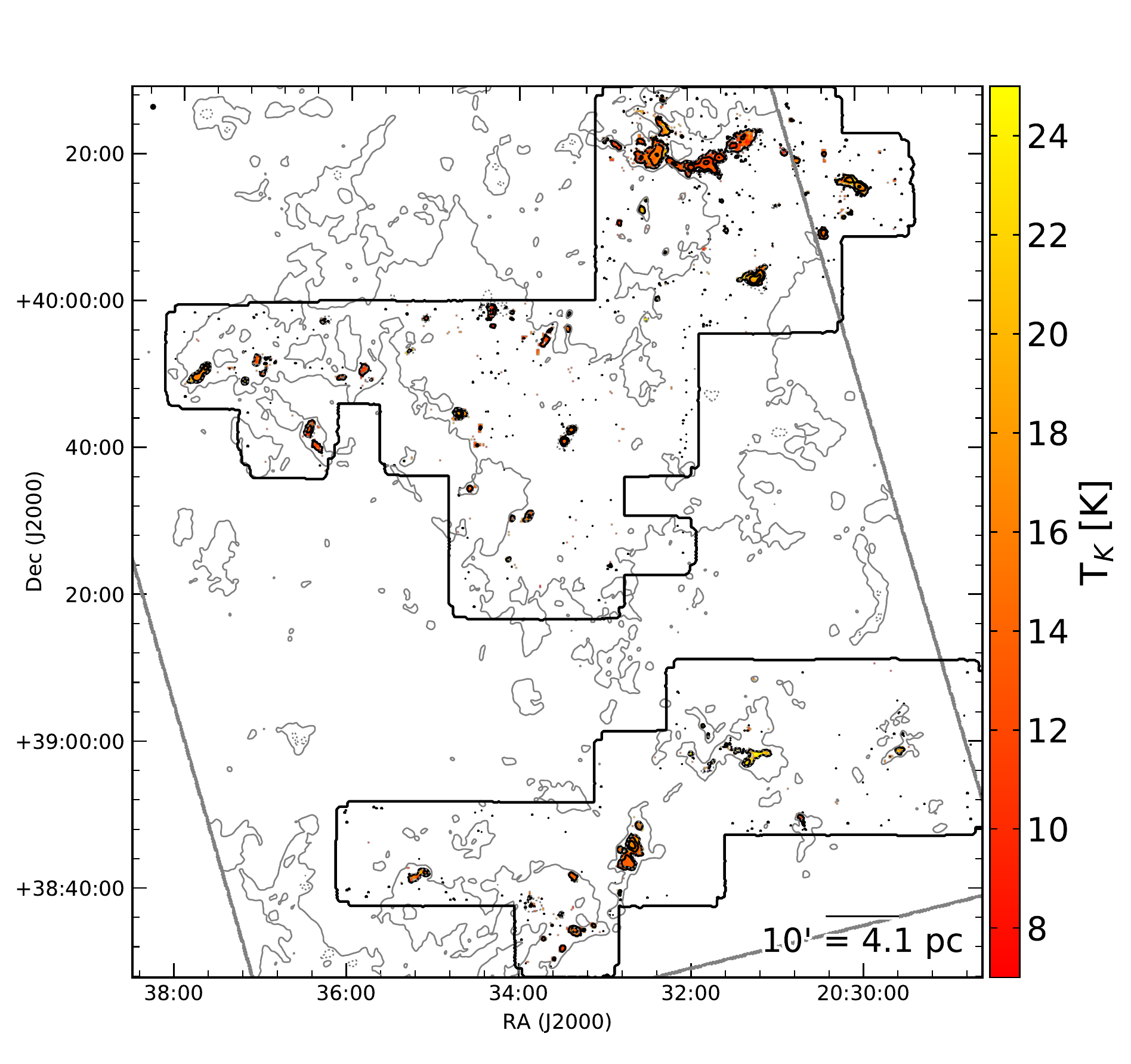}{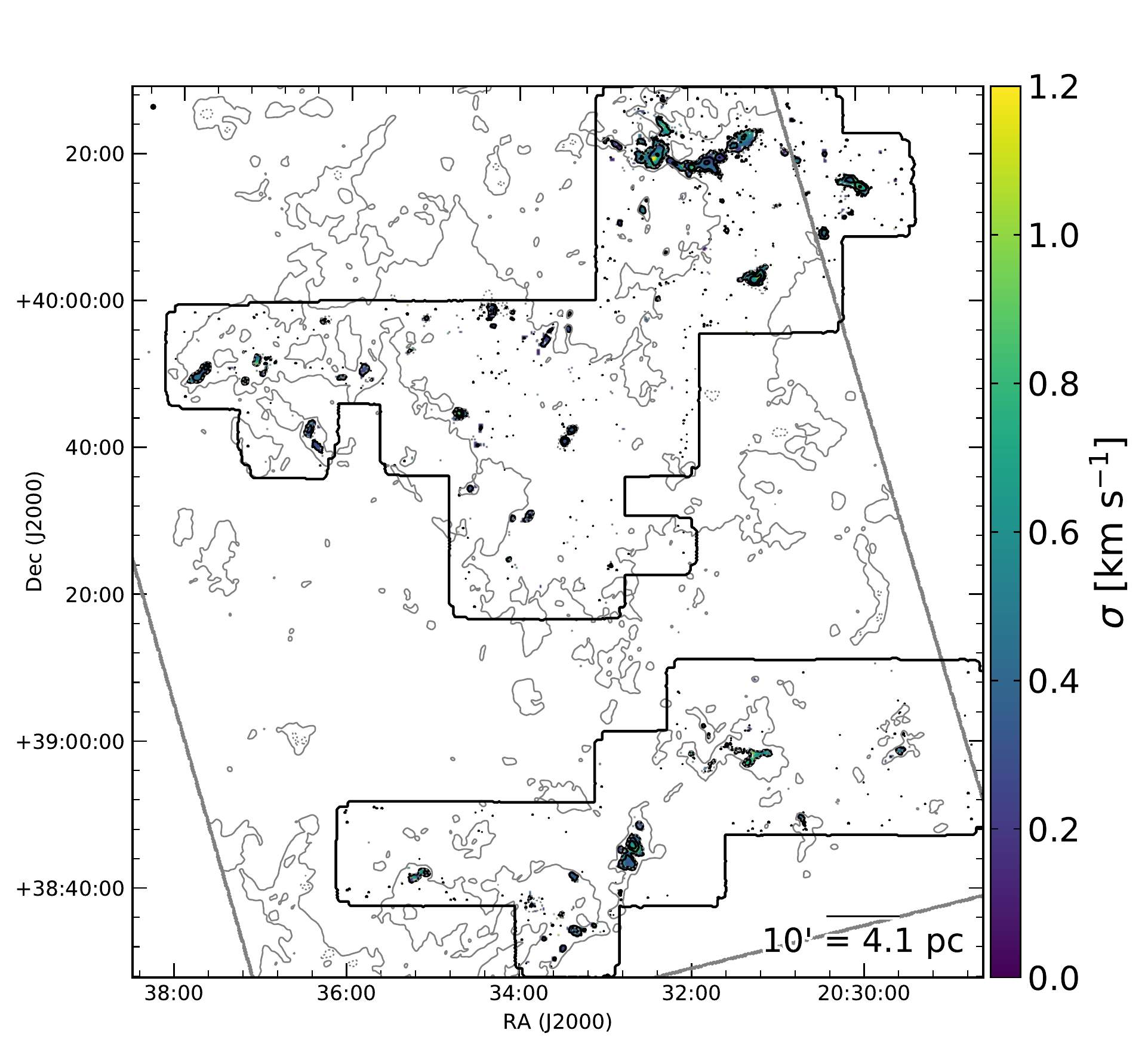}
\caption{Same as Figure \ref{W3_params} for Cygnus X South. The solid and dotted grey contours outline H$_2$ column densities of $4.7\times10^{21}$ cm$^{-2}$ and $9.4\times10^{21}$ cm$^{-2}$, respectively, which are equivalent to a total extinction in the V band of A$_{V}=5$ mag and 10 mag.}
\label{CygX_S_params}
\end{figure}

\begin{figure}[ht]
\epsscale{1.2}
\plottwo{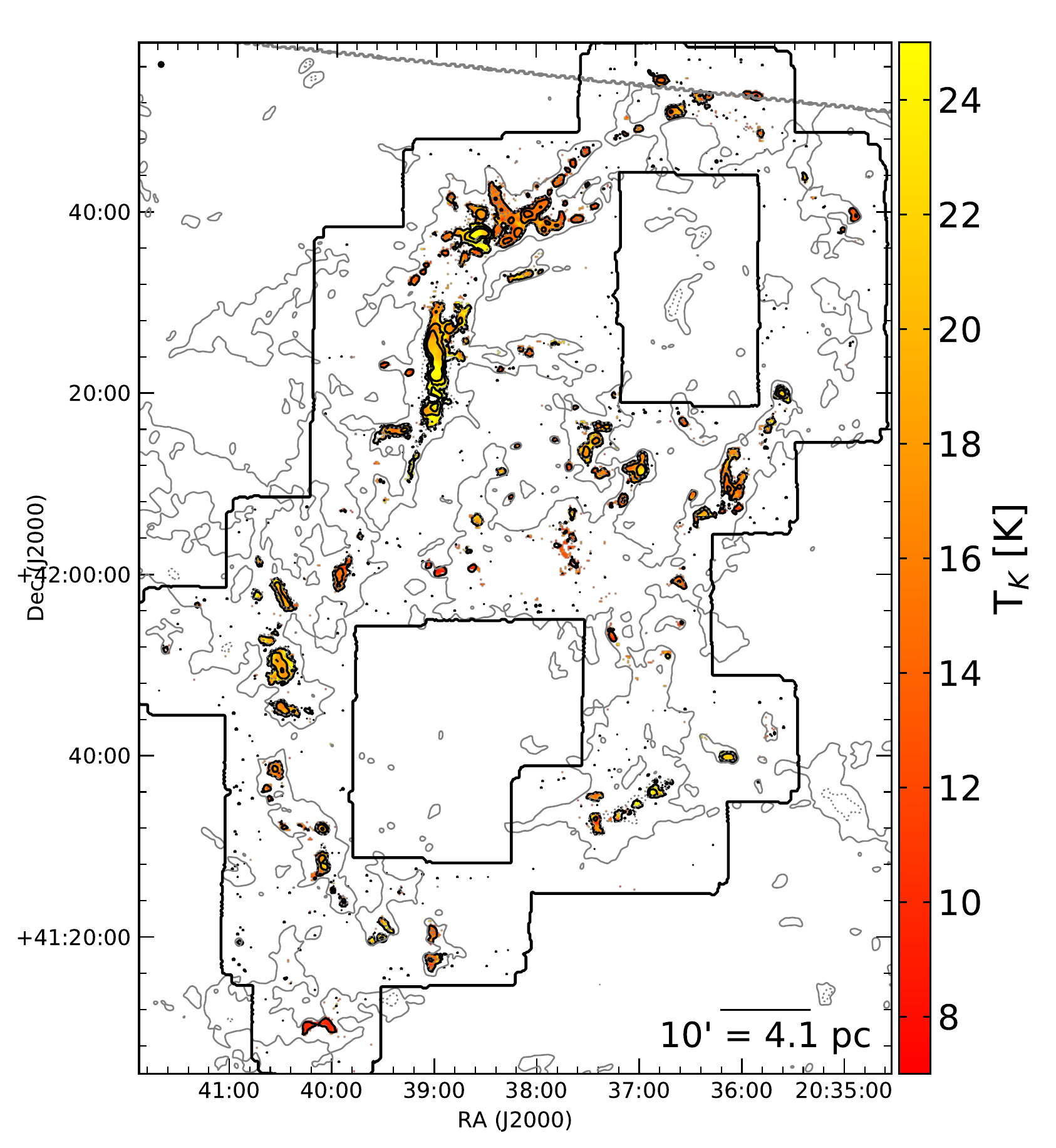}{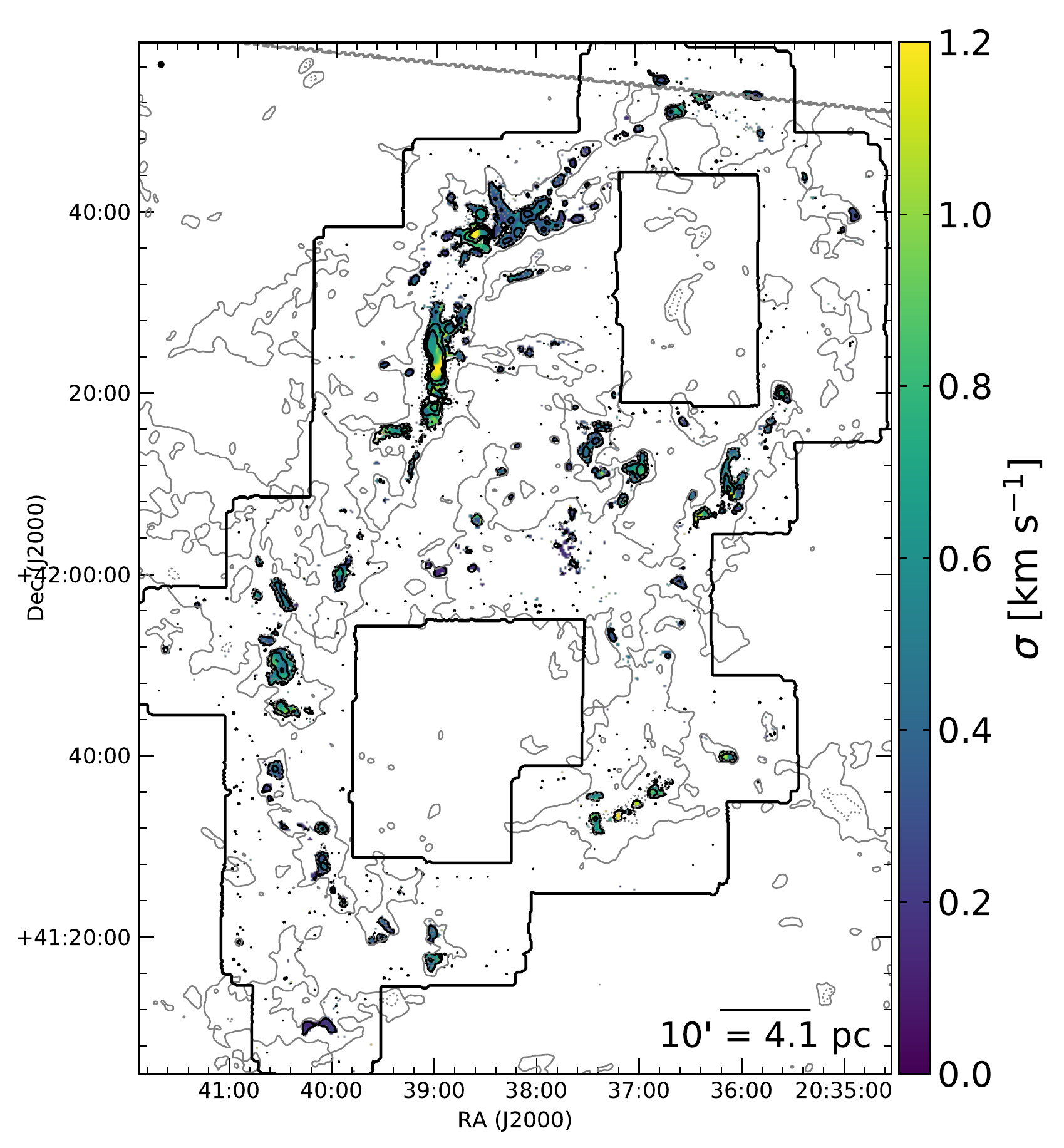}
\caption{Same as Figure \ref{CygX_S_params} for Cygnus X North.}
\label{CygX_N_params}
\end{figure}

\begin{figure}[ht]
\epsscale{0.8}
\plotone{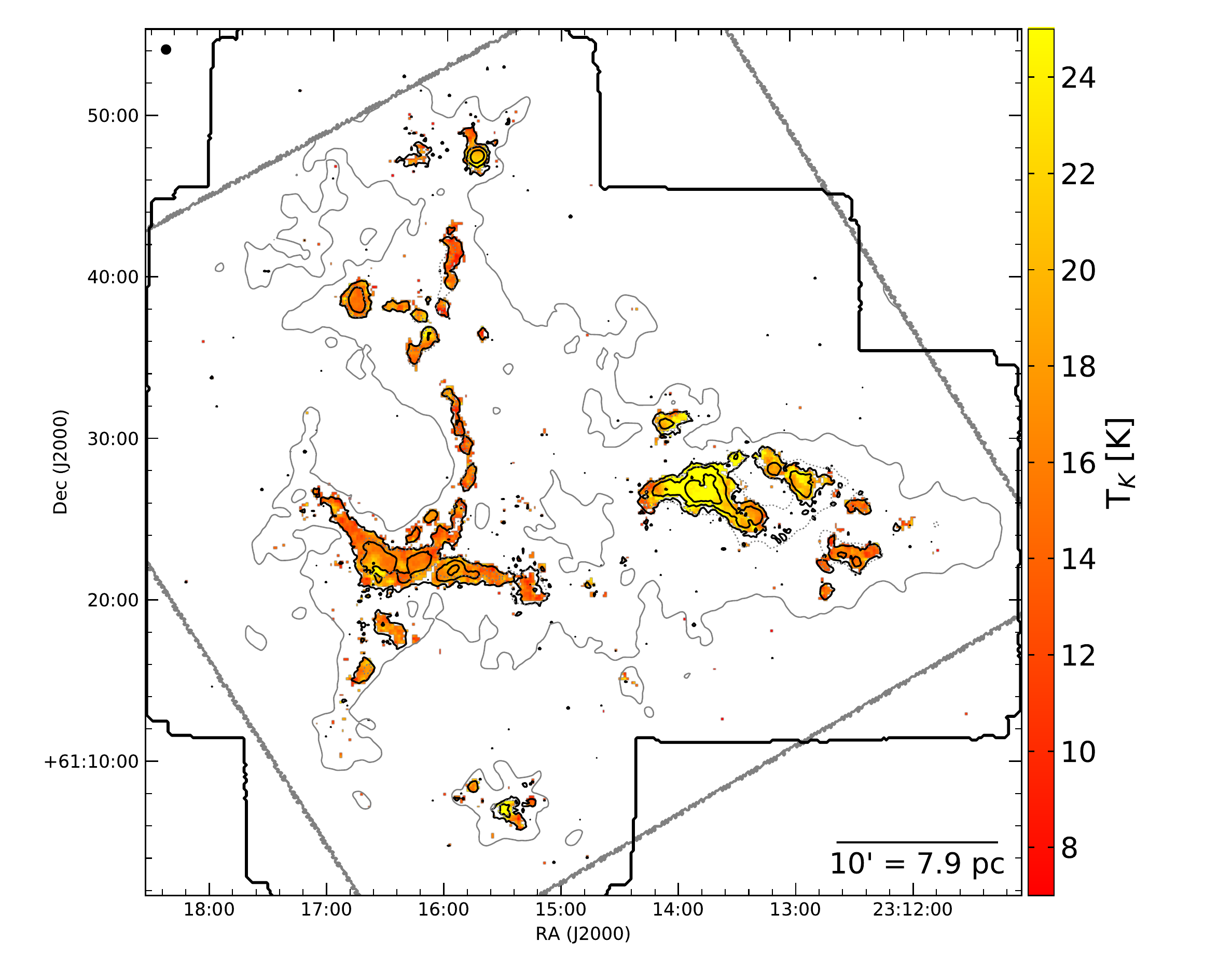}
\plotone{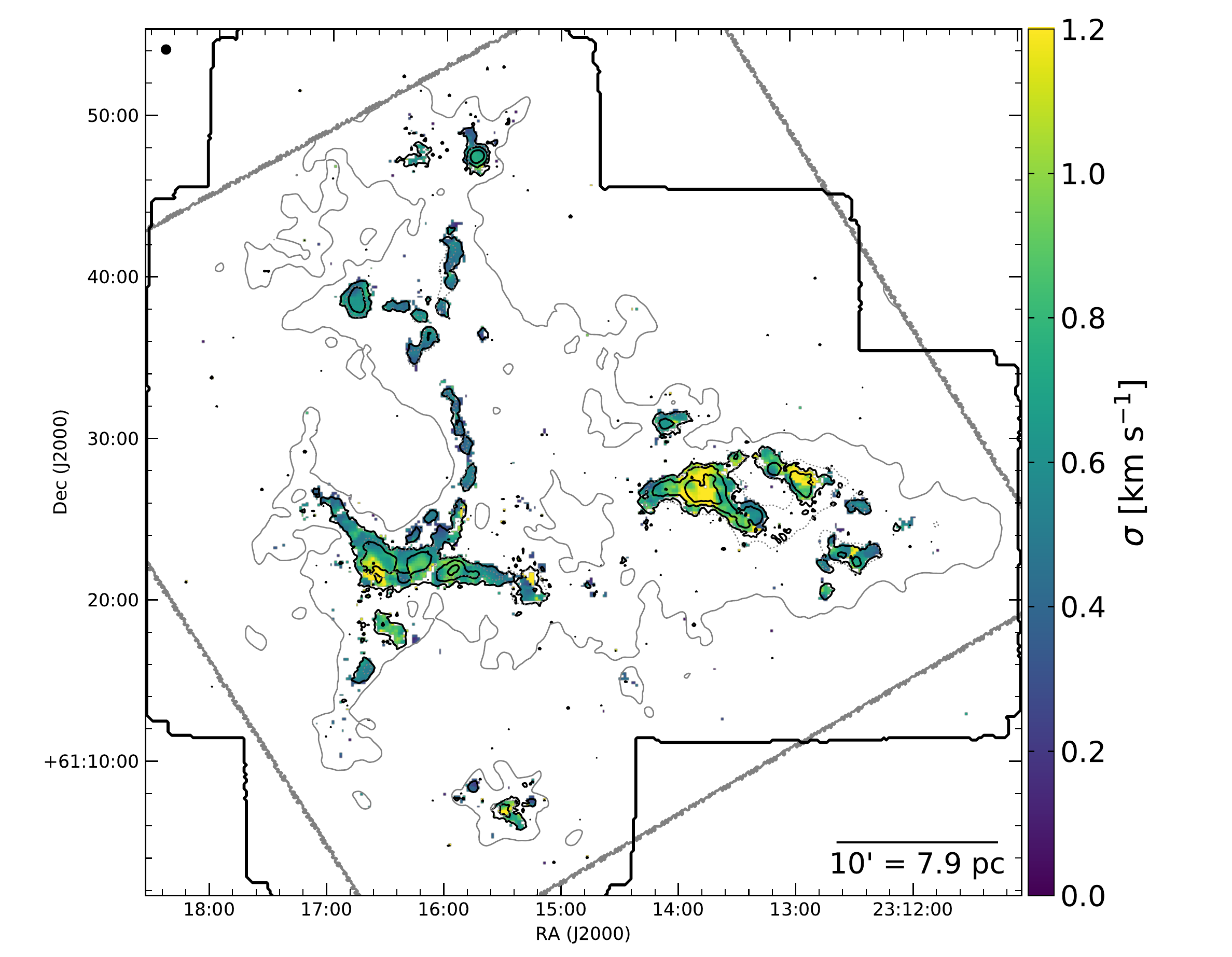}
\caption{Same as Figure \ref{W3_params} for NGC7538. }
\label{NGC7538_params}
\end{figure}

The final parameter maps for each region are shown in Figures \ref{W3_params}-\ref{NGC7538_params}.  To compare each region's ammonia emission to its dust continuum emission, we also plot \textit{Herschel} H$_2$ column density contours overtop the ammonia parameter maps presented in Figures \ref{W3_params}-\ref{NGC7538_params}.  The ammonia emission tends to occur where total extinction in the V band ($A_V$) is larger than $\sim6-8$ mag.

\begin{figure}[ht]
\epsscale{1.1}
\plottwo{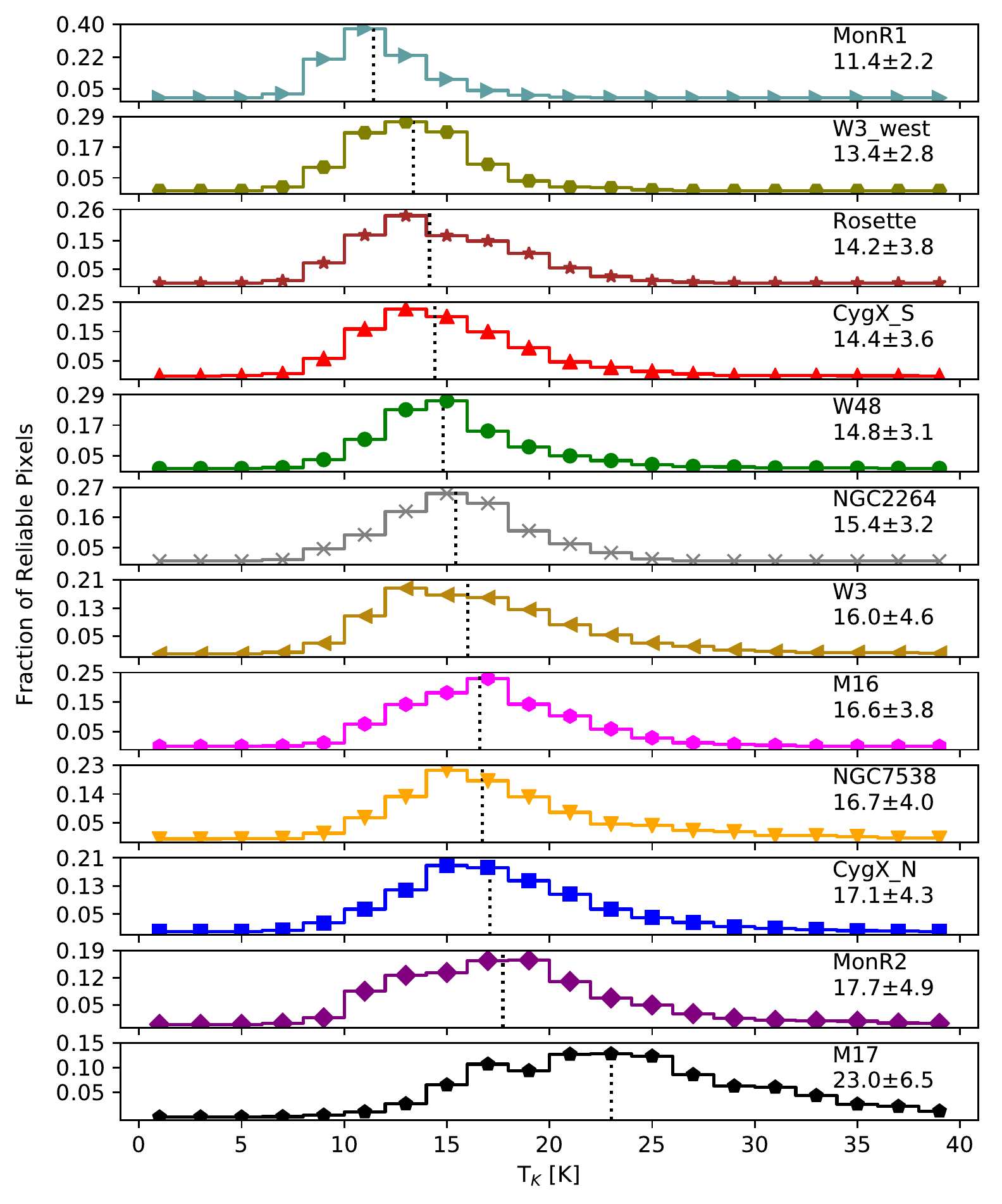}{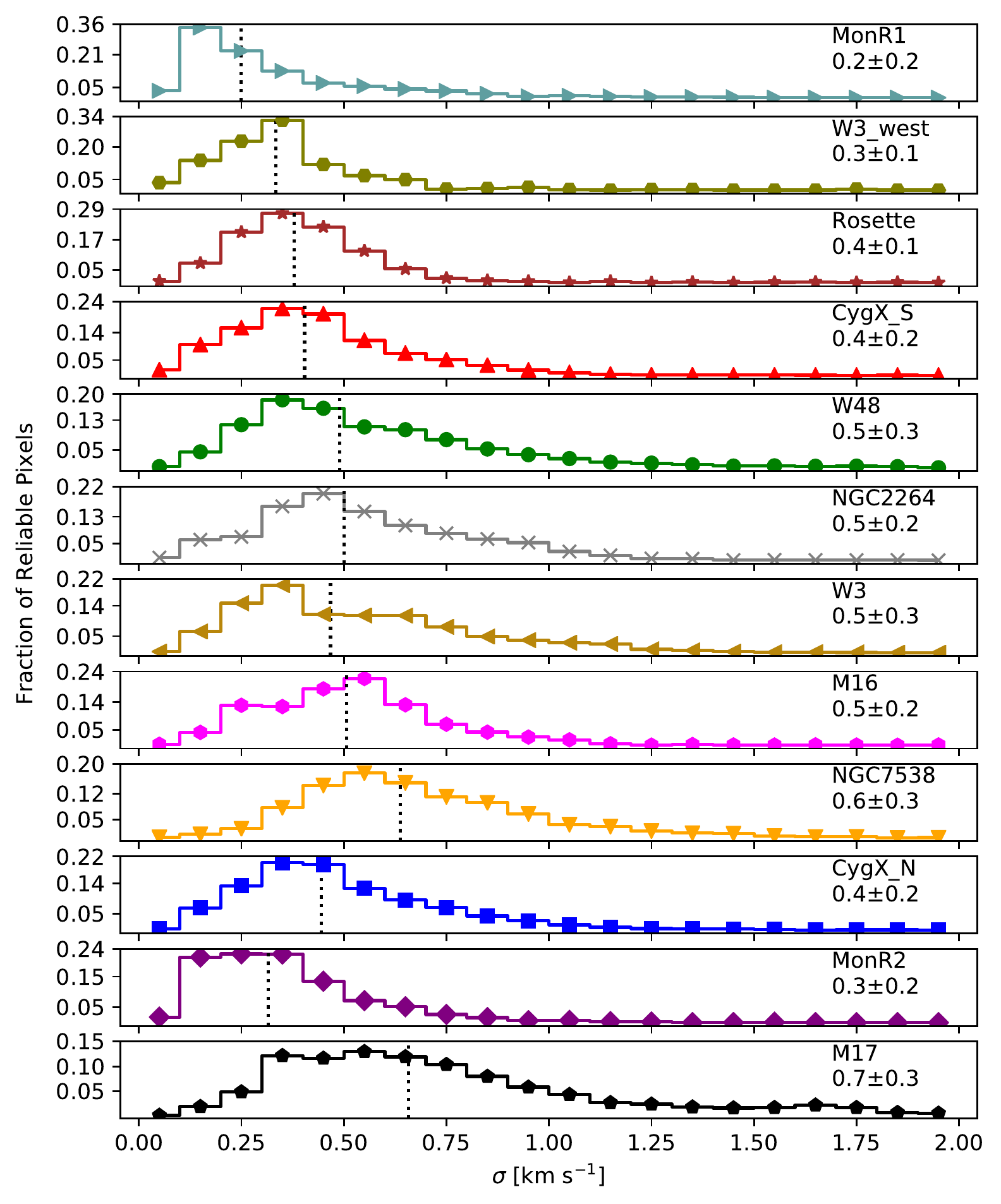}
\caption{Histograms of kinetic temperature (left) and NH$_3$ (1,1) velocity dispersion (right) for the reliably fit pixels in each region.  The median and median absolute deviation of each distribution is printed in the top right corner of each panel.  The clouds are ordered from top to bottom by increasing median kinetic temperature.}
\label{histograms}
\end{figure}

A comparison of the $T_K$ and $\sigma$ histograms for each region is presented in Figure \ref{histograms}.  Although the $T_K$ distributions are consistent for most of the regions, there are significant temperature differences between the regions with the lowest temperatures (MonR1 and W3-west) compared to the highest temperature regions (M17 and MonR2).  Similarly, the $\sigma$ distributions are fairly consistent across regions, with peak values of $0.3-0.7$ km s$^{-1}$.  There are several regions (NGC7538, W48, M17), however, that have a tail of pixels with large line widths $>$ 1 km s$^{-1}$.  These large line width tails are likely due to a higher fraction of pixels with strong outflows or multiple velocity components along the line of sight.  


\clearpage

\subsection{NH$_3$ (1,1) Integrated Intensity Maps}
The best-fit models from the NH$_3$ (1,1) line fitting described in Section 3.1 were used to identify the channels to integrate for producing NH$_3$ (1,1) integrated intensity maps.  Namely, the spectral channels in the best-fit models that were brighter than 0.0125 K were included in the integration.  This threshold was selected to include only the channels that are part of an emission line in the best-fit models.  Since the off-line channels in a \texttt{pyspeckit} model are slightly above zero due to machine precision, the threshold of 0.0125 K provides a conservative distinction between emission-line and off-line channels in the models.  For pixels that did not have any channels above that brightness criterion, we use the set of spectral channels centered on the mean cloud centroid velocity with a range defined by the mean cloud line width.  In addition, we blank all pixels within three pixels from the map edges since they have lower coverage by the KFPA and typically have higher noise.  Figures \ref{W3_leaves}-\ref{NGC7538_leaves} show the final NH$_3$ (1,1) integrated intensity maps for each region.

\subsection{Identifying NH$_3$ Structures with Dendrograms}
The hierarchical nature of molecular clouds \citep[e.g.,][]{Falgarone_1986, Lada_1992, Bonnell_2003} warrants a structure-identification method that handles features with different sizes, shapes, and spatial scales.  Dendrograms are a proven identification method that excel at identifying such hierarchical features in both continuum \citep[e.g.,][]{Kirk_2013_CSAR, Konyves_2015} and molecular line emission observations \citep[e.g.,][]{Rosolowsky_2008, Goodman_2009, Lee_2014, Seo_2015, Friesen_2017, Keown_2017} and simulations \citep{Boyden_2016, Koch_2017, Boyden_2018}.  This ability arises from the tree-diagram architecture of dendrogram algorithms, which first identifies the pixels in a map that represent local maxima.  Next, structures are assembled around the local maxima by joining nearby fainter pixels.  These top-level \textit{leaves} are grown until they either merge with another nearby leaf, at which point they are connected by a \textit{branch}, or reach a pre-defined noise threshold below which no more pixels are added to the structure.  The lowest-level structures above this noise threshold that are connected to branches are known as \textit{trunks}.  

Due to the hyperfine structure of NH$_3$ (1,1) emission, each hyperfine group would be detected as a distinct structure in a 3D dendrogram extraction of the KEYSTONE spectral cubes.  To ``remove'' the hyperfine structures of NH$_3$ (1,1), some authors have created a single-Gaussian cube from the line width, peak brightness temperature, and centroid velocity measured from the NH$_3$ (1,1) emission \citep[e.g.,][]{Friesen_2017, Keown_2017}.  Such a single-Gaussian cube attempts to represent how the ammonia emission would appear without hyperfine splitting.  Unless the ammonia emission is fit using a model with multiple velocity components along the line of sight, however, the output single-Gaussian cube does not account for emission with multiple velocity components.  Instead, a robust multiple velocity component line-fitting method would first need to be applied to the data to take full advantage of a 3D dendrogram extraction of NH$_3$ (1,1) cubes.  The multiple velocity component models would then allow for the creation of a ``multi-Gaussian'' cube that removes the hyperfine structures of NH$_3$ (1,1) while preserving the presence of multiple velocity components along the line of sight.  Although a multiple velocity component NH$_3$ (1,1) line-fitting method has been developed in another KEYSTONE paper (Keown et al. 2019, submitted), the analysis presented here neglects multiple velocity components along the line of sight.

Here, we instead perform a dendrogram analysis of the NH$_3$ (1,1) integrated intensity maps described in Section 3.2.  When the observed emission has only a single velocity component along the line of sight, a 2D dendrogram extraction of the integrated intensity maps will produce similar results as a 3D extraction of the full emission cube.  Since the majority of the KEYSTONE observations appear to lack multiple velocity components, a 2D analysis is warranted.  We defer a 3D dendrogram analysis of the ammonia data to a future KEYSTONE paper. 

The \texttt{astrodendro} Python package was applied to the integrated intensity map for each region.  For consistency with the ammonia dendrogram analyses by \cite{Friesen_2016} and \cite{Keown_2017}, we chose the following values for the dendrogram algorithm input parameters: 

\begin{itemize}
\item \texttt{min\char`_value} = 5 $\times$ RMS, where RMS is the rms noise measured in a region of the integrated intensity map where no emission was detected.  For clouds with highly variable noise in the integrated intensity map, the RMS was calculated using an emission-free region representative of the highest noise portion of the map.  While this conservative approach may leave some low brightness sources undetected, it reduces the amount of spurious noise sources detected by the dendrogram. \texttt{min\char`_value} is the lowest intensity a pixel can have to be joined to a neighboring structure.
\item \texttt{min\char`_delta} = 2 $\times$ RMS, where RMS is the same as described for \texttt{min\char`_value}. \texttt{min\char`_delta} is the minimum difference in brightness between two structures before they are merged into a single structure.
\item \texttt{min\char`_npix} = 10 pixels.  \texttt{min\char`_npix} is the minimum number of pixels a structure must contain to remain independent.  This parameter prevents noise spikes from being identified as sources.
\end{itemize}

After running the dendrogram algorithm on the maps, we cull \textit{leaf} sources from our final catalog that do not meet the following criteria: 

\begin{itemize}
\item The total area of the leaf, in terms of all the pixels associated with it, must be larger than the total area of the GBT beam.  This criterion ensures further that small noise spikes are excluded from our final catalog and analyses.   
\item The leaf contains at least one pixel that was reliably fit by the NH$_3$ line fitting method described in Section 3.1.  Here, a reliably fit pixel is one that passes the seven constraints listed in Section 3.1.
\end{itemize}

\begin{figure}[ht]
\plotone{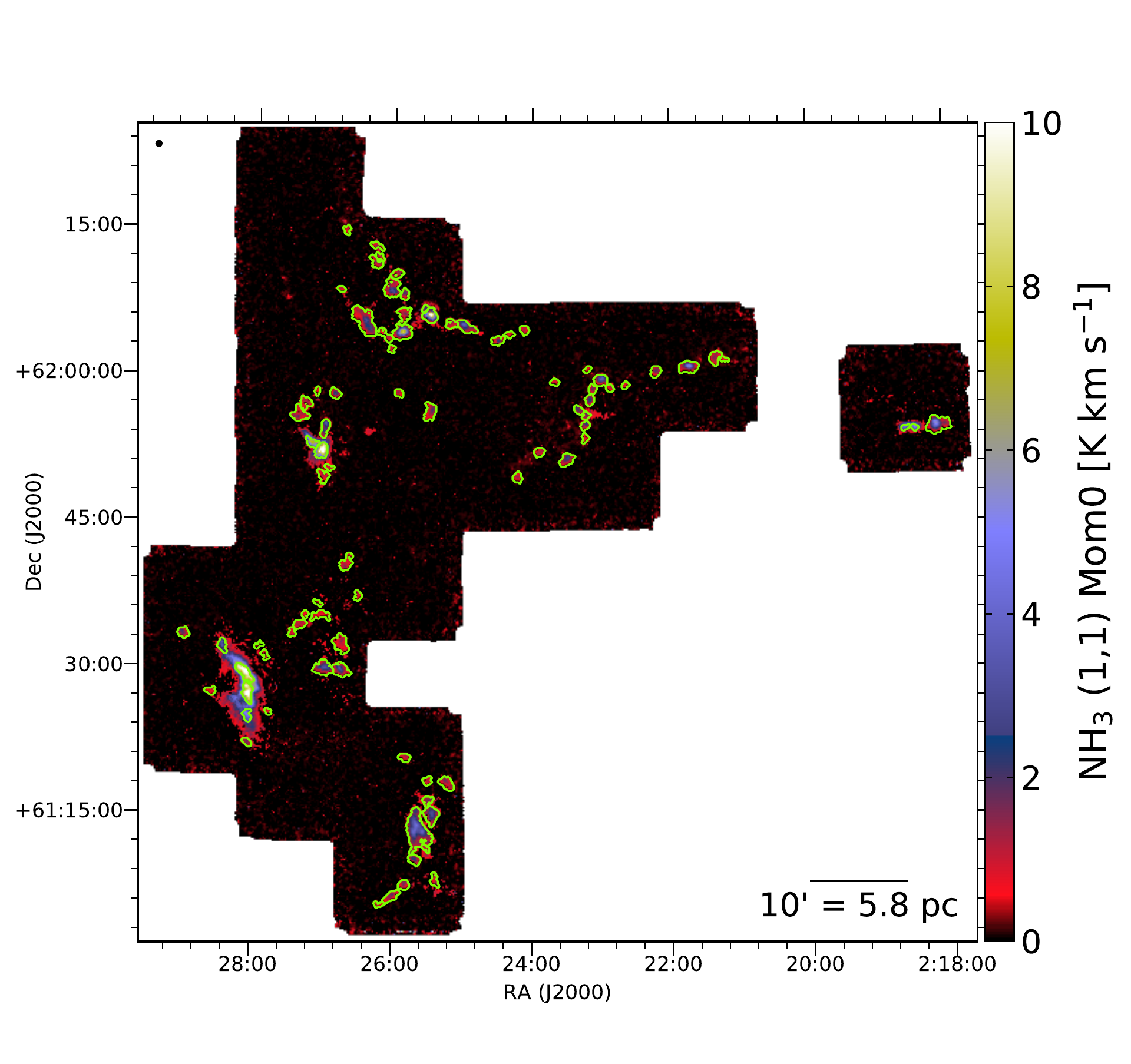}
\caption{NH$_3$ (1,1) integrated intensity map for W3.  Green contours outline leaves identified by a dendrogram analysis of the map that passed the culling criteria listed in Section 3.3. }
\label{W3_leaves}
\end{figure}

\begin{figure}[ht]
\plotone{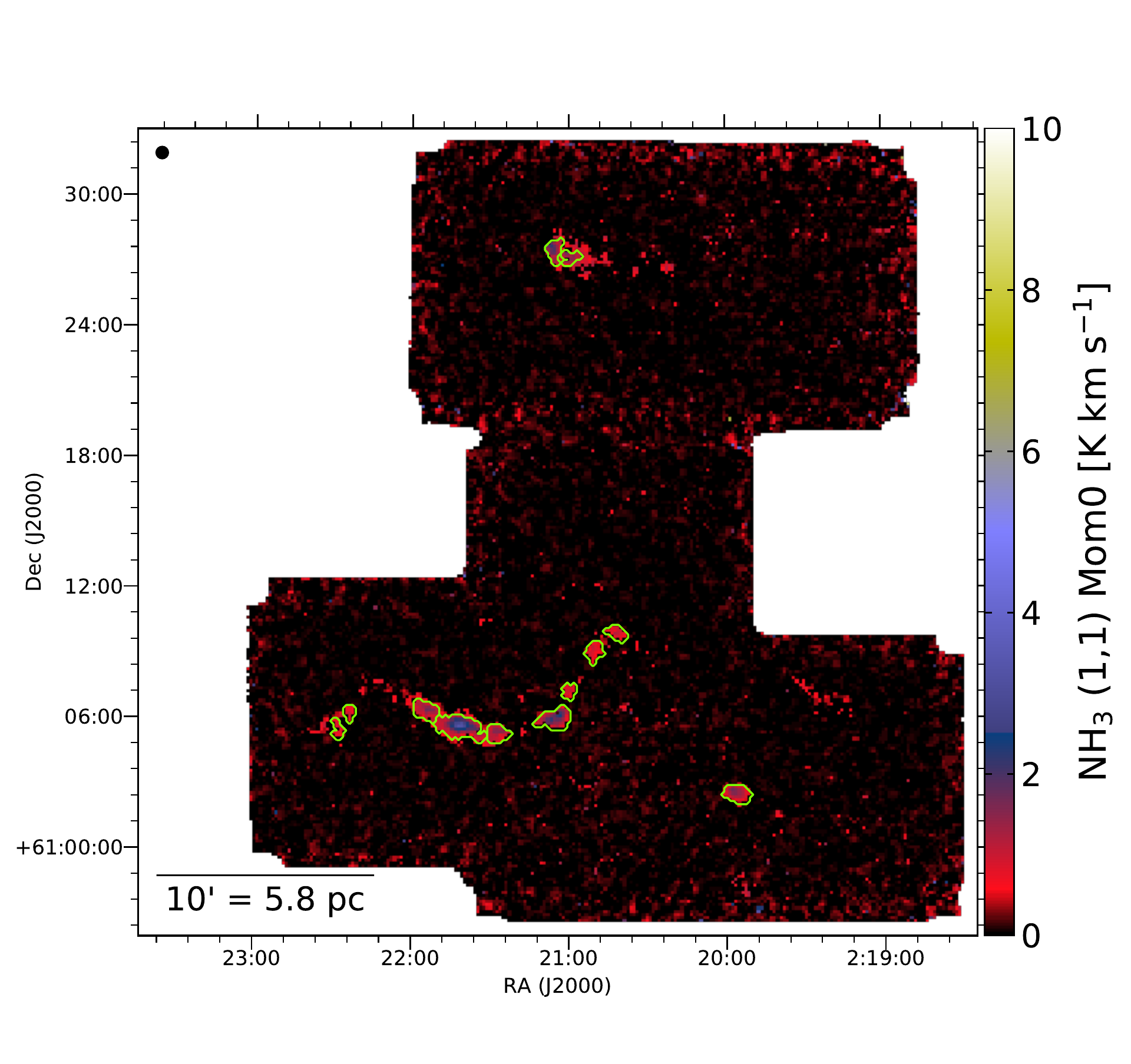}
\caption{Same as Figure \ref{W3_leaves} for W3-west.}
\label{W3_west_leaves}
\end{figure}

\begin{figure}[ht]
\plotone{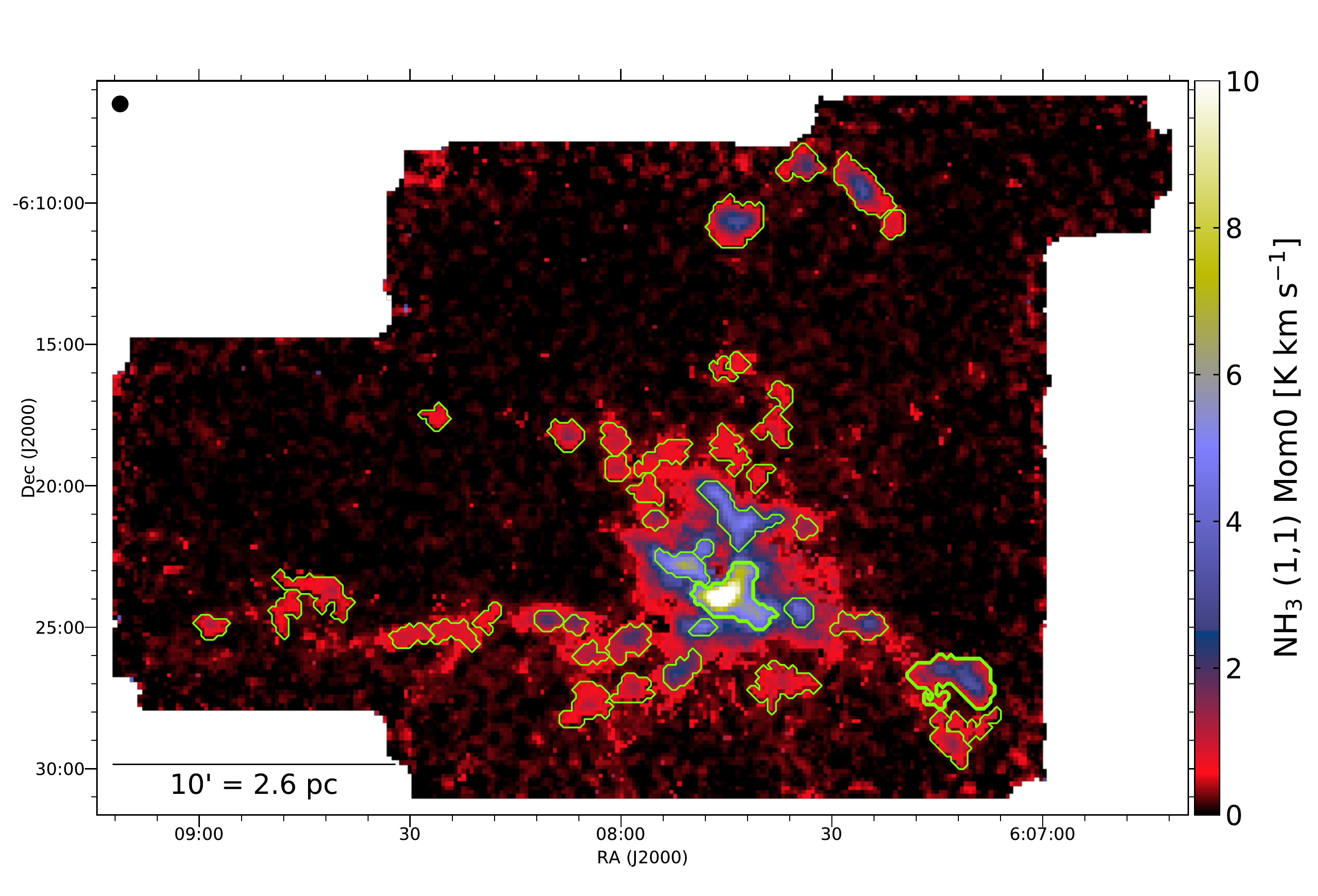}
\caption{Same as Figure \ref{W3_leaves} for MonR2.}
\label{MonR2_leaves}
\end{figure}

\begin{figure}[ht]
\plotone{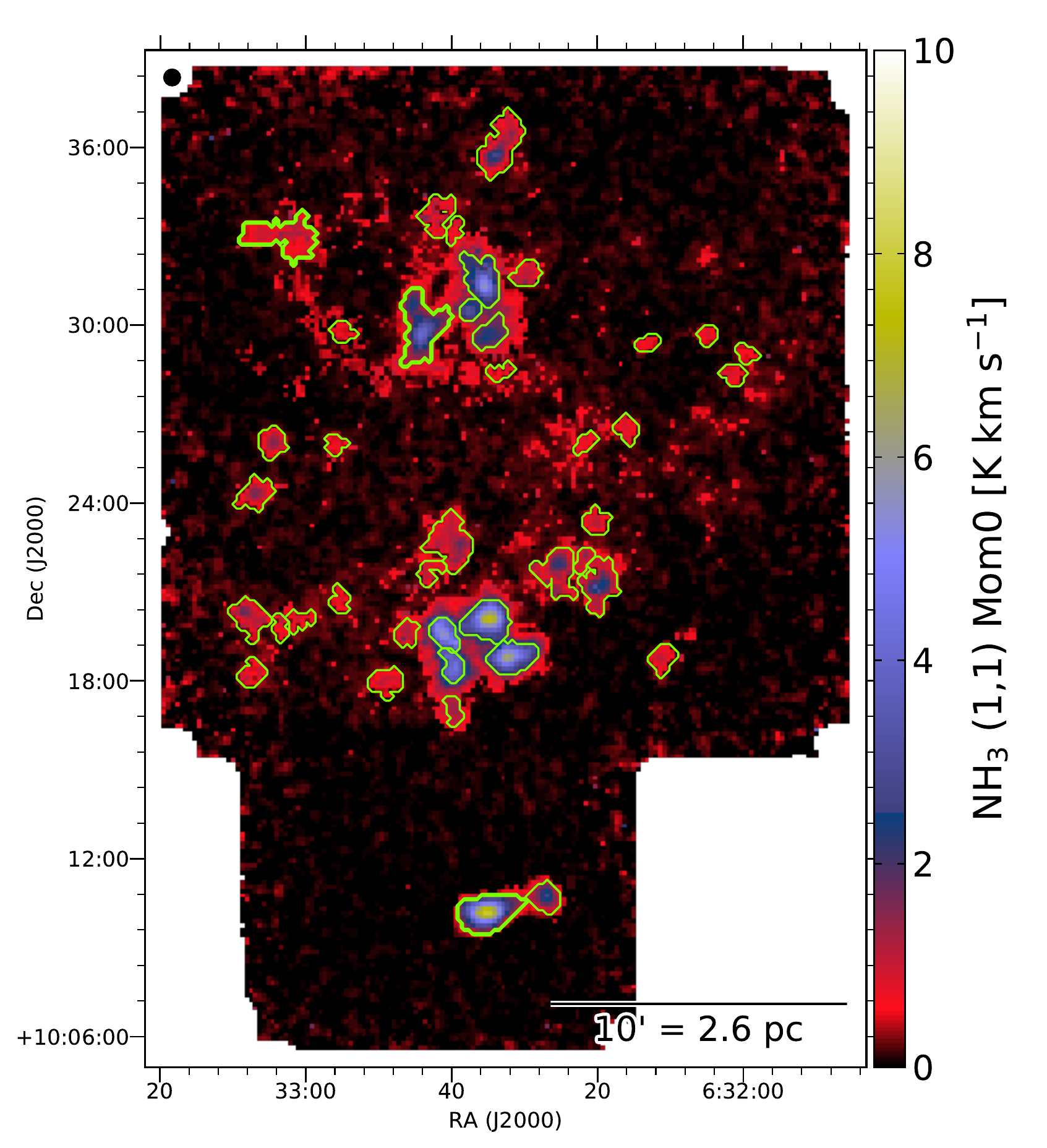}
\caption{Same as Figure \ref{W3_leaves} for MonR1.}
\label{MonR1_leaves}
\end{figure}

\begin{figure}[ht]
\plotone{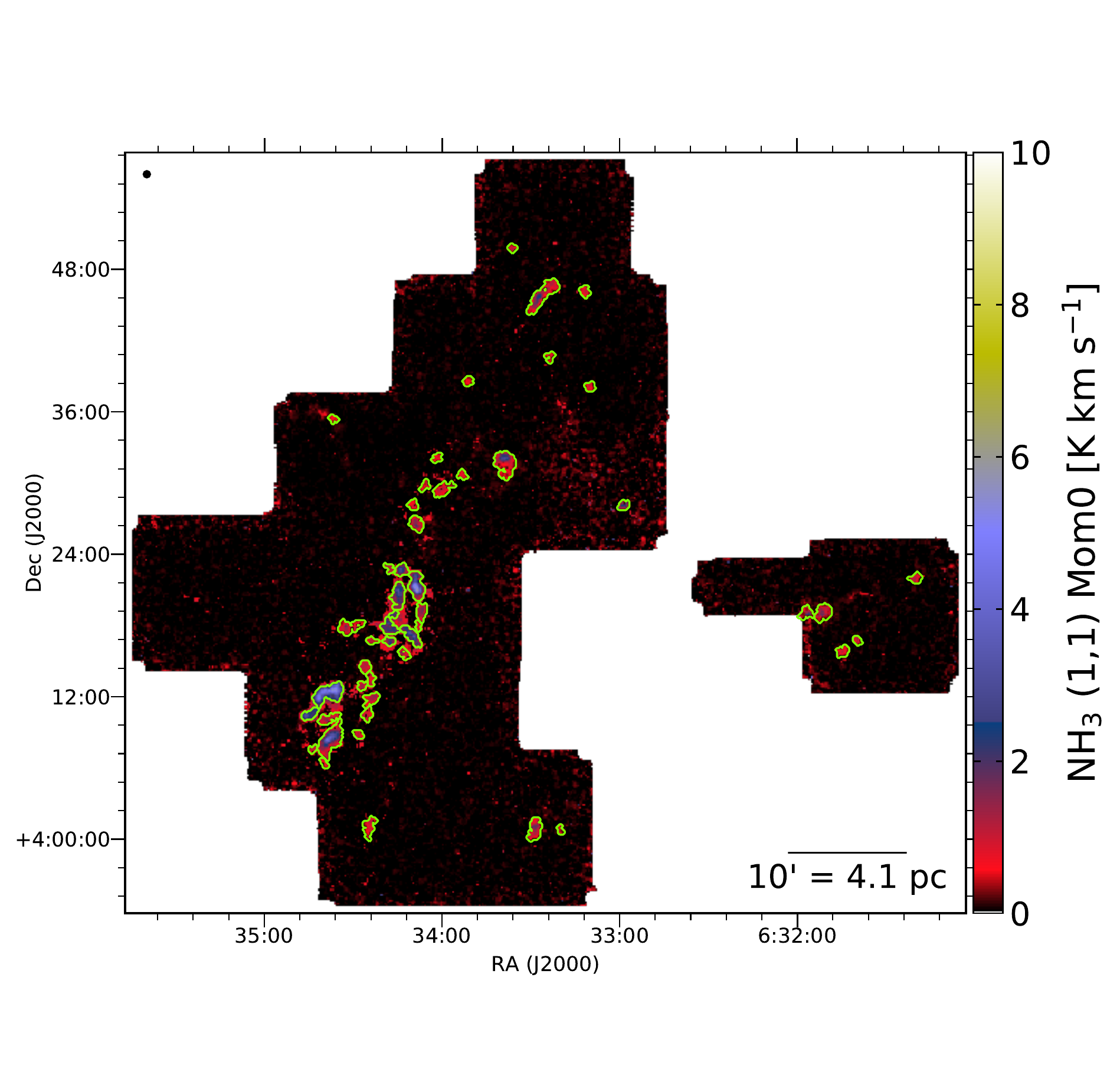}
\caption{Same as Figure \ref{W3_leaves} for Rosette.}
\label{Rosette_leaves}
\end{figure}

\begin{figure}[ht]
\plotone{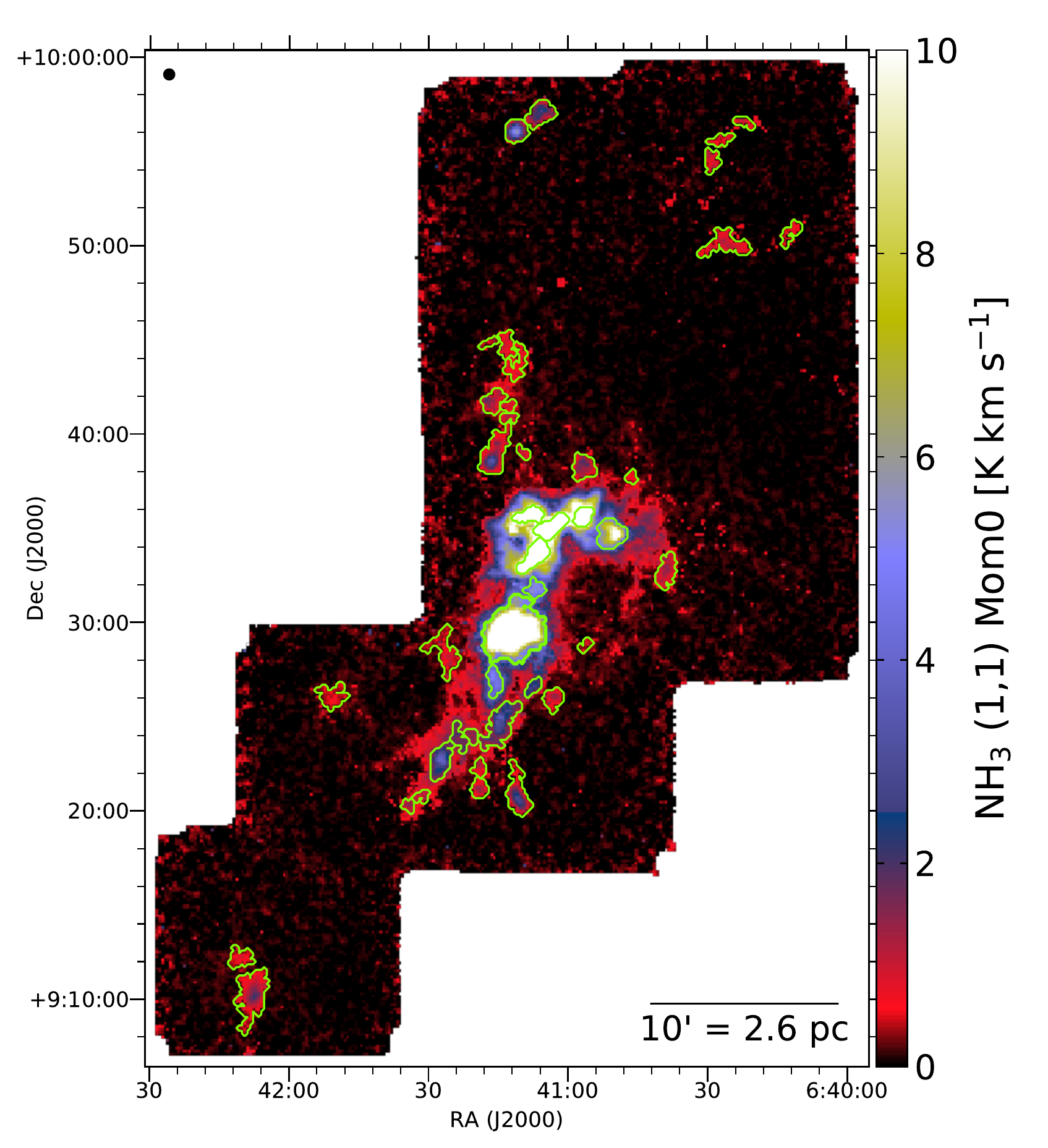}
\caption{Same as Figure \ref{W3_leaves} for NGC2264.}
\label{NGC2264_leaves}
\end{figure}

\begin{figure}[ht]
\plotone{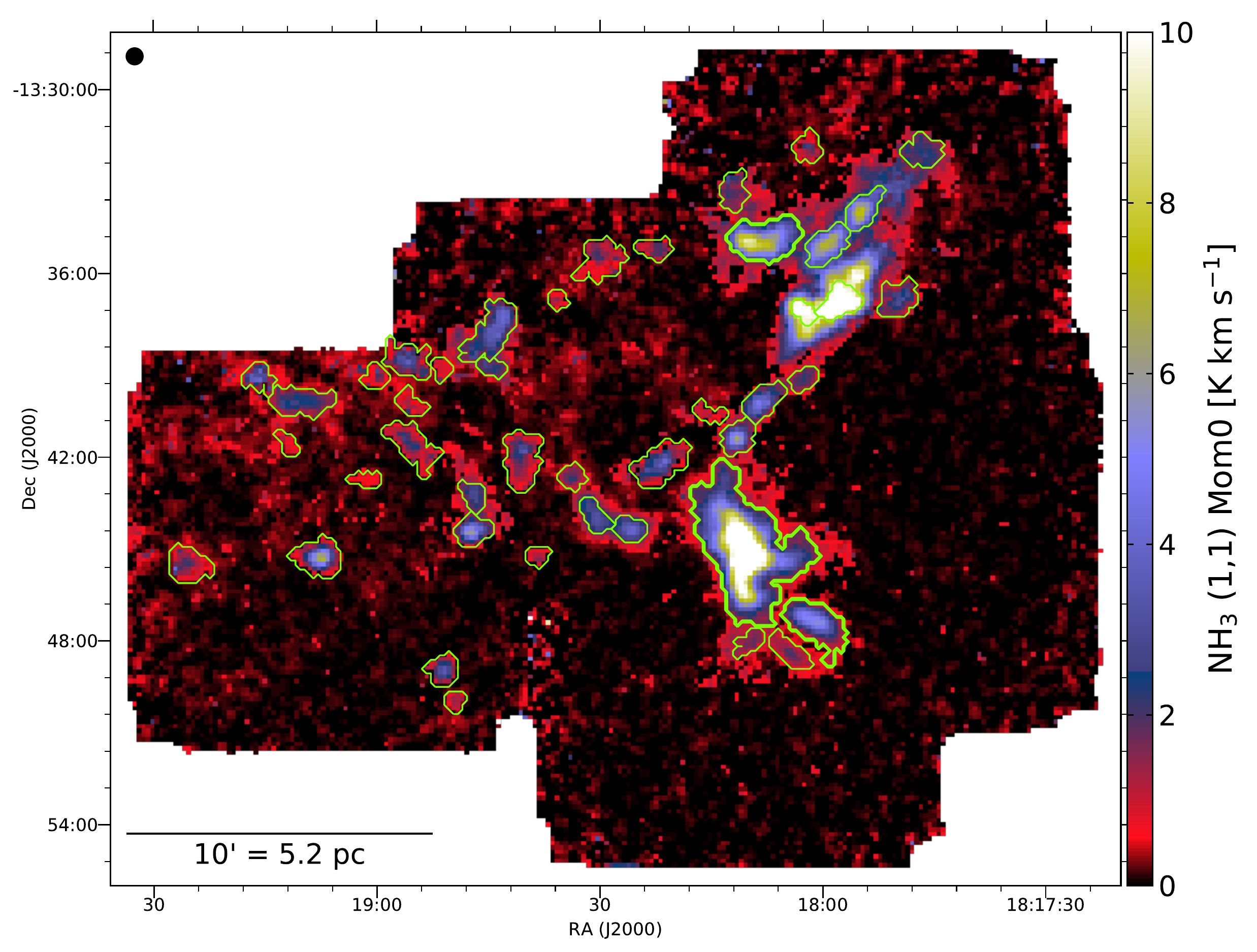}
\caption{Same as Figure \ref{W3_leaves} for M16.}
\label{M16_leaves}
\end{figure}

\begin{figure}[ht]
\plotone{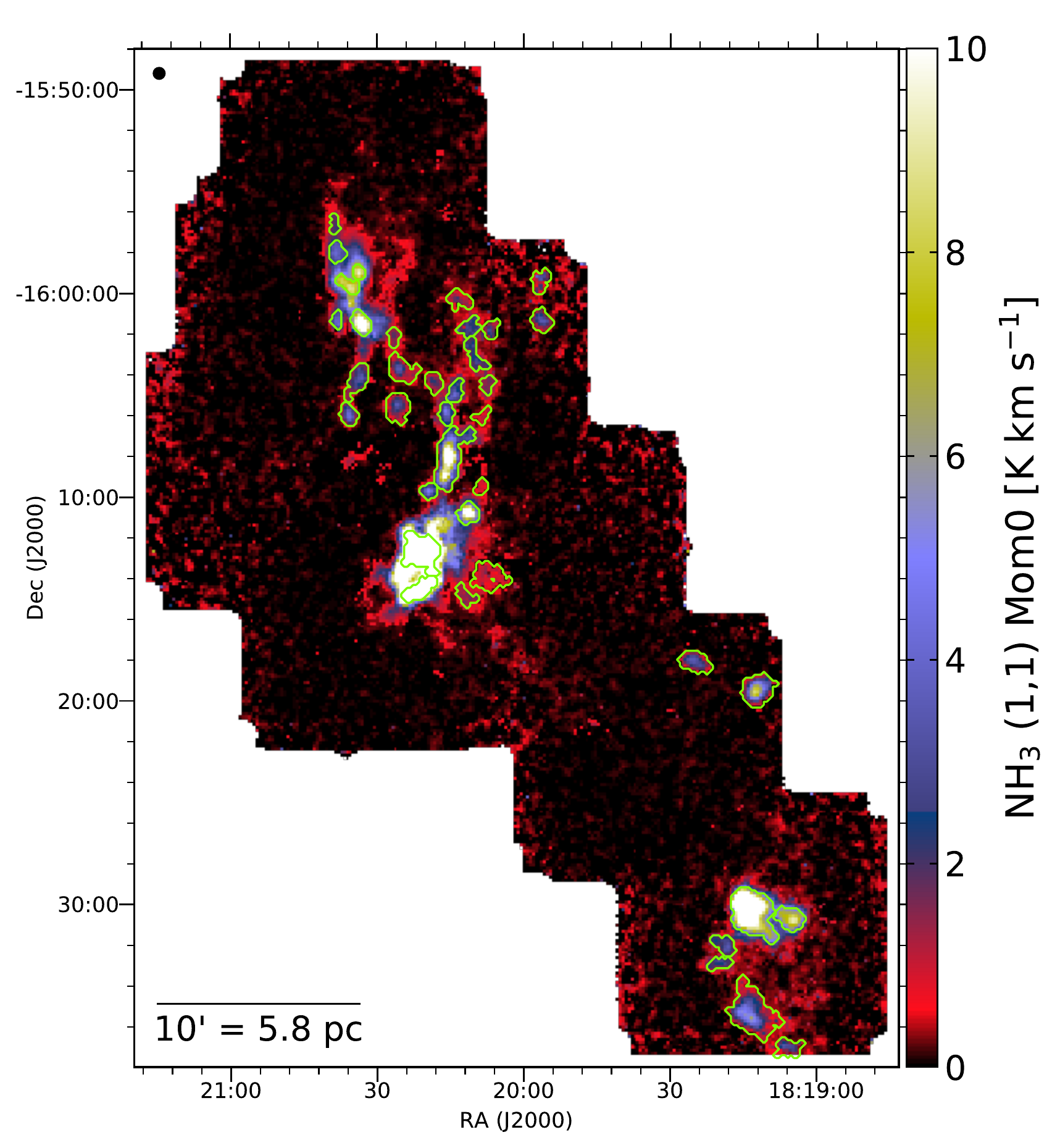}
\caption{Same as Figure \ref{W3_leaves} for M17.}
\label{M17_leaves}
\end{figure}

\begin{figure}[ht]
\epsscale{1.0}
\plotone{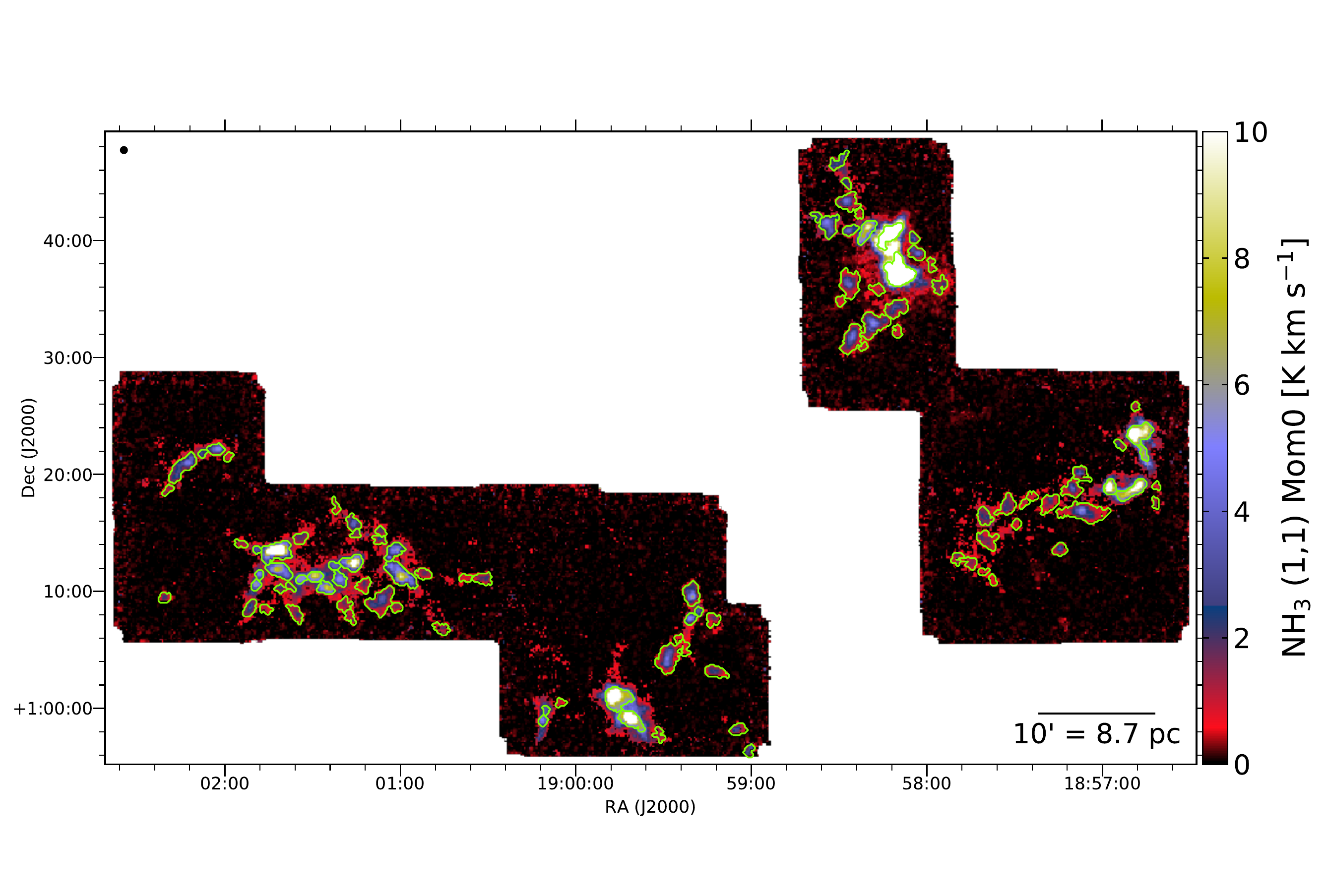}
\caption{Same as Figure \ref{W3_leaves} for W48.}
\label{W48_leaves}
\end{figure}

\begin{figure}[ht]
\epsscale{1.0}
\plotone{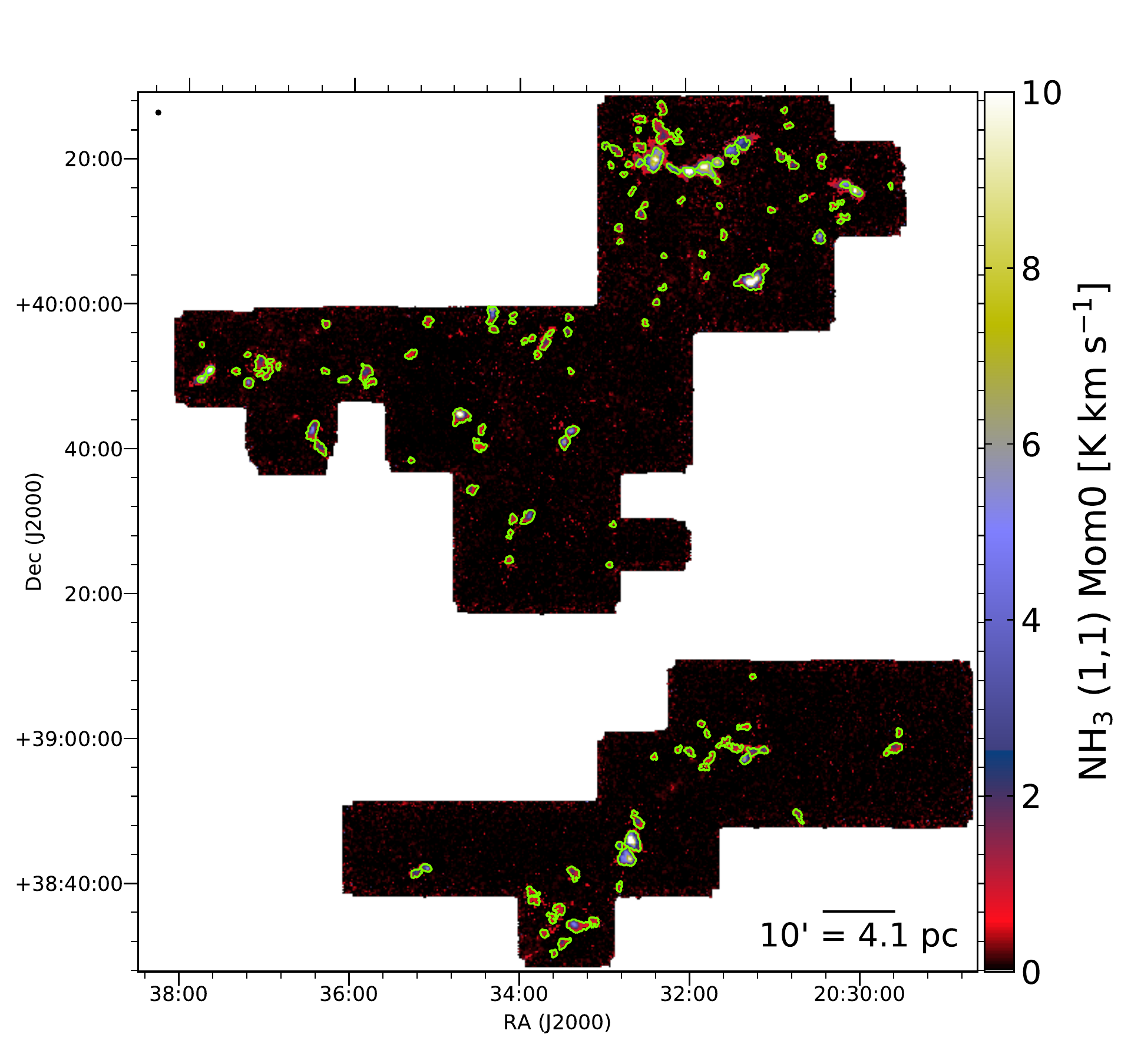}
\caption{Same as Figure \ref{W3_leaves} for Cygnus X South.}
\label{CygX_S_leaves}
\end{figure}

\begin{figure}[ht]
\epsscale{1.0}
\plotone{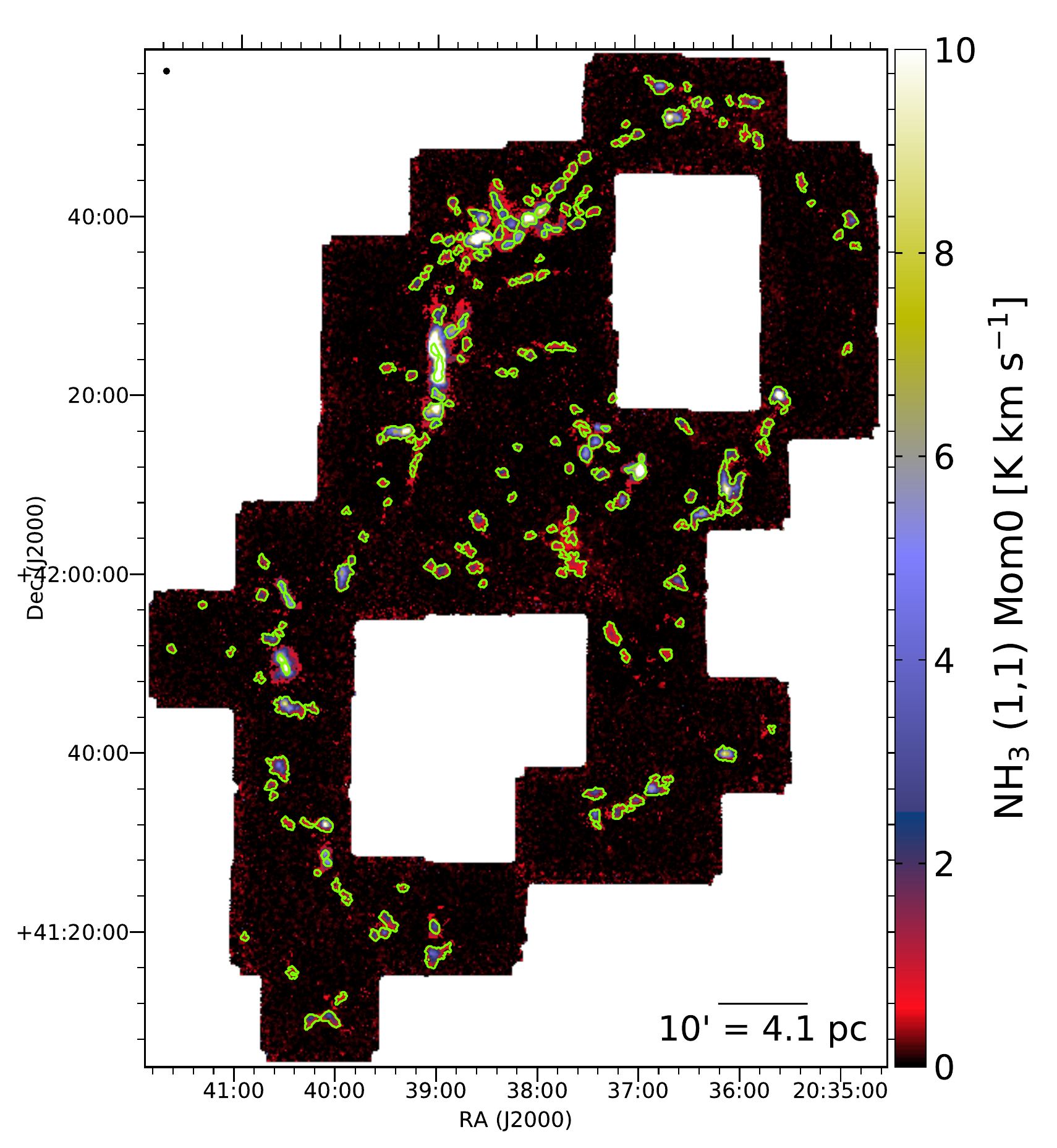}
\caption{Same as Figure \ref{W3_leaves} for Cygnus X North.}
\label{CygX_N_leaves}
\end{figure}

\begin{figure}[ht]
\epsscale{1.1}
\plotone{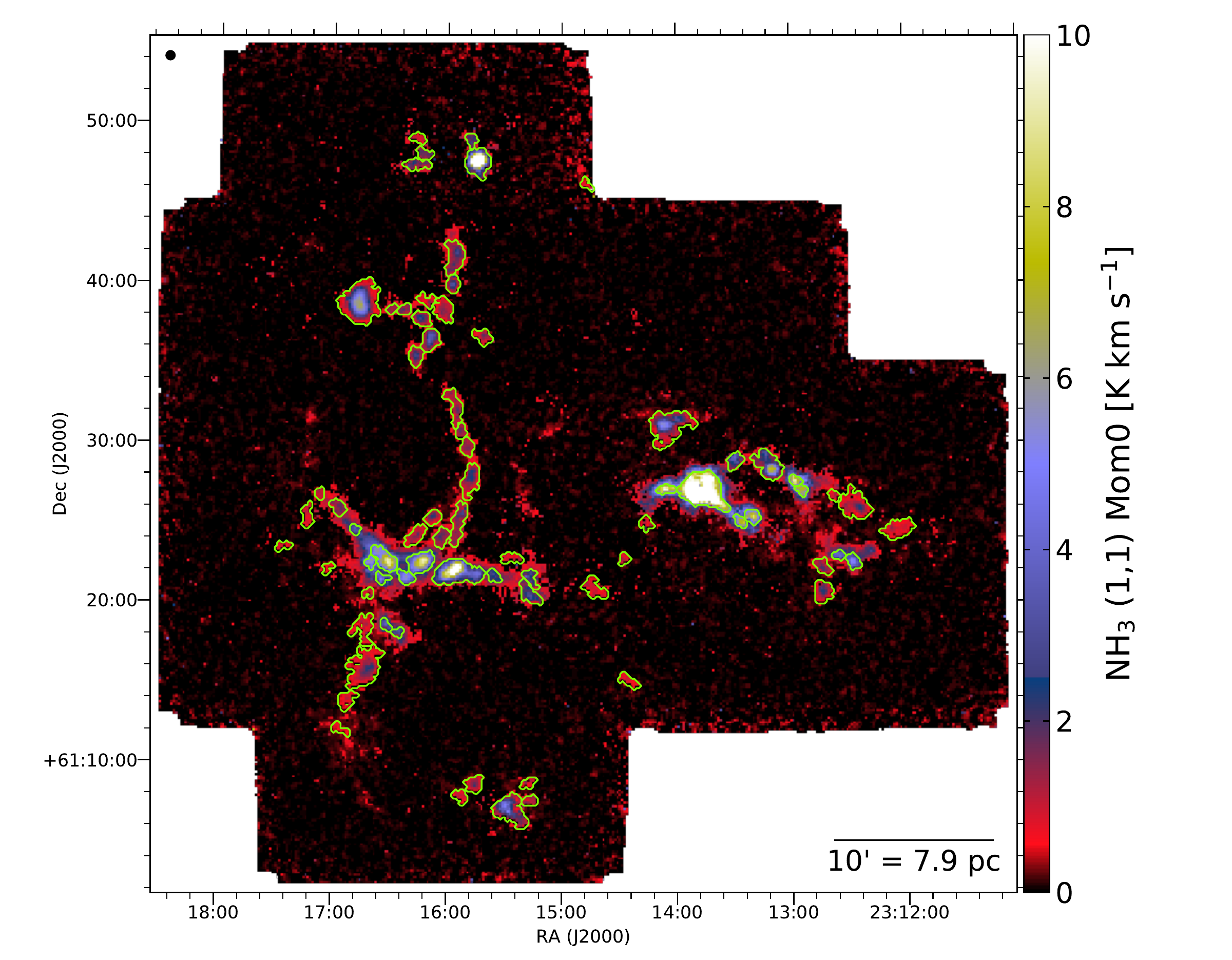}
\caption{Same as Figure \ref{W3_leaves} for NGC7538.}
\label{NGC7538_leaves}
\end{figure}

\begin{figure}[ht]
\epsscale{1.0}
\plotone{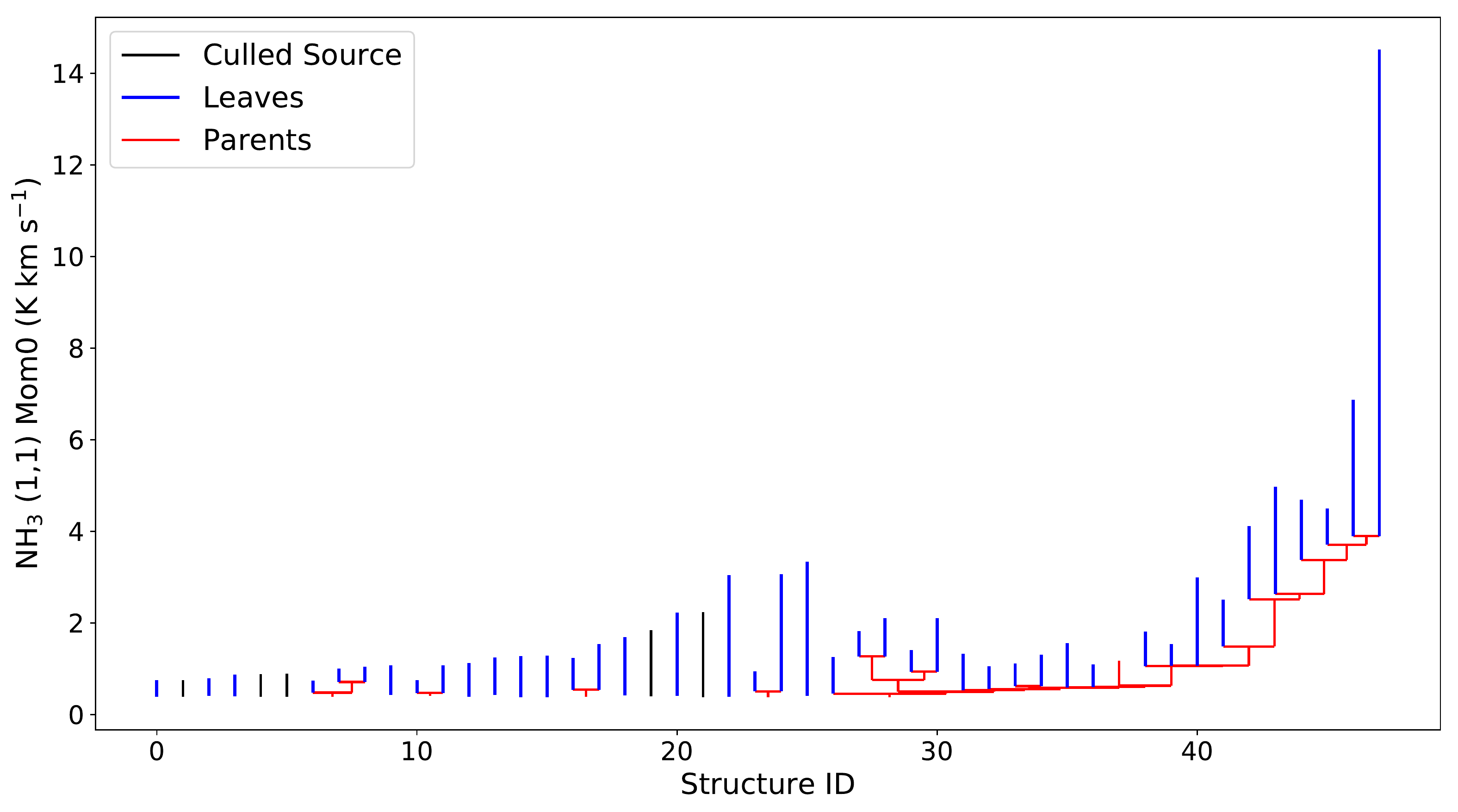}
\caption{Dendrogram tree diagram for MonR2 showing the peak intensity for each structure identified.  Leaves are shown in blue and branches are shown in red. Leaves that were culled based on the selection criteria described in Section 3.3 are shown in black.  Leaves that are red denote culled leaves that are part of a branch with an accepted leaf.}
\label{CygX_S_tree}
\end{figure}

Figure \ref{CygX_S_tree} shows an example tree diagram for the dendrogram extraction of the MonR2 region.  Leaves that do not pass our selection criteria are shown as black vertical lines, while robust leaves are shown in blue and parent structures are shown in red.  The tree diagram shows that our selection criteria preferentially cull isolated leaves that are not associated with larger-scale parent structures and are likely noise spikes in the map.  Table 4 provides a sample catalog of the leaves that pass our selection criteria in W3-west.  Similar catalogs for all eleven KEYSTONE regions are available online.  Of the 970 total leaves identified by the dendrograms in each region, the final catalog includes a total of 856 leaves ($\sim88\%$) that passed all of the culling criteria.  Figures \ref{W3_leaves}-\ref{NGC7538_leaves} show the final catalog leaf masks overlaid atop the NH$_3$ (1,1) integrated intensity map for each region.  These masks show the full extent of all pixels associated with each leaf.  For the remainder of the paper, we refer to leaves and clumps synonymously since the ammonia-identified leaves represent the dense gas structures from which new stars may form.

\clearpage

\subsection{Determining Leaf Radii and Masses}
The effective radii of the ammonia-identified leaves were estimated using the area of the leaf masks identified by the dendrogram analysis.  Following \cite{Kauffmann_2013}, we adopt $R_{eff}=(A/\pi)^{1/2}$ as the effective radius, where $A$ is the area of all pixels in the leaf's mask on the position-position plane.  \cite{Rosolowsky_2006} showed that this area-based radius formulation becomes inaccurate for structures with low SNR and sizes much larger or smaller than the beam size.  Although we do enforce that leaves are comprised of pixels with at least 5$\sigma$ detections (see Section 3.3) and limit structures to being larger than the beam size, $R_{eff}$ may still be susceptible to such biases.  To estimate the uncertainties on our measured radii, we use the method described by \cite{Chen_2018}, which uses the radii of the largest circle that fits inside the leaf boundary as the leaf's radii lower limit and the radii of the smallest circle that encompasses the leaf as its radii upper limit.  The corresponding uncertainties on $R_{eff}$ based on these upper and lower limits are listed in Table 4.  The uncertainties range from $0.1\%$ to $153\%$ of $R_{eff}$, with a median of $35\%$.

Masses for the ammonia-identified leaves were estimated by summing all the H$_2$ column density for the pixels inside each leaf's mask.  The integrated column densities are then converted to mass assuming the distances to each region listed in Table \ref{Table_regions} and a mean molecular weight per hydrogen molecule $\mu_H$ = 2.8.  In Cygnus X and MonR2, a small number of leaves (six in Cygnus X North, 14 in Cygnus X South, and one in MonR2) fall outside the boundaries of our H$_2$ column density maps.  Those sources are, therefore, excluded from our analyses that require a mass determination.  

Summing all the column density within the leaf boundaries is likely an upper limit on the mass of the structure.  Conversely, a lower limit on the structure's mass can be obtained by using the ``clipping'' technique described in \cite{Rosolowsky_2008} and \cite{Chen_2018}.  Namely, before summing the column density pixels within the leaf boundary, the lowest column density pixel's value is subtracted from all other pixels.  This method aims to remove contributions to the structure's observed column density from background sources, but is likely over-estimating the true background contribution in most cases.  As such, we adopt the regular integrated column density masses throughout this paper, but show the range the mass could be assuming the ``clipped'' mass is a lower limit.  The clipped masses are also displayed in Table 4 alongside the integrated column density masses.  The clipped masses are typically a factor of $\sim 5$ (median) lower than the integrated column density masses.

\begin{figure}[ht]
\epsscale{1.0}
\plotone{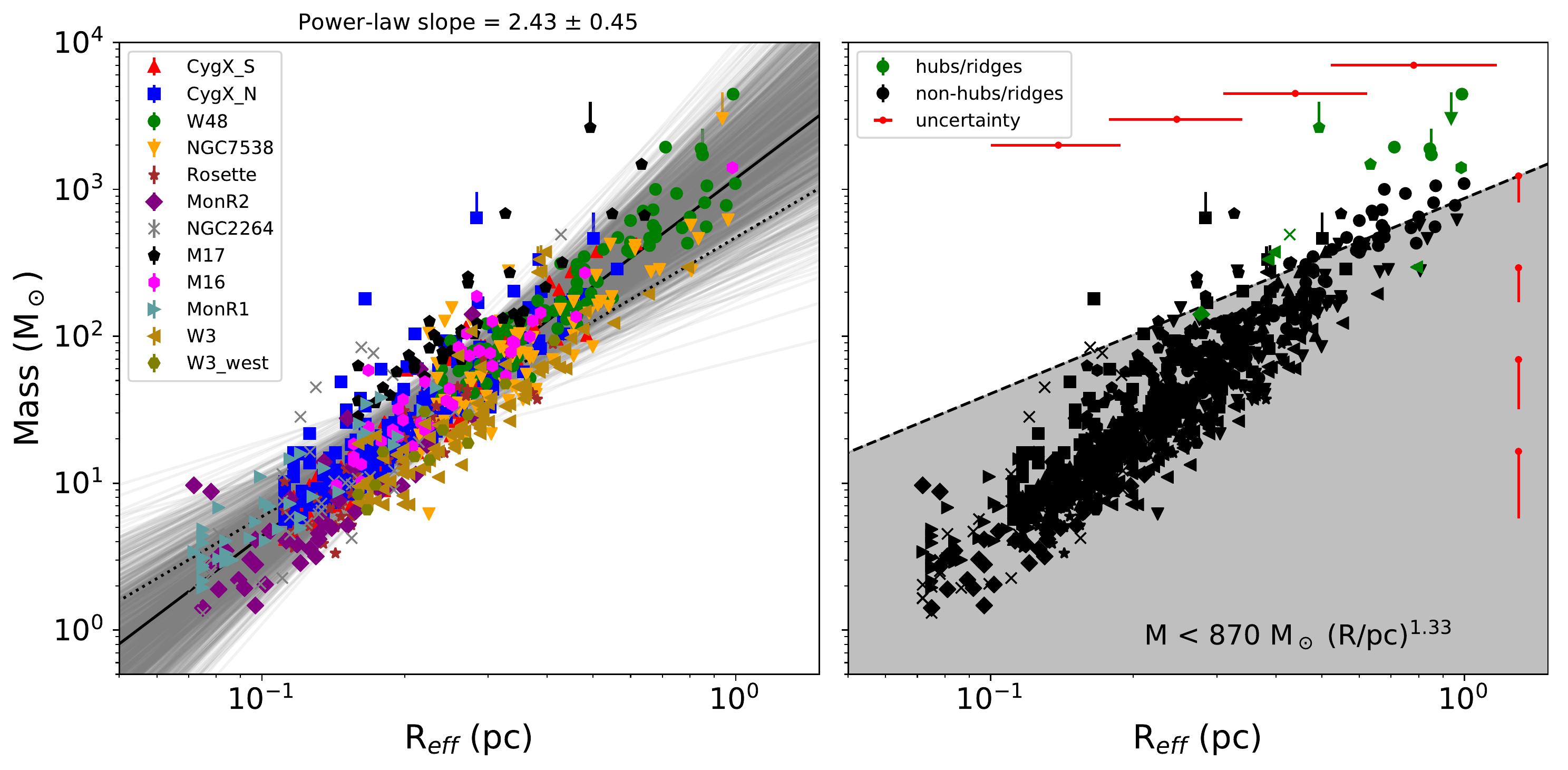}
\caption{Left: effective radius versus mass for the leaves identified in each region.  The solid black line shows the best power-law fit to the data using an MCMC sampler. The grey lines show 1000 random selections from the full MCMC chain.  The dotted line shows a power-law slope of 1.9 \citep{Larson_1981} for comparison.  Right: effective radius versus mass for hubs (green) and non-hubs (black) identified in each region (see Section 4.1 for discussion of hubs).  The dashed black line denotes the empirically-derived threshold for massive star formation determined by \cite{Kauffmann_2010}.  The red dots with errorbars show the median mass and radius uncertainties in different bins along each axis.  All other errorbars show leaves that have mass lower limits due to saturated pixels in the H$_2$ column density map (see Section 2.3).}
\label{Reff_mass}
\end{figure}

The left panel of Figure \ref{Reff_mass} shows the effective radii versus mass for all leaves in our final catalog.  A power-law fit to the radius versus mass distribution reveals a best-fit slope of 2.43 $\pm$ 0.45, which is consistent with the value of 2 expected for clumps of constant surface density and the value of 3 expected for clumps of constant volume density.  The data were fit using a Markov Chain Monte Carlo (MCMC) sampler\footnote{the \texttt{emcee} package: \url{https://emcee.readthedocs.io}}.  The MCMC sampling used an orthogonal least-squares likelihood function and uniform priors on the power-law slope and intercept. The best-fit model parameters were taken to be the medians of the accepted parameters in the MCMC chain, while the uncertainty on the best-fit parameters was taken to be the standard deviations of the accepted parameter distributions.



\subsection{Virial Analysis}
To estimate the virial stability of the ammonia-identified leaves, we adopt the virial analysis method described in \cite{Keown_2017}, which uses the ammonia-derived line widths to derive a virial mass ($M_{vir}$) for each structure given by:
\begin{equation}
M_{vir} = \frac{5\sigma^{2}R}{aG}
\end{equation} where $\sigma$ is the velocity dispersion of the core (including both the thermal and nonthermal components), \textit{R} is the core radius, \textit{G} is the gravitational constant, and 

\begin{equation}
a = \frac{1 - k/3}{1 - 2k/5}
\end{equation} is a term which accounts for the radial power-law density profile of a core, where $\rho(r) \propto r^{-k}$ \citep{Bertoldi_1992}.  $M_{vir}$ represents the mass that a structure with a given radius and internal kinetic energy would have if it were in virial equilibrium when considering only its gravitational potential and kinetic energies.  We also assume that the structure is in a steady state, spherical, isothermal, and has a radial power-law density profile of the form: $\rho(r) \propto r^{-1.5}$.  Our density profile assumption is motivated by recent observations that found $\rho(r) \propto r^{-1.5\pm0.3}$ for the inner regions of dense cores \citep[e.g.,][]{Kurono_2013, Pirogov_2009} and is likely a more accurate choice than the Gaussian density profile chosen in previous virial analyses \citep[e.g.,][]{Pattle_2017, Kirk_2017, Keown_2017}.  See \cite{Keown_2017} for a discussion of the implications of assuming a power-law density profile for sources in a virial analysis.  We also set $R$ in Equation 1 to be $R_{eff}$ and calculate the thermal plus nonthermal velocity dispersion as:

\begin{equation}
\sigma^2 = \sigma_{v}^{2} - \frac{k_{B}T}{m_{NH_3}} + \frac{k_{B}T}{\mu_p m_H}
\end{equation} where $k_B$ is Boltzmann's constant, $m_{NH_3}$ is the molecular mass of NH$_3$, $m_H$ is the atomic mass of hydrogen, and $\mu_p$ is the mean molecular mass of interstellar gas.  We use $\mu_p$ rather than $\mu_H$ for this analysis since $\mu_p$ considers the additional contributions of helium, assuming a hydrogen-to-helium abundance ratio of 10 and a negligible admixture of metals, that are required to calculate the thermal gas pressure accurately \citep[2.33; see, e.g., Appendix A in][]{Kauffmann_2008}.  $\sigma_{v}$ and $T$ are the average velocity dispersion and kinetic temperature, respectively, within the core boundaries measured from the NH$_3$ line-fitting parameter maps.  Both the $\sigma_{v}$ and $T$ averages are weighted by the NH$_3$ (1,1) integrated intensity map such that $\sigma_{v, avg} = w_1\sigma_1 + w_2\sigma_2 \cdots w_n\sigma_n$, where $w_n$ and $\sigma_n$ are the fraction of the source's integrated intensity and value of the velocity dispersion, respectively, for pixel $n$.  

\begin{figure}[ht]
\plotone{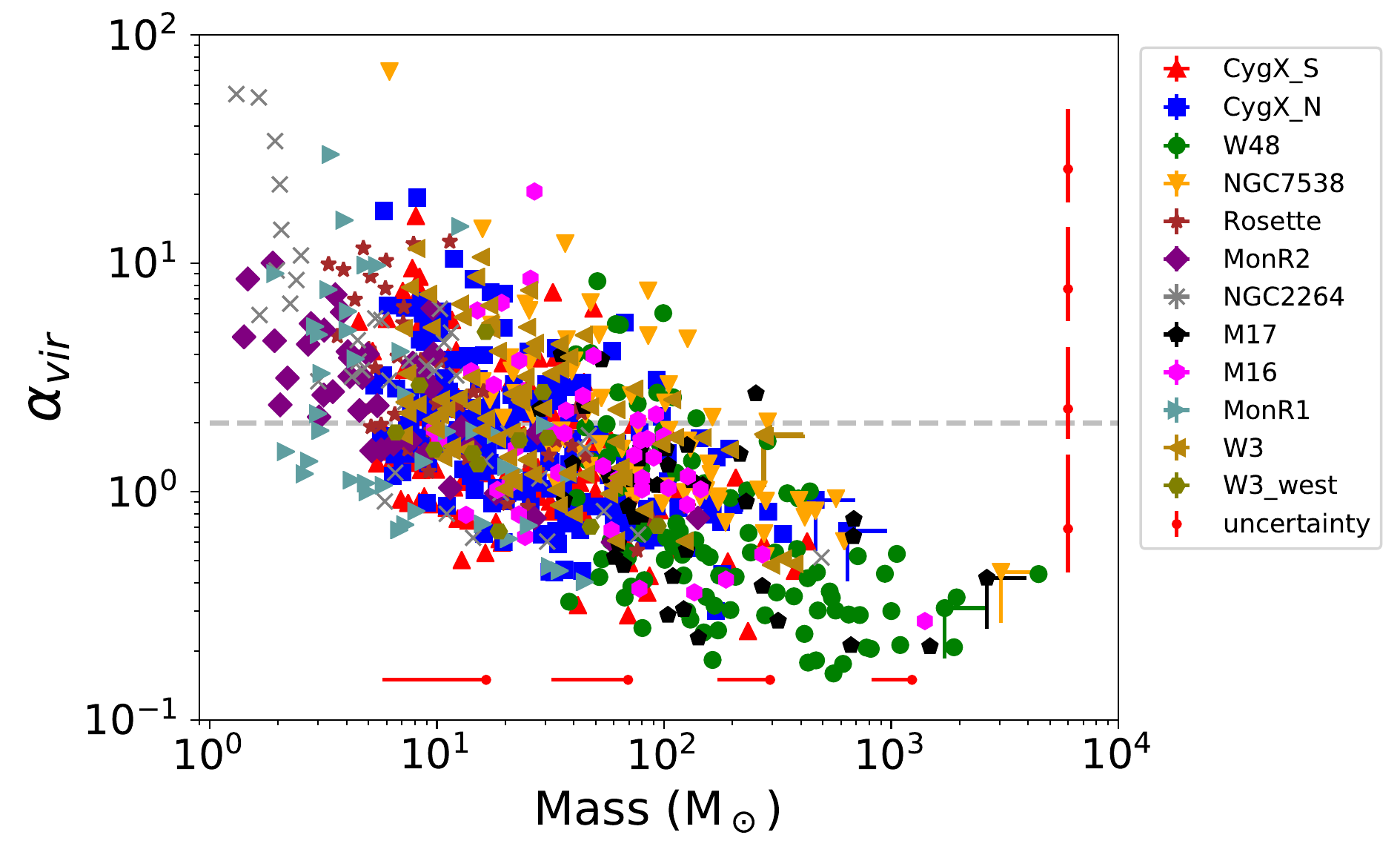}
\caption{Virial parameter ($\alpha_{vir}$) versus mass for the leaves identified in each region.  The dashed line denotes $\alpha_{vir}$ = 2 when assuming a power-law density profile for the structures.  Above this line, structures are deemed to be gravitationally unbound in the absence of magnetic fields or external pressure. The red dots with errorbars show the median mass and radius uncertainties in different bins along each axis.  All other errorbars show leaves that have mass lower limits due to saturated pixels in the H$_2$ column density map (see Section 2.3).}
\label{Mass_alpha}
\end{figure}

The ratio of $M_{vir}$ to the actual observed mass of the structure ($M_{obs}$) is known as the virial parameter ($\alpha_{vir}=M_{vir}/M_{obs}$).  This virial parameter can also be written as $\alpha_{vir}=a2\Omega_K / |\Omega_G|$, where $\Omega_K$, $\Omega_G$, and $a$ represent the structure's total kinetic energy, gravitational potential energy, and density profile (Equation 2), respectively \citep{Bertoldi_1992}.  Thus, the virial parameter neglects the surface term for the kinetic energy that considers the ambient gas pressure exerted on the structure by the cloud. When $\alpha_{vir} \geq 2$, the structure's internal kinetic energy is large enough to prevent it from being gravitationally bound.  Conversely, when $\alpha_{vir} < 2$, the structure is deemed gravitationally bound since its gravitational potential energy is large enough relative to its internal kinetic energy (neglecting the effects of magnetic fields and external pressure).  Figure \ref{Mass_alpha} shows observed mass versus virial parameter for all leaves in our final catalog.  Of the 835 leaves, 523 ($\sim 63 \%$) fall below the $\alpha_{vir}=2$ threshold to be considered gravitationally bound.  When looking at each cloud individually, the bound leaf fraction varies from $\sim$0.3 in MonR2 to $\sim$0.9 in M17 and W48.  Table 5 lists the virial parameters for the leaves identified in W3-west (similar tables for the other regions are provided online).  Table \ref{Table_clouds} shows the bound leaf fractions for each individual cloud, while Table \ref{Table_population} lists the bound fraction and other population statistics for the full leaf sample. 

We note that the variations in distance to each KEYSTONE target provide a variety of linear scales resolved by our observations.  These linear resolution effects are not accounted for with the analysis presented here.  In Appendix A, however, we show the impact on the virial parameters of NGC 2264, MonR1, and MonR2 ($d = 0.9$ kpc) if those maps were convolved and downsampled to the linear resolution of the W48 ($d = 3.0$ kpc) observations.  We show that there is indeed a tendency for the identified structures in the distance-adjusted analysis to be bound, which might be affecting the higher fraction of bound structures observed in W48 and M17 ($d = 2.0$ kpc).


\begin{table}
\let\center\empty
\let\endcenter\relax
\centering
\caption{W3-west NH$_3$ (1,1) Leaves Catalog 1}
\resizebox{.7\width}{!}{\begin{tabular}{rrrrrrlrrllrrr}
\toprule
ID & RA & decl. & PA & $\sigma_{major}$ & $\sigma_{minor}$ &  $R_{eff}$ &  $M_{obs}$ &  $M_{clip}$ &  $T_K$ &  $\sigma_{NH_3}$ &  $V_{LSR, NH_3}$ &  log($N_{para-NH_3}$) &  log($N_{H_2}$) \\& (deg) & (deg) & (deg) & ($\arcsec$) & ($\arcsec$) & (pc) & (M$_\odot$) & (M$_\odot$) & (K) & (km s$^{-1}$) & (km s$^{-1}$) & (cm$^{-2}$) & (cm$^{-2}$) \\\midrule
  0 &  34.9844 &  61.0428 &  166 &         16.6 &          9.9 &  0.27 $^{+0.07}_{-0.06}$ &  18.7 &           3.8 &  10.9 $\pm$ 2.4 &  0.13 $\pm$ 0.02 &         -15.2 &            13.7 &             21.4 \\
  1 &  35.4211 &  61.0938 &  167 &         30.7 &         12.8 &  0.41 $^{+0.27}_{-0.15}$ &  93.7 &          37.2 &  13.1 $\pm$ 1.7 &  0.36 $\pm$ 0.04 &         -49.9 &            13.9 &             21.5 \\
  2 &  35.3625 &  61.0890 &  180 &         14.6 &         10.8 &  0.24 $^{+0.08}_{-0.13}$ &  23.0 &           4.2 &  13.0 $\pm$ 2.4 &  0.36 $\pm$ 0.05 &         -49.7 &            14.1 &             21.5 \\
  3 &  35.6144 &  61.0918 &  101 &         14.5 &          7.0 &  0.17 $^{+0.09}_{-0.12}$ &   6.6 &           1.5 &  10.2 $\pm$ 3.6 &  0.21 $\pm$ 0.04 &         -49.4 &            13.6 &             21.3 \\
  4 &  35.2711 &  61.1001 &  196 &         22.1 &         12.1 &  0.33 $^{+0.22}_{-0.19}$ &  47.5 &          14.3 &  12.2 $\pm$ 2.5 &  0.27 $\pm$ 0.04 &         -49.7 &            14.0 &             21.5 \\
  5 &  35.5964 &  61.1042 &   90 &         10.8 &          6.2 &  0.16 $^{+0.05}_{-0.01}$ &   8.4 &           1.3 &  16.5 $\pm$ 3.0 &  0.34 $\pm$ 0.06 &         -50.1 &            13.4 &             21.3 \\
  6 &  35.4752 &  61.1064 &  152 &         17.8 &         10.6 &  0.27 $^{+0.07}_{-0.07}$ &  29.2 &           7.2 &  15.4 $\pm$ 2.6 &  0.52 $\pm$ 0.07 &         -50.0 &            13.8 &             21.5 \\
  7 &  35.2490 &  61.1213 &   63 &         10.2 &          8.1 &  0.17 $^{+0.02}_{-0.03}$ &   9.7 &           1.6 &  13.0 $\pm$ 3.9 &  0.23 $\pm$ 0.06 &         -49.6 &            13.9 &             21.5 \\
  8 &  35.2099 &  61.1515 &   68 &         13.8 &          9.7 &  0.23 $^{+0.08}_{-0.03}$ &  14.4 &           2.3 &  12.3 $\pm$ 3.2 &  0.25 $\pm$ 0.05 &         -48.8 &            13.1 &             21.5 \\
  9 &  35.1733 &  61.1655 &  155 &         14.4 &          8.2 &   0.21 $^{+0.1}_{-0.03}$ &  15.2 &           4.8 &  12.8 $\pm$ 2.8 &  0.25 $\pm$ 0.05 &         -48.7 &            13.8 &             21.5 \\
 10 &  35.2731 &  61.4578 &   86 &         16.1 &          9.5 &  0.22 $^{+0.13}_{-0.04}$ &  30.8 &           7.6 &  14.8 $\pm$ 2.5 &  0.46 $\pm$ 0.06 &         -51.0 &            13.8 &             21.5 \\
 11 &  35.2470 &  61.4522 &  191 &         13.4 &          9.0 &  0.18 $^{+0.08}_{-0.08}$ &  16.4 &           1.2 &  12.6 $\pm$ 3.5 &  0.67 $\pm$ 0.11 &         -51.4 &            14.0 &             21.5 \\
\bottomrule
\end{tabular}
} \\
\raggedright Columns show the following values for each leaf: (1) Leaf ID, (2-3) Mean Right Ascension and Declination in J2000 coordinates, (4) Position angle of the major axis, measured in degrees counterclockwise from the west on sky, (5-6) Major and minor axis measured by \textit{astrodendro} based on the intensity weighted second moment in the direction of greatest elongation, (7) Effective radius defined as $R_{eff}=(A/\pi)^{1/2}$, where $A$ is the area of all pixels in the leaf's mask on the position-position plane, (8) Observed mass of leaf from the sum of its H$_2$ column density, (9) lower limit mass of leaf calculated using the ``clipping'' technique (see text), (10-13) Average kinetic gas temperature, velocity dispersion, NH$_3$ (1,1) centroid velocity, and para-NH$_3$ column density for leaf, all weighted by the NH$_3$ (1,1) integrated intensity map, along with their 1-sigma uncertainties, (14) Median H$_2$ column density for leaf measured from the spatially-filtered column density map and used as $N$ in Equation 7.  Both column densities are shown in logarithmic scale. Similar tables for all other KEYSTONE regions are available online.  Although the intensity-weighted major and minor axes of some sources are less than the $32\arcsec$ beam size of the observations, our culling criteria ensure that their total areas when considering all their associated pixels are larger than $32\arcsec$.
\end{table}

\begin{table}
\let\center\empty
\let\endcenter\relax
\centering
\caption{W3-west NH$_3$ (1,1) Leaves Catalog 2}
\resizebox{.75\width}{!}{\begin{tabular}{rrrrrrrrllrr}
\toprule
 ID &  $\alpha_{vir}$ &  $M_{vir}$ &  $\sigma_{nt}$ & log$|\Omega_G|$ & log$\Omega_K$ & log$|\Omega_{Pw}|$ &  log$|\Omega_{Pt}|$ & on-filament & hub & $N_{proto}$ & Bad N(H$_2$) Pixels \\& & (M$_\odot$) & (km s$^{-1}$) & (erg) & (erg) & (erg) & (erg) & & & \\\midrule
  0 &   0.67 &              12.6 &      0.10 &  43.8 &     43.4 &     44.2 &         NaN &         True &  False &           0 &         0 \\
  1 &   0.71 &              66.3 &      0.36 &  45.0 &     44.7 &     44.8 &         NaN &         True &  False &           4 &         0 \\
  2 &   1.68 &              38.6 &      0.36 &  44.1 &     44.1 &     44.1 &         NaN &         True &  False &           1 &         0 \\
  3 &   1.82 &              12.0 &      0.20 &  43.1 &     43.2 &     43.5 &         NaN &         True &  False &           0 &         0 \\
  4 &   0.70 &              33.3 &      0.26 &  44.5 &     44.2 &     44.5 &         NaN &         True &  False &           2 &         0 \\
  5 &   2.92 &              24.5 &      0.33 &  43.4 &     43.6 &     43.5 &         NaN &         True &  False &           1 &         0 \\
  6 &   2.72 &              79.6 &      0.51 &  44.2 &     44.4 &     44.3 &         NaN &         True &  False &           0 &         0 \\
  7 &   1.53 &              14.9 &      0.21 &  43.4 &     43.4 &     43.7 &         NaN &         True &  False &           1 &         0 \\
  8 &   1.47 &              21.2 &      0.24 &  43.7 &     43.6 &     44.0 &         NaN &         True &  False &           0 &         0 \\
  9 &   1.32 &              20.0 &      0.24 &  43.8 &     43.7 &     43.9 &         NaN &         True &  False &           1 &         0 \\
 10 &   1.72 &              52.9 &      0.45 &  44.3 &     44.4 &     44.0 &         NaN &         True &  False &           1 &         0 \\
 11 &   5.01 &              82.0 &      0.67 &  43.9 &     44.4 &     43.8 &         NaN &         True &  False &           0 &         0 \\
\bottomrule
\end{tabular}
} \\
\raggedright Columns show the following values for each leaf: (1) Leaf ID, (2) virial parameter defined as M$_{vir}$/M$_{obs}$, (3) virial mass calculated using Equation 1, (4) non-thermal component of the velocity dispersion, (5) gravitational energy density calculated using Equation 5, (6) kinetic energy density calculated using Equation 6, (7) cloud weight pressure energy density calculated using Equation 4, (8) turbulent pressure energy density calculated using Equation 4, with NaN representing a lack of C$^{18}$O data for that leaf, (9-10) whether or not the leaf is on-filament or a hub, (11) number of 70 $\mu$m point sources within the leaf's boundary, (12) fraction of pixels in the leaf that were saturated in the H$_2$ column density map.   Similar tables for all other KEYSTONE regions are available online.
\end{table}

\subsection{Identifying filaments and candidate YSOs}
\begin{figure}[ht]
\epsscale{1.05}
\plotone{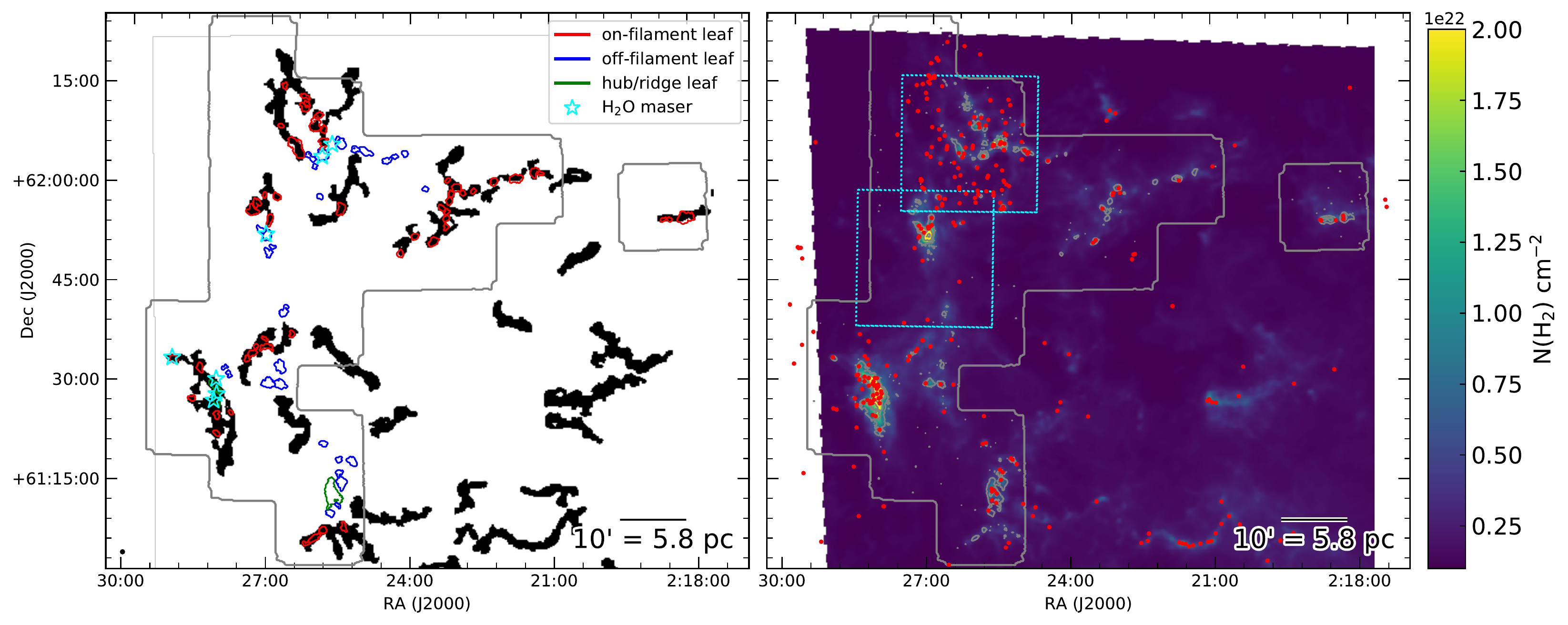}
\epsscale{1.0}
\plotone{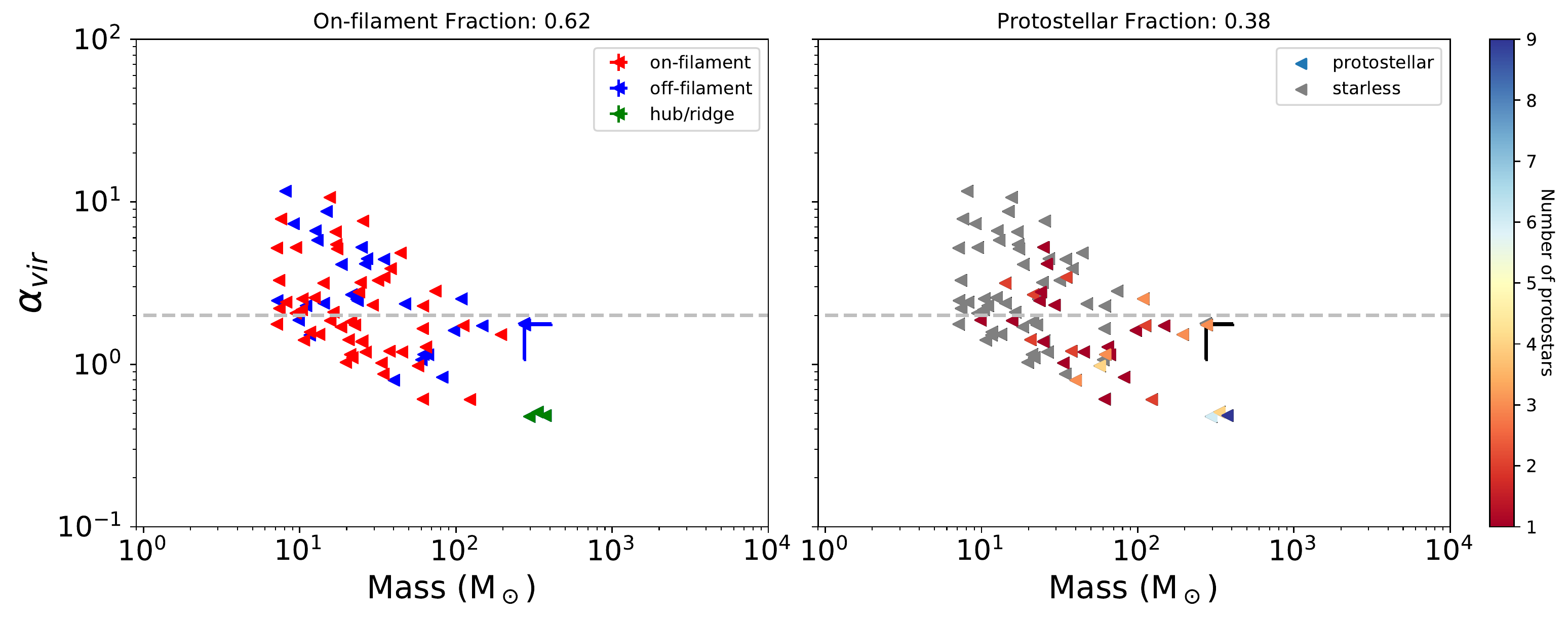}
\caption{Top Right: \textit{Herschel} H$_2$ column density map of W3 with positions of candidate YSOs identified by \texttt{getsources} at 70 $\mu$m overlaid as red dots.  The outer grey outline denotes the area mapped by KEYSTONE.  The grey contours show NH$_3$ (1,1) integrated intensity at 1.0 K km s$^{-1}$, 3.5 K km s$^{-1}$, and 10 K km s$^{-1}$.  The cyan dotted line outlines the area observed in C$^{18}$O $(3-2)$ by the JCMT, which we fit to derive external pressure terms for the subset of leaves falling within those observations (see Section 4.5). Top Left: masks of filaments identified in the \textit{Herschel} H$_2$ column density map by \textit{getfilaments}, reconstructed up to scales of 145$\arcsec$.  The positions of our ammonia-identified leaves are overlaid, with red denoting an ``on-filament" leaf, blue representing an ``off-filament" leaf, and green showing ``hubs/ridges" that have uncharacteristically larger masses than the majority of leaves in their respective cloud and tend to be located at filament intersections (see Section 4.1).  Cyan stars show the positions of H$_2$O maser emission. Bottom row: virial parameters versus mass for the protostellar and starless leaves (right) as well as the on-filament, off-filament, and hub sources shown in the top left panel (left). The data point shape in these plots is the same as in Figure \ref{rms_hists}. Errorbars show leaves that have mass lower limits due to saturated pixels in the H$_2$ column density map (see Section 2.3).}
\label{W3_filaments}
\end{figure}

\begin{figure}[ht]
\epsscale{1.05}
\plotone{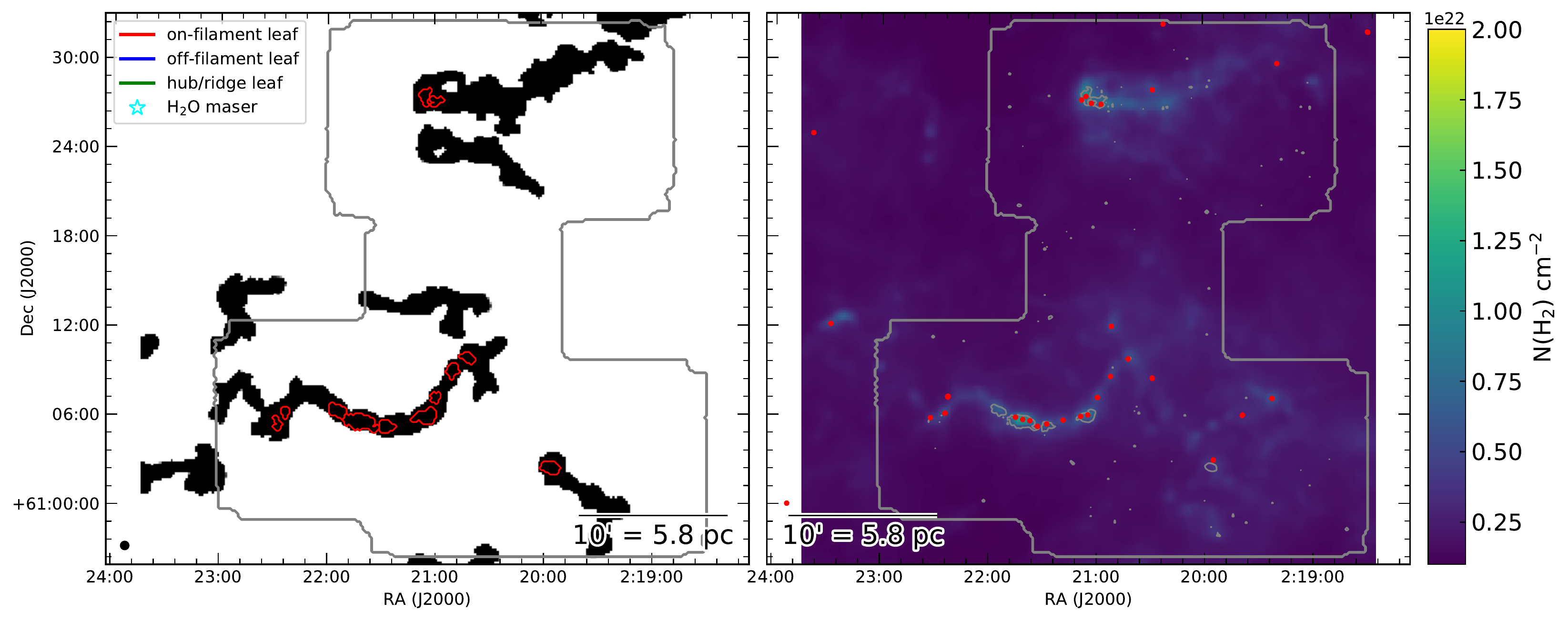}
\epsscale{1.0}
\plotone{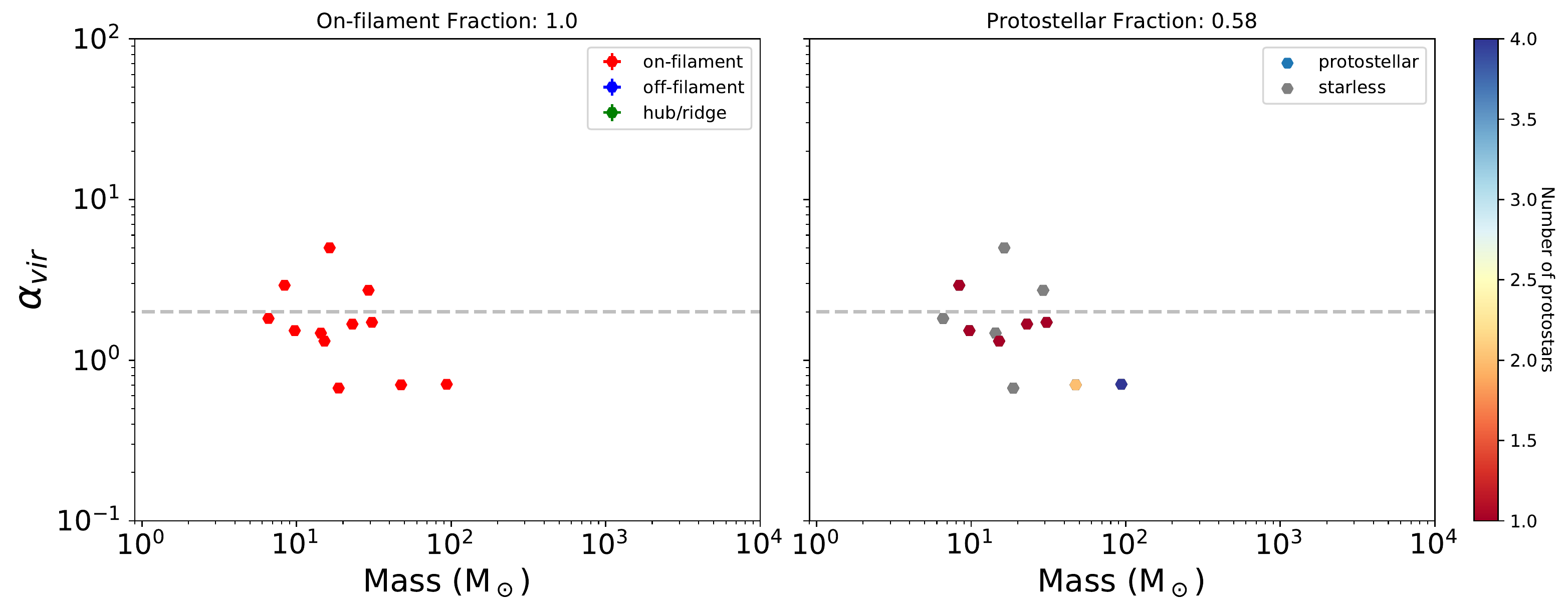}
\caption{Same as Figure \ref{W3_filaments} for W3-west.}
\label{W3-west_filaments}
\end{figure}

\begin{figure}[ht]
\epsscale{1.05}
\plotone{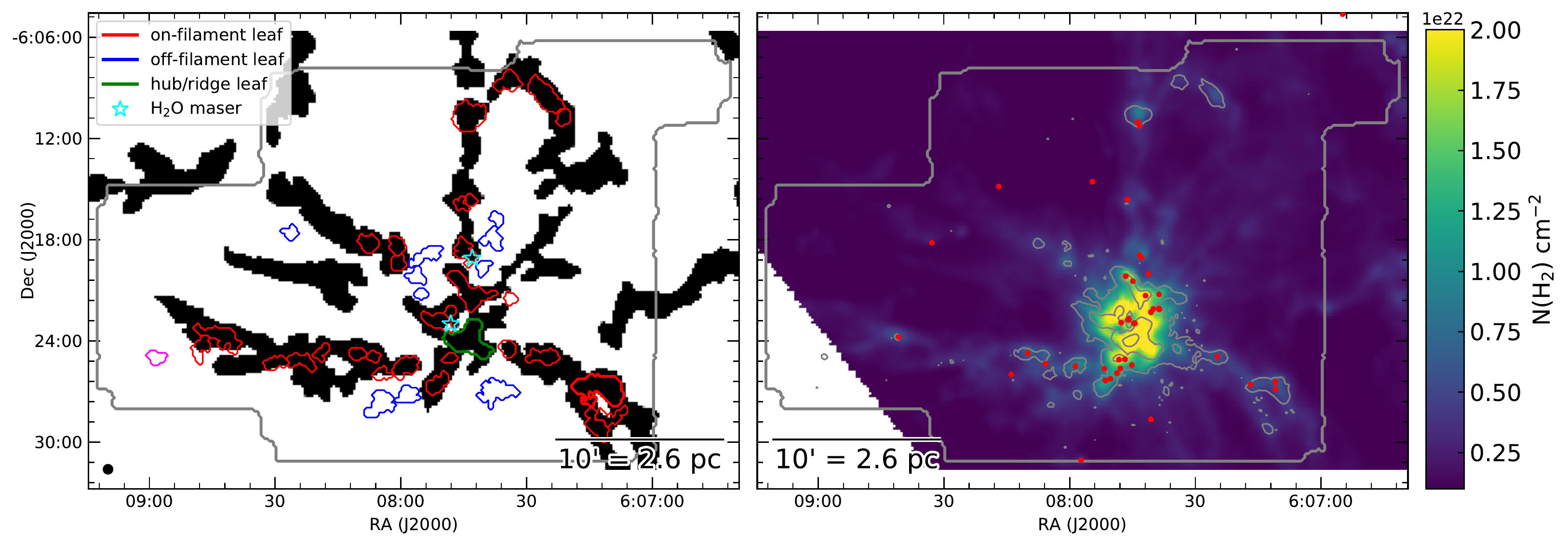}
\epsscale{1.0}
\plotone{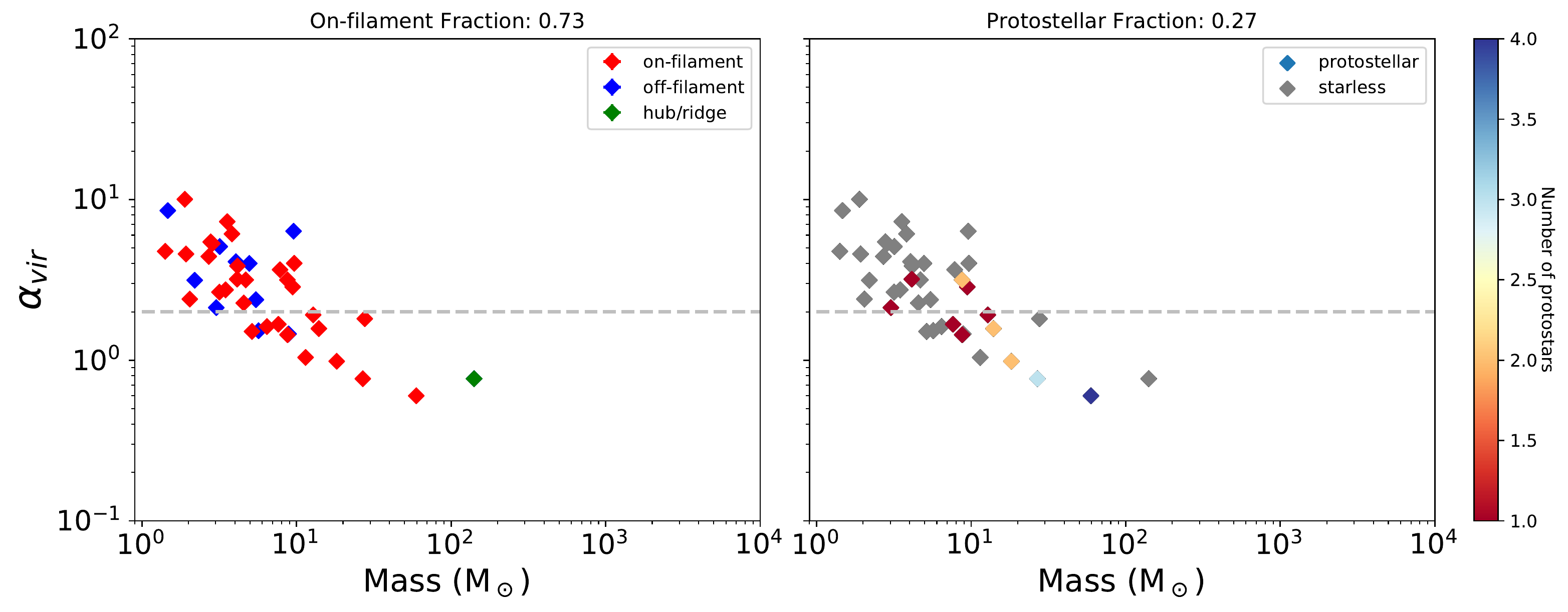}
\caption{Same as Figure \ref{W3_filaments} for MonR2. Magenta contours on the left side of the upper left panel denote leaves that were not included in our virial analysis since their masses could not be estimated due to their locations being outside our H$_2$ column density map boundaries.}
\label{MonR2_filaments}
\end{figure}

\begin{figure}[ht]
\epsscale{1.05}
\plotone{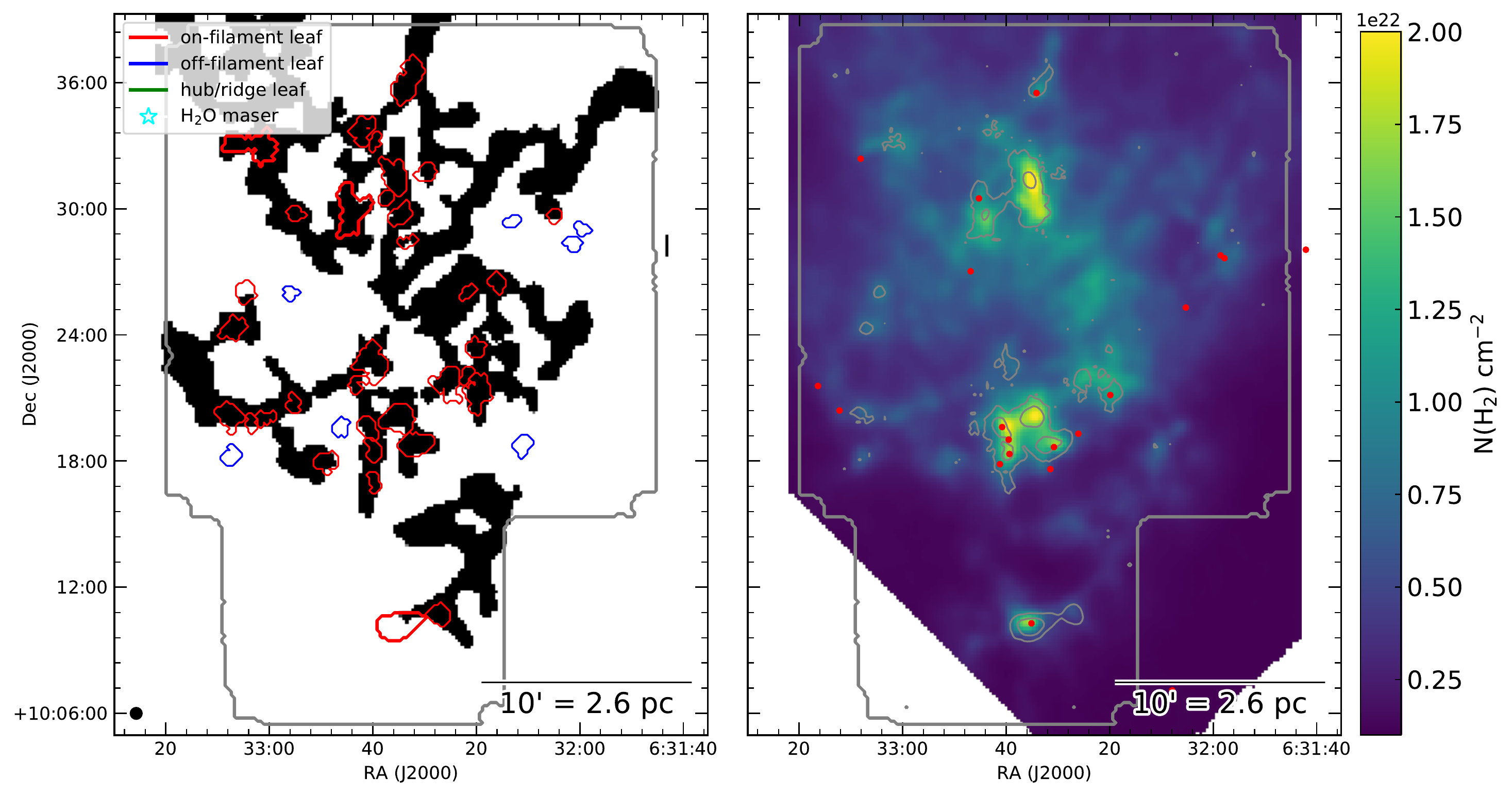}
\epsscale{1.0}
\plotone{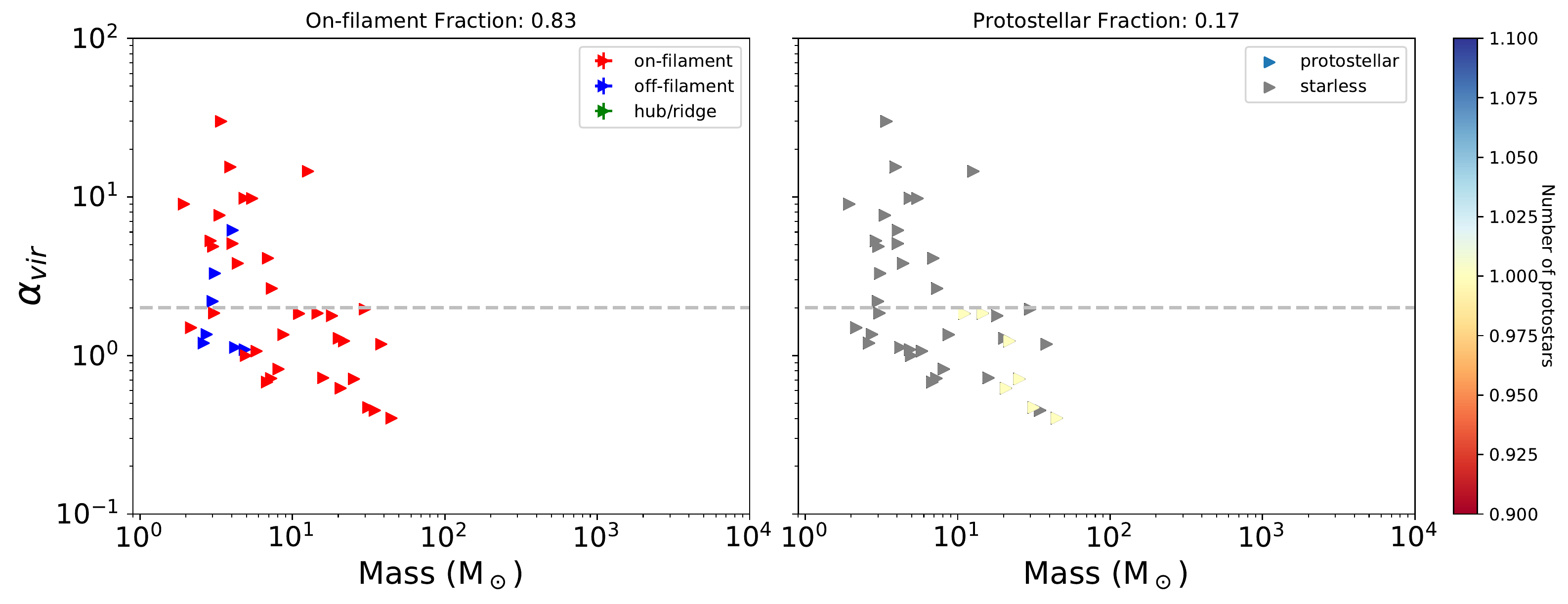}
\caption{Same as Figure \ref{W3_filaments} for MonR1.}
\label{MonR1_filaments}
\end{figure}

\begin{figure}[ht]
\epsscale{1.05}
\plotone{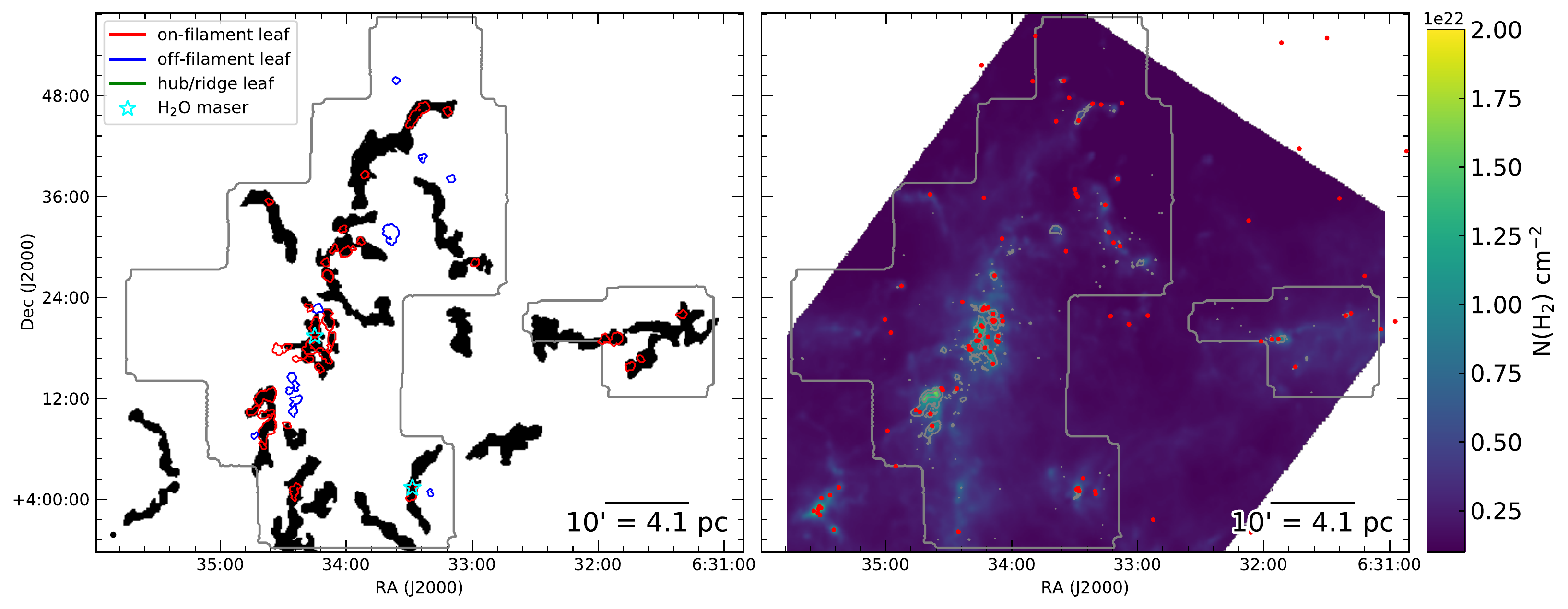}
\epsscale{1.0}
\plotone{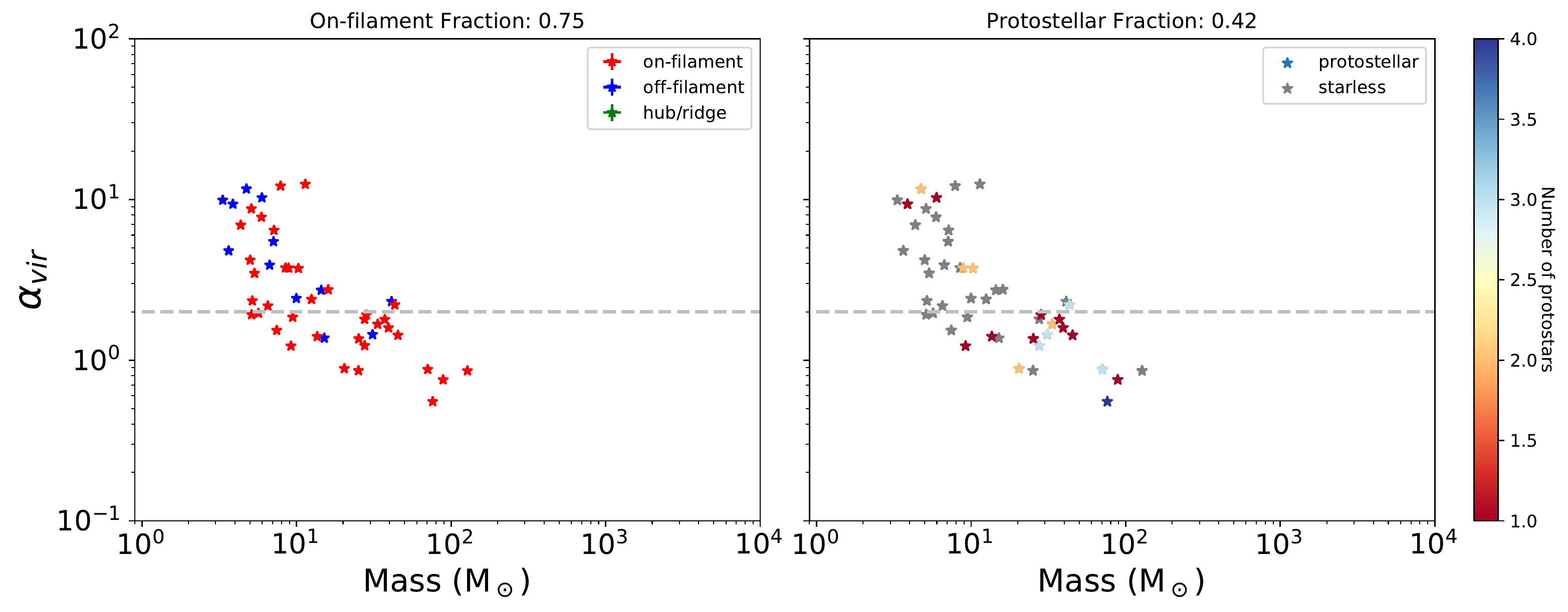}
\caption{Same as Figure \ref{W3_filaments} for Rosette.}
\label{Rosette_filaments}
\end{figure}

\begin{figure}[ht]
\epsscale{1.05}
\plotone{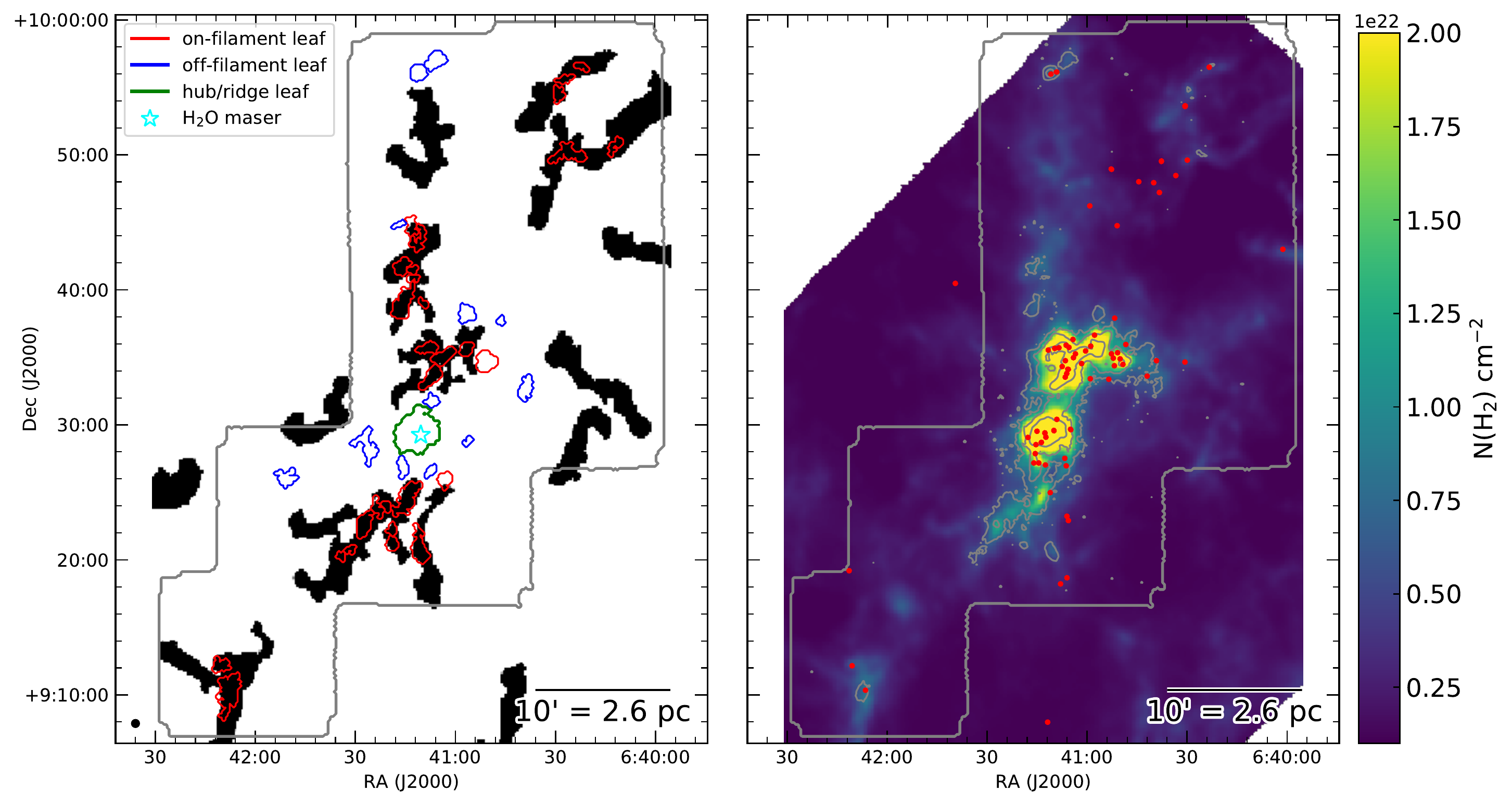}
\epsscale{1.0}
\plotone{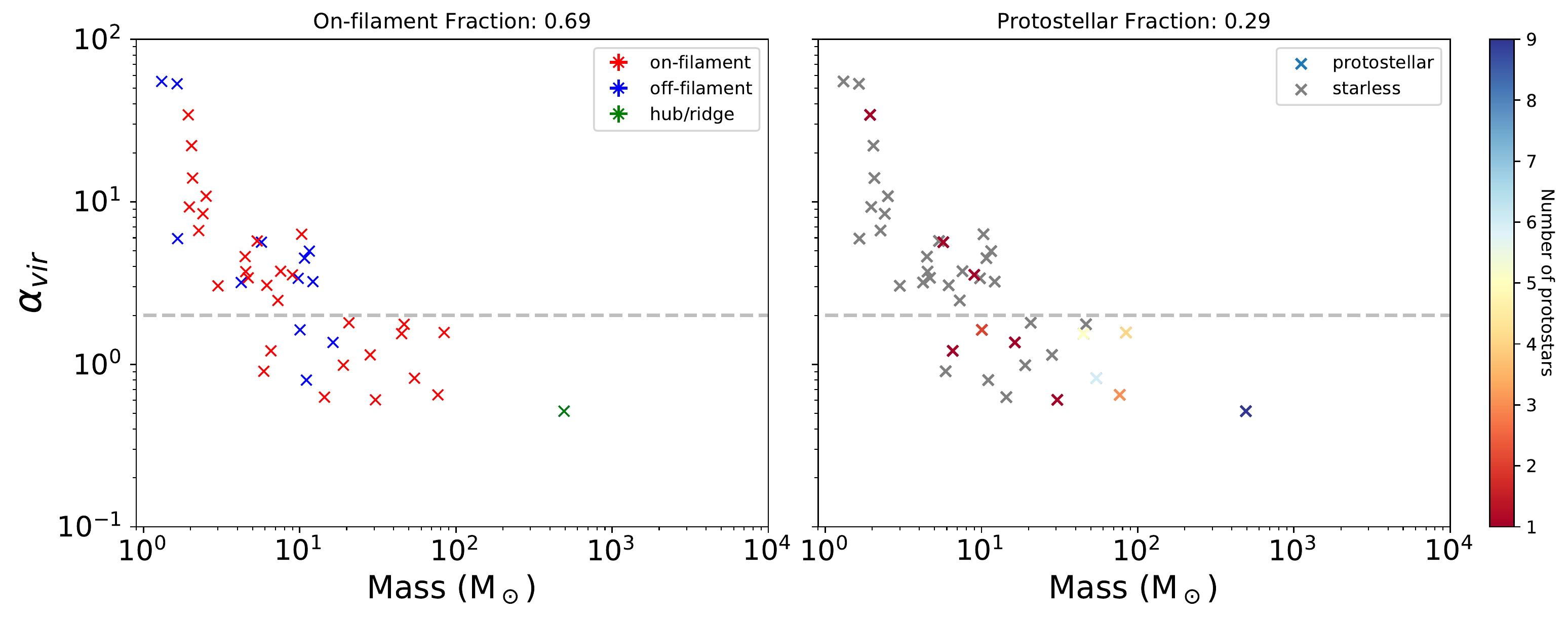}
\caption{Same as Figure \ref{W3_filaments} for NGC2264.}
\label{NGC2264_filaments}
\end{figure}

\begin{figure}[ht]
\epsscale{1.05}
\plotone{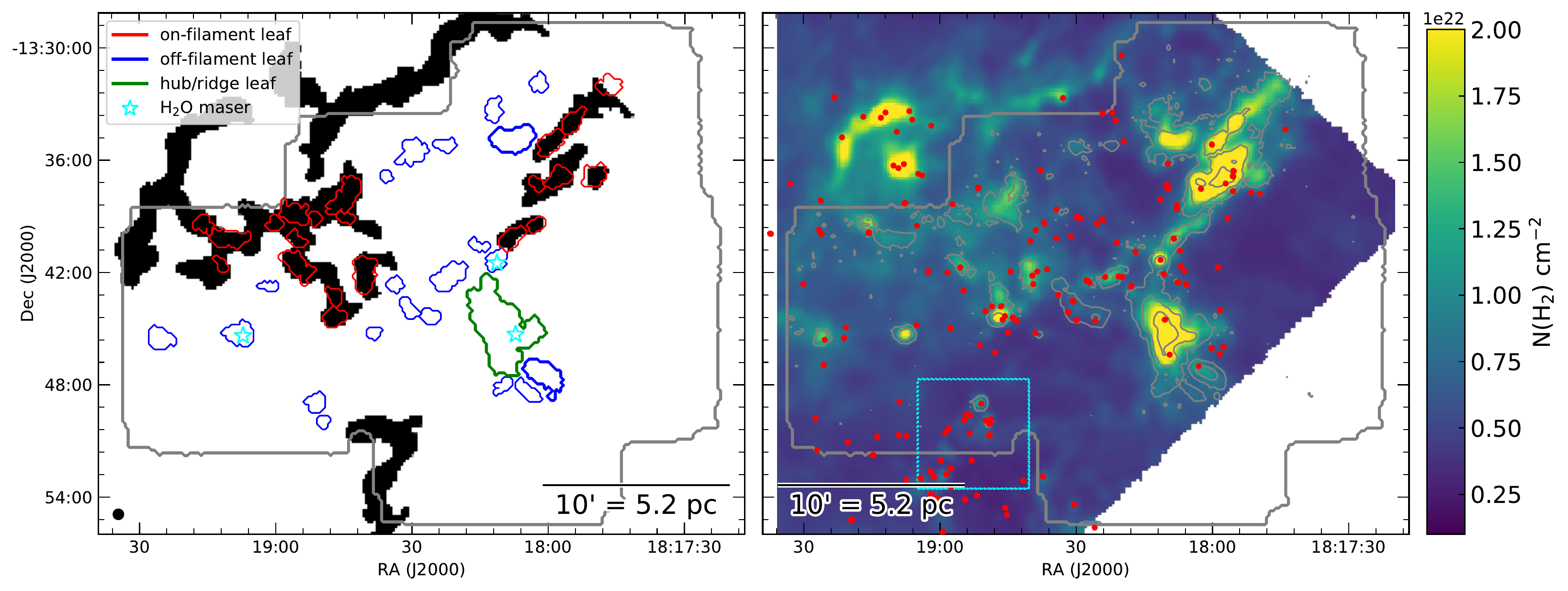}
\epsscale{1.0}
\plotone{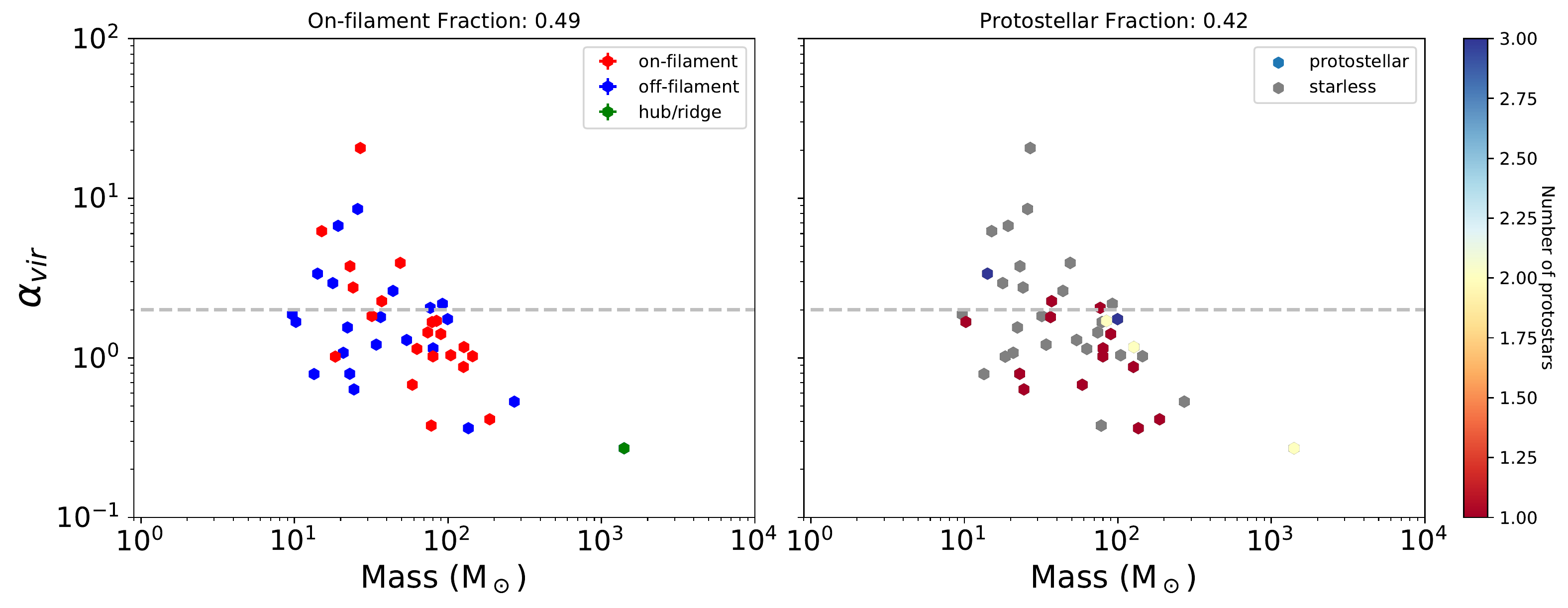}
\caption{Same as Figure \ref{W3_filaments} for M16.}
\label{M16_filaments}
\end{figure}

\begin{figure}[ht]
\epsscale{1.05}
\plotone{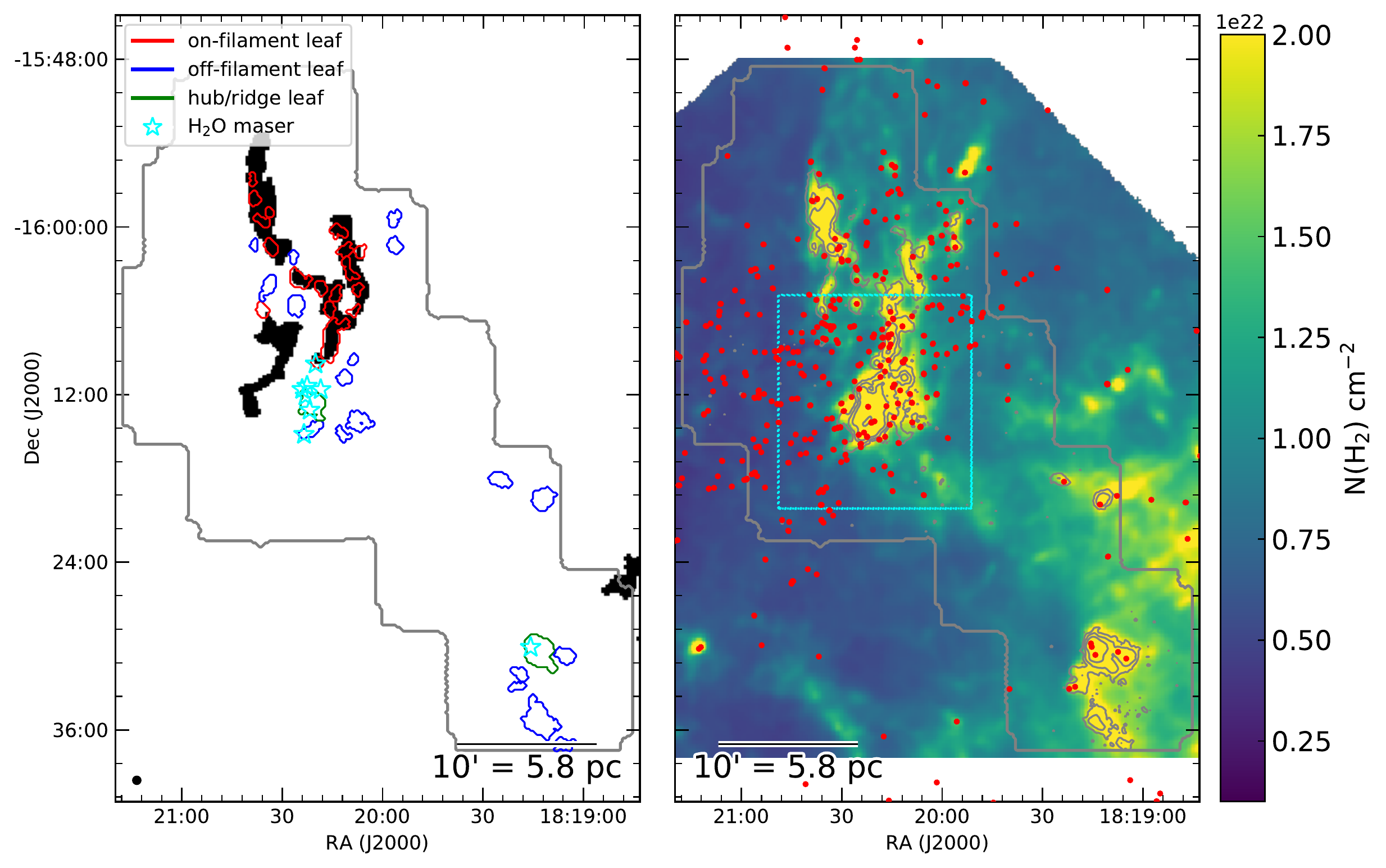}
\epsscale{1.0}
\plotone{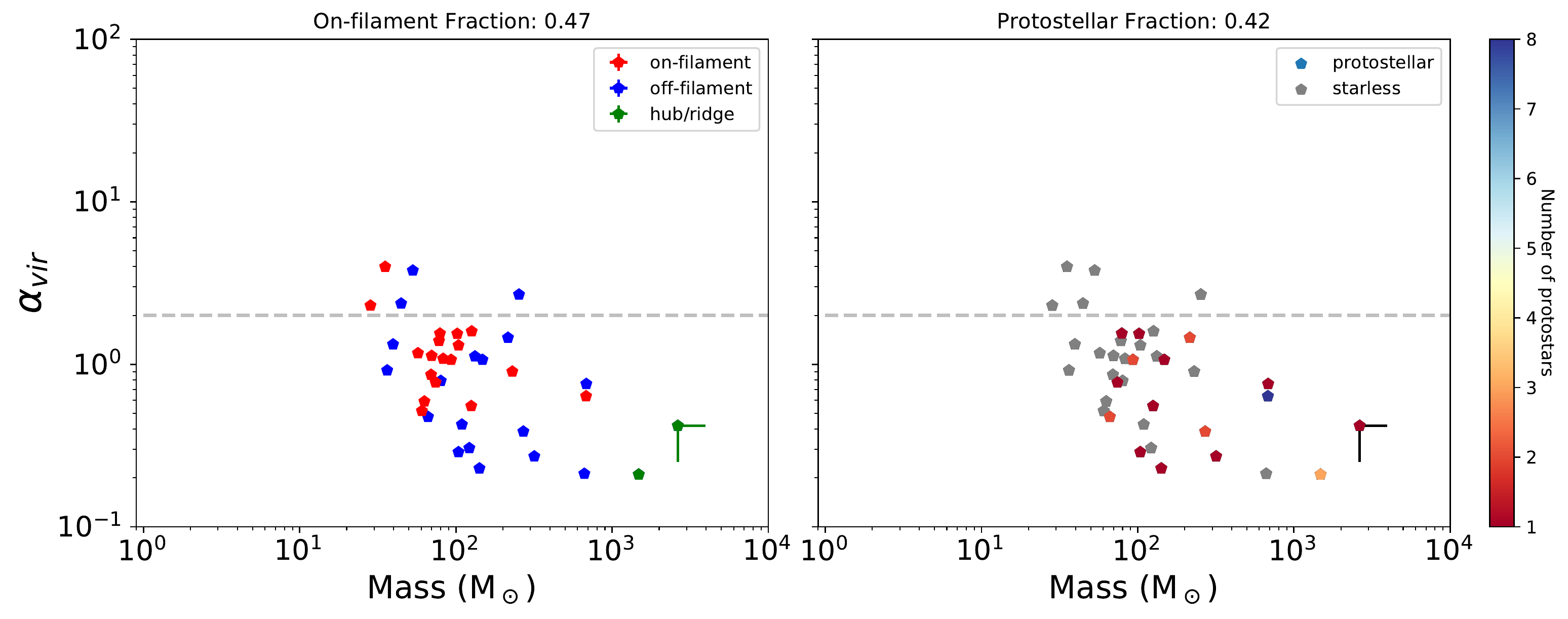}
\caption{Same as Figure \ref{W3_filaments} for M17.}
\label{M17_filaments}
\end{figure}

\begin{figure}[ht]
\epsscale{1.05}
\plotone{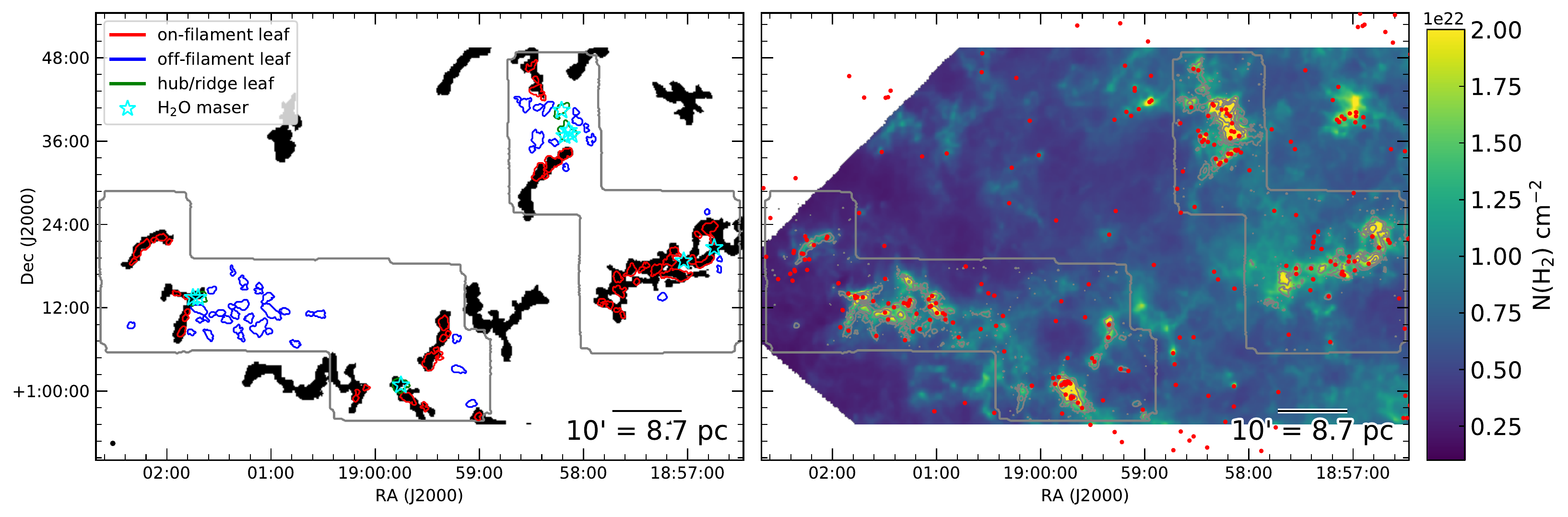}
\epsscale{1.0}
\plotone{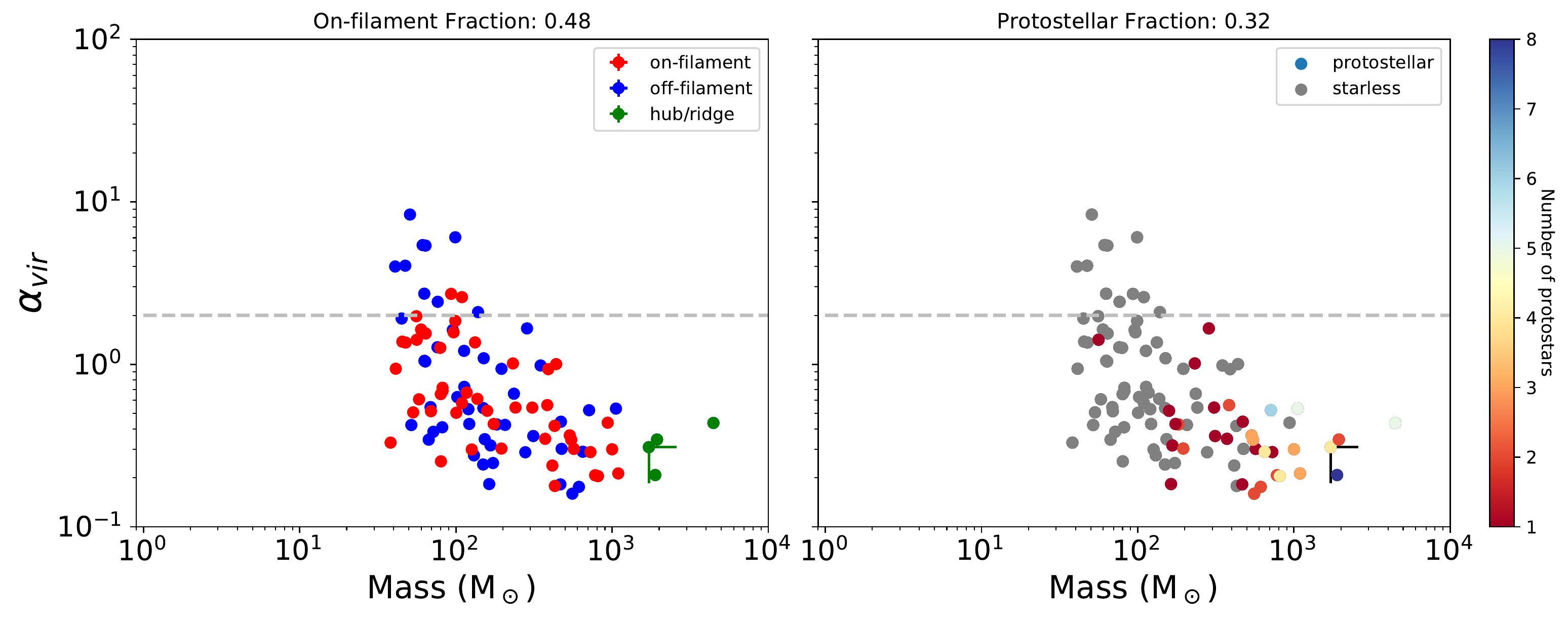}
\caption{Same as Figure \ref{W3_filaments} for W48.}
\label{W48_filaments}
\end{figure}

\begin{figure}[ht]
\epsscale{1.05}
\plotone{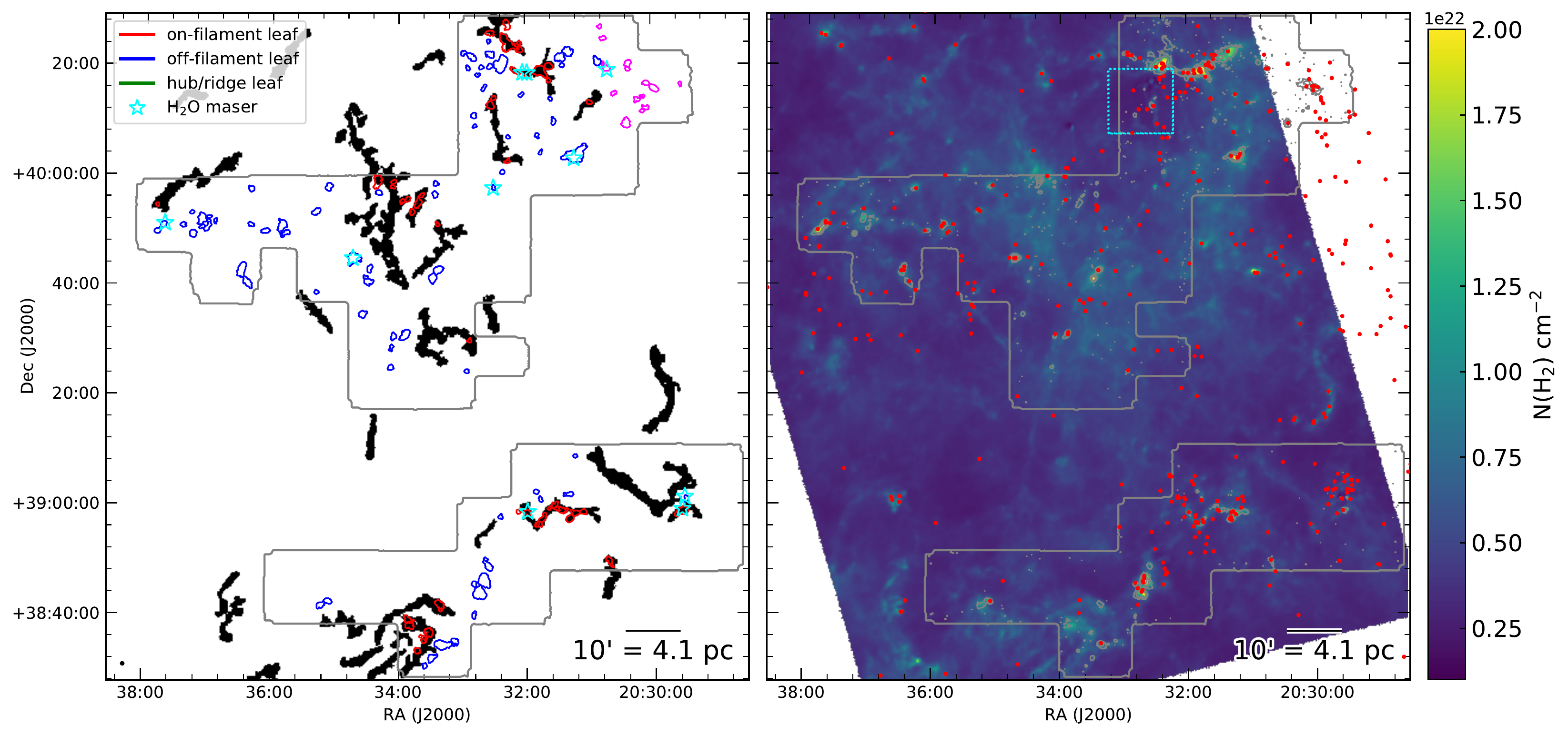}
\epsscale{1.0}
\plotone{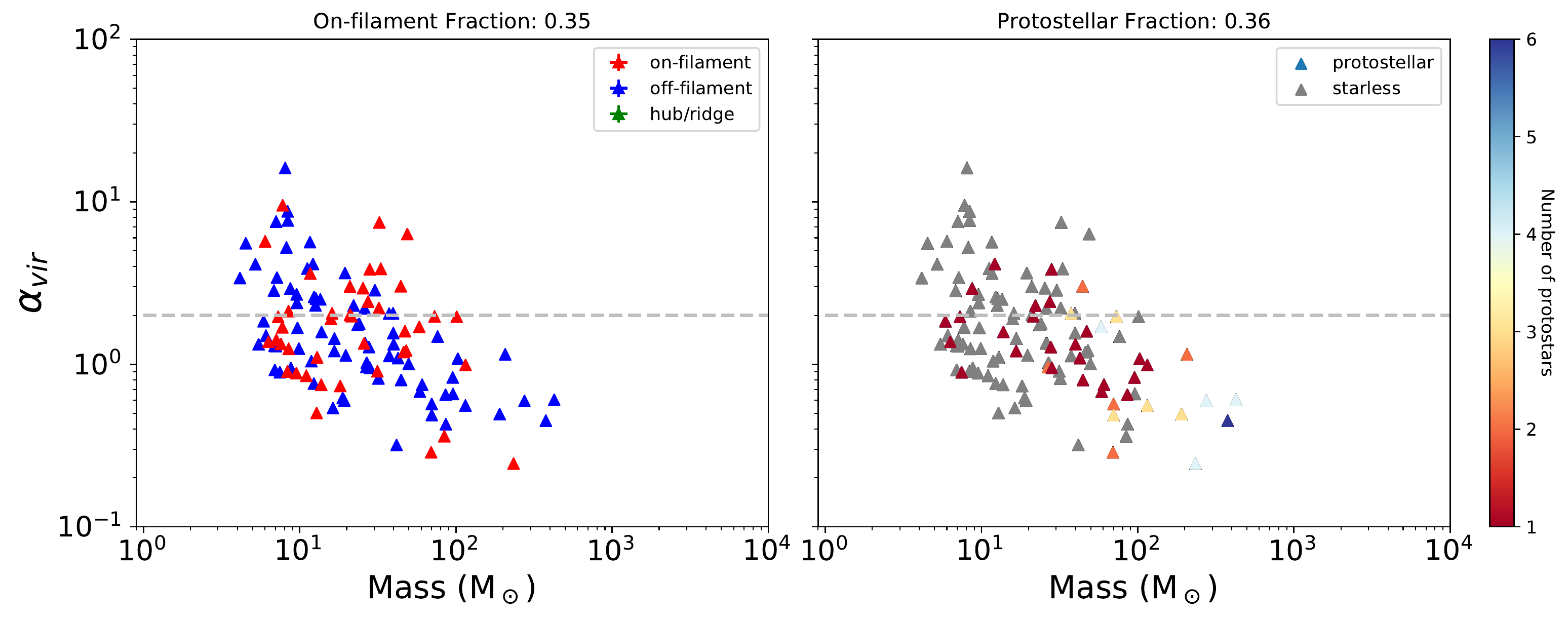}
\caption{Same as Figure \ref{MonR2_filaments} for Cygnus X South.}
\label{CygX_S_filaments}
\end{figure}

\begin{figure}[ht]
\epsscale{1.05}
\plotone{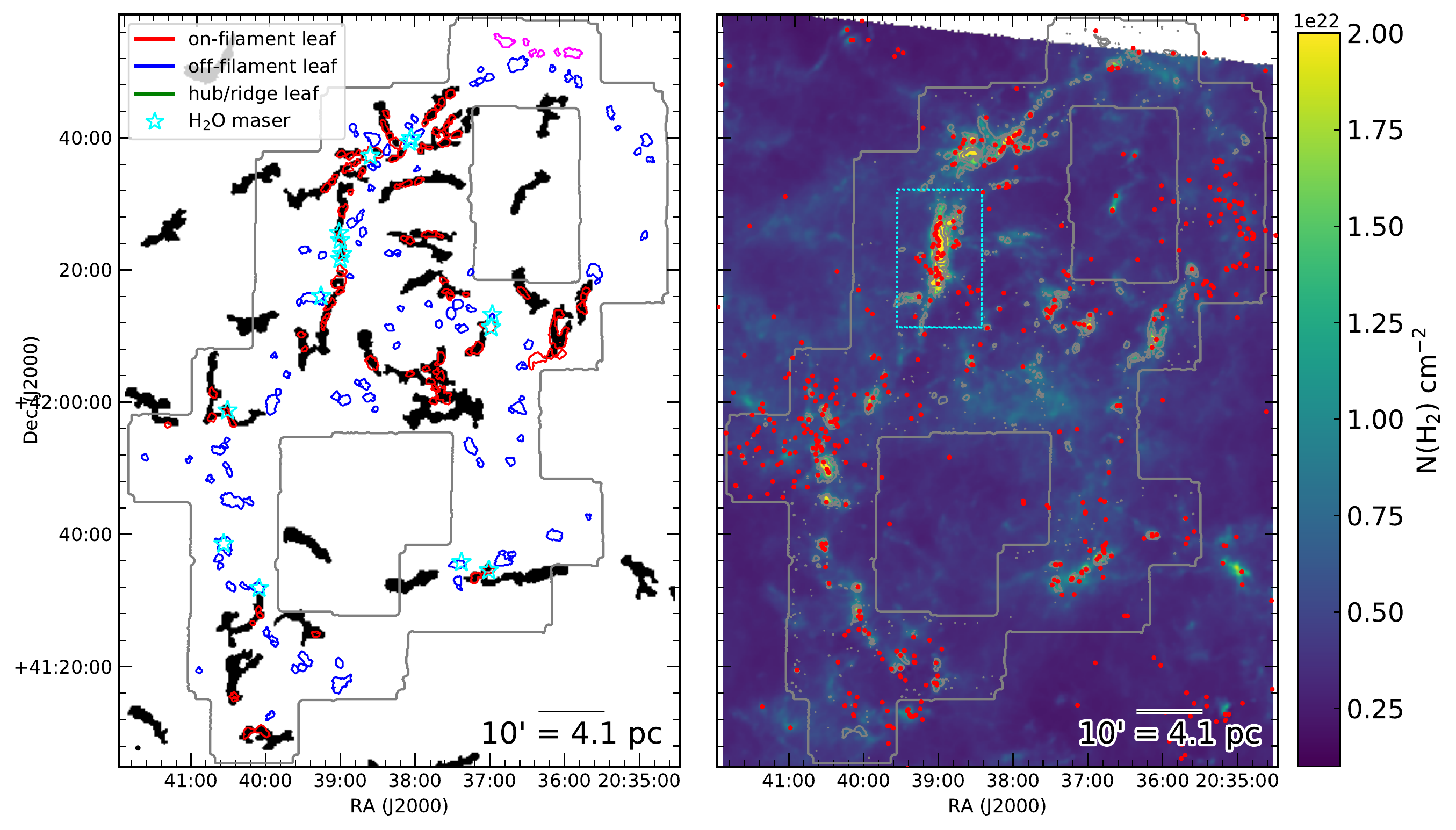}
\epsscale{1.0}
\plotone{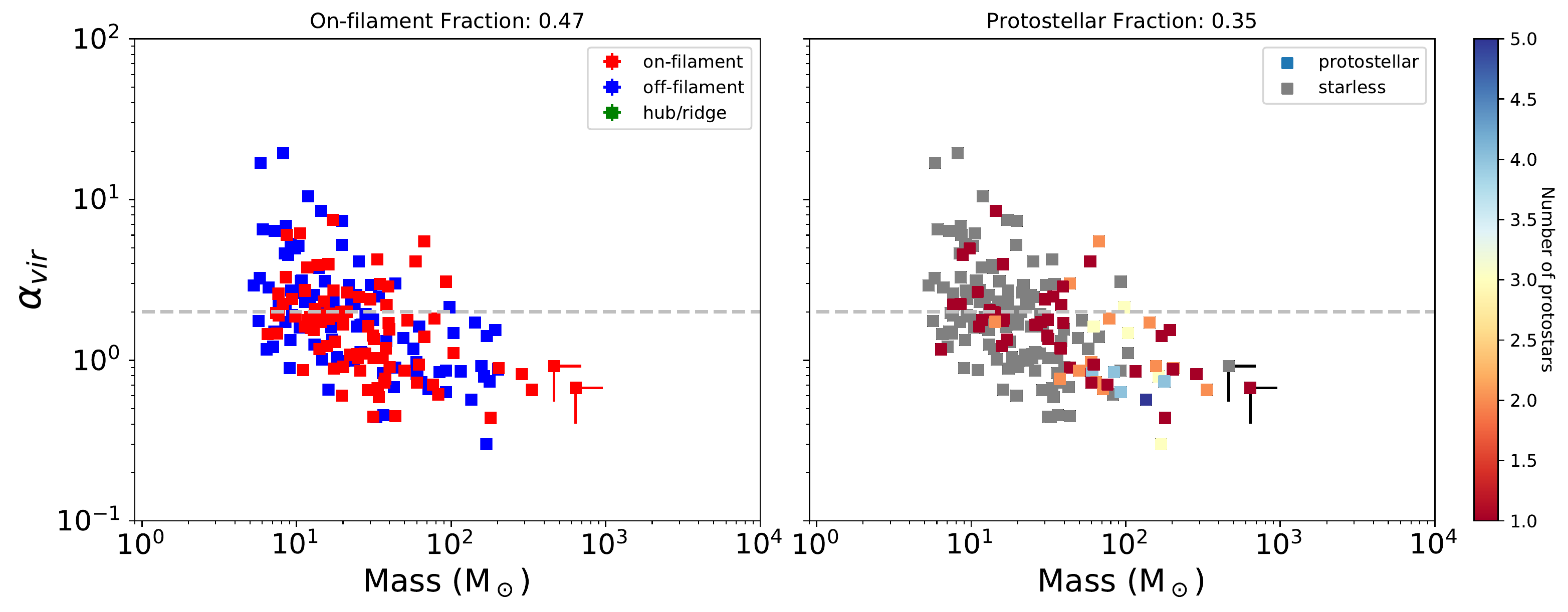}
\caption{Same as Figure \ref{MonR2_filaments} for Cygnus X North.}
\label{CygX_N_filaments}
\end{figure}

\begin{figure}[ht]
\epsscale{1.05}
\plotone{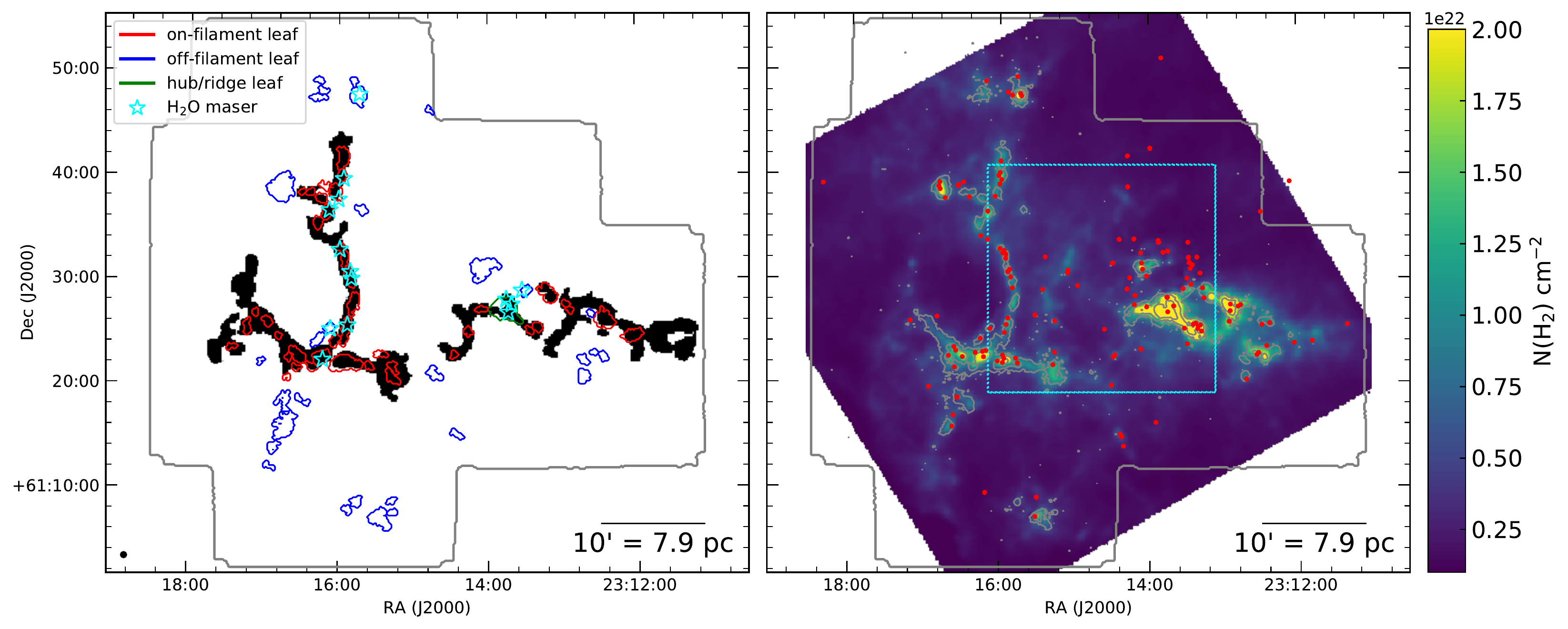}
\epsscale{1.0}
\plotone{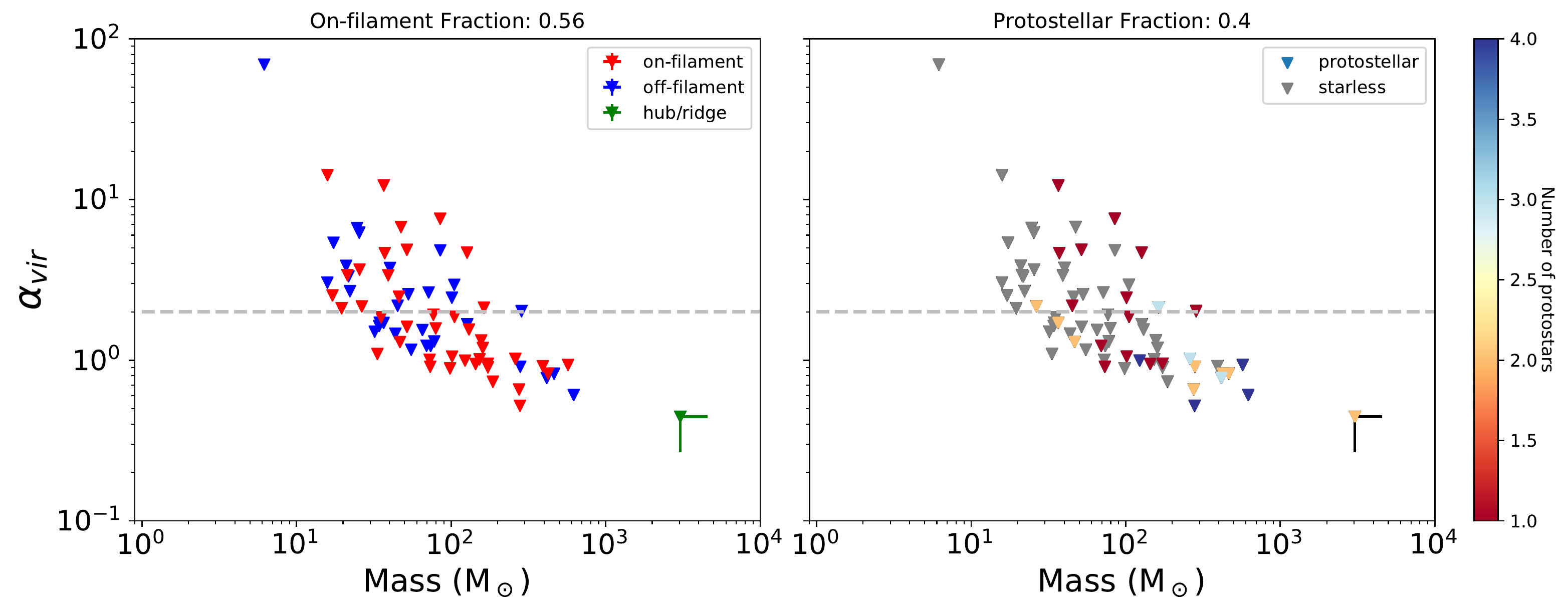}
\caption{Same as Figure \ref{W3_filaments} for NGC7538.}
\label{NGC7538_filaments}
\end{figure}

Although dendrograms are able to identify the hierarchical parent structures in which leaves are embedded, they are not optimized for isolating the elongated, filamentary structures that are commonly observed in molecular clouds.  To understand how the ammonia-identified leaves in this paper relate to surrounding filamentary structures, we employ a dedicated filament extraction algorithm called \texttt{getfilaments} \citep{Menshchikov_2013} to identify filaments in each region's H$_2$ column density map.  The \texttt{getfilaments} algorithm is a multi-scale extraction approach designed to identify filamentary background structures in \textit{Herschel} maps \citep[e.g.,][]{Konyves_2015, Rivera_Ingraham_2016, Marsh_2016, Bresnahan_2018}.  As such, it performs far better than dendrograms at identifying filaments.  \texttt{getfilaments} was run on all the \textit{Herschel} H$_2$ column density maps using the standard extraction parameters for the algorithm (see \cite{Menshchikov_2013} for the extensive list of \texttt{getfilaments} parameters).  The top left panels in Figures \ref{W3_filaments}-\ref{NGC7538_filaments} display the final filament masks, reconstructed up to spatial scales of 145$\arcsec$.

A leaf that has at least one of its pixels corresponding to at least one pixel in a filament mask is designated ``on-filament'' and all other leaves are termed ``off-filament.''  The ``on-filament fraction,'' i.e., the fraction of leaves in a cloud that are on-filament, is listed in Table \ref{Table_clouds} and ranges from 0.35 in Cygnus X South to 1.0 in W3-west.  For the full sample of 835 leaves with $\textit{Herschel}$ observations, 454 are on-filament (on-filament fraction of $\sim 54 ~\%$).

To compare the number of star-forming leaves in each KEYSTONE region, we identify young, embedded protostars using the \textit{Herschel} $70~\mu$m maps observed for each cloud.  \texttt{getsources} \citep{Menshchikov_2012}, a multi-scale source extraction algorithm designed to identify dense cores and protostars in \textit{Herschel} observations, was employed to extract point sources at 70 $\mu$m only.  We adopt the \textit{Herschel} Gould Belt Survey selection criteria for candidate young stellar objects (YSOs) described in Section 4.5 of \cite{Konyves_2015}.  The final candidate YSOs are shown as red circles in Figures \ref{W3_filaments}-\ref{NGC7538_filaments}.  We note that at the distances of the KEYSTONE clouds (0.9 kpc $< d <$ 3.0 kpc), there may be significant incompleteness plus insufficient resolution to separate close sources in the \textit{Herschel} 70 $\mu$m maps.  For some regions, there may also be contamination by photodissociation regions that can appear as 70 $\mu$m point sources.  Since we are only using these 70 $\mu$m point sources to indicate which leaves are currently star-forming, rather than using them as a complete catalog of YSOs, the extraction is sufficient for our goals.

We perform a cross-match between the candidate YSO catalogs and leaf catalogs to determine which leaves are protostellar.  Leaves with at least one candidate YSO falling within their dendrogram-identified boundary are designated ``protostellar.''  Conversely, leaves without a candidate YSO are termed ``starless.''  The protostellar leaf fraction is listed in Table \ref{Table_clouds} and ranges from 0.17 in MonR1 to 0.58 in W3-west.  For the 835 leaves with $\textit{Herschel}$ observations, 288 are protostellar ($\sim 34 ~\%$).

\clearpage

\subsection{Cloud Population Statistics}

To understand the impact environment has upon the star formation in the KEYSTONE clouds, we search for relationships between ten variables of interest: leaf on-filament fraction, ``bound'' leaf fraction (i.e., the fraction of leaves with $\alpha_{vir} < 2$), protostellar leaf fraction, total dense gas mass, total protostar count, total leaf count, dense gas surface mass density, surface protostar density, median cloud kinetic temperature, and cloud distance.  The dense gas mass, total protostar count, dense gas surface mass density, and surface protostar density are calculated over the KEYSTONE-mapped boundaries of the cloud where the NH$_3$ (1,1) integrated intensity is greater than 1.0 K km s$^{-1}$.  The dense gas mass is defined as the integrated H$_2$ column density within the NH$_3$ (1,1) 1.0 K km s$^{-1}$ integrated intensity contour, while the total protostar count is defined as the number of candidate YSOs identified by \texttt{getsources} within that same contour.  The threshold of NH$_3$ (1,1) integrated intensity above 1.0 K km s$^{-1}$ was chosen since it typically highlights the extent of pixels that were robustly fit during our line-fitting procedure (see, e.g., the lowest NH$_3$ (1,1) integrated intensity contours in Figures \ref{W3_params}-\ref{NGC7538_params}).  Median values of H$_2$ column density within the NH$_3$ (1,1) 1.0 K km s$^{-1}$ integrated intensity contours are typically around $1\times10^{22}$ cm$^{-2}$, with a minimum of $4.2\times10^{21}$ cm$^{-2}$ measured in W3-west and a maximum of $2.3\times10^{22}$ cm$^{-2}$ measured in M17. 

\begin{figure}[ht]
\epsscale{1.0}
\plotone{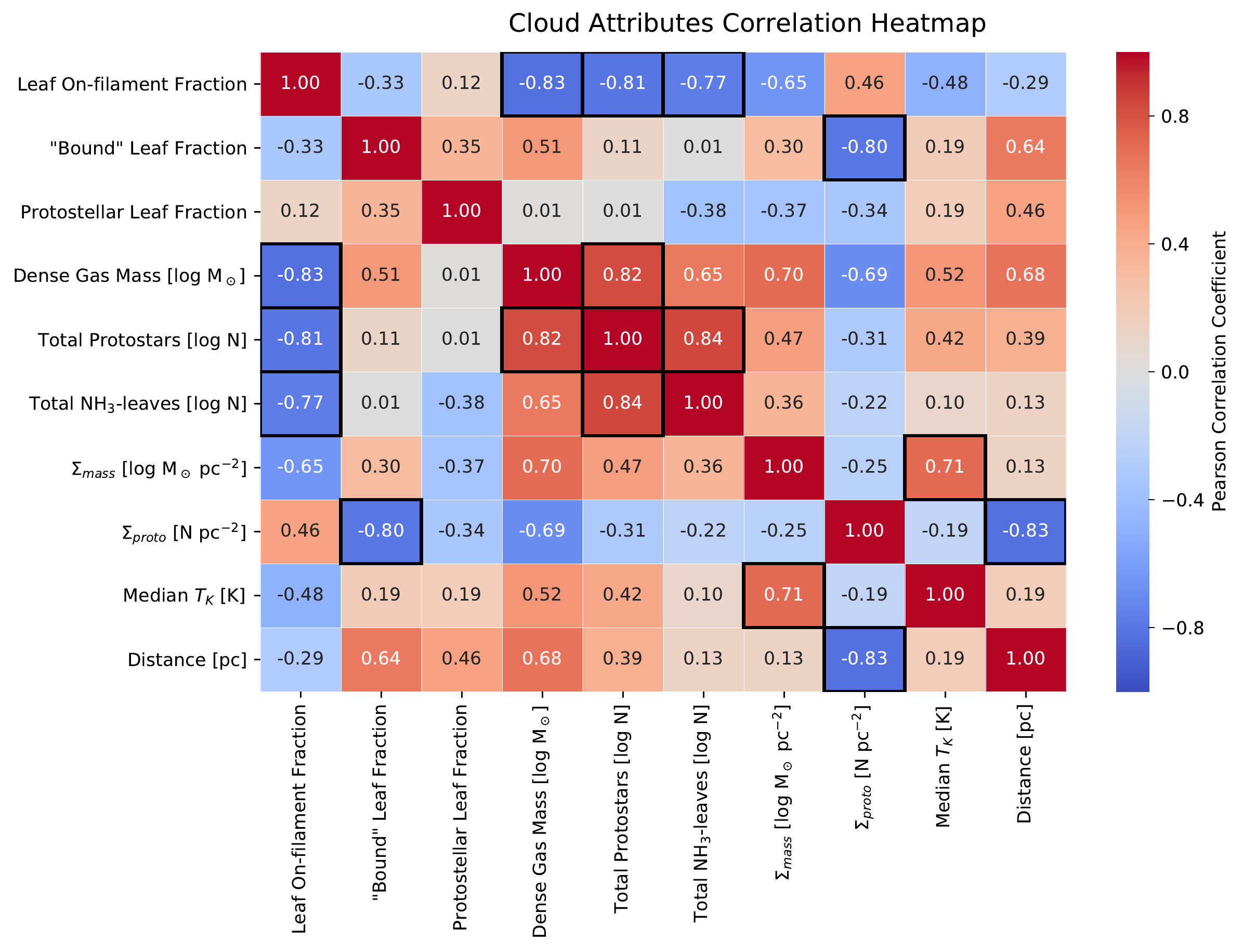}
\caption{Heatmap of Pearson correlation coefficients for the ten variables of interest examined: leaf on-filament fraction, ``bound'' leaf fraction (i.e., the fraction of leaves with $\alpha_{vir} < 2$), protostellar leaf fraction, total dense gas mass, total protostar count, total leaf count, dense gas surface mass density, surface protostar density, median cloud kinetic temperature, and cloud distance.  The colorbar ranges from $-1$ (perfect anti-correlation) to 1 (perfect positive correlation).  Panels with a black outline denote statistically significant correlations that are displayed in Figure \ref{scatters}.}
\label{correlations}
\end{figure}

\begin{figure}[ht]
\epsscale{0.75}
\plottwo{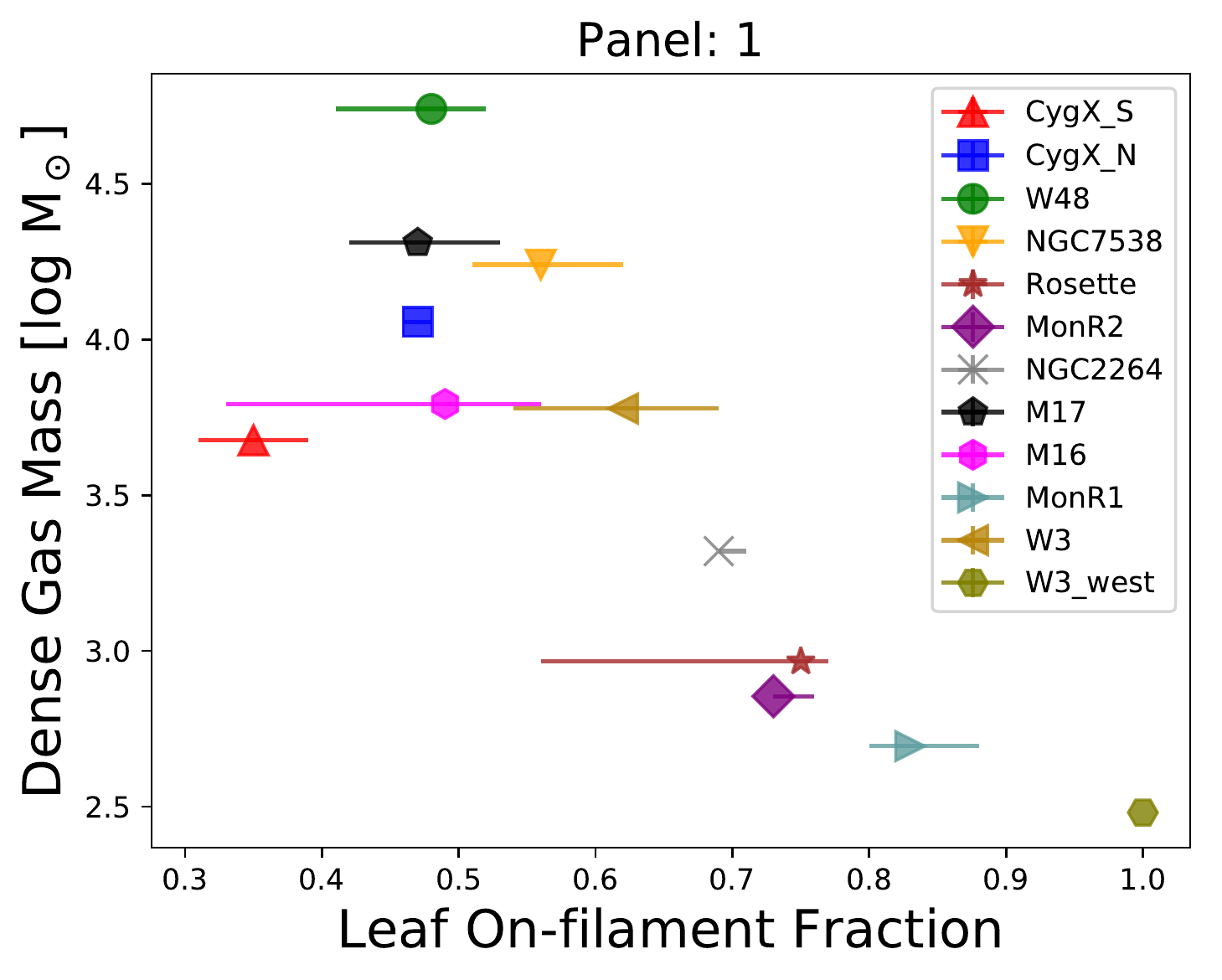}{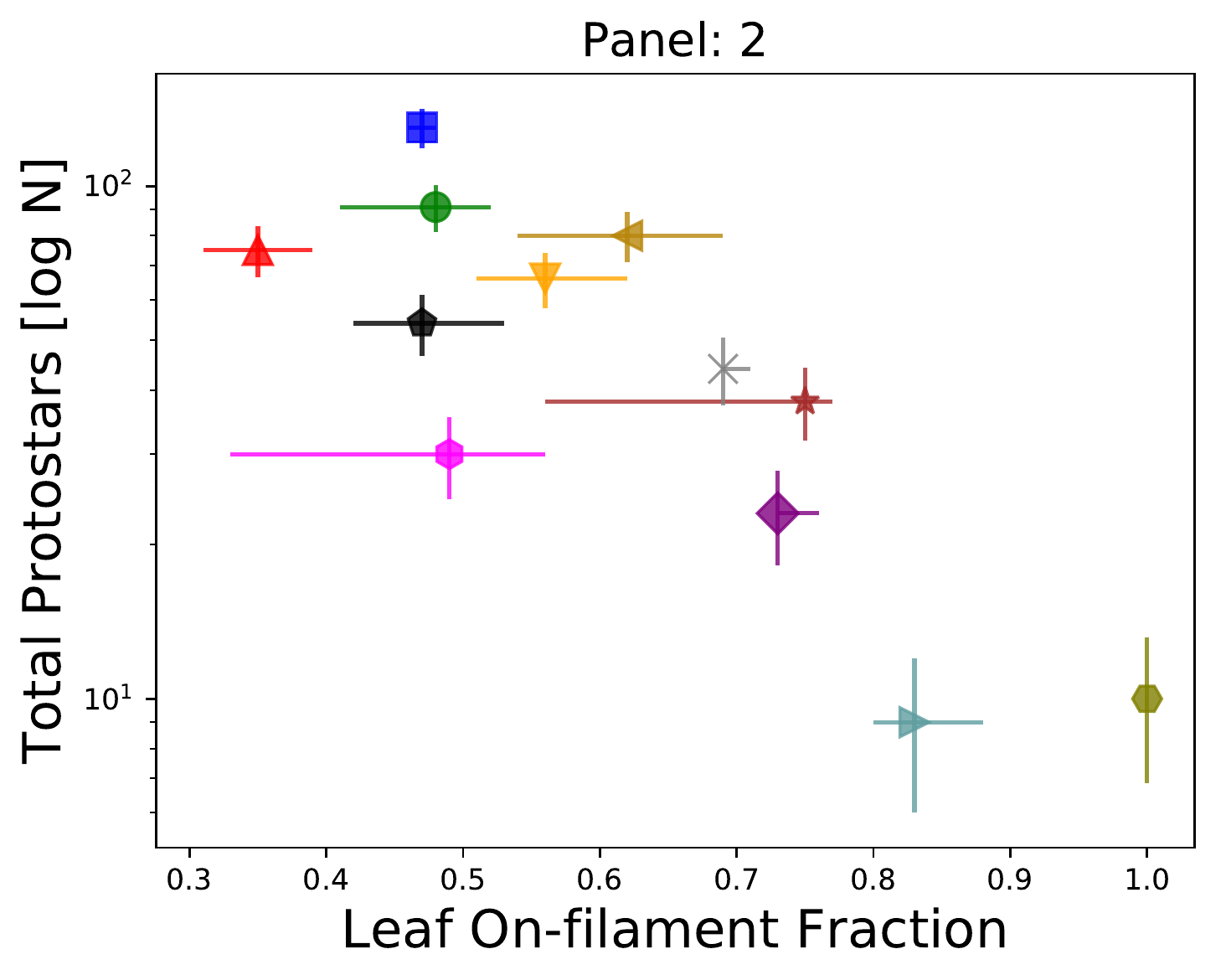}
\plottwo{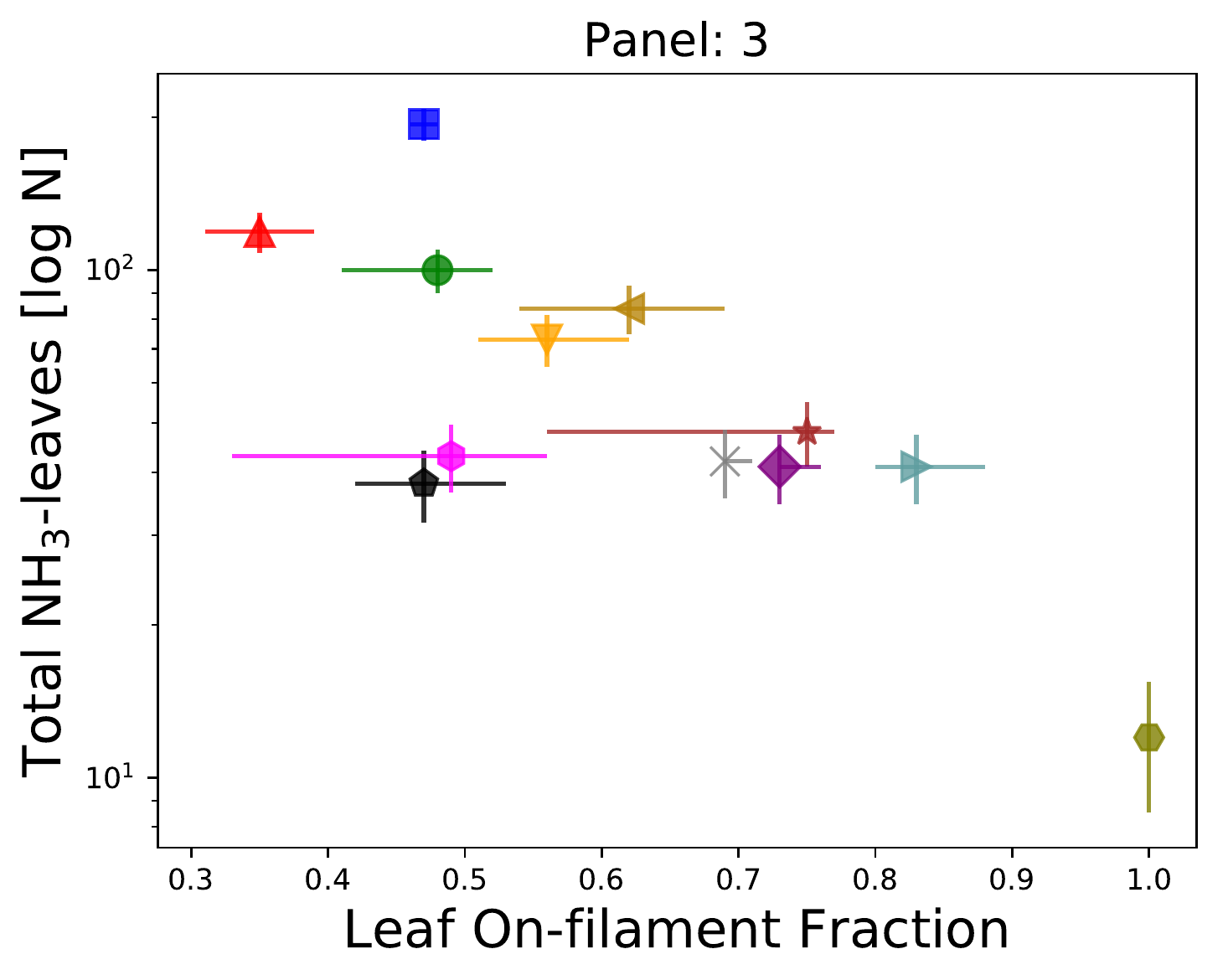}{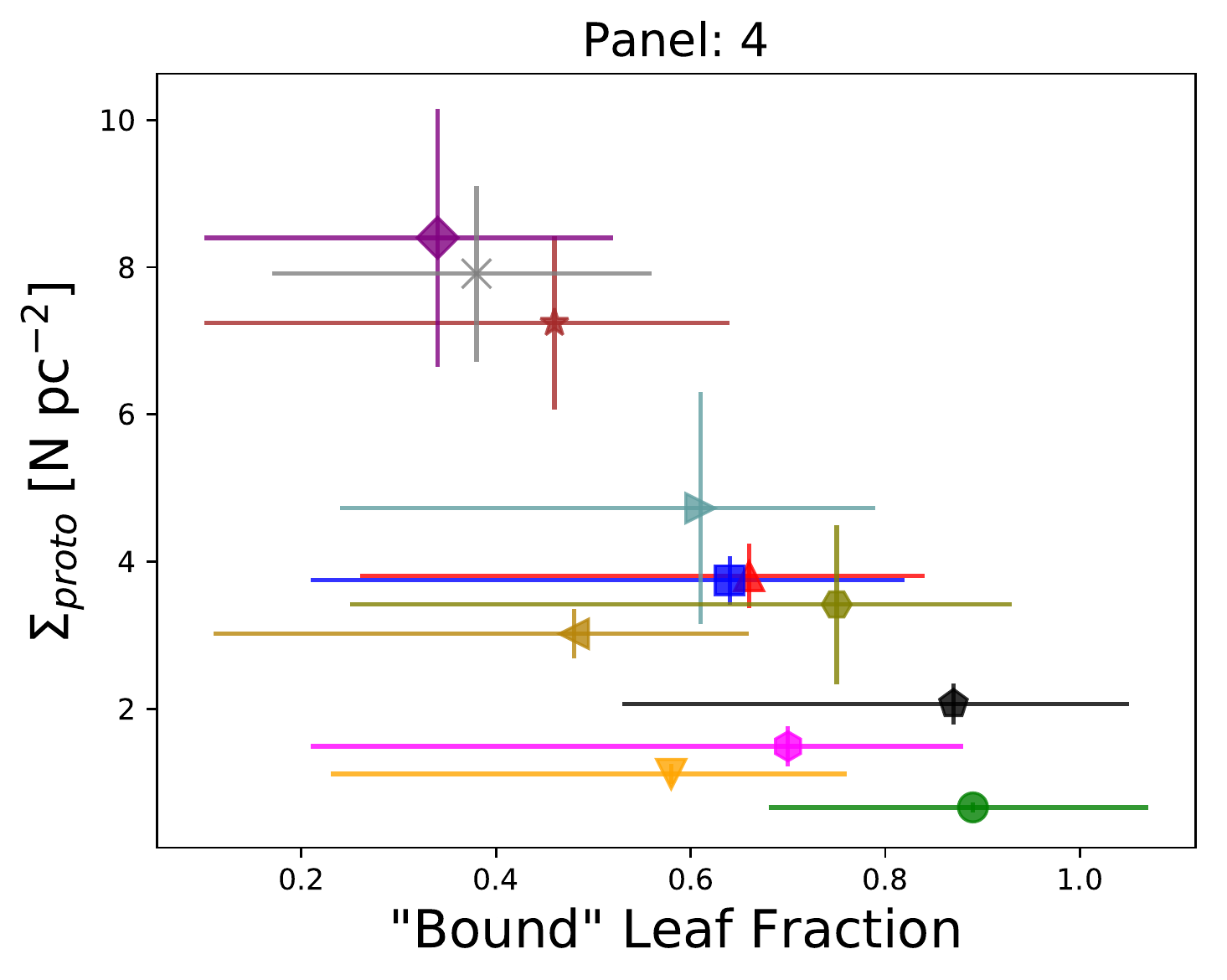}
\plottwo{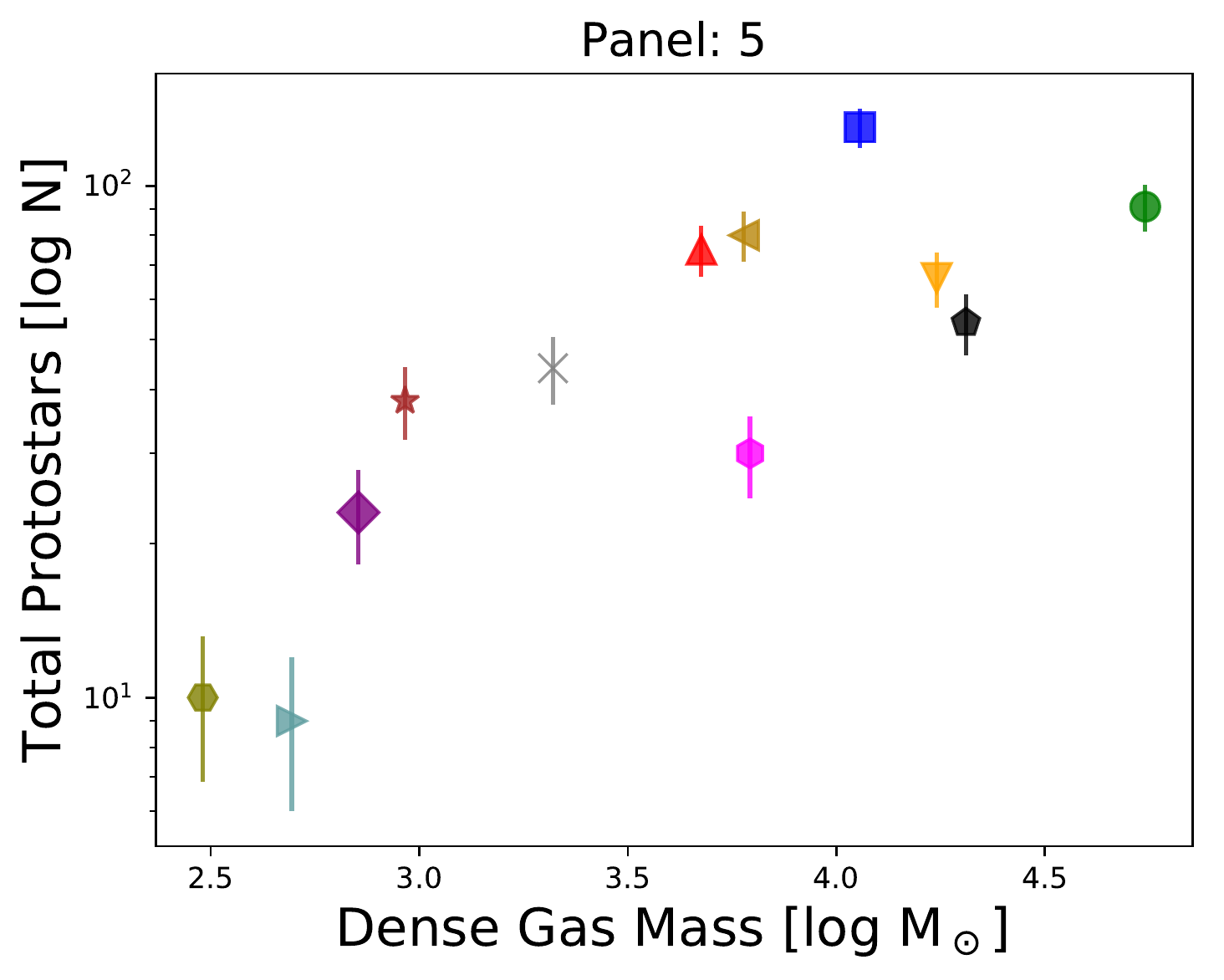}{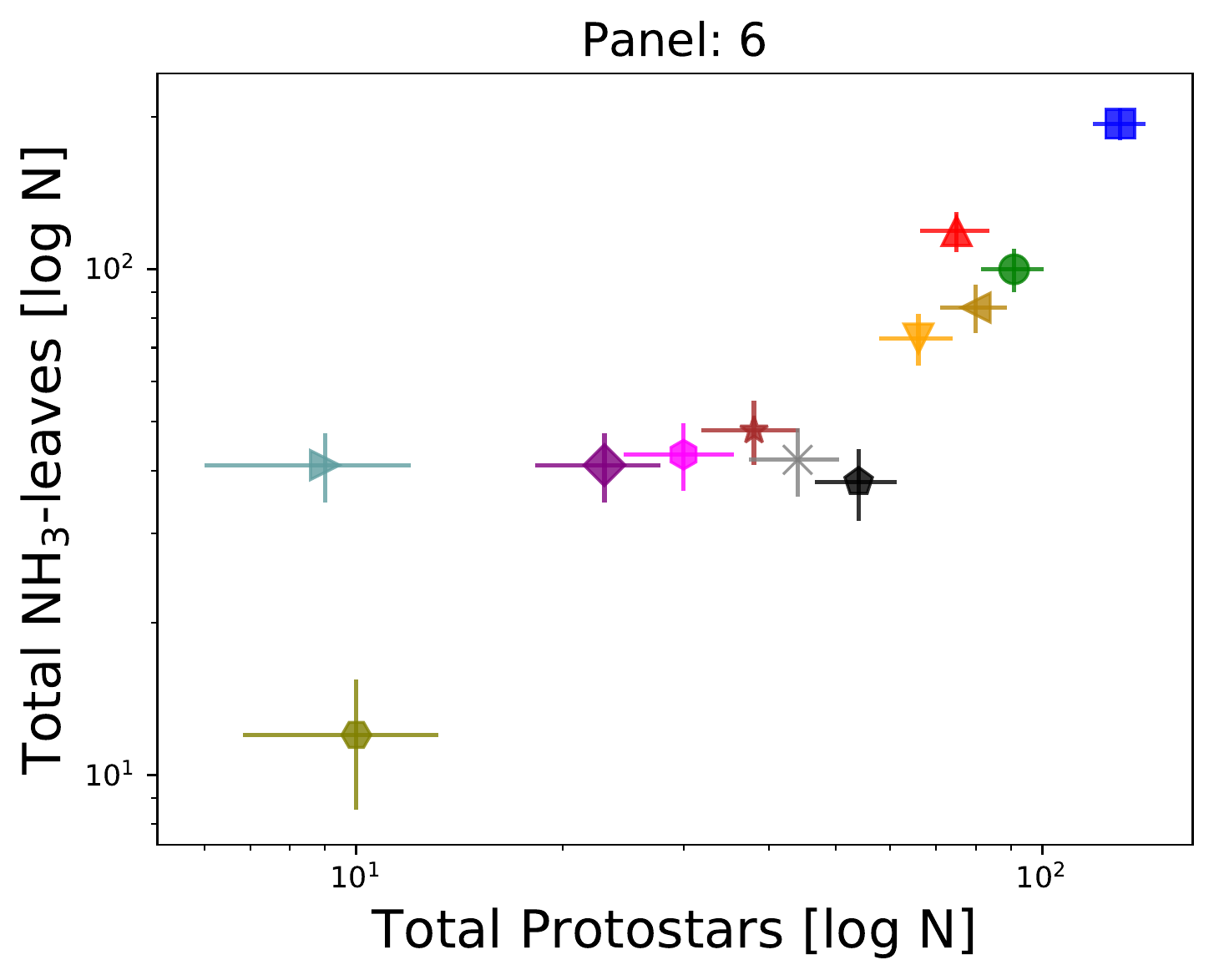}
\centering
\plottwo{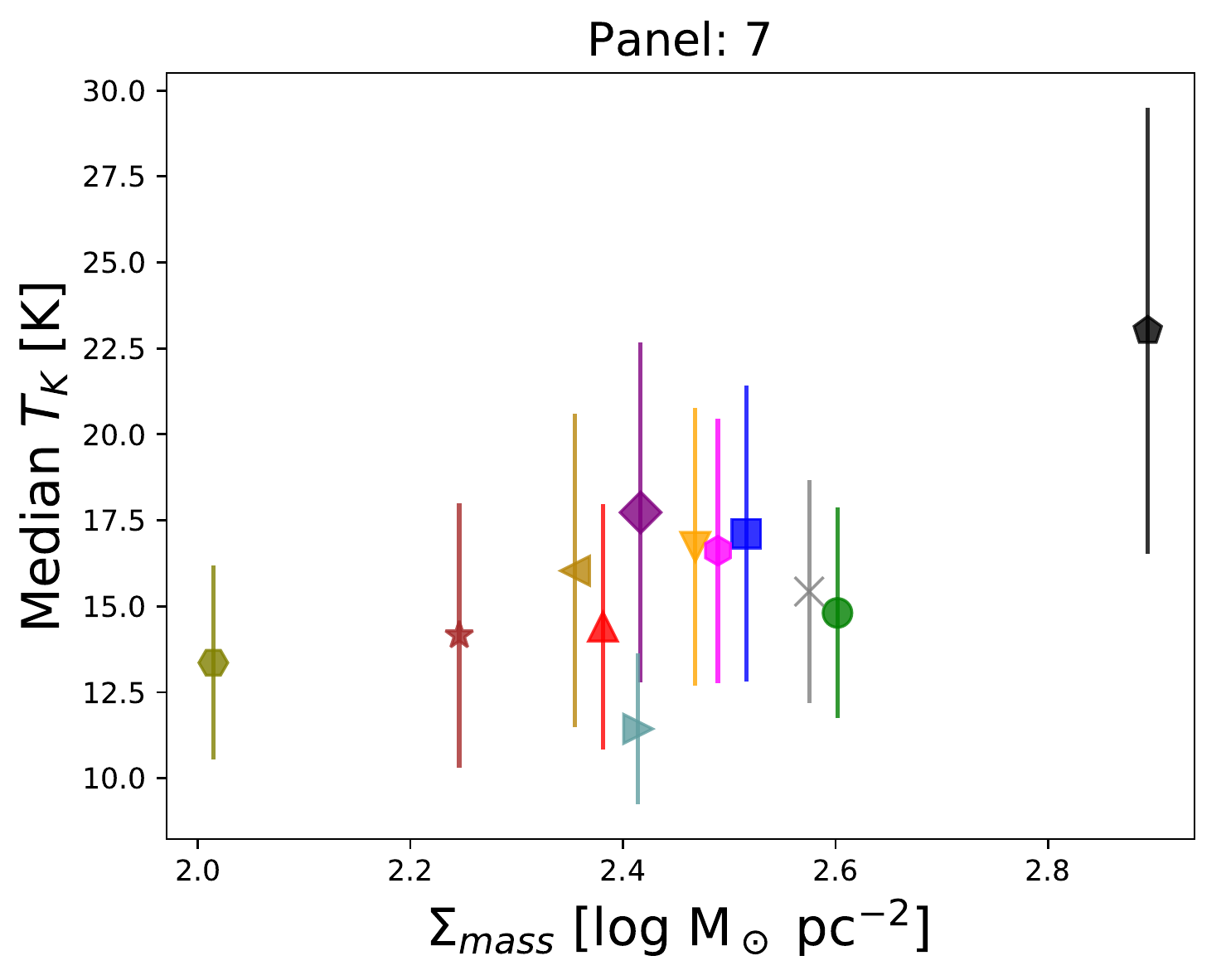}{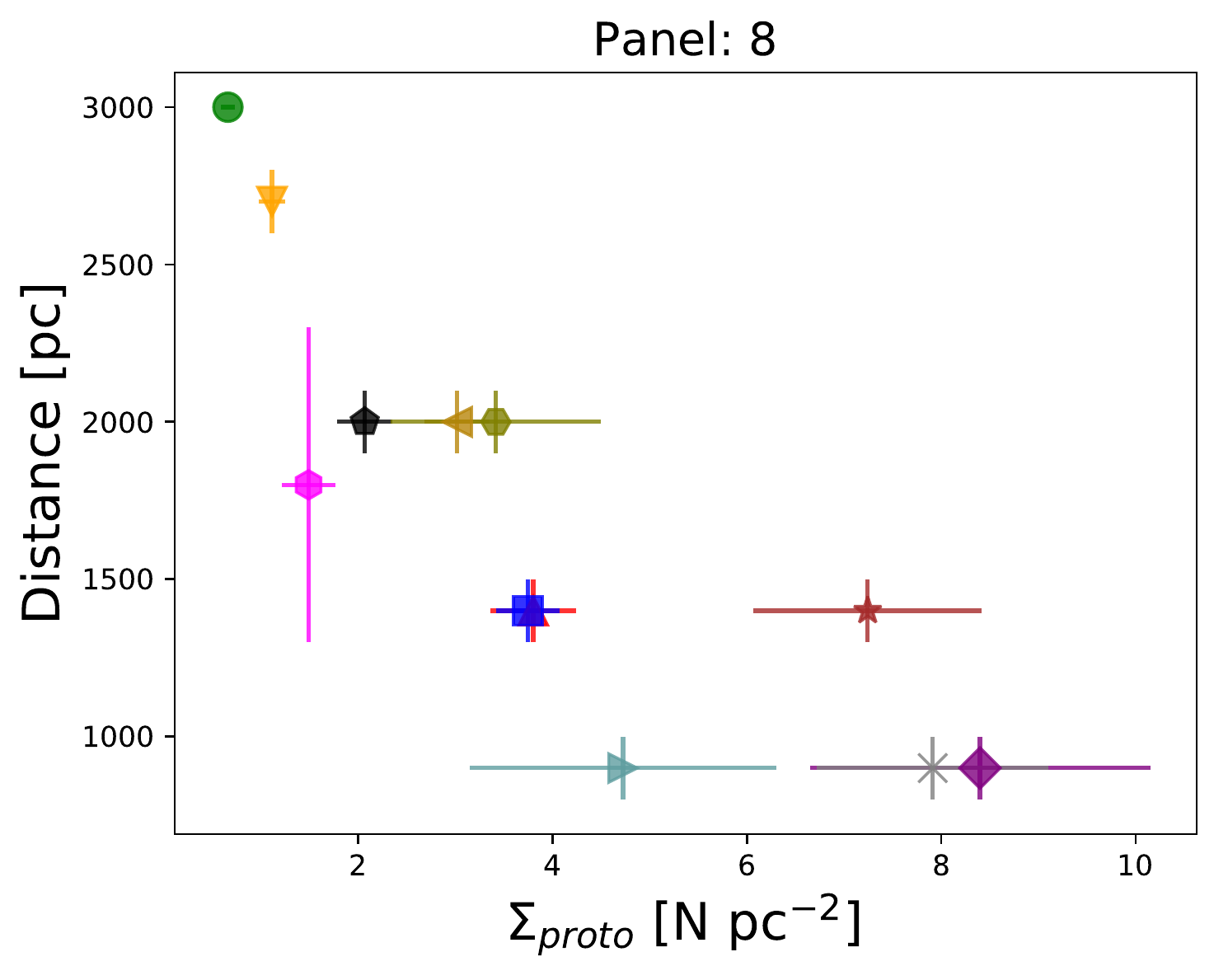}

\caption{Statistically significant correlations found from Figure \ref{correlations} when neglecting data errorbars.  The dense gas mass, number of protostars, and surface protostar density are calculated within the boundaries of the KEYSTONE-mapped regions in each cloud where NH$_3$ (1,1) integrated intensity was greater than 1.0 K km s$^{-1}$.  The errorbars on each data point are calculated as described in Section 3.7.}
\label{scatters}
\end{figure}

Figure \ref{correlations} shows the Pearson correlation coefficients for the matrix of ten variables.  The Pearson correlation coefficient can range from $-1$ to 1, where $-1$ and 1 signify the data fall on a straight line with a negative or positive correlation, respectively. A Pearson correlation coefficient of zero indicates there is no correlation between the variables.  For 12 data points and using the Student's $t$-distribution to test for statistical significance, the null hypothesis that the data have no relationship is rejected at the $99.5\%$ confidence level when the Pearson correlation coefficient is greater than $\sim |0.71|$.  As such, correlations that meet this threshold are outlined by black in Figure \ref{correlations} and the corresponding scatter plots are shown in Figure \ref{scatters}. 



Eight statistically significant correlations were found in the data: 1) decreasing leaf on-filament fraction with increasing dense gas mass, 2) decreasing leaf on-filament fraction with increasing number of protostars, 3) decreasing leaf on-filament fraction with increasing number of leaves, 4) decreasing bound leaf fraction with increasing protostellar surface density, 5) increasing dense gas mass with increasing number of protostars, 6) increasing number of protostars with increasing number of leaves, 7) increasing surface mass density with increasing temperature, and 8) decreasing protostellar surface density with increasing cloud distance.

Since the Pearson correlation coefficient does not take into consideration the uncertainties on each data point, we visualize the scatter of each parameter in Figure \ref{scatters} by adding errorbars as follows: The lower and upper leaf on-filament fraction errorbars represent the on-filament fractions obtained when using the \texttt{getfilaments} filament masks reconstructed up to spatial scales of $72\arcsec$ and $290\arcsec$, respectively, rather than the $145\arcsec$ scale map.  Errorbars for the bound leaf fraction reflect the fractions obtained when assuming the virial parameters for each leaf are at the extremes of their individual uncertainty range.  Errorbars for the total protostars, protostellar surface density, and NH$_3$-leaves represent the $\sqrt{N}$ counting uncertainty. Errorbars for the median kinetic temperature represent the median absolute deviation of each cloud's $T_K$ distribution.

Many of these correlations can be explained with our current understanding of the star formation process.  For instance, the positive correlation between dense gas mass and protostars (Panel 5 in Figure \ref{scatters}) is related to the relationship between star formation rate (SFR) and dense gas mass that is well-established \citep[e.g.,][]{Gao_2004, Wu_2005, Wu_2010, Lada_2012, Stephens_2016, Burkhart_2018}.  Similarly, the correlation we observe between protostars and total ammonia-identified leaves (Panel 6 in Figure \ref{scatters}) is also related to the SFR-Mass relation since the leaves are tracing the dense gas mass in each cloud.  Moreover, the negative relationship observed between leaf on-filament fraction and dense gas mass (Panel 1 in Figure \ref{scatters}) may also be loosely related to the SFR-Mass relation.  Since the lower mass clouds in our sample tend to have higher leaf on-filament fractions, this trend may suggest that the star formation in those environments is more heavily dependent on filaments creating the high densities required to form ammonia leaves.  Clouds with higher dense gas mass, however, can form clumps and stars even when filaments are not present due to their more widespread dense gas.  The same argument can be applied to the anti-correlations observed between leaf on-filament fraction versus protostars and ammonia-identified leaves (Panels 2 and 3 in Figure \ref{scatters}).  The SFR-Mass relation could also explain the positive relationship between temperature and surface mass density displayed in Panel 7 of Figure \ref{scatters} since a higher SFR could lead to higher gas temperatures.  Panel 7 has the lowest Pearson coefficient absolute value (0.71) of all the relationships shown in Figure \ref{scatters} and is dominated by two data points, however, which suggests it is not as robust as the other relationships presented.  

In addition, the negative trend observed between bound leaf fraction and protostellar surface density (Panel 4 in Figure \ref{scatters}) may be related to the heating and turbulence injected into the cloud by protostars \citep[e.g.,][]{Krumholz_2008, Hansen_2012, Offner_2017, Offner_2018, Cunningham_2018}.  As the protostellar density increases, the virial parameters of the leaves may increase due to the higher velocity dispersions and temperatures caused by nearby protostellar (or cluster) feedback (e.g., radiation and outflows).  Such a scenario is also suggested by the magneto-hydrodynamic simulations of \cite{Offner_2017}, which showed that cores become unbound soon after ($<0.1$ Myr) the onset of protostar formation due to outflow-induced turbulence.  Lastly, the negative correlation between protostar surface density and cloud distance shown in Panel 8 is likely related to the larger areas observed for the more distant clouds, which would lower their protostar surface densities.  Since protostellar surface density is the only parameter significantly correlated with distance, the distance dependency of the other parameters shown in Figure \ref{correlations} is likely minimal.

\clearpage

\clearpage

\section{Discussion}
\subsection{Leaf/Filament Relationship}
A relationship between dense cores and filamentary structures in dust continuum observations has been noted in several nearby star-forming environments.  \cite{Polychroni_2013} show that $67\%$ of the dense cores they identified with CuTEx, a Gaussian fitting and background subtraction source extraction algorithm developed by \cite{Molinari_2011}, were coincident with filamentary structures in L1641 of Orion A.  Using the same filament identification algorithm adopted in this paper, \cite{Konyves_2015}, \cite{Marsh_2016}, and \cite{DiFrancesco_prep} also found that $75\%$, $40\%$, and $40-80\%$ of starless dense cores in Aquila, Taurus-L1495, and five regions of the Cepheus Flare, respectively, are coincident with filamentary structures.  In addition, hydrodynamical simulations of molecular clouds \citep[e.g.,][]{Offner_2013} also show a strong correspondence between cores and filamentary structures \citep{Mairs_2014}.  Although the KEYSTONE clouds analyzed in this paper are located at much farther distances ($0.9-3.0$ kpc) than L1641 (400 pc), Aquila ($250-450$ pc), Taurus-L1641 (140 pc), and Cepheus ($\sim300$ pc), we find consistent values for the on-filament fraction ($\sim0.4-1.0$ for KEYSTONE clouds).  


Despite the apparent relationship observed between leaves and filaments, we find no significant variations between the virial parameters of the on- and off-filament leaf populations in any of the clouds.  For the 454 on-filament leaves,  294 ($\sim 65 ~\%$) have $\alpha_{vir} < 2$, which is consistent with the bound fraction for the entire leaf population ($\sim 63 ~\%$).  Furthermore, Figure \ref{leaf_hists} shows that the mass, effective radius, average kinetic temperature, and average velocity dispersion for the on-filament and off-filament leaves are essentially identical.  Although the more distant clouds (e.g., W48 and NGC7538) tend to have larger masses and radii than the nearest clouds in our sample (e.g., MonR1, MonR2, and NGC2264; see Appendix A for a discussion of the distance dependency of our results), the similarities between the on-filament and off-filament leaf parameter distributions appear to hold for the individual clouds as well. These similarities indicate that star formation away from filaments might be equally as likely as star formation on filaments in high-mass GMCs since dense gas may be more widespread in those environments.  Such a scenario is also suggested by the anti-correlation found between leaf on-filament fraction and dense gas mass discussed in Section 3.7.  As dense gas becomes more widespread in high-mass GMCs, the fraction of star formation taking place on filaments may decrease since dense gas is equally as likely to be found away from filaments as it is to be found within filaments.  


We also note the existence of a group of ammonia-identified leaves with uncharacteristically larger masses ($10^{2}-10^3$ M$_\odot$) than the majority of leaves in their respective clouds.  Deemed ``hubs'' or ``ridges'' based on the nomenclature suggested in \cite{Myers_2009}, these high-mass leaves are shown by the color green in Figures \ref{W3_filaments}-\ref{NGC7538_filaments}.  The hubs are located at the intersection of multiple filaments (e.g., MonR2, NGC2264, eastern and northern regions in W48) and the ridges are massive filaments (e.g., NGC7538, M16, southern region of W48).  Due to their high masses, these structures all have low virial parameters ($\alpha_{vir} = 0.2-0.5$) and are likely gravitationally bound or collapsing.  As such, the hubs and ridges may be a result of mass build-up at the locations where filaments are transporting mass from other parts of the cloud.

In addition to the hubs and ridges being coincident with filaments, their mass and size indicate they are likely the precursors of massive young stellar objects and stellar clusters.  For example, the right panel of Figure \ref{Reff_mass} shows the Mass-Radius plot for the hub/ridge and non-hub/ridge leaves in relation to the \cite{Kauffmann_2010} empirically-derived threshold for massive star formation: m(r) $>$ 870 M$_\odot$ (r/pc)$^{1.33}$.  The hubs and ridges tend to be above this threshold, indicating that they will likely form high-mass stars.  As shown in Figure \ref{leaf_hists}, the hubs and ridges tend to have higher masses (median log($M_{obs}$) = $3.2 \pm 0.5$), larger radii (median $R_{eff}$ = $0.7 \pm 0.3$), warmer temperatures (median $T_{K,avg}$ = $19.5 \pm 5.1$), and larger velocity dispersions (median $\sigma = 0.7 \pm 0.1$) than the starless leaf population.  Instead, the hub and ridge masses, radii, and virial parameters are more similar to the massive star-forming clumps identified by \cite{Urquhart_2015}, highlighting further their propensity to form massive stars.  Furthermore, the hubs and ridges tend to align with the positions of H$_2$O maser emission (identified by eye and shown as cyan stars in Figures \ref{W3_filaments}-\ref{NGC7538_filaments}) also detected by KEYSTONE (White et al., in prep.), which is frequently associated with massive young stellar objects.  If dense gas hubs and ridges are indeed the current, or future, sites of massive young stellar object and stellar cluster formation, it would explain the high correspondence observed between those objects and filament intersections \citep[e.g.,][]{Myers_2009, Schneider_2012, Hennemann_2012, Li_2016, Motte_2017}.  

\subsection{Leaf/Protostar Relationship}

\begin{figure}[ht]
\epsscale{1.15}
\plottwo{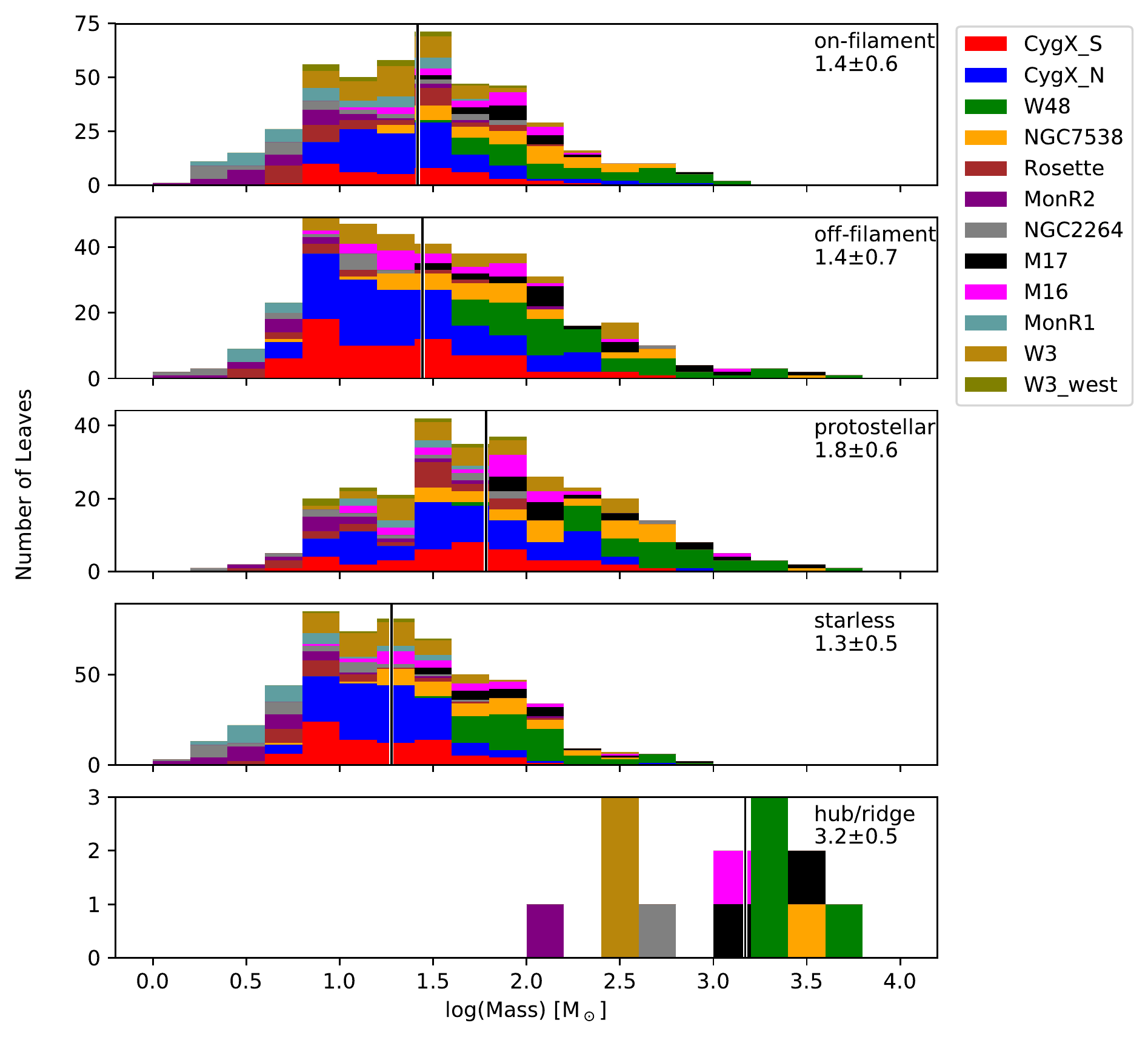}{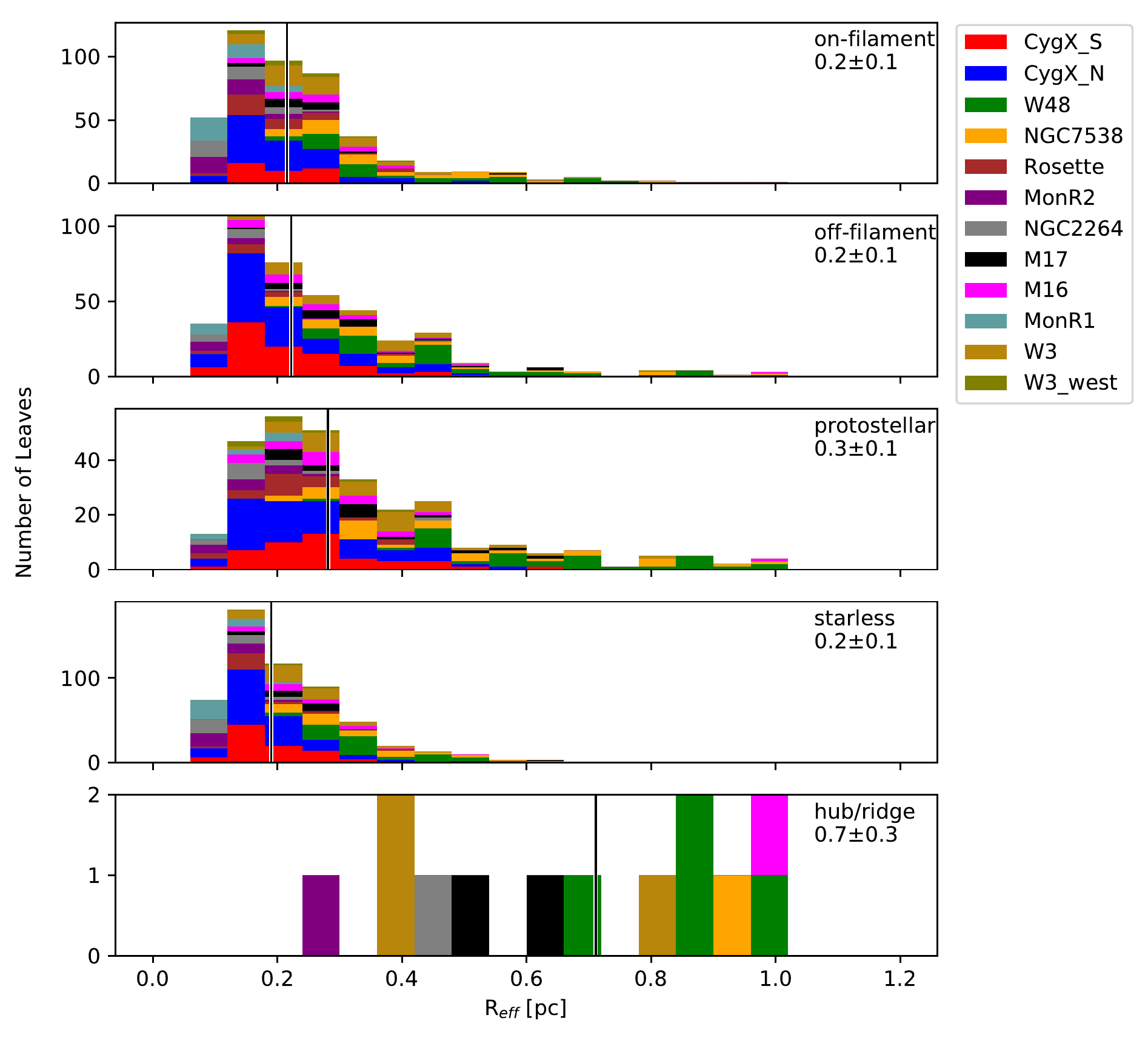}
\plottwo{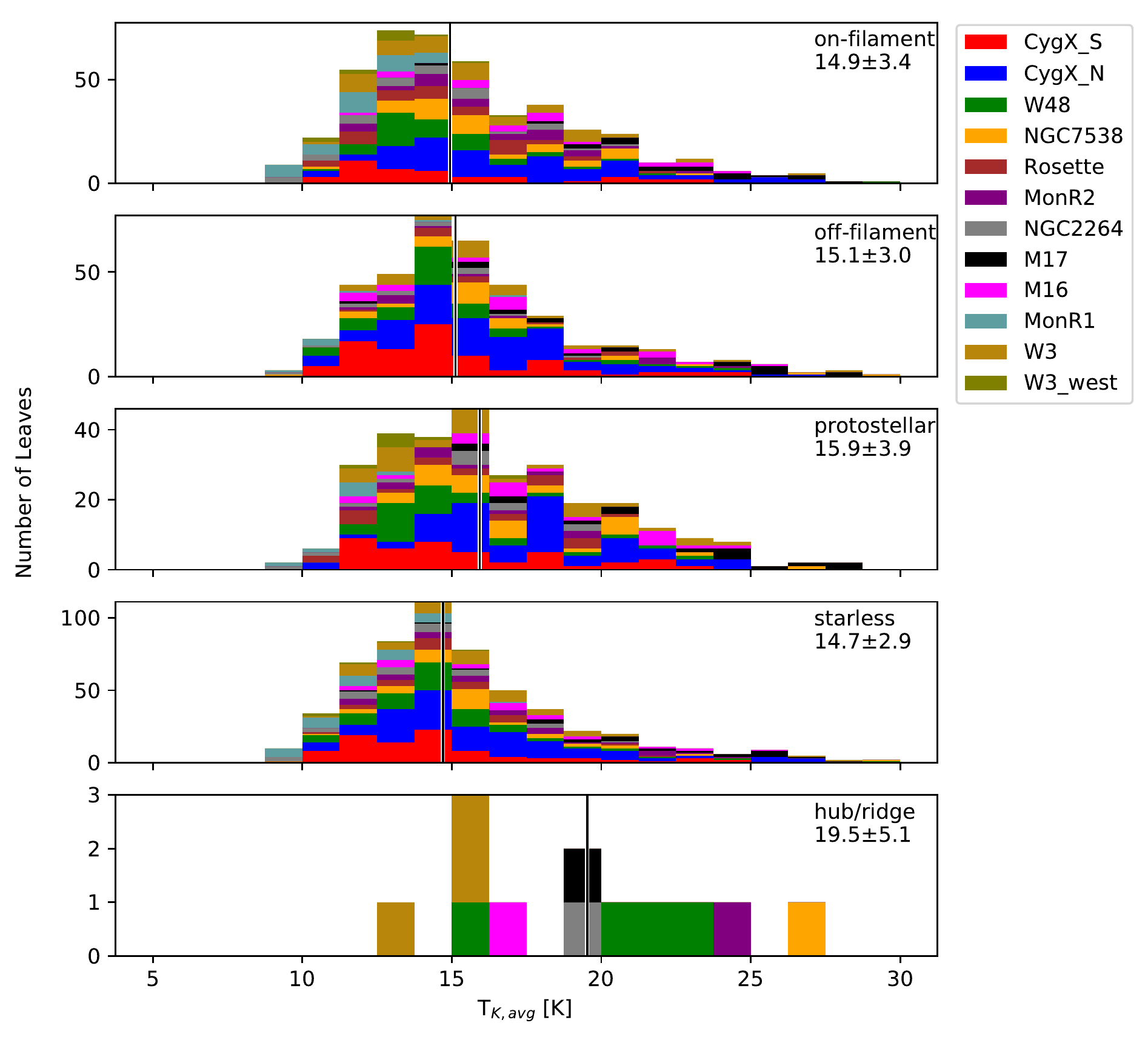}{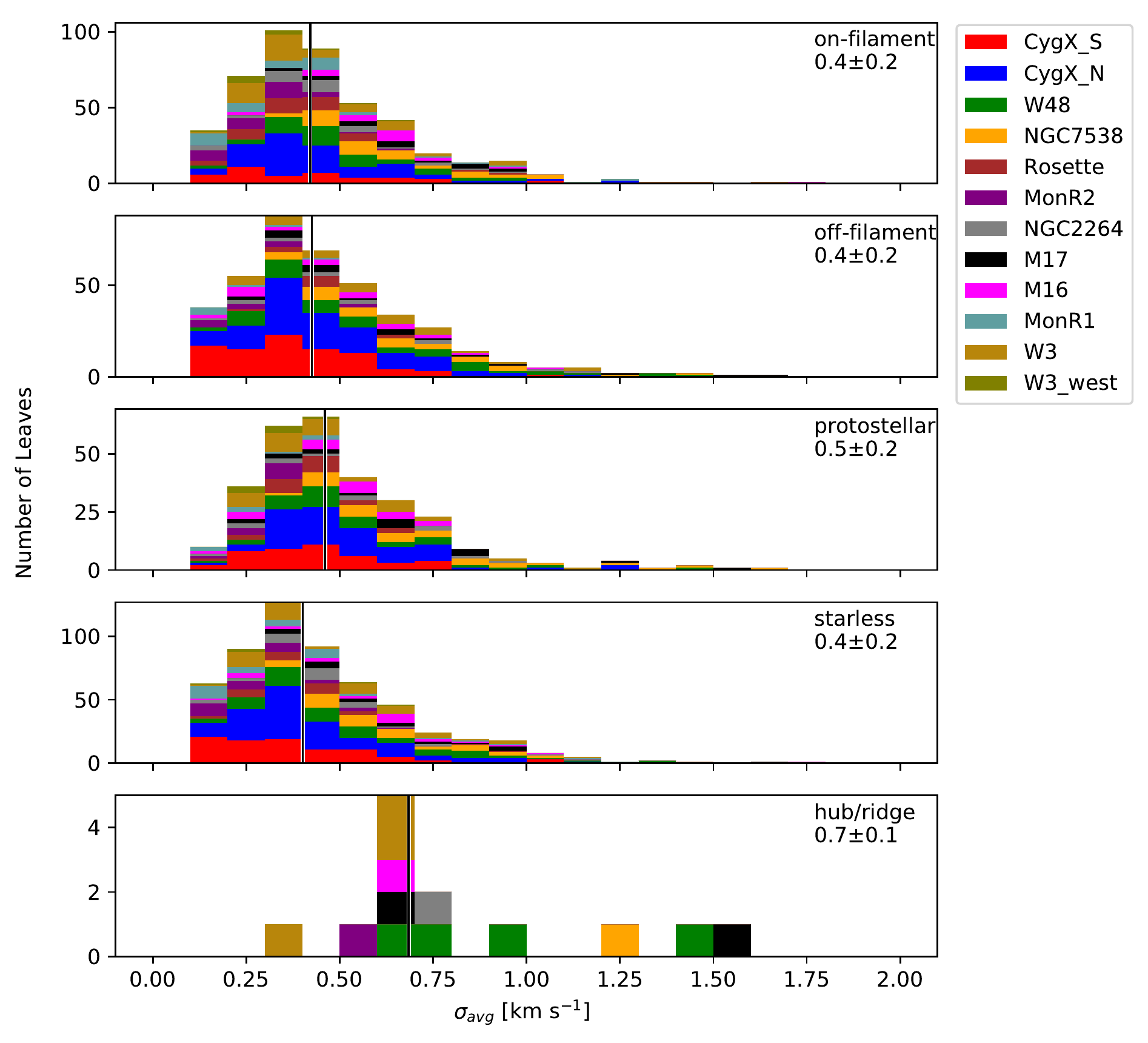}
\caption{Stacked histograms of mass (upper left), effective radius (upper right), average kinetic temperature (lower left), and average velocity dispersion (lower right) for the leaves identified as on-filament, off-filament, protostellar, starless, and hub/ridge.  The median and median absolute deviation of the distributions are shown in the upper right corner of each panel.  The black vertical lines show the distribution median.}
\label{leaf_hists}
\end{figure}

Figures \ref{W3_filaments}-\ref{NGC7538_filaments} also show the virial parameters of the leaves identified as protostellar and starless in each of the clouds observed by KEYSTONE.  In many regions (e.g., MonR1, Rosette, Cygnus X North, MonR2, W48), the protostellar core population clearly has larger masses and lower virial parameters than the starless core population.  For the 288 protostellar leaves identified, 229 ($\sim 80 ~\%$) have $\alpha_{vir} < 2$.  In comparison, 294 of the 547 starless leaves identified ($\sim 54 ~\%$) have $\alpha_{vir} < 2$.  As shown in Figure \ref{leaf_hists}, the protostellar population's mass distribution peaks at higher values (Median log(Mass) = 1.8 $\pm$ 0.6 M$_\odot$) than that of the starless population (Median log(Mass) = 1.3 $\pm$ 0.5 M$_\odot$), which could explain the lower protostellar virial parameters.  In addition, the hubs and ridges identified in the previous section tend to host multiple protostars (see, e.g., NGC7538, NGC2264, W48, W3, M16).  In several regions, the hubs and ridges host over six protostars.  Since the 70 $\mu$m maps are typically confusion-limited in the hubs and ridges, their protostar counts are likely under-estimated. This attribute highlights the exceptional environment hubs and ridges provide for cluster formation.  

\clearpage


\begin{deluxetable}{cccccccccc}
\tabletypesize{\tiny}
\tablewidth{0pt}
\tablecolumns{10}
\tablecaption{Cloud Statistics}
\tablehead{  \colhead{Cloud} &  T$_{med}$ & $\sigma_{med}$ &  Protostars &   Area &   Mass &  Total Leaves &  Filament Leaves &  Proto Leaves  &  Bound Leaves \\
& K & km s$^{-1}$ &  & pc$^2$ & M$_\odot$ & & fraction & fraction & fraction}
\startdata
      W3 &       16.0 $\pm$        4.6 &             0.5 $\pm$             0.3 &        80 &   26.5 &   5999 &          84 &                0.62 &                 0.38 &          0.48 \\
 W3-west &       13.4 $\pm$        2.8 &             0.3 $\pm$            0.1 &        10 &    2.9 &    303 &          12 &                1.00 &                 0.58 &          0.75 \\
  MonR2 &       17.7 $\pm$        4.9 &             0.3 $\pm$             0.2 &        23 &    2.7 &    715 &          41 &                0.73 &                 0.27 &          0.34 \\
   MonR1 &       11.4 $\pm$        2.2 &             0.2 $\pm$            0.2 &         9 &    1.9 &    494 &          41 &                0.83 &                 0.17 &          0.61 \\
 Rosette &       14.2 $\pm$       3.8 &             0.4 $\pm$             0.1 &        38 &    5.2 &    924 &          48 &                0.75 &                 0.42 &          0.46 \\
 NGC2264 &       15.4 $\pm$        3.2 &             0.5 $\pm$             0.2 &        44 &    5.6 &   2091 &          42 &                0.69 &                 0.29 &          0.38 \\
     M16 &       16.6 $\pm$        3.8 &             0.5 $\pm$            0.2 &        30 &   20.1 &   6207 &          43 &                0.49 &                 0.42 &          0.70 \\
     M17 &       23.0 $\pm$        6.5 &             0.7 $\pm$             0.3 &        54 &   26.1 &  20439 &          38 &                0.47 &                 0.42 &          0.87 \\
     W48 &       14.8 $\pm$       3.1 &             0.5 $\pm$             0.3 &        91 &  137.6 &  55013 &         100 &                0.48 &                 0.32 &          0.89 \\
  Cygnus X South &       14.4 $\pm$        3.6 &             0.4 $\pm$             0.2 &        75 &   19.7 &   4744 &         119 &                0.35 &                 0.36 &          0.66 \\
  Cygnus X North &       17.1 $\pm$        4.3 &             0.4 $\pm$             0.2 &       130 &   34.7 &  11383 &         194 &                0.47 &                 0.35 &          0.64 \\
 NGC7538 &       16.7 $\pm$       4.0 &             0.6 $\pm$             0.3 &        66 &   59.2 &  17393 &          73 &                0.56 &                 0.40 &          0.58 \\

\enddata
\tablecomments{Columns show the following: (1) cloud name, (2-3) median kinetic gas temperature and velocity dispersion for all reliably fit ammonia pixels, (4) number of protostars identified where NH$_3$ (1,1) integrated intensity is greater than 1 K km s$^{-1}$, (5) area of map where NH$_3$ (1,1) integrated intensity is greater than 1 K km s$^{-1}$, (6) total mass where NH$_3$ (1,1) integrated intensity is greater than 1 K km s$^{-1}$, (7) total number of ammonia leaves identified in cloud, (8) fraction of total leaves that are on-filament, (9) fraction of total leaves that are protostellar, (10) fraction of total leaves that are bound ($\alpha_{vir} < 2$).}

\label{Table_clouds}
\end{deluxetable}

\begin{deluxetable}{cc}
\tablewidth{0pt}
\tablecolumns{2}
\tablecaption{Leaf Population Statistics}
\tablehead{\colhead{Leaf Statistic} & \colhead{Fraction}}
\startdata
Bound & 523 of 835 ($\sim 63\%$)\\  
Starless & 547 of 835 ($\sim 66\%$)\\
Protostellar & 288 of 835 ($\sim 34\%$)\\
On-filament & 454 of 835 ($\sim 54\%$)\\
Bound Starless & 294 of 547 ($\sim 54\%$) \\ 
Bound Protostellar & 229 of 288 ($\sim 80\%$) \\   
Bound On-filament & 294 of 454 ($\sim 65\%$) \\
Sub-virial (cloud weight pressure) & 573 of 835 ($\sim 69\%$) \\              
\enddata
\tablecomments{~Leaf population statistics quoted throughout the paper.  All fractions are in relation to the sample of leaves with mass estimates.}
\label{Table_population}
\end{deluxetable}

\subsection{Virial Stability in Low- and High-Mass Star-Forming Regions}
Our final sample of 835 ammonia-identified leaves, for which virial parameters have been measured in a consistent manner, forms one of the largest current samples of dense gas structure virial parameters in high-mass star-forming regions.  While many studies have derived virial parameters for single molecular clouds or sub-samples of clumps/cores \citep[e.g.,][]{Dunham_2010, Schneider_2010_cyg, Urquhart_2011, Tan_2013, Li_2013, Urquhart_2015, Friesen_2016, Kirk_2017, Billington_2019}, few have viewed virial stability across many clouds.  \cite{Wienen_2012} observed ammonia emission from a sample of 862 clumps in the inner Galactic disk, but only $\sim300$ of those had known distance measurements required to estimate masses and virial parameters.  Similarly, \cite{Kauffmann_2013} compiled a catalog of 1325 virial parameter estimates from previously published catalogs of sources in both high- and low-mass star-forming regions.  The \cite{Kauffmann_2013} catalog, however, featured virial parameters that had been measured using a variety of molecular tracers ($^{13}$CO, NH$_3$, N$_2$D$^+$), mass estimation methods (dust continuum and NIR extinction), and probed varying scales (clouds, clumps, and cores).  Nevertheless, we can usefully compare the \cite{Kauffmann_2013} catalog to our data since it deals primarily with high-mass star-forming regions, uses the same formulation for source effective radius, and uses dust continuum emission to derive the masses for most sources.

Overall, our virial parameters are consistent with those found for the high-mass cores, clumps, and clouds included in the \cite{Kauffmann_2013} compilation, which included the clumps observed by \cite{Wienen_2012}.  $\alpha_{vir}$ ranges from $\sim 10^{-1}$ to $10^{2}$ and roughly half of the sources fall below the $\alpha_{vir}$=2 threshold for both the \cite{Kauffmann_2013} sources and those presented in this paper.  This result is in contrast to the virial analyses of \cite{Friesen_2016} in Serpens South and \cite{Keown_2017} in Cepheus-L1251, which found that nearly all their ammonia-identified leaves had $\alpha_{vir}<2$.  Serpens South and Cepheus-L1251 are closer ($d\sim250-450$ pc and 300 pc, respectively) than the clouds observed by KEYSTONE and are thus probing smaller scale structures, which may explain the higher rate of gravitationally bound leaves in those papers.  These results are supported by those of \cite{Ohashi_2016_2} and \cite{Chen_2019}, which used ALMA observations of infrared dark clouds to show that $\alpha_{vir}$ decreases with decreasing spatial scales from filaments to clumps to cores, and may be showing that gravity is more important in the stability of structures at small scales.  Alternatively, the KEYSTONE observations may indicate that ammonia is more widespread throughout GMCs than it is in low-mass clouds, producing detectable ammonia emission in both bound and unbound sources.  Such a scenario is supported by the observations of \cite{Henshaw_2013}, which found N$_2$H$^+$ ($1-0$) to be more extended in the infrared dark cloud G035.39-00.33 than in low-mass star-forming environments. 

Observations of more distant cloud clumps ($\sim 2 - 11$ kpc) by the Bolocam Galactic Plane Survey \citep[BGPS;][]{Rosolowsky_2010, Ginsburg_2013}, which mapped 1.1 mm dust continuum emission across the Galactic plane, have also indicated low virial parameters ($\alpha_{vir} < 2$) for clumps.  For instance, \cite{Svoboda_2016} combined NH$_3$ observations with BGPS clump detections to calculate virial parameters for 1640 clumps and found that $76\%$ of starless candidates and $86\%$ of protostellar candidates had $\alpha_{vir}<2$.  Similarly, \cite{Dunham_2011} found that a separate sample of 456 BGPS clumps had a median $\alpha_{vir}=0.74$. We note, however, that the NH$_3$ observations used for those analyses were targeted follow-ups to previously identified clumps in the BGPS data.  Thus, they may not be tracing the faint NH$_3$ emission probed by KEYSTONE and included in our leaf catalog.  These low brightness NH$_3$ sources comprise the lower mass leaves in our sample that have $\alpha_{vir}>2$ and may be the reason we detect unbound sources that are lacking in the BGPS data. 

A similar selection bias for high brightness sources is likely also impacting clump virial analyses using \textit{Herschel} Hi-GAL \citep{Molinari_2010} and \textit{APEX Telescope} ATLASGAL \citep{Schuller_2009} observations. For example, \cite{Merello_2019} combined Hi-GAL clump detections with NH$_3$ catalogs to derive virial parameters for 1068 clumps at distances typically between $\sim 2$ kpc and $\sim 15$ kpc.  $72\%$ of the 1068 clumps had $0.1 < \alpha_{vir} < 1$, with a median $\alpha_{vir}$ of 0.3.  A similar virial analysis of 213 Hi-GAL clumps by \cite{Traficante_2018} found that $76\%$ have $\alpha_{vir} < 1$.  Using a sample of 1244 ATLASGAL clump detections with masses much larger (typically $>10^3 M_\odot$) than the leaves presented in this paper, \cite{Contreras_2017} found a median $\alpha_{vir}$ of 1.1. Since these analyses rely on clump catalogs identified by dust continuum observations, however, they likely exclude the faint NH$_3$ sources detected by KEYSTONE. 

\subsection{Cloud Weight Pressure}
\begin{figure}[ht]
\epsscale{1.0}
\plottwo{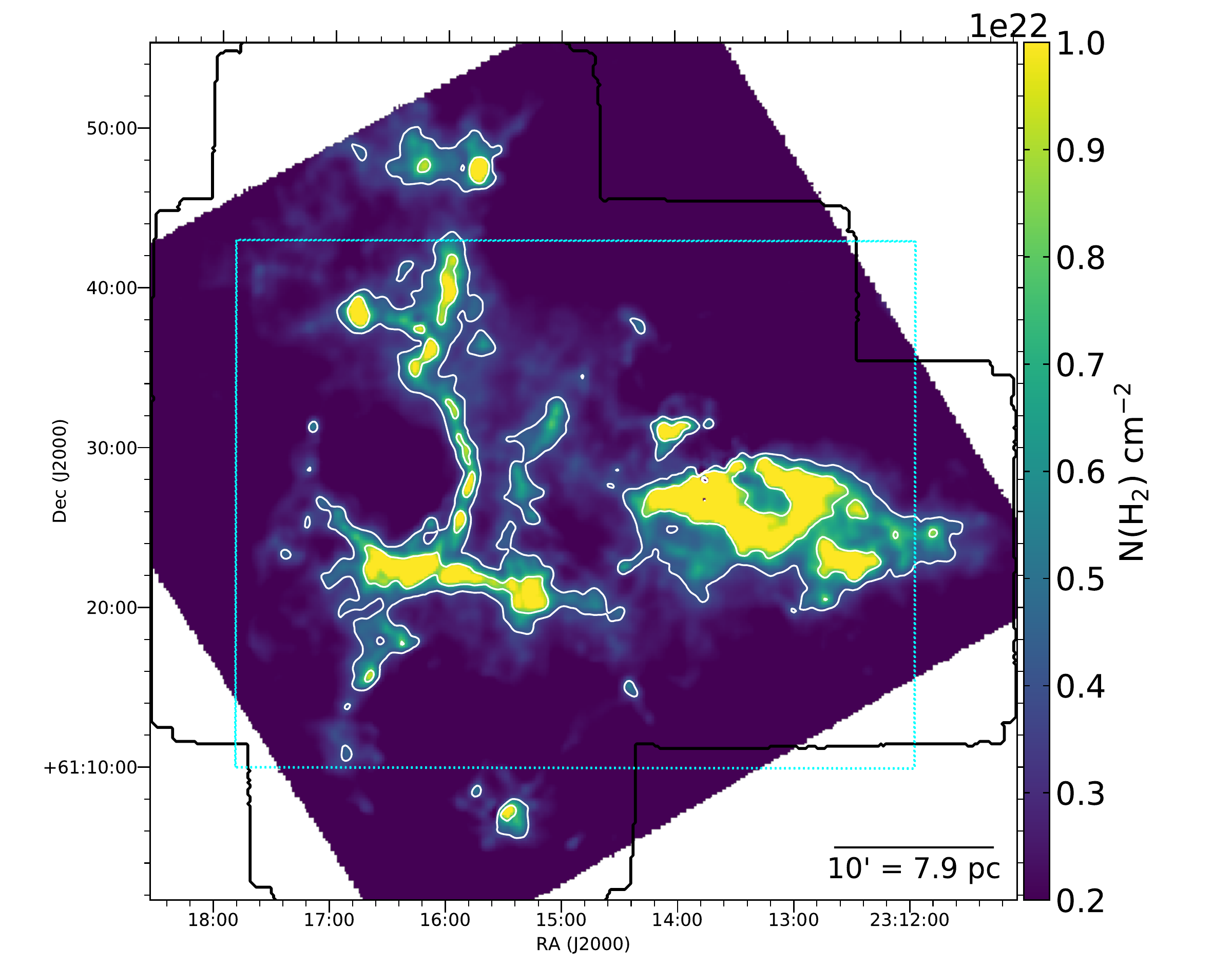}{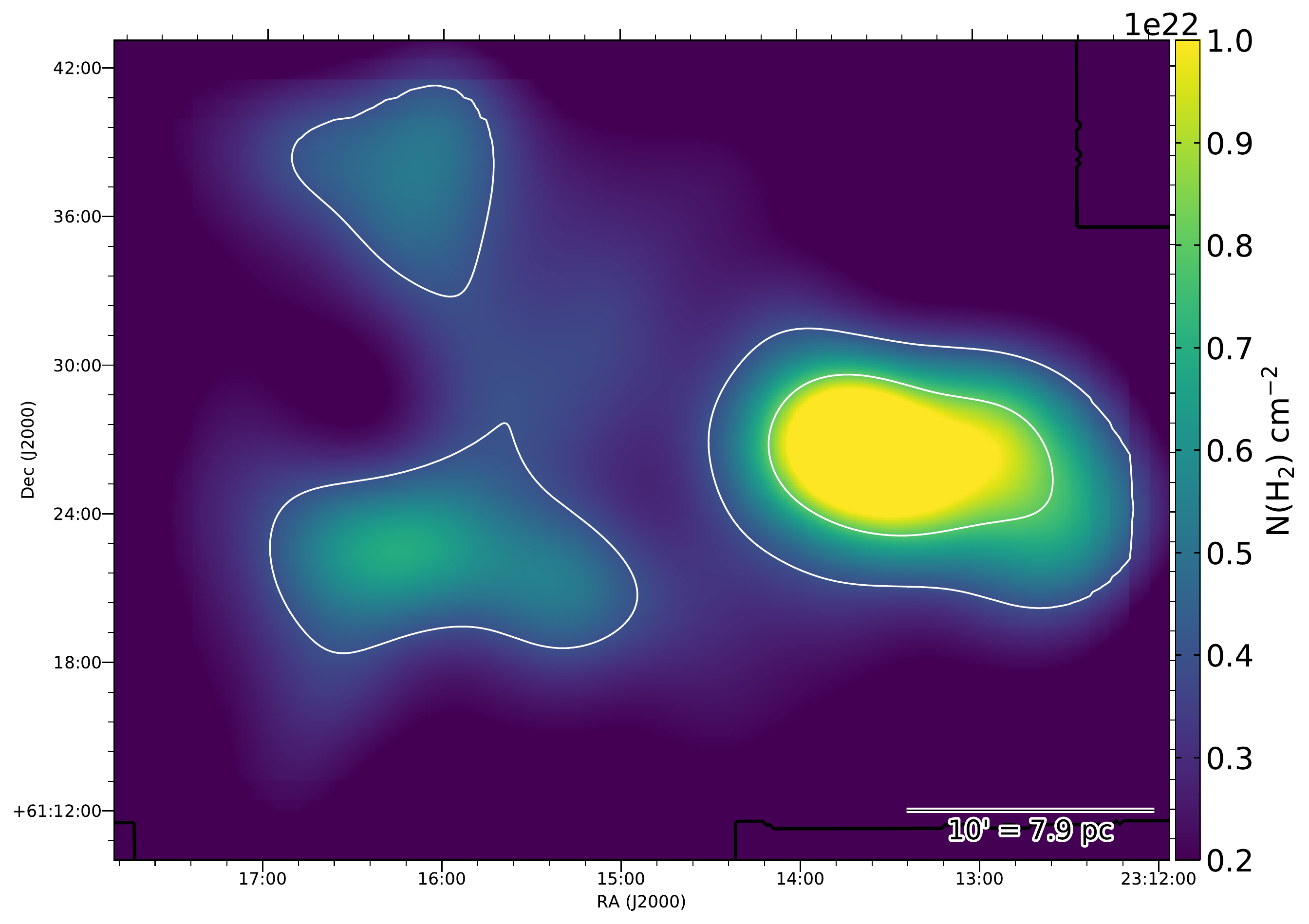}
\caption{Left: H$_2$ column density map of NGC7538 with white contours at $4\times 10^{21}$ cm$^{-2}$ and $8\times 10^{21}$ cm$^{-2}$.  Black outlines the extent of the KEYSTONE mapping of the region.  Right: Spatially filtered H$_2$ column density map over the cyan dotted area overlaid in the left panel.  The map includes spatial scales larger than 16 pixels, which is equivalent to $\sim96\arcsec$ or $\sim1.3$ pc at the distance of NGC7538.}
\label{NGC7538_atrous}
\end{figure}

Although the virial analysis presented in Section 3.5 compares the gravitational energy of the ammonia-identified leaves with their kinetic energy, it excludes the possible influence of external pressure applied by the leaves' surroundings \citep{Field_2011}.  For instance, the weight of the molecular cloud in which the leaves are embedded can contribute to their confinement \citep[e.g.,][]{Pattle_2015, Pattle_2017, Kirk_2017}.  Here, we add the cloud weight pressure energy density ($\Omega_{Pw}$) to the virial equation using the technique described in \cite{Keown_2017} and \cite{Kirk_2017}.  The three energy densities in the virial equation considered in our analysis are given by the following expressions:

\begin{equation}
\Omega_{Pw} = -4 \pi P_w R^3
\end{equation}

\begin{equation}
\Omega_G = \frac{-1}{2\sqrt{\pi}}\frac{GM^2}{R}
\end{equation}

\begin{equation}
\Omega_K = \frac{3}{2}M\sigma^2
\end{equation} where $M$ is the observed structure mass, $R$ is the effective radius, $G$ is the gravitational constant, $\sigma^2$ is the same as Equation 3, and $P_w$ is cloud weight pressure:

\begin{equation}
P_w = \pi G \bar{N} N (\mu_H m_H)^2
\end{equation} where $\bar{N}$ is the mean cloud column density and $N$ is the column density at the structure \citep[e.g., ][]{Mckee_1989, Kirk_2006, Kirk_2017}.  Both $\bar{N}$ and $N$ are measured from a spatially filtered column density map to determine the cloud's mass contribution from large-scale structures.  Following other recent virial analyses incorporating turbulent pressure \citep{Kirk_2017, Keown_2017, Kerr_2019}, the \textit{a trous} transform\footnote{The \textit{atrous.pro} IDL script developed by Erik Rosolowsky, which is available at https://github.com/low-sky/idl-low-sky/blob/master/wavelet/atrous.pro, was used for this analysis.} is used to filter spatially each column density map to spatial scales larger than 2$^n$ pixels, where we have chosen $n$=4 or 16 pixels for all regions.  Figure \ref{NGC7538_atrous} shows an example spatially filtered column density map for NGC 7538, where 16 pixels is equivalent to $\sim 96\arcsec$ or $\sim1.3$ pc.  $\bar{N}$ is calculated as the mean of the spatially filtered column density map, while $N$ represents the mean spatially filtered column density within each leaf's dendrogram-identified boundary. 

Although we have chosen a single scale for the spatial filtering, a recent virial analysis by \cite{Kerr_2019} investigated the impact that varying this scale has upon the cloud weight pressure term.  In their analysis of L1688, B18, and NGC1333, Kerr et al. find that increasing or decreasing the spatial filtering scale by a factor of two results in less than a factor of two difference in the average $P_w$ values for those clouds.  As shown below, such a factor is not enough to change our overall conclusions about the virial stability of the observed structures.   

\begin{figure}[ht]
\epsscale{1.1}
\plottwo{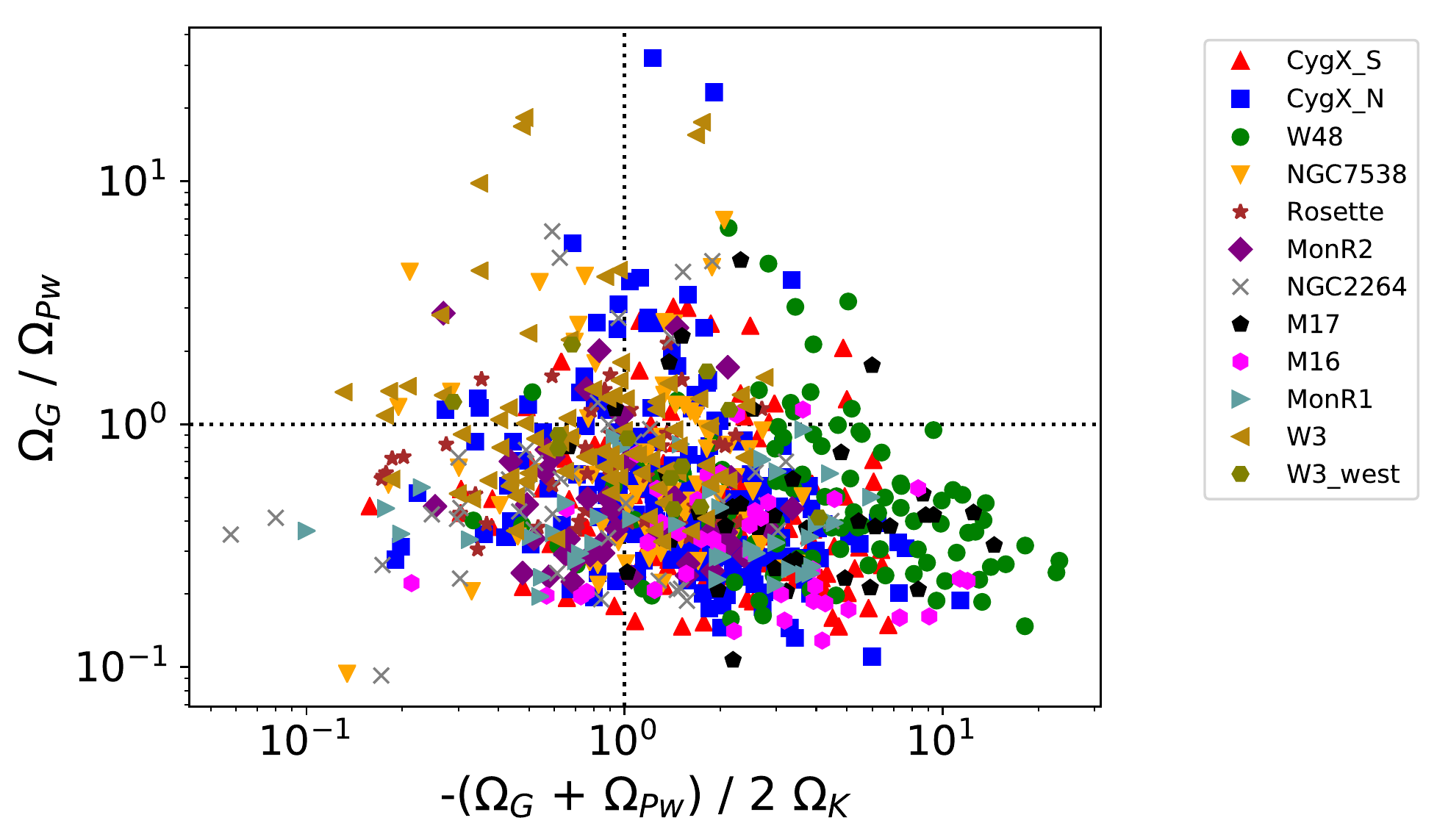}{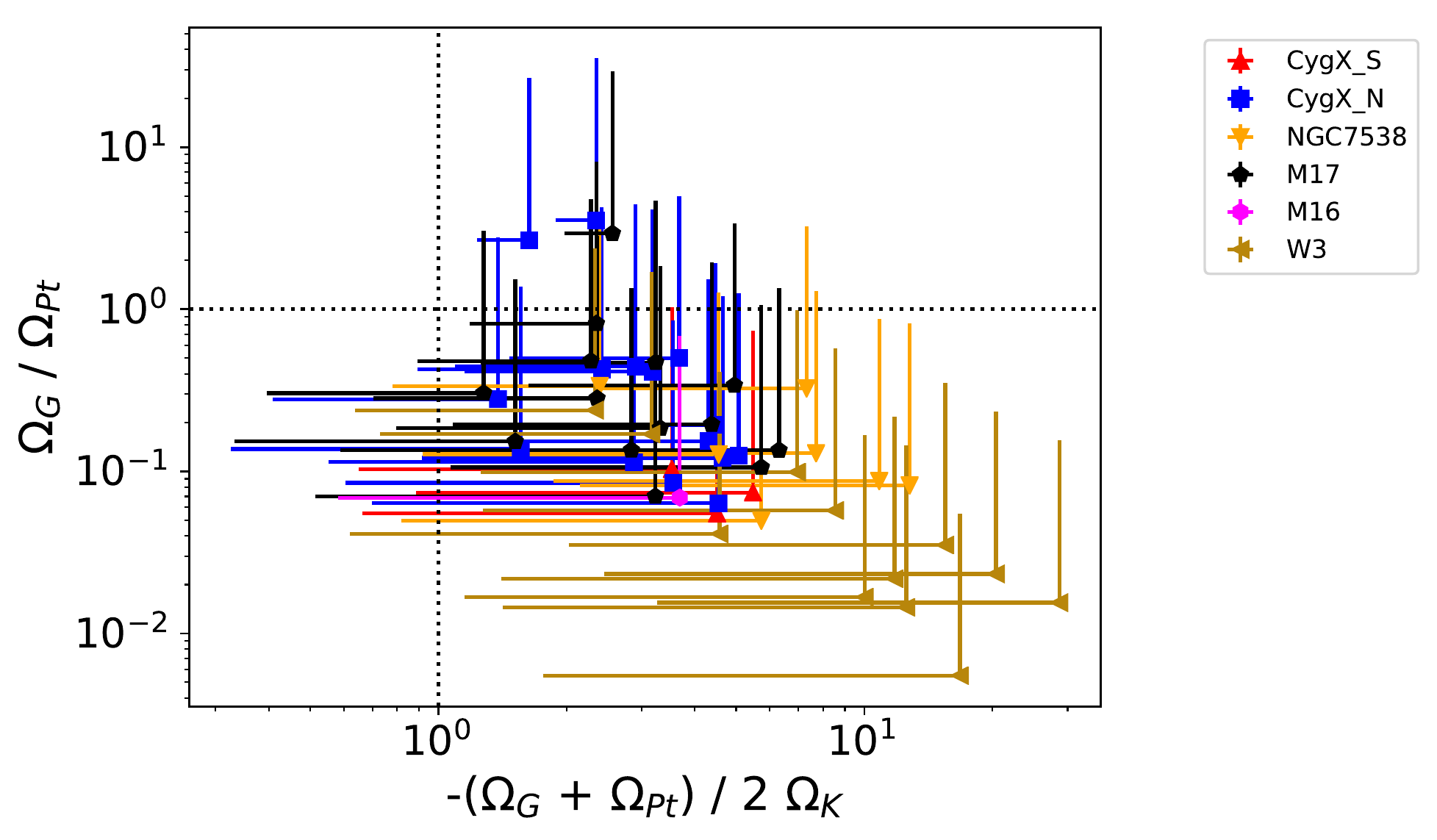}
\caption{Left: Virial plane for all ammonia-identified leaves displayed in Figure \ref{Mass_alpha} showing the balance between three energy densities in the virial equation: Gravitational ($\Omega_G$), Cloud Weight Pressure ($\Omega_{Pw}$), and Kinetic ($\Omega_K$). Sources to the right of the vertical line are sub-virial, while sources to the left are super-virial.  Sources above the horizontal line are gravitationally-dominated, while sources below the line are pressure-dominated. Right: Virial plane using instead the Turbulent Pressure energy density ($\Omega_{Pt}$) for all ammonia-identified leaves with reliable C$^{18}$O $(3-2)$ velocity dispersion measurements.  The data points are calculated using the critical density ($3\times10^{4}$ cm$^{-3}$) of C$^{18}$O ($3-2$) as the volume density traced by its emission.  The errorbars show where the source would fall on the plot if the density traced by the C$^{18}$O were a factor of ten lower ($3\times10^{3}$ cm$^{-3}$).}
\label{virial_plane}
\end{figure}

The left panel of Figure \ref{virial_plane} shows the balance between cloud weight pressure ($\Omega_{Pw}$), kinetic energy ($\Omega_K$), and gravitational potential energy ($\Omega_G$) for the 835 leaves with mass estimates.  Leaves to the right of the vertical dotted line are deemed ``sub-virial'' since their gravitational potential and external pressure are enough to overcome their internal kinetic energy in the absence of magnetic fields.  Leaves to the left of the vertical line are ``super-virial'' since they are not bound by their gravitational and external pressure energy densities.  Below the horizontal dotted line are ``pressure-dominated'' sources that have a higher external pressure energy density than their gravitational energy density.  Conversely, ``gravity-dominated'' sources lie above the horizontal line. 

As previous virial analyses in nearby low-mass star-forming regions have shown \citep{Pattle_2015, Pattle_2017, Kirk_2017, Keown_2017, Chen_2018, Kerr_2019}, most of the leaves are sub-virial and pressure-dominated.  In particular, \cite{Kerr_2019} showed that 79 of 134 ($\sim 59\%$) of the cores in their combined sample from the NGC1333, L1688, and B18 clouds were sub-virial when considering both cloud weight pressure and turbulent pressure.  Similarly, we find that $\sim 69\%$ of the leaves presented in this paper are deemed sub-virial when considering only the cloud weight pressure in the pressure energy density.  This implies that the larger H$_2$ column densities and total cloud masses observed for GMCs are sufficient to create virially bound structures without the necessity of large levels of turbulent pressure that appear to be required in low-mass star-forming environments (see Section 4.5 below for a discussion of turbulent pressure).  

Although many of the structures are pressure-dominated, the large surrounding reservoirs of dense gas in GMCs may facilitate their evolution into the gravitationally-dominated regime.  For instance, some recent simulations have shown that even though dense cores may appear as pressure-confined and stable structures at various stages in their evolution, they are still likely to be gaining mass by accreting material from their surroundings and will eventually undergo gravitational collapse \citep[e.g., ``global hierarchical collapse,''][]{Naranjo_2015, Vazquez-Semadeni_2017, Ballesteros-Paredes_2018, Vazquez-Semadeni_2019}.  As such, the observed pressure-dominated state of our ammonia-identified leaves may be a common stage in their evolution from dense gas structures to protostars and clusters.  Furthermore, the dearth of structures in the sub-virial and gravitationally-dominated regime may indicate that clumps spend the majority of their time being pressure-dominated, with a quick transition to being gravitationally-dominated and subsequently collapsing to form protostars.  

\subsection{Turbulent Pressure}
In addition to cloud weight, pressure due to cloud-scale turbulence may have a significant impact on the virial stability of dense cores \citep[e.g.,][]{Kerr_2019, Pattle_2015, Pattle_2017, Kirk_2017, Keown_2017, Chen_2018}.  Here, we calculate the turbulent pressure energy density ($\Omega_{Pt}$) for a subset of our ammonia-identified leaves following the method described by \cite{Keown_2017} (see also \cite{Pattle_2017} and \cite{Kirk_2017} for detailed discussions of turbulent pressure).  $\Omega_{Pt}$ is calculated from Equation 4, with P$_w$ being replaced with P$_t$ given by:

\begin{equation}
P_T=\mu_H m_H\times \rho_{C^{18}O}\times \sigma ^2_{C^{18}O} ~,
\end{equation} where $\sigma_{C^{18}O}$ is the velocity dispersion measured from C$^{18}$O $(3-2)$, a moderate density tracer, and $\rho_{C^{18}O}$ is the volume density at which the C$^{18}$O $(3-2)$ emission originates.  Here, we assume the C$^{18}$O $(3-2)$ emission is tracing a volume density of $3 \times 10^4$ cm$^{-3}$, which is the critical density ($n_{cr}(u-l) = A_{ul}/\gamma_{ul}$, where $A_{ul}$ is the Einstein A coefficient and $\gamma_{ul}$ is the collisional rate coefficient for collisions with H$_2$) of C$^{18}$O $(3-2)$ at 20 K \citep[calculated using collisional rate coefficients accessed from the Leiden Atomic and Molecular Database,][]{Schoier_2005}.  



Six of the eleven KEYSTONE regions were found to have partial JCMT observations of C$^{18}$O $(3-2)$ (see Section 2.4 for a discussion of these data and our reduction techniques).  A Gaussian model with three free parameters (peak brightness temperature, centroid velocity, and velocity dispersion) was fit to all pixels in the C$^{18}$O $(3-2)$ data cubes with SNR $>$ 6 using the non-linear least squares curve-fitting method in the \texttt{scipy.optimize.curvefit} Python package.  SNR is measured from the ratio of the peak brightness temperature in each spectra to the standard deviation of the off-line spectral channels.  The conservative cutoff of SNR $>$ 6 was chosen to remove low SNR spectra from consideration since they often have higher uncertainties in the fitted parameters. The initial parameter guesses of each fit are based on the brightness and velocity of the peak brightness channel, with a set guess of 1.5 km s$^{-1}$ for the velocity dispersion.  After the line fitting, pixels must meet the following criteria to be included in our final parameter maps:

\begin{enumerate}
\item $\sigma >$ 0.05 km s$^{-1}$ (below 0.05 km s$^{-1}$ is unrealistic since our channel width is only $\sim$ 0.11 km s$^{-1}$;
\item $T_{peak, err} <$ 1 K;
\item $\sigma_{err} <$ 0.5 km s$^{-1}$;
\item $V_{LSR, err} <$ 0.75 km s$^{-1}$.
\end{enumerate} 

\begin{figure}[ht]
\epsscale{1.0}
\plottwo{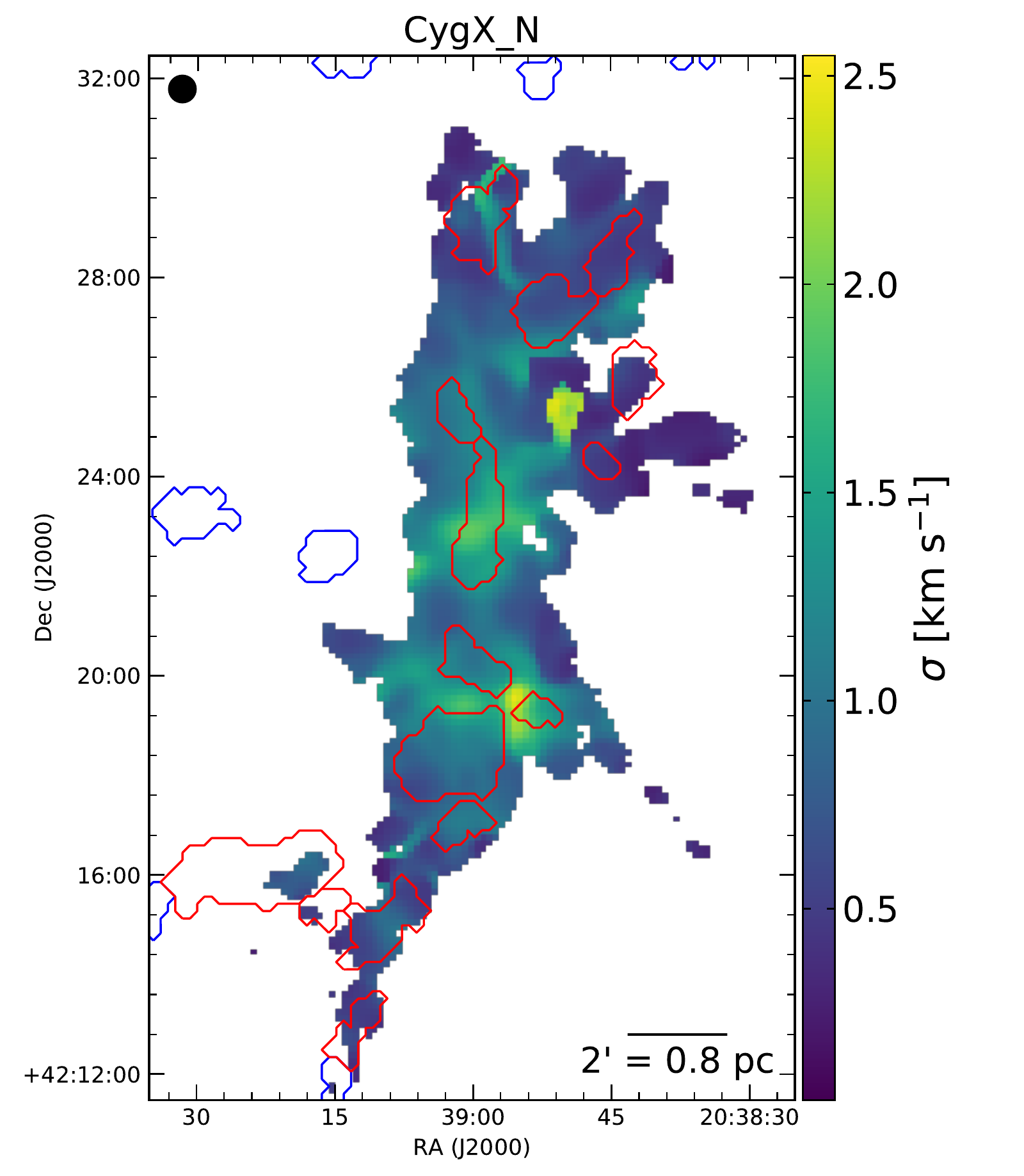}{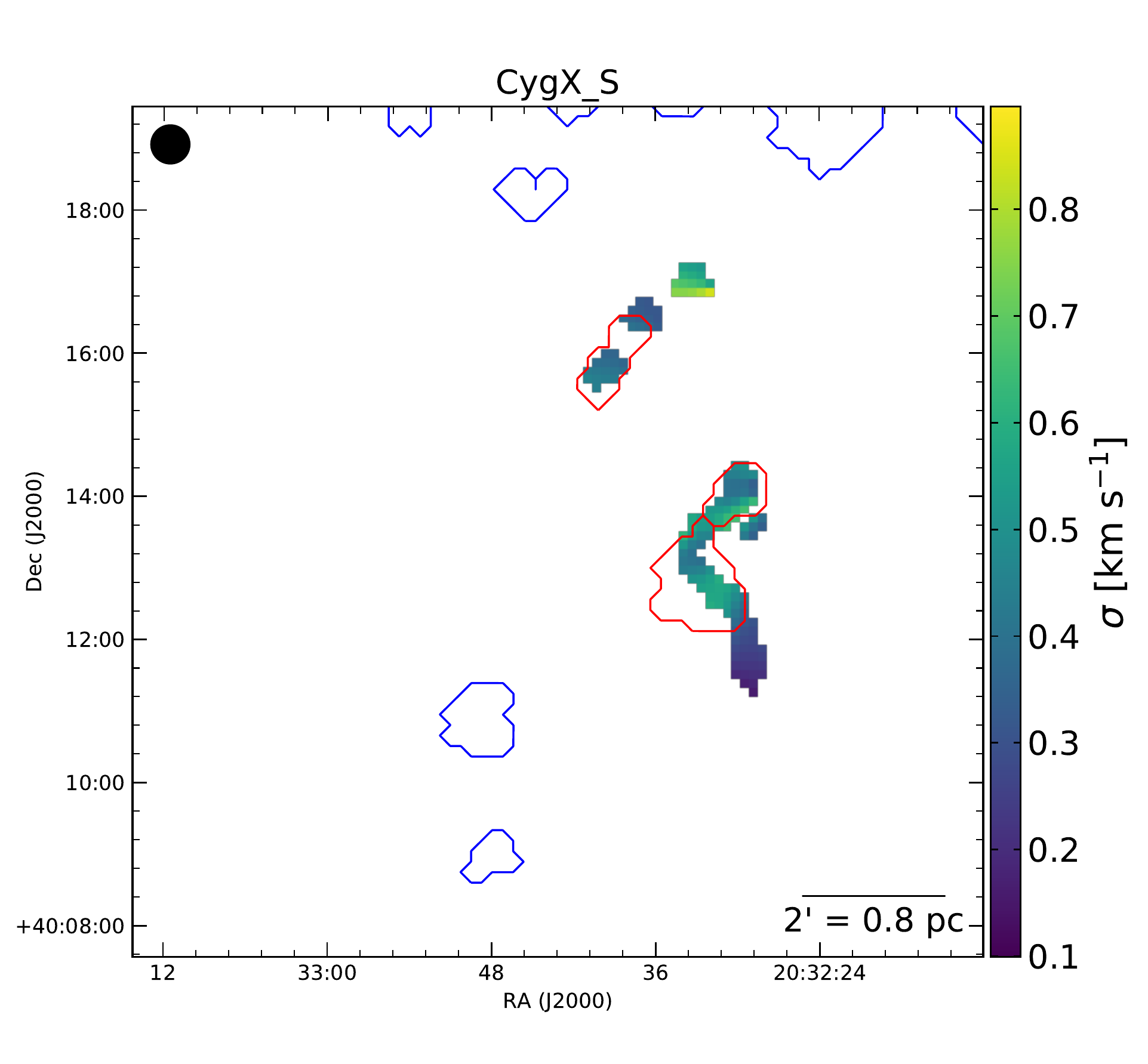}
\plottwo{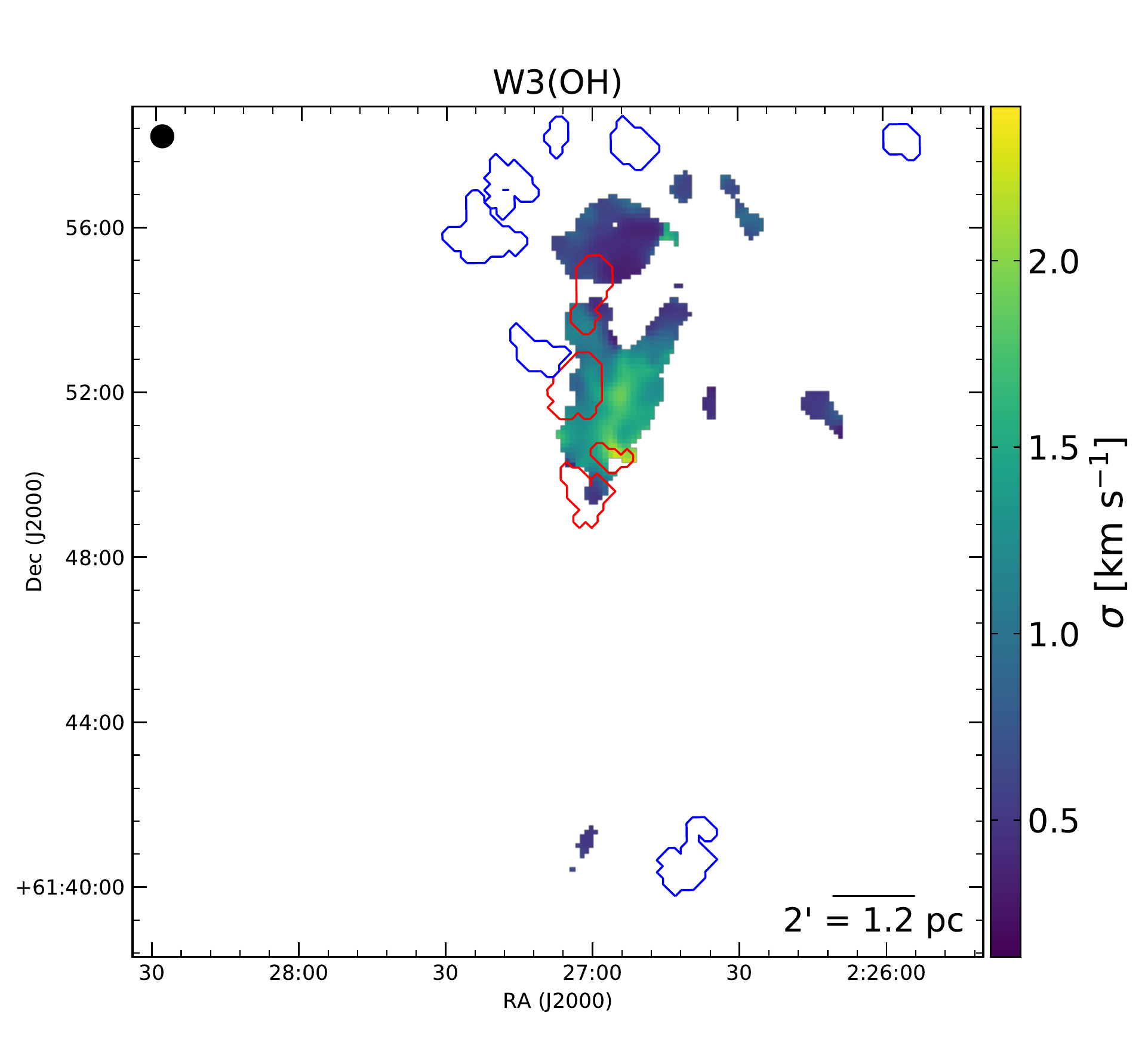}{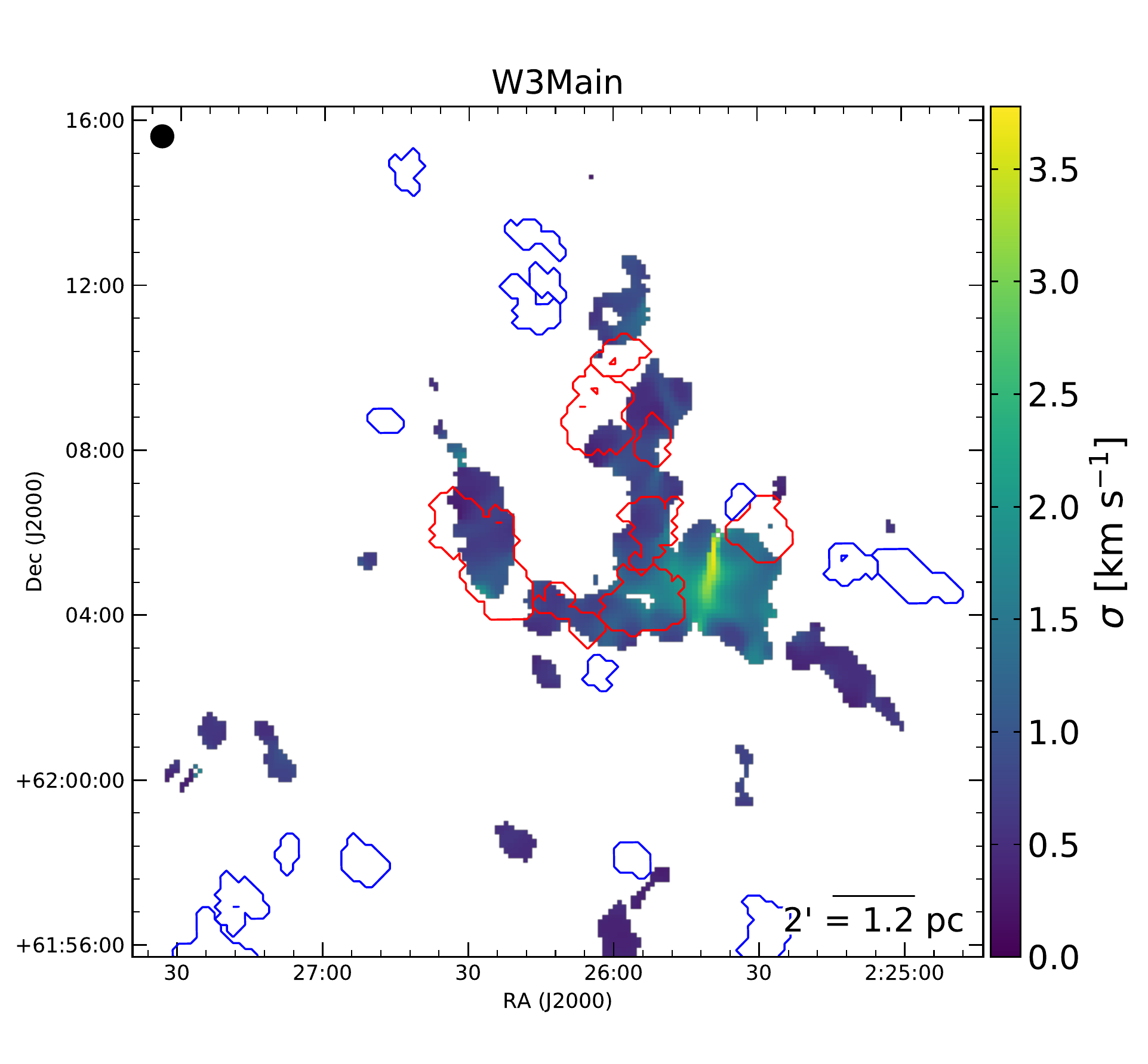}
\caption{C$^{18}$O $(3-2)$ velocity dispersion measured from Gaussian fits to JCMT observations of DR21 in Cygnus X North (top left), G79.34 in Cygnus X South (top right), W3(OH) (bottom left), and W3-Main (bottom right).  Red contours denote ammonia-identified leaves that have at least one reliably fit C$^{18}$O $(3-2)$ pixel and were included in our turbulent pressure virial analysis.  Blue contours show all other ammonia-identified leaves in the field of view.}
\label{Cygnus_C18O}
\end{figure}

\begin{figure}[ht]
\epsscale{0.45}
\plotone{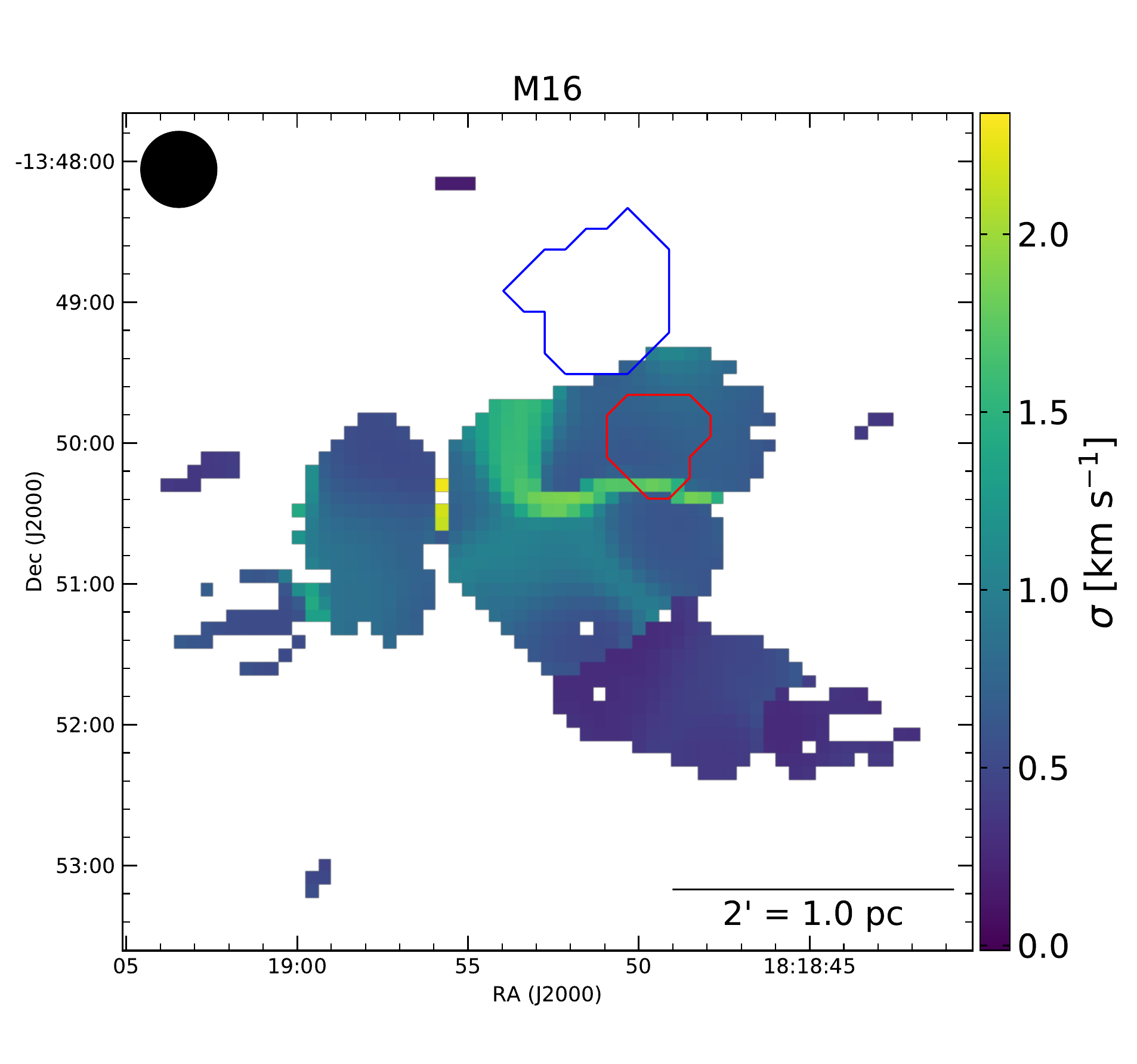}
\plotone{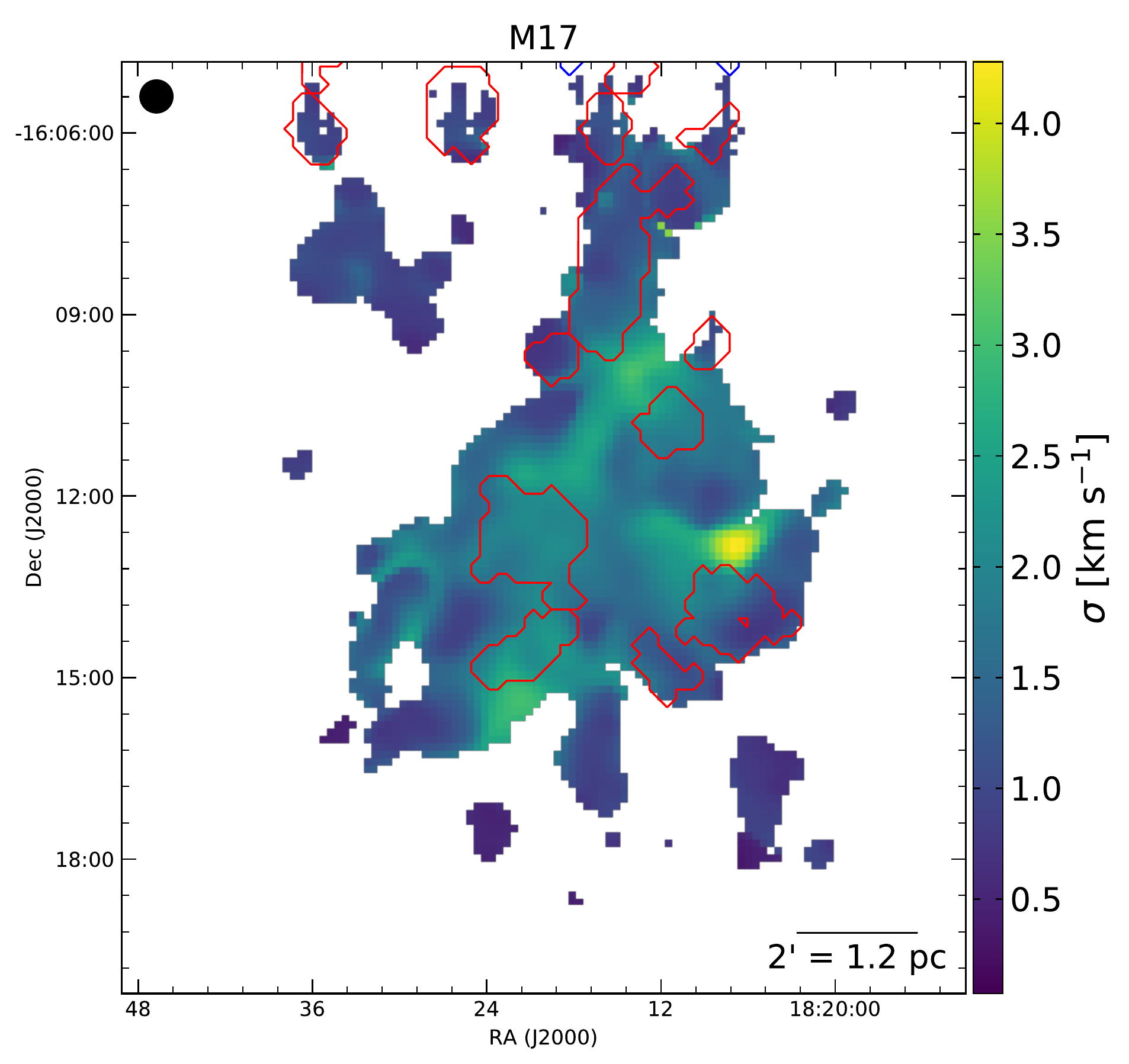}
\plotone{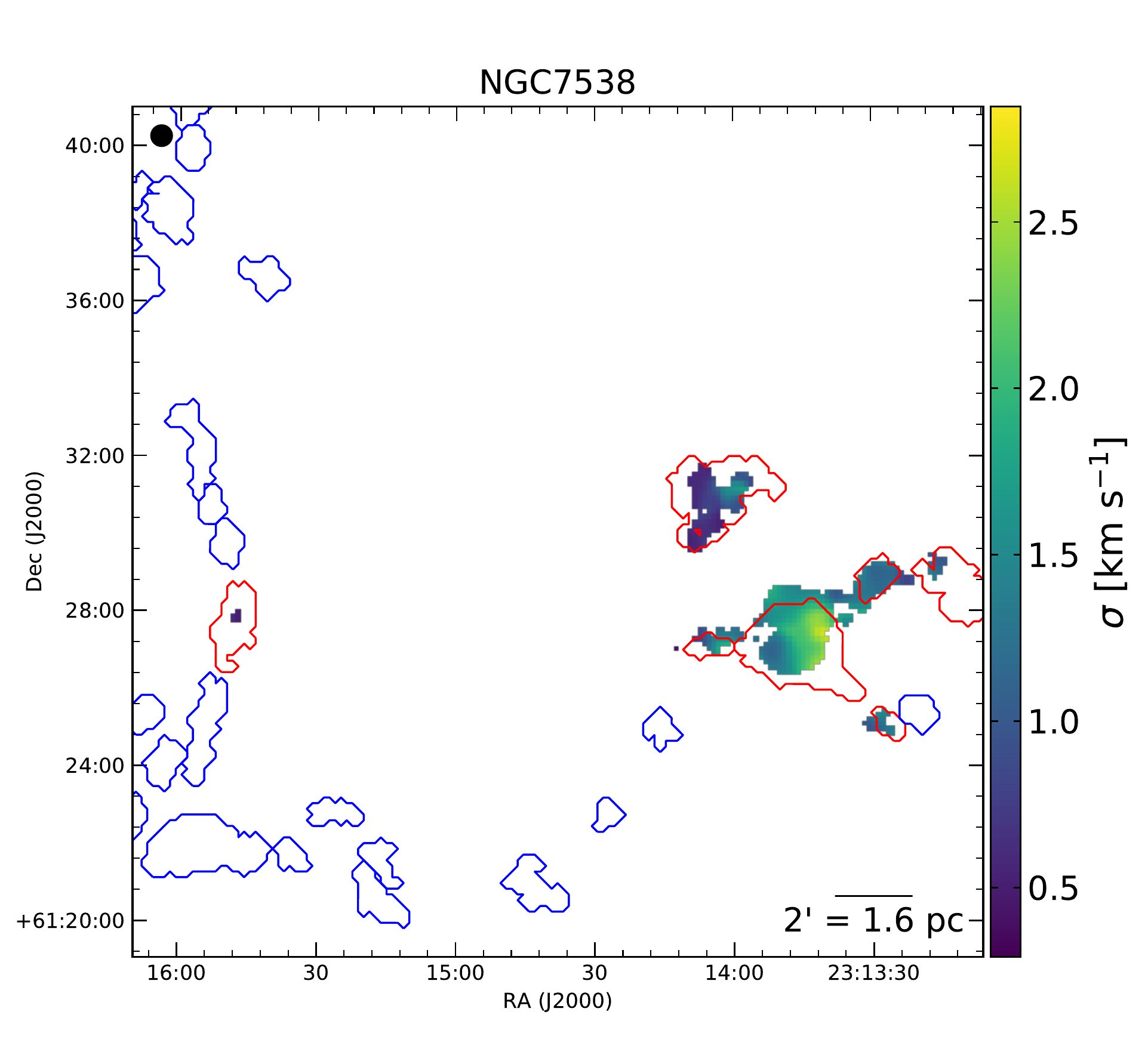}
\caption{Same as Figure \ref{Cygnus_C18O} for M16 (top left), M17 (top right), and NGC7538 (bottom). }
\label{M16_C18O}
\end{figure}

The final parameter maps for the velocity dispersion, $\sigma_{C18O}$, are shown in Figures \ref{Cygnus_C18O}-\ref{M16_C18O}.  The 52 leaves with at least one reliably fit C$^{18}$O $(3-2)$ pixel are shown by red contours in Figures \ref{Cygnus_C18O}-\ref{M16_C18O}.  The median value of $\sigma_{C18O}$ is calculated within the boundaries of each leaf and converted into a turbulent pressure using Equations 5 and 9. Although most of the C$^{18}$O $(3-2)$ spectra are well-characterized by a single Gaussian, some do show wings that may be due to outflows or multiple velocity components along the line of sight.  These areas can be seen in the velocity dispersion maps as sharp increases in $\sigma_{C18O}$.  Since most of the leaves do not overlap with these sharp transitions, our single Gaussian fit is likely sufficient for our velocity dispersion estimates. 

The right panel of Figure \ref{virial_plane} shows the virial plane for structures with C$^{18}$O $(3-2)$ measurements when using $\Omega_{Pt}$ as the pressure term in the virial equation.  All of these structures fall within the pressure-dominated, sub-virial quadrant.  The ratio of $\Omega_{Pt}$/$\Omega_{Pw}$ for these structures ranges from $\sim 1$ to $\sim 200$, with a median of $\sim 7$.  This would suggest that cloud-scale turbulence or global collapse, rather than cloud weight, is the dominant contributor to the pressure term in the virial equation for these ammonia-identified leaves. 

Since the turbulent pressure calculation is sensitive to the assumed value of $\rho_{C^{18}O}$, we also calculate $P_T$ using a factor of ten lower value ($3 \times 10^3$ cm$^{-3}$) for $\rho_{C^{18}O}$ and show the difference as errorbars in Figure \ref{virial_plane}.  This lower density is more characteristic of the effective excitation density of C$^{18}$O $(3-2)$, which is often 1-2 orders of magnitude lower than critical densities \citep{Shirley_2015}.  Under this assumption, the ratio of $\Omega_{Pt}$/$\Omega_{Pw}$ correspondingly drops by a factor of ten, with a range from $\sim 0.1$ to $\sim 20$ and a median of $\sim 0.7$.  In contrast to the scenario using the higher $\rho_{C^{18}O}$ assumption, this new estimation would suggest that cloud weight is dominant over turbulent pressure.

Several recent virial analyses of nearby star-forming regions such as Ophiuchus, B18, and NGC1333 have shown that turbulent pressure tends to be larger than cloud weight pressure \cite[e.g.,][]{Kerr_2019, Pattle_2015}.  Conversely, cloud weight pressure appears to be larger than turbulent pressure in Orion A \citep[][]{Kirk_2017}.  We note, however, that these analyses included turbulent pressure measurements across entire clouds.  This approach is in contrast to the analysis we present here, which has turbulent pressure measurements for only a small subset of leaves that are generally concentrated on the most active star formation sites in each cloud observed (e.g., DR21 in Cygnus X, W3(OH) and W3 Main in W3, and M17SW in M17) and tend to qualify as hubs or ridges.  As such, this biased sample cannot be used to draw generalizations for the full ammonia-identified leaf catalog presented in this paper.  Instead, widespread C$^{18}$O mapping across the KEYSTONE clouds is required to investigate further the role of turbulent pressure on cloud structure and core dynamics. 

\clearpage

In addition to external pressure, the magnetic field may also play an important role in the virial stability of the ammonia-identified leaves.  For instance, both observations \cite[e.g.,][]{Tan_2013, Pillai_2015} and simulations \citep[e.g.][]{Peters_2011} suggest that magnetic fields are equally as important as turbulence and gravity for high-mass star formation.  Although we currently lack large-scale magnetic field measurements for the KEYSTONE clouds, \cite{Auddy_2019} have recently developed a ``core field structure'' model that predicts the magnetic field strength and fluctuation profile using dense gas kinematics.  We reserve a detailed analysis of the clump magnetic fields, however, to a future KEYSTONE paper.




\section{Summary}
We have presented initial observations and results from KFPA Examinations of Young STellar Object Natal Environments (KEYSTONE), a survey of filamentary dense gas structure in eleven GMCs (Cygnus X North, Cygnus X South, M16, M17, MonR1, MonR2, NGC2264, NGC7538, Rosette, W3, and W48) at distances of $0.9 - 3.0$ kpc.  We identified 856 dense gas clumps, traced by NH$_3$ (1,1) emission, across all the observed clouds using a dendrogram analysis.  Simultaneous line fitting of the NH$_3$ (1,1) and (2,2) emission provided estimates of kinetic gas temperature, centroid velocity, velocity dispersion, and para-NH$_3$ column density for the dense gas.  These parameter maps and the NH$_3$ (1,1) and (2,2) data cubes are publicly available.\footnote{\url{https://doi.org/10.11570/19.0074}}

The ammonia parameter maps were used to derive virial stability parameters for each dense gas structure identified, providing insight into whether or not the gravitational potential energy of the structures is enough to overcome their internal kinetic energies in the absence of magnetic fields or external pressure.  HOBYS \textit{Herschel Space Observatory} observations of dust continuum emission were utilized to create H$_2$ column density maps, identify young protostellar candidates, and identify filamentary structures in each region.  James Clerk Maxwell Telescope observations of C$^{18}$O $(3-2)$ emission were accessed to determine the turbulent pressure applied by the ambient molecular cloud upon a subset of the dendrogram-identified dense gas structures.  Our main results are listed below:

\begin{enumerate}

\item Significant variations in kinetic gas temperature are observed between clouds, with median $T_K = 11.4 \pm 2.2$ K in the coldest region (MonR1) and $T_K = 23.0 \pm 6.5$ K in the warmest (M17).  The velocity dispersion distributions are more similar between clouds, with characteristic median values of $0.3-0.7$ km s$^{-1}$. 

\item Of the 835 ammonia-identified clumps with mass estimates, 523 ($\sim 63 ~\%$) have virial parameters less than two, suggesting the mass of those structures is gravitationally bound and more susceptible to gravitational collapse when neglecting the effects of magnetic fields or external pressure. Similar analyses in nearby, low-mass star-forming clouds have found much higher rates of gravitationally bound ammonia-identified cores, which may suggest ammonia is more widespread in GMCs than in nearby clouds or that gravity is more important to structure stability at small scales.
 
\item The fraction of ammonia-identified clumps that are spatially coincident with filaments identified in the H$_2$ column density maps ranges from 0.35 in Cygnus X South to 1.0 in W3-west.  These values are consistent with core on-filament fractions found from dust continuum observations of nearby star-forming regions, which tend to be from $\sim0.4-0.8$ depending on the cloud and core class considered.

\item  On- and off-filament clumps show no substantial differences in their virial parameter, mass, radius, temperature, and velocity dispersion distributions.  We do find, however, a tendency for clouds with low dense gas mass to have a higher fraction of on-filament clumps.  These findings may indicate that filaments play a lesser role in the star formation process of high-mass GMCs.  In those environments, dense gas may be more widespread allowing for clump formation to be equally as likely on and off filaments.  In lower mass environments where dense gas is less widespread, however, clump formation may be limited to the filaments that harbor the main supply of dense gas. 

\item In several regions there are ``hubs'' or ``ridges'' of dense gas that have much higher masses and lower virial parameters than the other clumps in their respective cloud.  These hubs and ridges tend to be located at the intersections of multiple filaments or located near/within a single filament, are often associated with H$_2$O maser emission, and typically host multiple protostars.  Based on these characteristics, hubs may be the sites of future cluster formation.  


\item When considering the external pressure exerted on the clumps, most are considered sub-virial and pressure-dominated structures.  This characteristic state may indicate that high-mass clumps spend the majority of their lifetime confined by external pressure.  Over time, as the clumps accrete mass from their surroundings, they may gain enough mass to be gravitationally dominated and undergo gravitational collapse or fragmentation. 

\end{enumerate}

\section{Future Work}
Although it was not the focus of this paper, a key use-case of the KEYSTONE data is the analysis of filament kinematics in GMCs.  For instance, in regards to the observed spatial relationship between massive young stellar objects and stellar clusters with filament intersections, the KEYSTONE data could be used to determine: 1) whether or not the observed clumps and filaments are truly velocity coherent structures \citep{Pineda_2010, Chen_2018}, and 2) if the mass flow rates along them are large enough to produce MYSOs.  Several independent studies have already measured gas motions in individual filaments such as Serpens South \citep{Kirk_2013, Friesen_2013}, G035.39-00.33 \citep{Henshaw_2013}, DR21 \citep{Schneider_2010_cyg}, W43-MM1 \citep{NL_2013}, M17SW \citep{Chen_2019}, and eight other high-mass filaments \citep{Lu_2018}, noting that the observed mass flow rates could supply the mass required to assemble the stellar clusters at their centres.  Similarly, observations of the Large Magellanic Cloud by \cite{Fukui_2015} showed two separate instances of MYSOs forming at the centers of adjoining filaments that have gas flowing into the central junction.  \cite{Svoboda_2016} and \cite{Motte_2017} also contend that such mass flow onto sites of cluster formation is prominent throughout Galactic high-mass star-forming regions, providing the mass build-up and compression necessary to form stellar clusters.  These studies, however, have focused primarily on regions that have already formed clusters/MYSOs and do not address the conditions of the parental clumps, i.e., the dense gas out of which stellar clusters form, prior to the onset of star formation.  As such, the gas velocity patterns of those regions are susceptible to distortions due to stellar feedback, which raises questions about the applicability of their kinematic measurements.  Furthermore, observations after the onset of star formation cannot be used to decipher whether clumps form before or after filaments collide to form intersections.  Using the clump catalog presented in this paper, in conjunction with an analysis of the adjacent filament kinematics, it is now possible to investigate these questions across multiple high-mass star-forming environments.  

In addition to dense gas kinematics, the KEYSTONE data can provide insight into temperature and chemistry variations as a function of environment within GMCs.  For example, the KEYSTONE observations of the NH$_3$ (1,1), (2,2), (3,3), (4,4), and (5,5) transitions probe a range of gas temperatures up to $\sim$ 200 K.  Since dust temperatures derived from \textit{Herschel} data become increasingly uncertain above 20 K as the spectral energy distribution peak moves toward wavelengths shorter than 160 $\mu$m \citep{Chen_2016}, ammonia-derived gas temperatures will be important for understanding how the OB associations are impacting the star formation in the KEYSTONE target clouds.  Furthermore, the KEYSTONE observations will constrain the abundances of NH$_3$ in GMCs. These abundances, in combination with the gas temperatures, will provide a way to understand how temperature impacts the formation/destruction of ammonia in GMCs. 



\section*{Acknowledgments}
JK, JDF, ER, and MCC acknowledge the financial support of a Discovery Grant from NSERC of Canada.  SB and NS acknowledge support by the French ANR and the German DFG through the project ``GENESIS'' (ANR-16-CE92-0035-01/DFG1591/2-1).  The Green Bank Observatory is a facility of the National Science Foundation operated under cooperative agreement by Associated Universities, Inc.  \textit{Herschel} is an ESA space observatory with science instruments provided by European-led Principal Investigator consortia and with important participation from NASA. The James Clerk Maxwell Telescope is operated by the East Asian Observatory on behalf of The National Astronomical Observatory of Japan; Academia Sinica Institute of Astronomy and Astrophysics; the Korea Astronomy and Space Science Institute; Center for Astronomical Mega-Science (as well as the National Key R$\&$D Program of China with No. 2017YFA0402700). Additional funding support is provided by the Science and Technology Facilities Council of the United Kingdom and participating universities in the United Kingdom and Canada. This research made use of astrodendro, a Python package to compute dendrograms of Astronomical data (\url{http://www.dendrograms.org/}), Astropy (\url{http://www.astropy.org}), a community-developed core Python package for Astronomy \citep{Astropy_2013}, and pyspeckit (\url{https://pyspeckit.readthedocs.io/en/latest/}), a Python spectroscopic analysis and reduction toolkit \citep{Ginsburg_2011}. 

{\it Facility:} \facility{GBT}, \facility{Herschel}, \facility{JCMT}

\section*{Appendix}
\begin{appendix}
\section{Distance Dependence of Virial Parameters}
The virial analysis presented in this paper used the native KEYSTONE resolution for all clouds analyzed, despite the cloud distances ranging from 0.9 kpc to 3 kpc.  Such distance variations provide a factor of $\sim 3$ range of linear spatial resolutions, which may lead to different types of structures (in terms of size and mass) being identified in each cloud.  To test whether or not the distance dependence has any influence on the main results of this paper, we convolved the NH$_3$ (1,1) and (2,2) cubes for NGC 2264, MonR1, and MonR2 ($d = 0.9$ kpc, $\theta \sim 0.13$ pc), the closest clouds in KEYSTONE, to the linear resolution of W48 ($d = 3.0$ kpc, $\theta \sim 0.45$ pc), the most distant cloud observed.  This process degrades the resolution of the observations by a factor of $3.0 / 0.9 \sim 3.3$ to a final resolution of $\sim103\arcsec$.  Following a similar distance bias analysis by \cite{Baldeschi_2017} on \textit{Herschel} maps, we also downsampled the cubes by the same factor along each spatial axis.  Finally, Gaussian noise with a mean of zero and standard deviation of $s_N \sqrt{1-d_0/d_1}$ is added to the cubes, where $s_N$ is the median of the original cube's RMS map, $d_0$ is 0.9 pc and $d_1$ is 3.0 kpc.  This noise addition attempts to return the native RMS level to the convolved and resampled cubes, which have lower noise levels due to the spatial averaging. 

These processes allow us to simulate what might be observed if NGC 2264, MonR1, and MonR2 were at a distance of 3 kpc.  The convolved, downsampled, noise-injected cubes were then run through the line-fitting pipeline to produce a new set of ammonia parameter maps and integrated intensity maps.  Finally, we repeat the virial analysis presented in Section 3.5 using the new set of maps and assuming their distance is 3.0 kpc.

\begin{figure}[ht]
\epsscale{1.0}
\plottwo{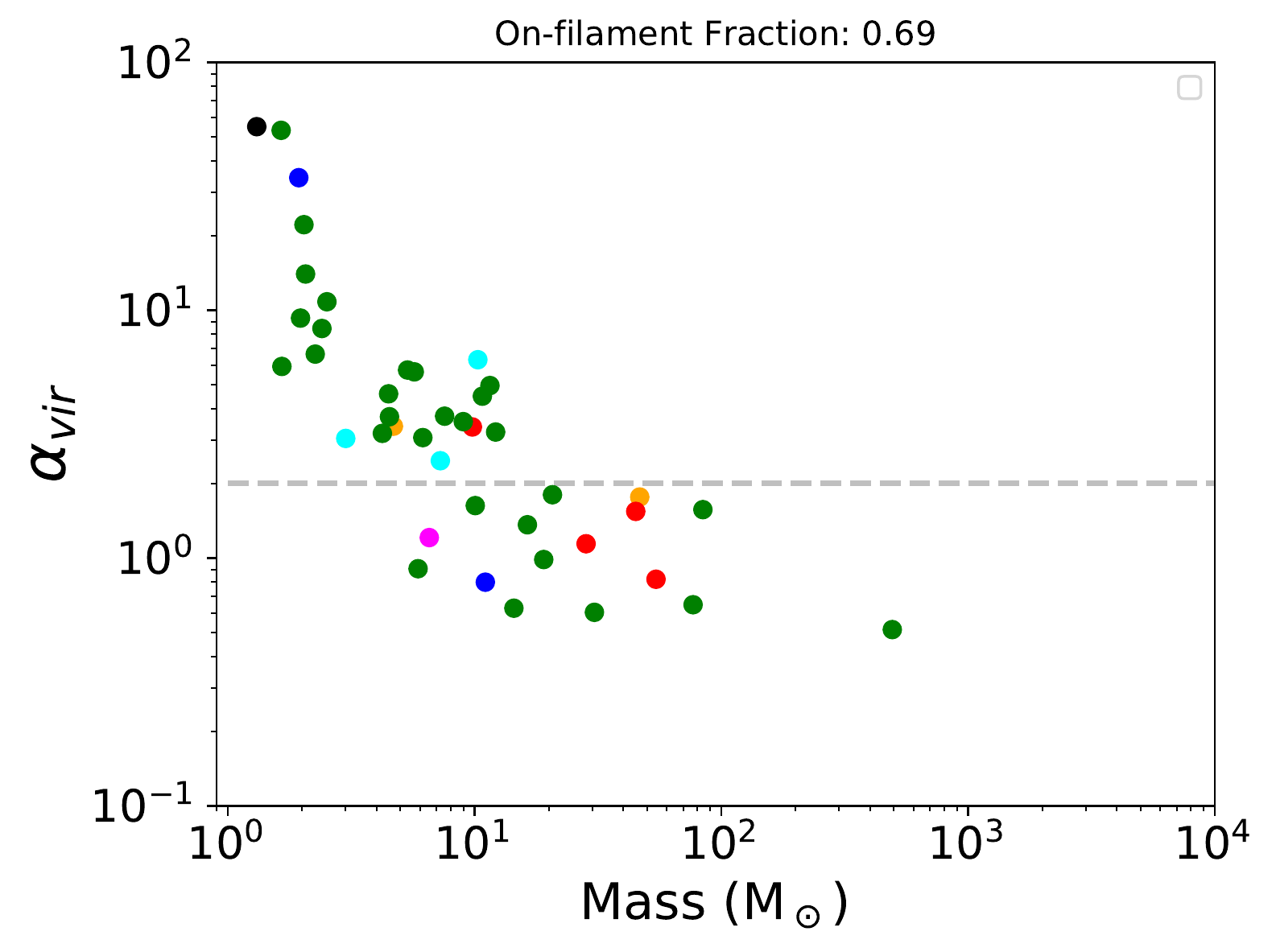}{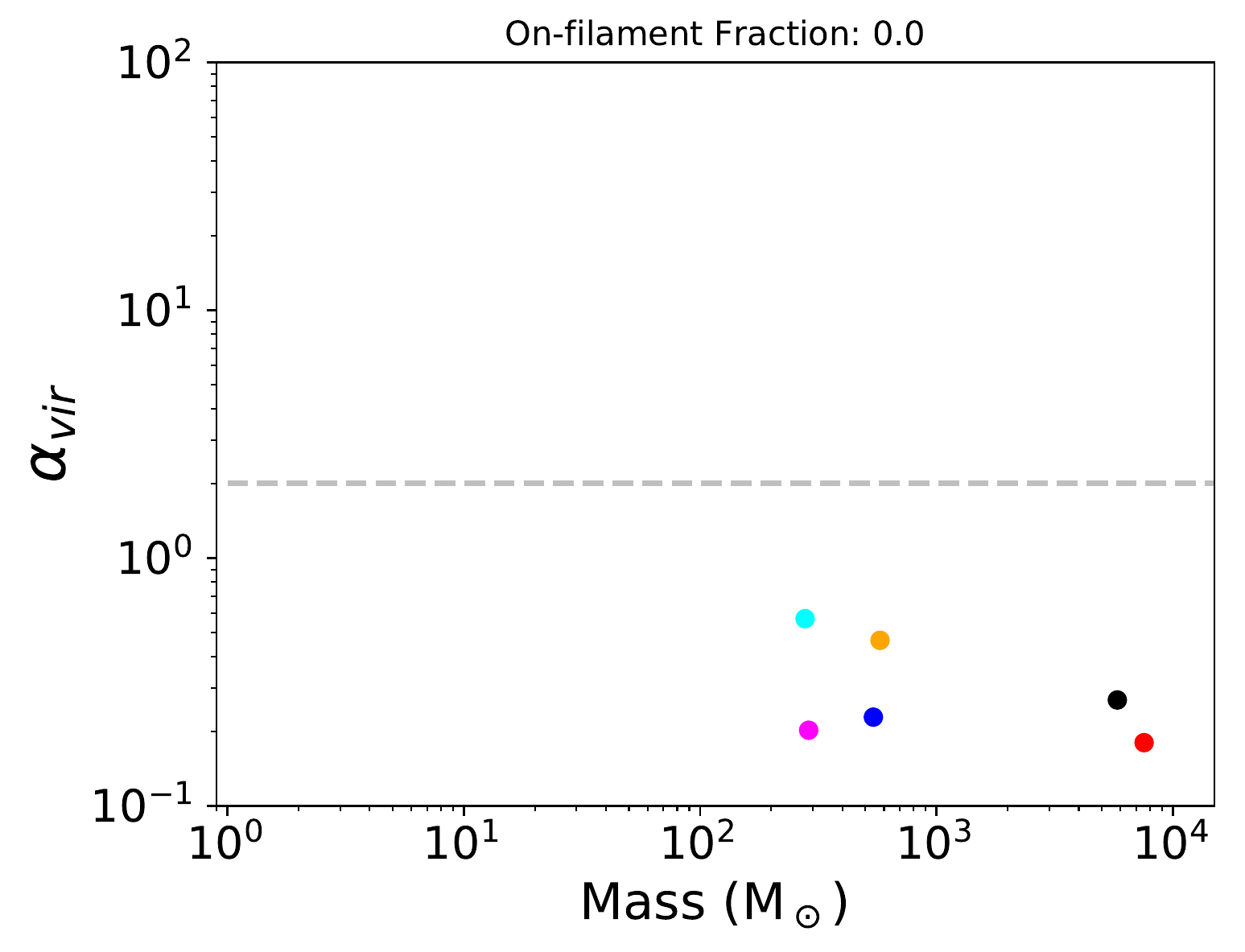}
\epsscale{0.6}
\plotone{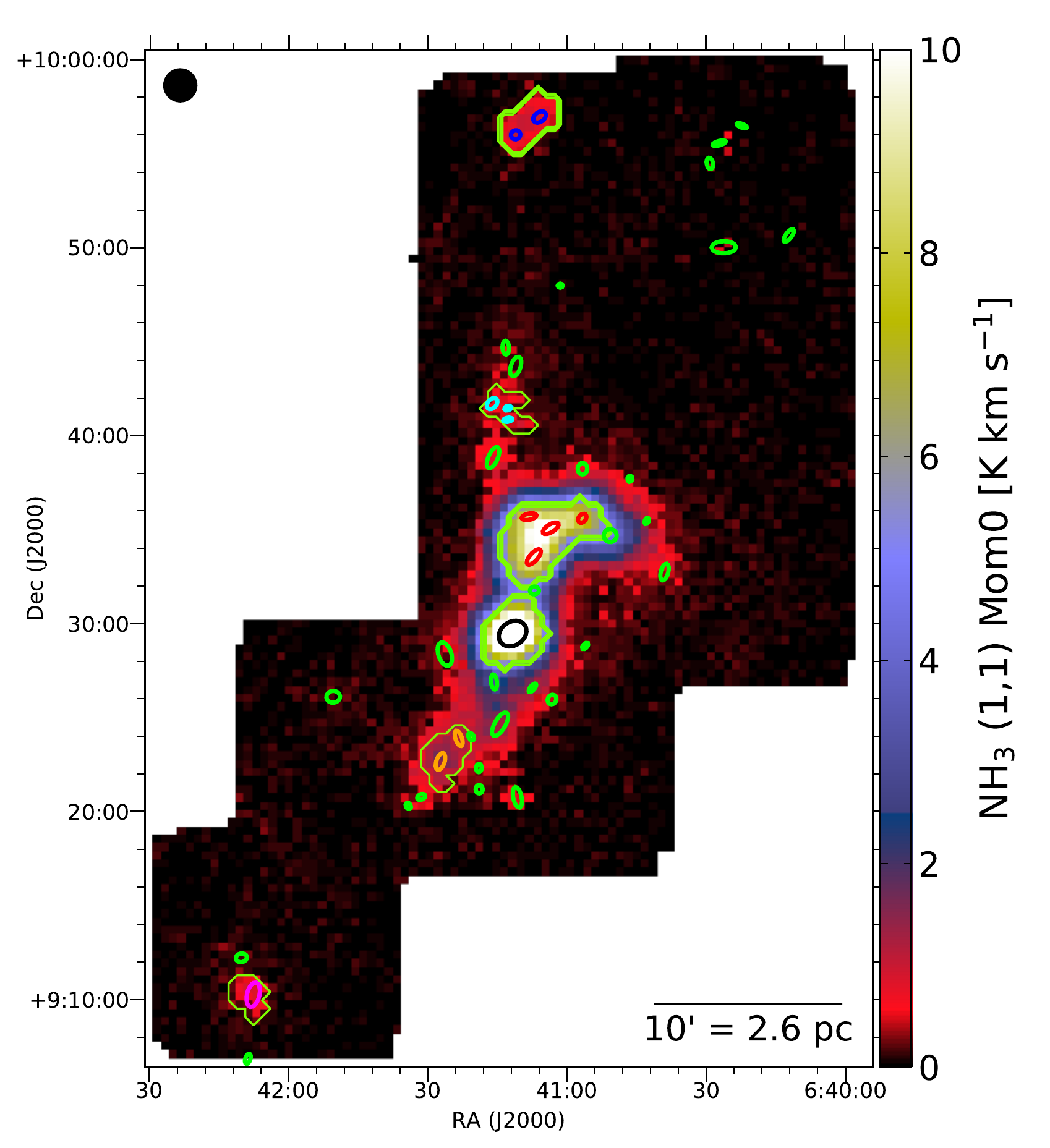}
\caption{Top row: Leaf virial parameters in NGC2264 using native KEYSTONE resolution (left) and after convolving the NH$_3$ (1,1) and (2,2) cubes to the linear resolution of W48 ($\theta \sim 0.45$ pc), downsampling the number of pixels, adding white noise, and re-running the full analysis.  Bottom: Integrated intensity map obtained using the convolved NH$_3$ (1,1) cube.  Ellipses represent the peaks of leaves identified when using the native resolution data.  The green dendrogram masks show the extent of the leaves identified using the convolved data.  Ellipses falling within a dendrogram mask were tagged as a colored pair in the top row plots.}
\label{convolved}
\end{figure}

\begin{figure}[ht]
\epsscale{1.0}
\plottwo{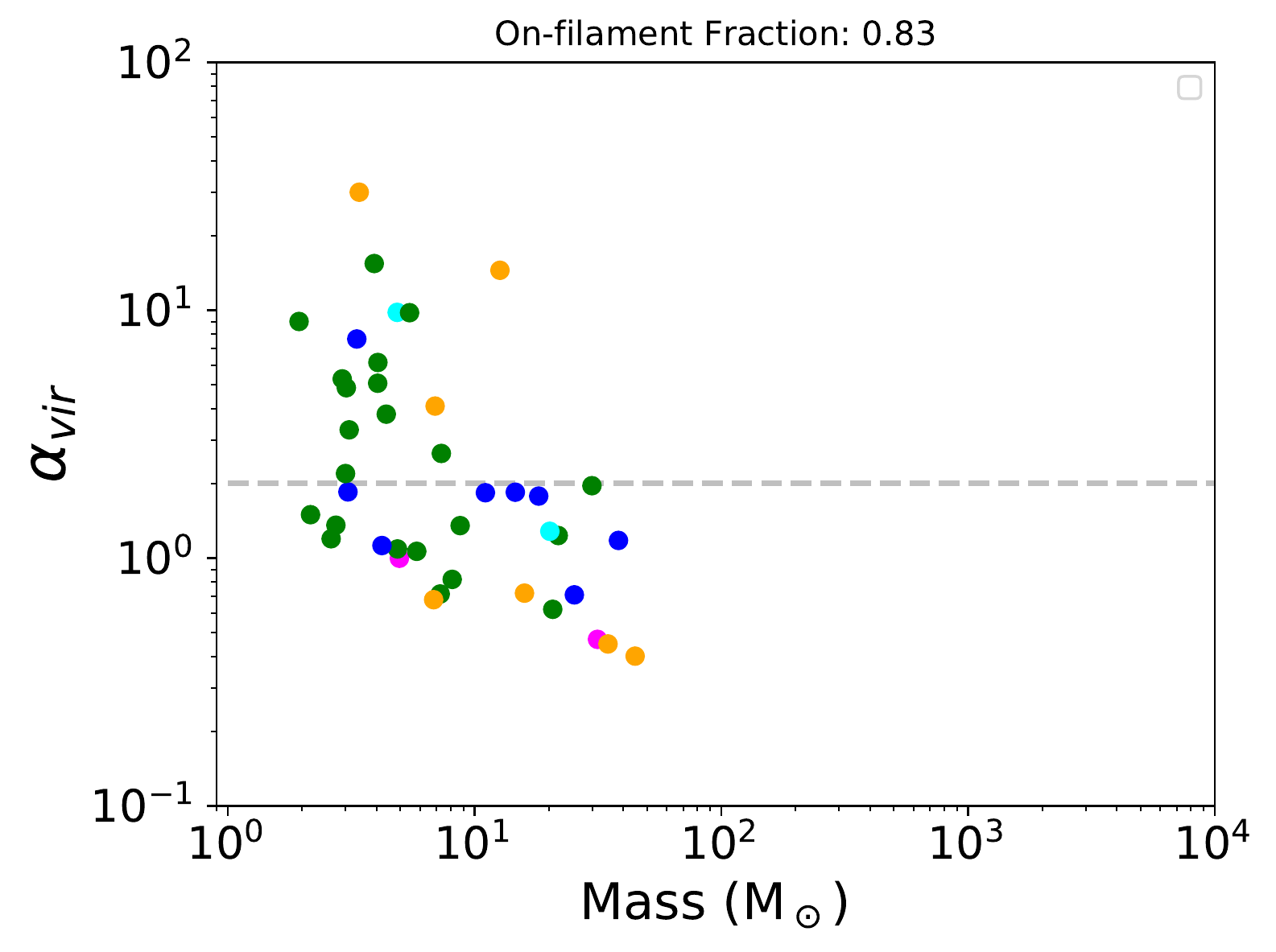}{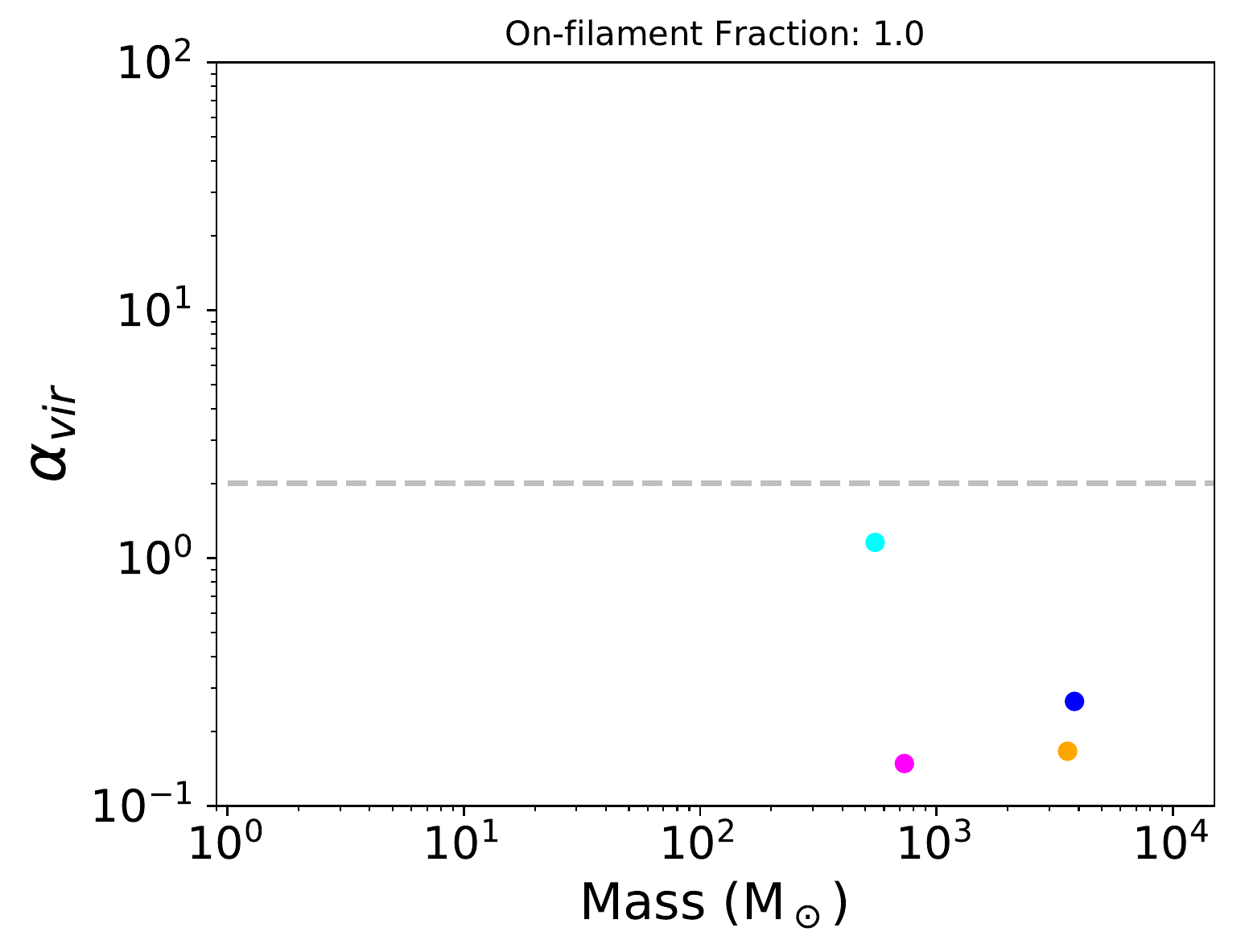}
\epsscale{0.6}
\plotone{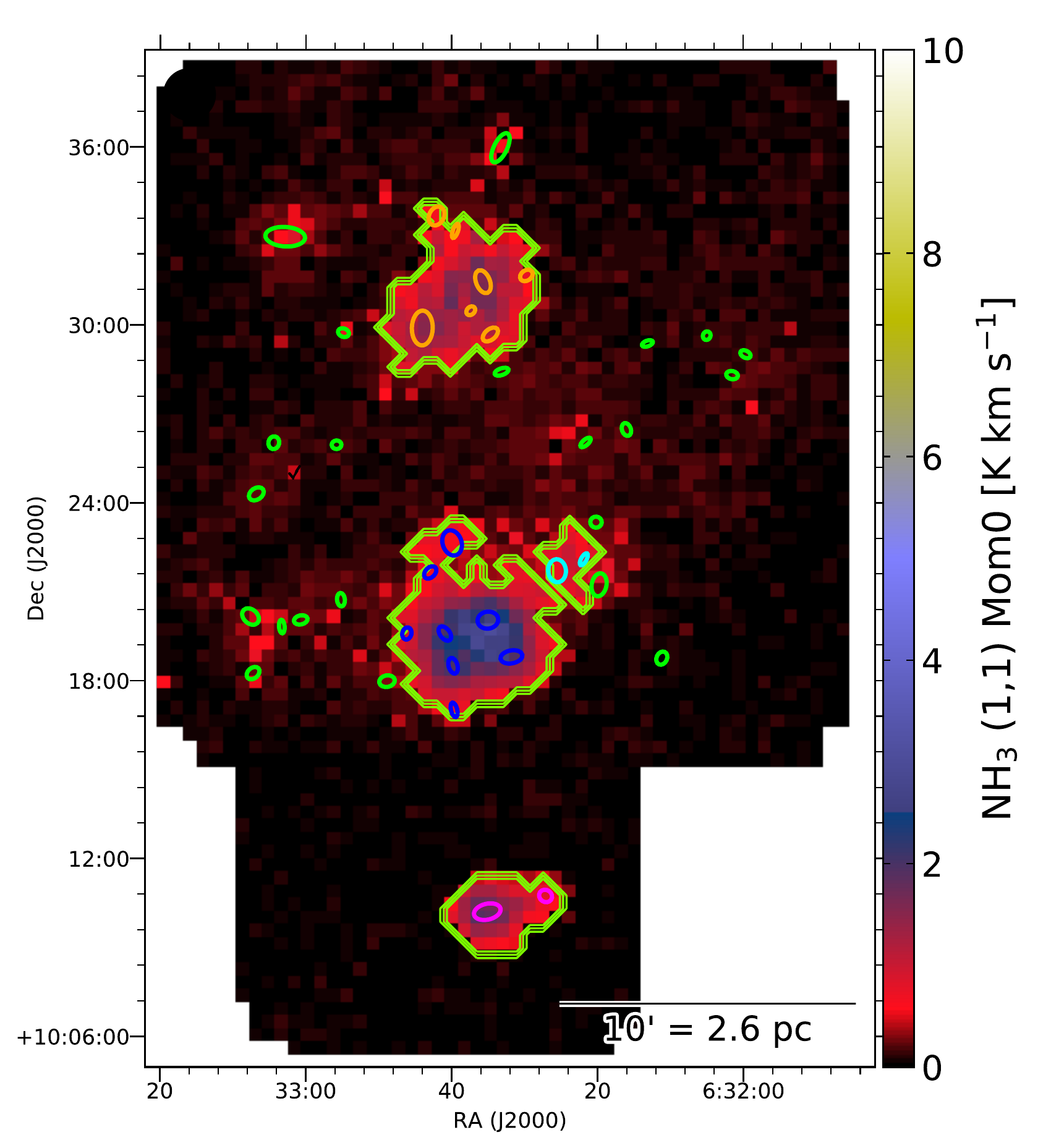}
\caption{Same as Figure \ref{convolved} for MonR1.}
\label{MonR1_convolved}
\end{figure}

\begin{figure}[ht]
\epsscale{1.0}
\plottwo{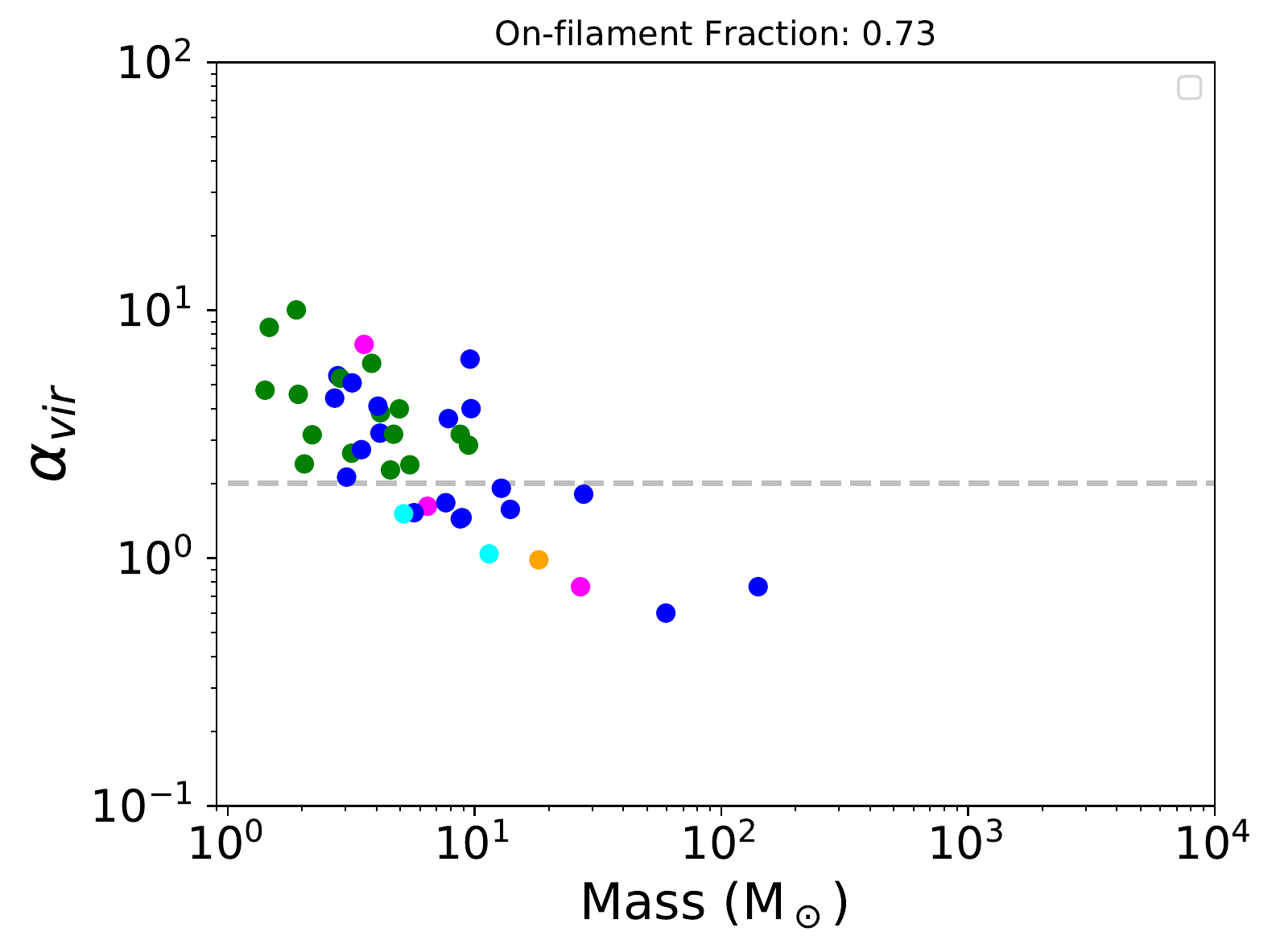}{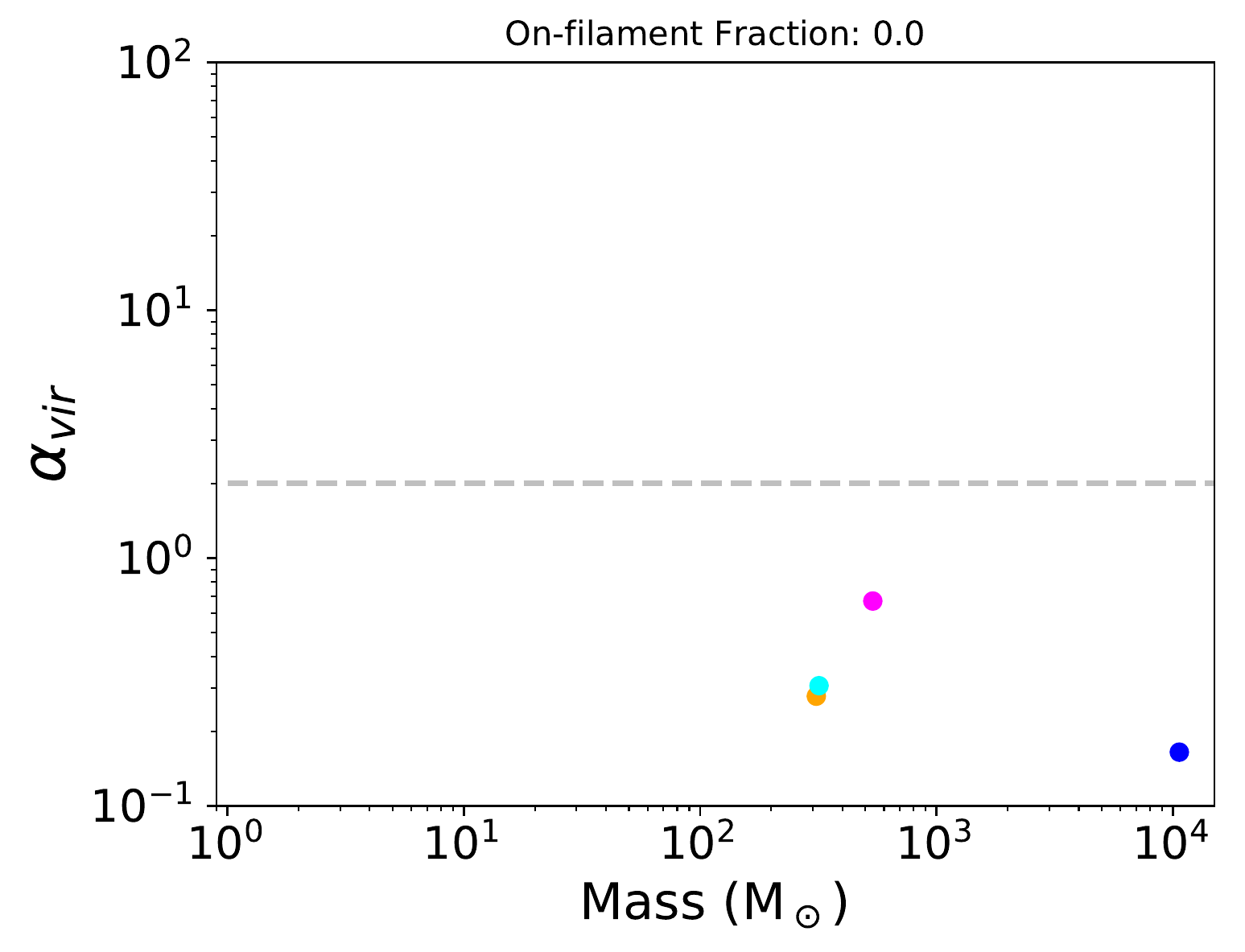}
\epsscale{0.6}
\plotone{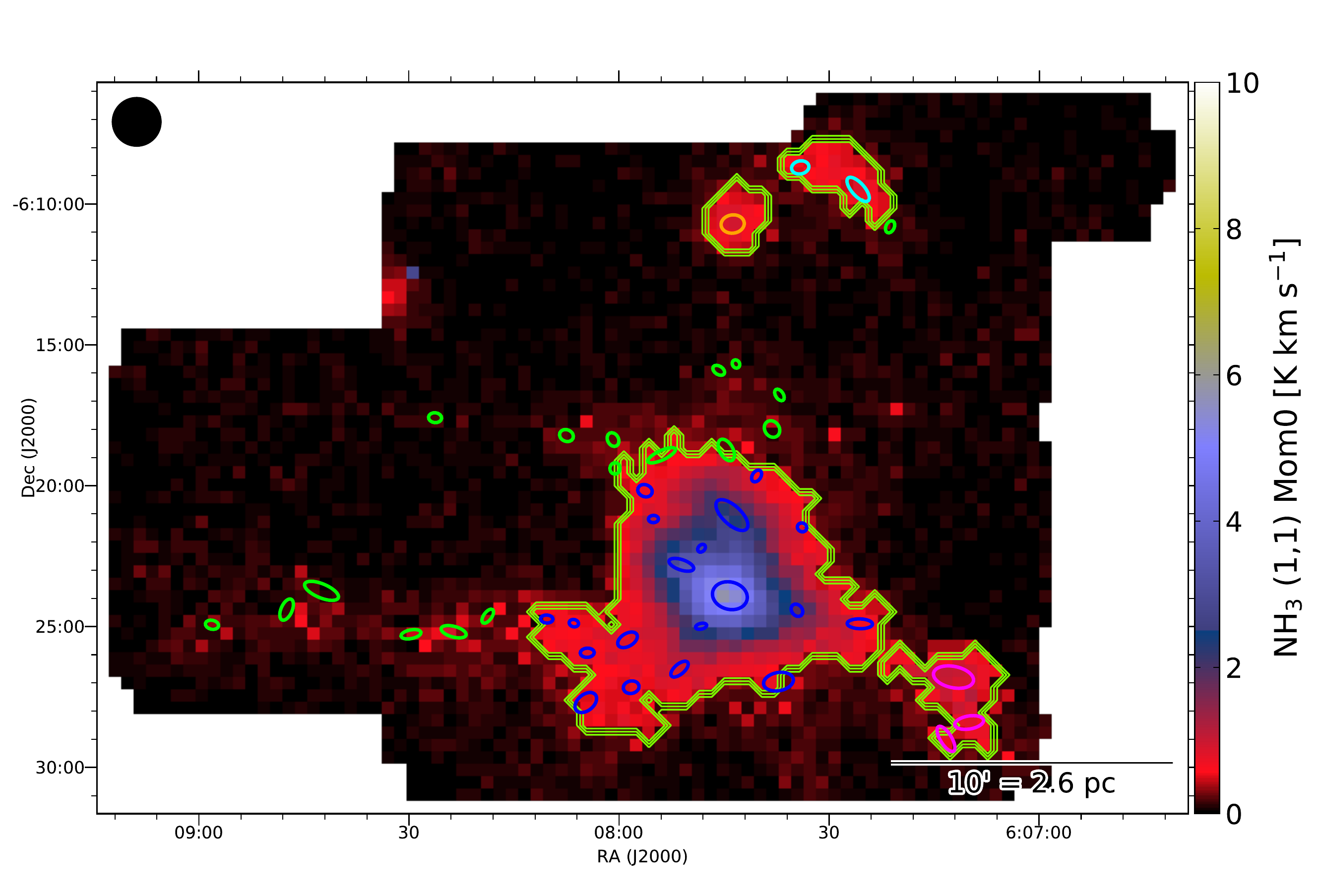}
\caption{Same as Figure \ref{convolved} for MonR2.}
\label{MonR2_convolved}
\end{figure}

In Figures \ref{convolved}-\ref{MonR2_convolved}, we compare the results of the original virial analyses in NGC 2264, MonR1, and MonR2 to their distance-adjusted virial analysis.  Leaves from the original analysis that fall within a leaf identified in the distance-adjusted analysis are tagged with a specific color in each plot of Figures \ref{convolved}-\ref{MonR2_convolved}.  As expected, much fewer structures are identified by the dendrogram in the distance-adjusted analysis.  Many of the small structures in the original analysis are lumped together into a single large structure in the distance-adjusted analysis due to the lower resolution.  In terms of virial parameters, the structures in the distance-adjusted analysis are all gravitationally bound.  This distance bias is likely why more bound structures are observed in W48 than these closer clouds.  At 900 pc, NGC2264, MonR1, and MonR2 represent the extreme examples of the distance bias in our sample.  The impact will undoubtedly be less for the moderate distance clouds in our sample (e.g., Cygnus X, W3, NGC7538, etc.), for which the distance difference is lower.  This statement is echoed by the fact that many of the cloud attributes analyzed in this paper (e.g., dense gas mass, leaf on-filament fraction, number of leaves, etc.) do not depend strongly on cloud distance (see Section 3.7 and Figure \ref{correlations}).  Nevertheless, we are indeed tracing different scales in the star formation hierarchy by including a range of cloud distances in our analysis.  Future high-resolution observations of the more distant KEYSTONE targets with the Karl G. Jansky Very Large Array (VLA) or Next Generation VLA (ngVLA) could provide the means to compare cloud structures between clouds on more similar spatial scales.  




\end{appendix}

\bibliographystyle{apj}
\bibliography{arxiv_27Aug2019}




\end{document}